# A Novel One Parameter Unit Distribution: Median Based Unit Rayleigh: Properties and Estimations

Iman M. Attia *

[Imanattiathesis1972@gmail.com](mailto:Imanattiathesis1972@gmail.com) ,[imanattia1972@gmail.com](mailto:imanattia1972@gmail.com)

*Department of Mathematical Statistics, Faculty of Graduate Studies for Statistical Research, Cairo University, Egypt*

## Abstract:

### Background and aim

In this paper, the author introduces the Median-Based Unit Rayleigh (MBUR) distribution, a newly formulated statistical distribution defined exclusively on the interval (0, 1). The development of such distributions is essential for enhancing our comprehension of phenomena modeled through ratios and proportions.

### Methods

The paper provides a detailed derivation of the probability density function (PDF) for the MBUR distribution, thoroughly articulating each phase of the derivation process. The analysis extends to a rigorous examination of the MBUR distribution's properties, encompassing related functions crucial for statistical evaluation, including the cumulative distribution function (CDF), survival function, hazard rate function, and quantile function. These functions are integral to elucidating the distribution's behavior and characteristics.In addition to theoretical insights, the author examines various methodologies for parameter estimation relevant to the MBUR distribution. A detailed overview of the statistical techniques used for parameter estimation is provided, highlighting their respective strengths and weaknesses.

### Results

To underpin these methodologies, extensive simulation studies will be conducted, demonstrating the efficacy and robustness of the proposed estimation techniques. These simulations will facilitate a comparative analysis to evaluate the fit of the MBUR distribution across diverse datasets.

### Discussion

Additionally, the paper incorporates real data analyses to showcase the empirical utility of the MBUR distribution. This will involve a systematic comparison of the MBUR distribution's performance against well-established unit distributions, such as the beta and Kumaraswamy distributions, highlighting its advantages and adaptability in modeling practical scenarios. This thorough exploration aims to provide significant contributions to the expanding domain of statistical distributions.

### Conclusion:

MBUR is an advanced statistical model designed to effectively manage a wide variety of skewed data distributions. It has been shown to outperform traditional distributions, such as the beta and Kumaraswamy distributions, in the analysis of certain real-world datasets. One of the main advantages of MBUR is its capability to estimate parameters using just a single parameter, which offers both flexibility and efficiency in modeling complex data patterns. This characteristic not only simplifies the estimation process but also enhances its applicability across different fields where skewed data is prevalent.

**Keywords:**

**Median Based Unit Rayleigh (MBUR) distribution, new distribution, unit distribution, Maximum product of spacing, MLE.**

## Introduction

Fitting data to a statistical distribution is essential for understanding the underlying processes that generate the data. Researchers have developed many distributions to describe complex real-world phenomena. Before 1980, the primary techniques for generating distributions included solving systems of differential equations, using transformations, and applying quantile function strategies. Since 1980, the methods have largely focused on either adding new parameters to existing distributions or combining already known distributions. These approaches provide researchers with a wide range of tractable and flexible distributions capable of accommodating various types of asymmetrical data, as well as outliers in data sets. Fitting distributions to data enhances modeling in analyses involving regression, survival analysis, reliability analysis, and time series analysis.

Quantile regression models are utilized by many researchers to model time-to-event response variables that exhibit skewness, long tails, and violations of normality and homogeneity assumptions. These models are robust to outliers, skewness, and heteroscedasticity as they specify the entire conditional distribution of the response variable, rather than merely the conditional mean. Many authors have applied quantile regression to investigate the effects of covariates on time-duration response variables at different quantiles. For instance, Flemming et al. (2017) [1] studied the association between time to surgery and survival among patients with colon cancer. Faradmal et al. (2016) [2] employed censored quantile regression to examine the overall factors affecting survival in breast cancer. Xue et al. (2018) [3] conducted an in-depth exploration of the censored quantile regression model for analyzing time-to-event data.

Numerous real-world phenomena can be represented as proportions, ratios, or fractions over the bounded interval (0,1). Various disciplines, such as biology, finance, mortality rates, recovery rates, economics, engineering, hydrology, health, and measurement sciences, have modeled these types of data using continuous distributions. Some of these distributions include: the Johnson SB distribution [4], Beta distribution [5], Unit Johnson distribution [6], Topp-Leone distribution [7], Unit Gamma distribution [8],[9],[10],[11], Unit Logistic distribution [12], Kumaraswamy distribution [13], Unit Burr-III distribution [14], Unit Modified Burr-III distribution [15], Unit Burr-XII distribution [16], Unit-Gompertz distribution [17], Unit-Lindley distribution [18], Unit-Weibull distribution [19], Unit-Birnbaum-Saunders [20] and Unit Muth distribution [21].

The unit distribution is primarily derived through variable transformation, which can take various forms: $y = e^{-w}$, $y = \frac{1}{1+w}$, $y = \frac{w}{1+w}$ or $y = \frac{e^w}{1+e^w}$

Table 1 shows some of the differences between Beta, Kumaraswamy and MBUR distributions. Jones (2009) [22] mentioned some of the differences between Beta and Kumaraswamy.

As shown from the differences, the new MBUR distribution has one parameter but with that single parameter the pdf has shapes that are increasing, decreasing, unimodal and bathtub distributions like the two-parameters Beta and Kumaraswamy distributions. Therefore, the new MBUR is tractable and flexible to accommodate and fit a wide range of data shapes. The new MBUR has explicit closed form of the CDF and subsequently the quantile function enables the distribution to be used in the median based quantile regression models like the Kumaraswamy distribution. The MBUR has simple formula for moments especially the mean which makes it candidate for mean based regression models like the Beta distribution when the data does not show extreme skewness. This is in contrast to Kumaraswamy which does not have that simple formula for the mean hence the distribution is not a candidate for mean-based regression models.

Table (1): some differences between Beta, Kumaraswamy and the new distribution MBUR:

| | Beta distribution | Kumaraswamy distribution | MBUR distribution |
|---|---|---|---|
| Parameters | Two parameters | Two parameters | One parameter |
| PDF shapes (depends on parameters) | Unimodal, Uni-antimodal (bathtub), Increasing & left skew, J-shape, decreasing & right skew, constant. | Unimodal, Uni-antimodal (bathtub), Increasing & left skew, J-shape, decreasing & right skew, constant. | Unimodal, Uni-antimodal (bathtub), Increasing & left skew, J-shape, decreasing & right skew. |
| Mode | Explicit expression | Explicit expression | Explicit expression |
| Behavior of Skewness & kurtosis | Good behavior as function of parameters | Good behavior as function of parameters | Good behavior as function of parameter |
| CDF | Involves special function. No explicit closed form | Simple explicit closed formula not involving any special functions | Simple explicit closed formula not involving any special functions |
| Quantile function | No explicit closed formula | Simple closed explicit formula | Closed explicit formula |
| R.N. generator | No simple formula | Simple formula | Simple formula |
| Moments | Simple formula | No simple closed formula | Simple formula |
| regression | Mean based regression | Median-based quantile regression | Mean and Median-based quantile regression |
| One-parameter subfamily symmetric distribution | Exist ( if both shape parameters are equal to 2, this gives symmetric distribution around 0.5) | Not exist ( if both shape parameters are equal to one, this gives uniform distribution) | Exist ( if alpha parameter equals to one, this gives symmetric distribution around 0.5) |
| Moments of order statistics | No simple formula | Simple formula | Simple formula |

Most of the previously mentioned unit distributions exhibit flexibility to fit wide range of data shapes, especially skewed data, but with more than one parameter and varying tractability. They differ considering the closeness of the CDF and subsequently the lack of special function in the definition of the quantile function. The simpler and the closer formula for the CDF and the quantile function is, the better the distribution is to suit for quantile regression models. The unit Lindley distribution, although it is one parameter unit distribution, the quantile function requires Lambert function evaluation which is a special function. Topp Loene distribution has PDF that does not express the bathtub appearance. The new (MBUR) offers a significant advantage due to its simplicity and parsimony, requiring the estimation of only one parameter. This distribution is versatile, as its probability density function (PDF) can exhibit a variety of shapes, including increasing, decreasing, unimodal, and bathtub configurations. Additionally, the cumulative distribution function (CDF) has a straightforward, closed-form expression, which means that its quantile function does not necessitate the use of complex special functions. This simplicity in modeling enhances usability, making MBUR a valuable contribution to the family of unit distributions. This is particularly important considering the absence of a consensus on the most suitable distribution for datasets that display skewness. This paper is organized into the following sections. Methods are discussed in Section 1, where the author will explain the methodology for obtaining the new distribution, and in Section 2 where the author will elaborate on its probability density function (PDF), cumulative distribution function (CDF), survival function, hazard function, reversed hazard function, and quantile function. Results are evaluated in Section 3 where the author will discuss methods of estimation, accompanied by a simulation study. Discussion is expounded in Section 4 where the author will explore real data analysis along with an elucidation. Matlab 2014R was used in all calculations. Finally, Conclusions are explicated in section 5 with illumination on suggestions for future works.

## Methods: Section 1:

### Derivation of the MBUR Distribution:

By utilizing the PDF of the median order statistics for a sample size of n=3, the author derives a new distribution based on a Rayleigh distribution as the parent distribution, as illustrated below. Equation (1.A) defines the PDF of order statistics:

$$f_{i:n}(w) = \frac{n!}{(i-1)!\,(n-i)!} \{F(w)\}^{i-1} \{1 - F(w)\}^{n-i} f(w), w > 0 \ldots\ldots\ldots\ldots\ldots\ldots\ldots(1.A)$$

For a sample size n=3, to calculate the median order statistics, replace n=3 and i=2 in equation (1.A) to obtain (1.B):

$$f_{2:3}(w) = \frac{3!}{(2-1)!\,(3-1)!} \{F(w)\}^{2-1} \{1 - F(w)\}^{3-2} f(w) \ldots\ldots\ldots\ldots\ldots\ldots\ldots(1.B)$$

substitute the PDF and CDF of Rayleigh distribution as a parent distribution in equation (1). Equation (2) defines both the PDF and CDF of the random variable w distributed as Rayleigh distribution. This yields a new distribution called Median-Based Rayleigh (MBR) distribution with PDF shown in equation (3).

$$F(w) = 1 - e^{\frac{-w}{\alpha^2}}, \quad f(x) = \frac{2w}{\alpha^2} e^{\frac{-w^2}{\alpha^2}} \quad \ldots\ldots\ldots\ldots\ldots\ldots\ldots\ldots\ldots\ldots(2)$$

$$f_{2:3}(w) = 3! \left\{1 - e^{\frac{-w^2}{\alpha^2}}\right\}^{2-1} \left\{e^{\frac{-w^2}{\alpha^2}}\right\}^{3-2} \frac{2w}{\alpha^2} e^{\frac{-w^2}{\alpha^2}}$$

$$f_{2:3}(w) = \frac{12w}{\alpha^2}\left[1 - e^{\frac{-w^2}{\alpha^2}}\right]\left[e^{\frac{-w^2}{\alpha^2}}\right]e^{\frac{-w^2}{\alpha^2}} = \frac{12w}{\alpha^2}\left[1 - e^{\frac{-w^2}{\alpha^2}}\right]\left[e^{\frac{-2w^2}{\alpha^2}}\right]$$

$$f_{2:3}(w) = \frac{12w}{\alpha^2}\left[1 - \left(e^{-w^2}\right)^{\frac{1}{\alpha^2}}\right]\left[\left(e^{-w^2}\right)^{\frac{2}{\alpha^2}}\right], \quad w > 0 \ldots\ldots\ldots\ldots\ldots\ldots\ldots\ldots(3)$$

Applying the following transformation on equation (3) to obtain the new unit distribution: let $y = e^{-w^2}, 0 < y < 1$

take the log on both sides: $\quad -ln(y) = w^2$

take square root of both sides : $[-ln(y)]^{.5} = w$

take the Jacobian: $\frac{dw}{dy} = \frac{1}{2}[-ln(y)]^{-.5} \left(\frac{-1}{y}\right)$

Replace the absolute value of the Jacobian and the above transformation in equation (3) to derive the new Median Based Unit Rayleigh (MBUR) Distribution shown in equation (4).

$$f(y) = \frac{12[-ln(y)]^{.5}}{\alpha^2}\left[1 - (y)^{\frac{1}{\alpha^2}}\right]\left[(y)^{\frac{2}{\alpha^2}}\right]\left|\frac{1}{2}[-ln(y)]^{-.5}\left(\frac{-1}{y}\right)\right|, 0 < y < 1 \dots \dots \dots \dots \dots (4)$$

After some algebraic manipulations, the PDF of the MBUR is shown is equation (5).

## Section 2: Some of the properties of the new distribution (MBUR):

### 2.1. Basic functions (PDF, CDF, Sf, HR, rHR) :

The following are the probability density function (PDF), cumulative distribution function (CDF), survival function, hazard function, and reversed hazard function as shown in equation (5-9) respectively:

$$f(y) = \frac{6}{\alpha^2}\left[1 - y^{\frac{1}{\alpha^2}}\right]y^{\left(\frac{2}{\alpha^2}-1\right)}, \quad 0 < y < 1, \quad \alpha > 0 \dots \dots \dots \dots \dots \dots \dots (5)$$

$$F(y) = 3y^{\frac{2}{\alpha^2}} - 2y^{\frac{3}{\alpha^2}}, \quad 0 < y < 1, \quad \alpha > 0 \dots \dots \dots \dots \dots \dots \dots \dots (6)$$

$$S(y) = 1 - F(Y) = 1 - \left(3y^{\frac{2}{\alpha^2}} - 2y^{\frac{3}{\alpha^2}}\right), \quad 0 < y < 1, \alpha > 0 \dots \dots \dots \dots \dots \dots (7)$$

$$h(y) = \frac{f(y)}{S(y)} = \frac{\frac{6}{\alpha^2}\left(1 - y^{\frac{1}{\alpha^2}}\right)y^{\left(\frac{2}{\alpha^2}-1\right)}}{1 - \left(3y^{\frac{2}{\alpha^2}} - 2y^{\frac{3}{\alpha^2}}\right)}, \quad 0 < y < 1, \alpha > 0 \dots \dots \dots \dots \dots \dots (8)$$

$$rh(y) = \frac{f(y)}{F(y)} = \frac{\frac{6}{\alpha^2}\left(1 - y^{\frac{1}{\alpha^2}}\right)y^{\left(\frac{2}{\alpha^2}-1\right)}}{3y^{\frac{2}{\alpha^2}} - 2y^{\frac{3}{\alpha^2}}}, \quad 0 < y < 1, \alpha > 0 \dots \dots \dots \dots \dots \dots (9)$$

### 2.2. Quantile Function:

The CDF of the MBUR can be written as a third degree polynomial

$$u = F(y) = 3y^{\frac{2}{\alpha^2}} - 2y^{\frac{3}{\alpha^2}} = -2\left(y^{\frac{1}{\alpha^2}}\right)^3 + 3\left(y^{\frac{1}{\alpha^2}}\right)^2$$

The inverse of this CDF is used to obtain y, the real root of this 3rd polynomial function is $y = F^{-1}(y)$ as shown in (equation 10):

$$y = \left\{-.5\left(\cos\left[\frac{\cos^{-1}(1-2u)}{3}\right] - \sqrt{3}\sin\left[\frac{\cos^{-1}(1-2u)}{3}\right]\right) + .5\right\}^{\alpha^2} \dots \dots \dots \dots \dots (10)$$

Figures 1-9 illustrate the specified functions for various values of alpha for a random variable X distributed as MBUR.

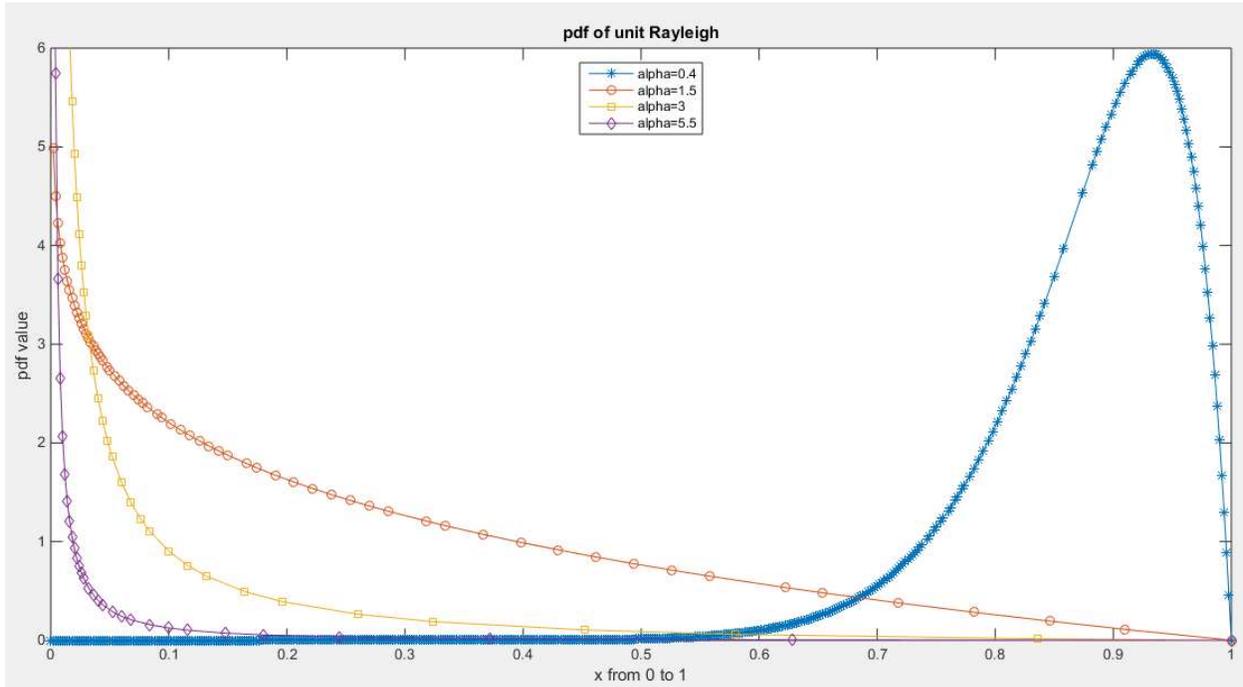

Fig. 1: PDF of Median Based Unit Rayleigh (MBUR) distribution.

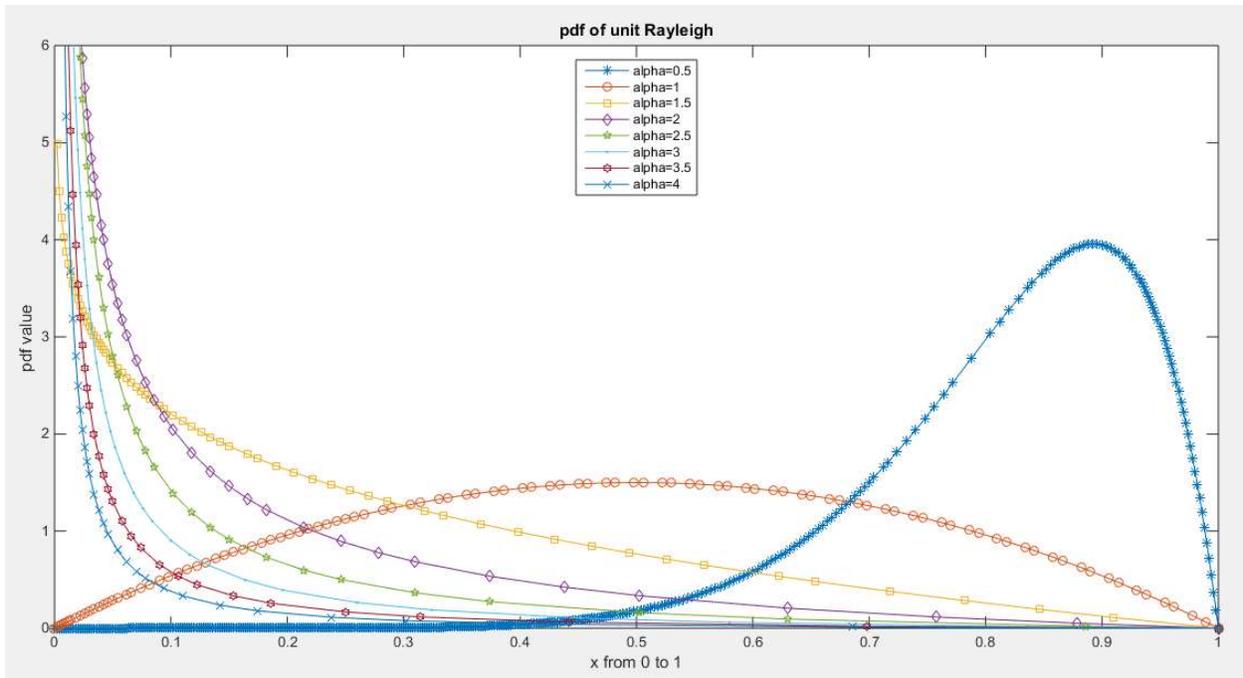

Fig. 2: PDF of Median Based Unit Rayleigh (MBUR) distribution.

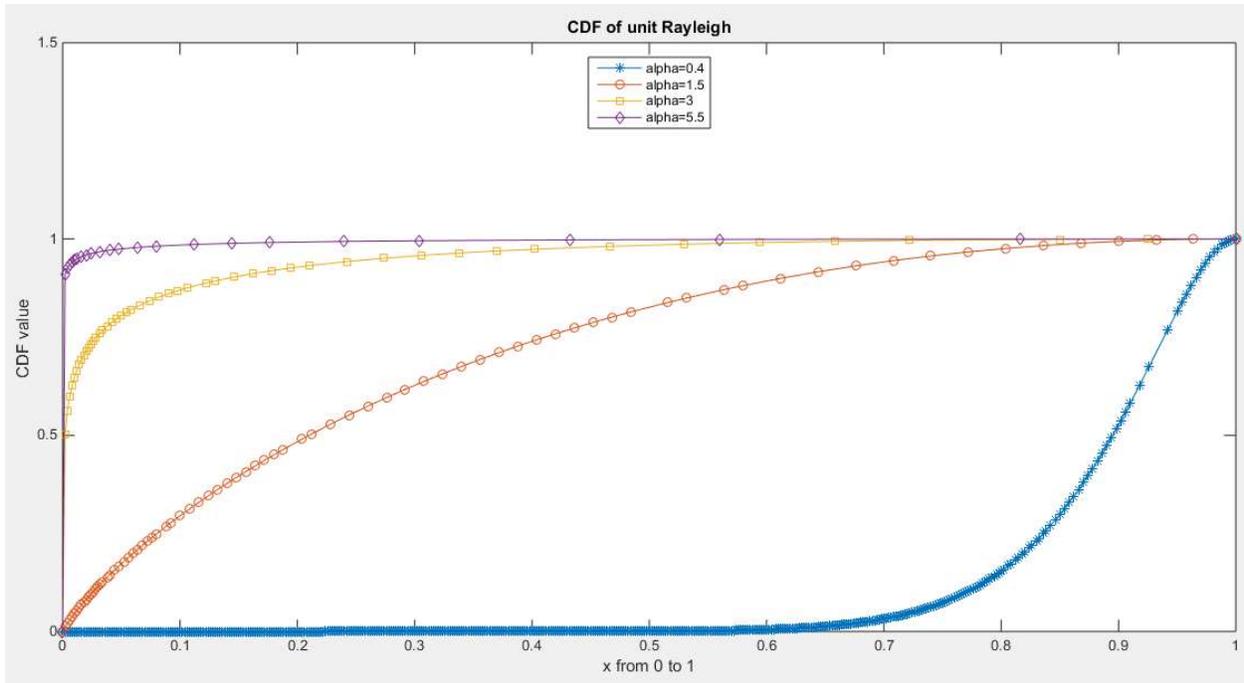

Fig. 3: CDF of Median Based Unit Rayleigh (MBUR) Distribution.

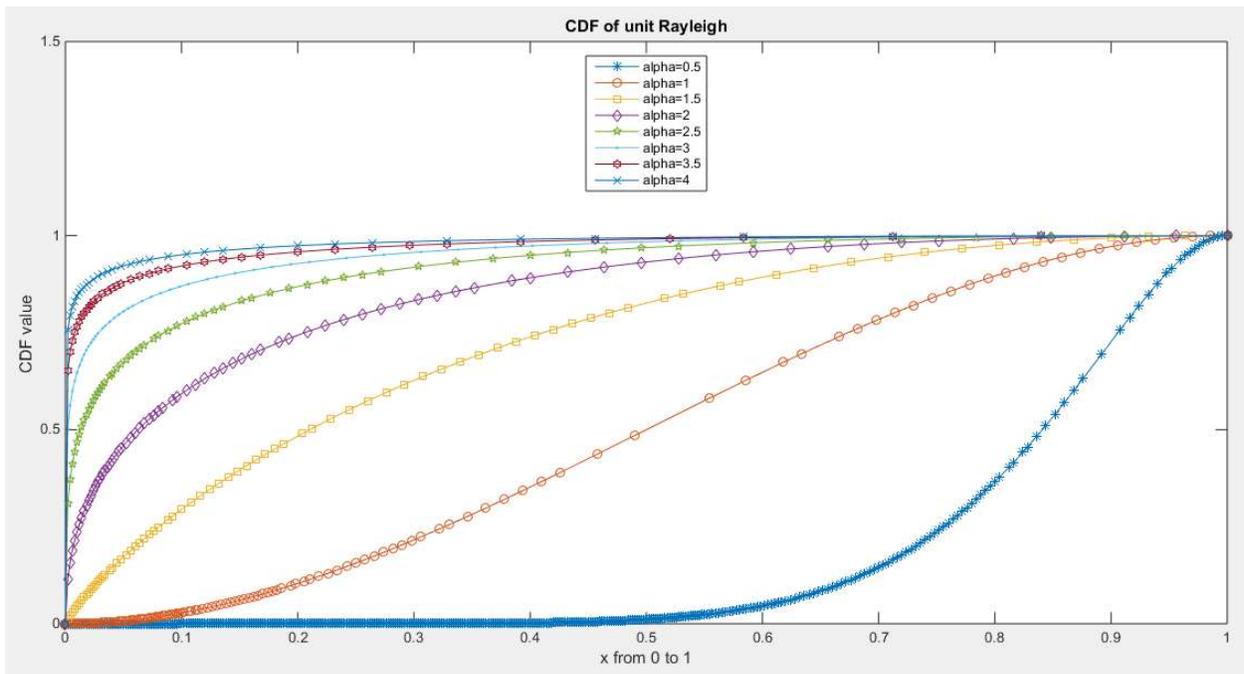

Fig. 4: CDF of Median Based Unit Rayleigh (MBUR) Distribution

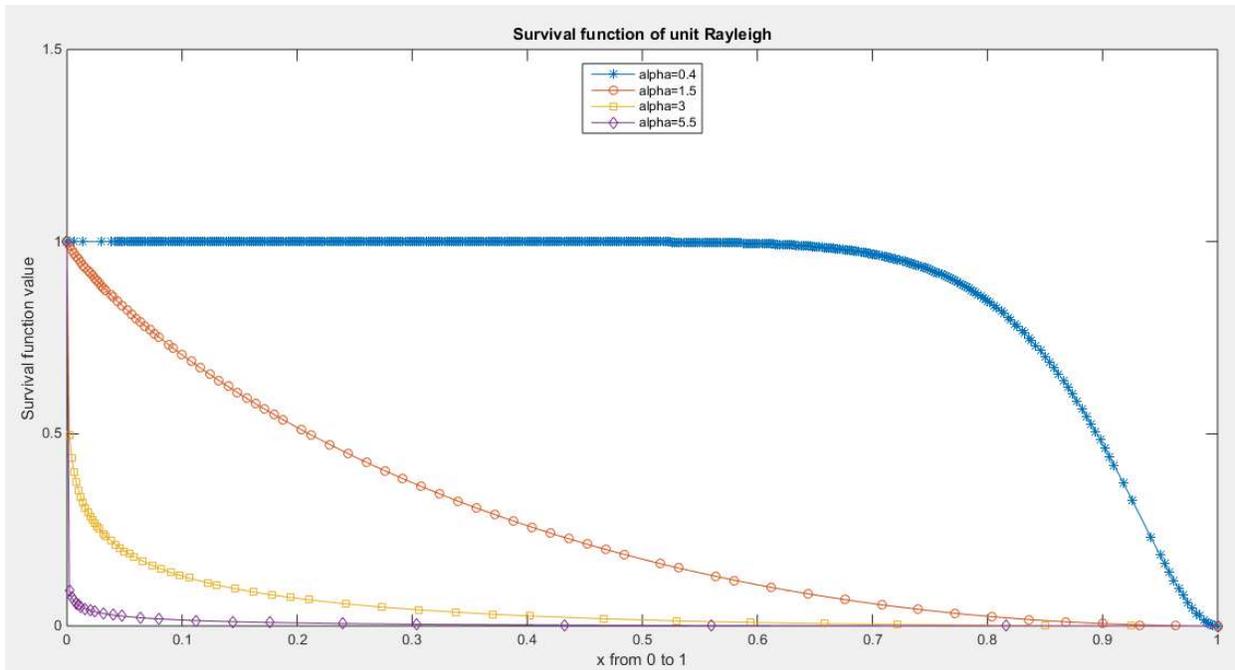

Fig. 5: Survival function of MBUR Distribution.

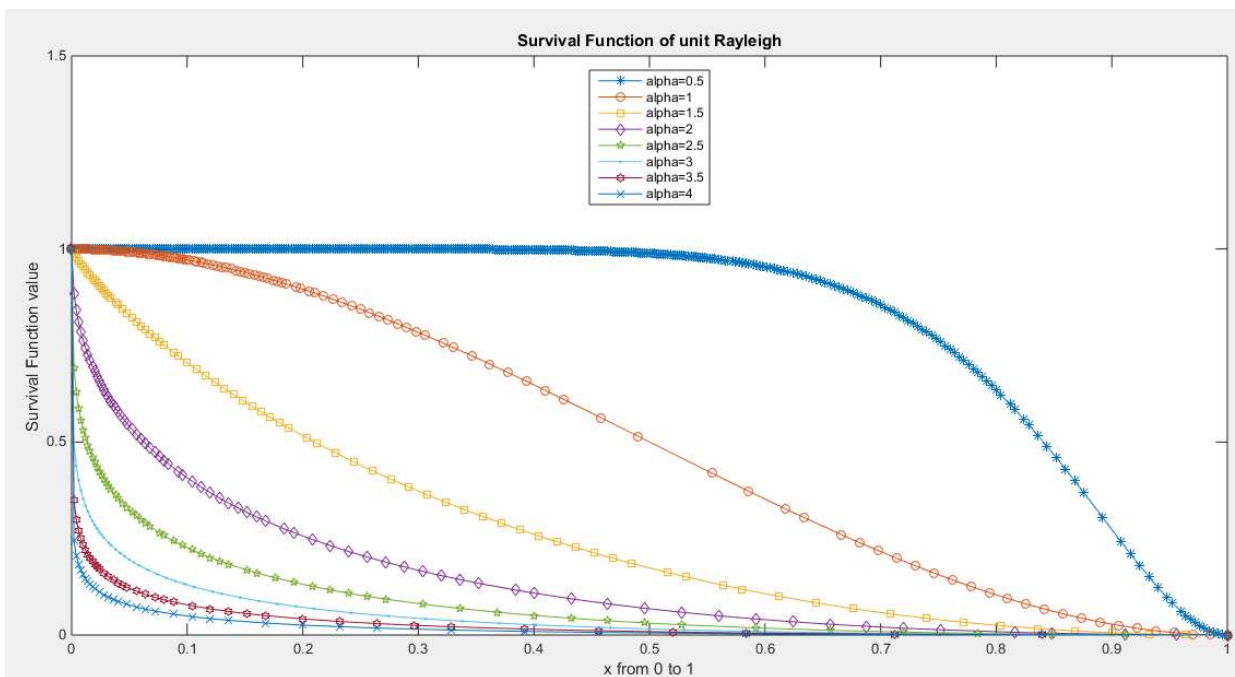

Fig. 6: Survival function of MBUR Distribution.

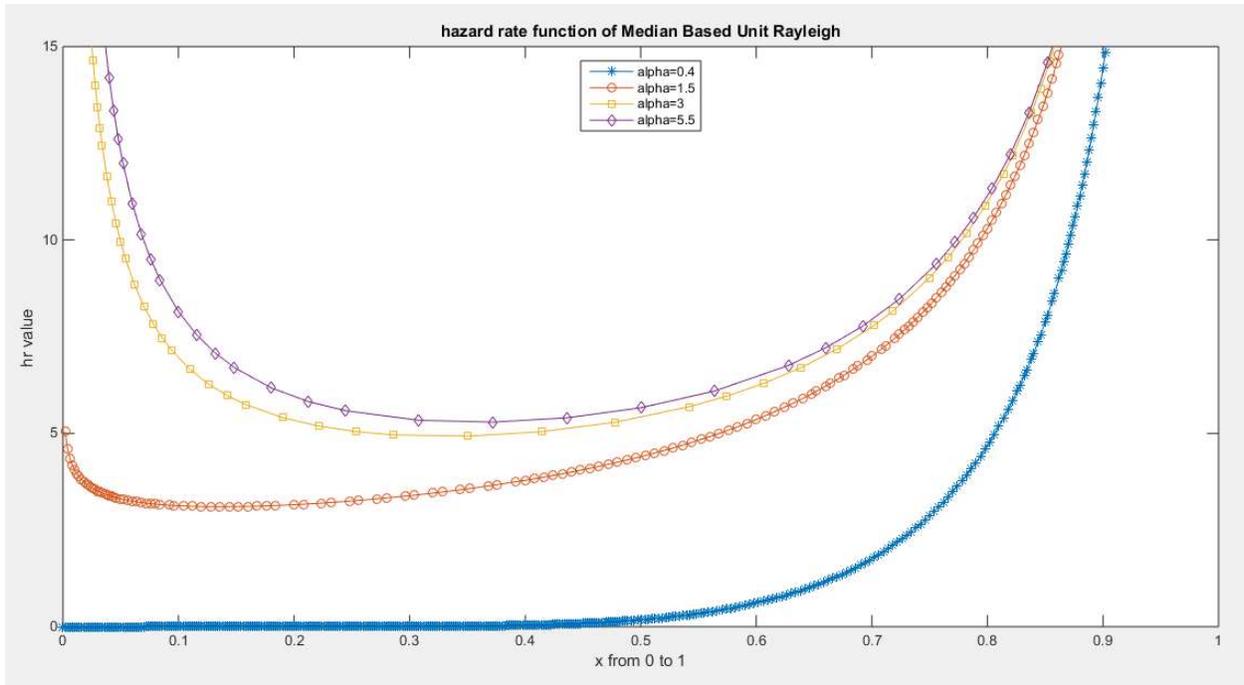

Fig. 7: hazard rate function of MBUR Distribution

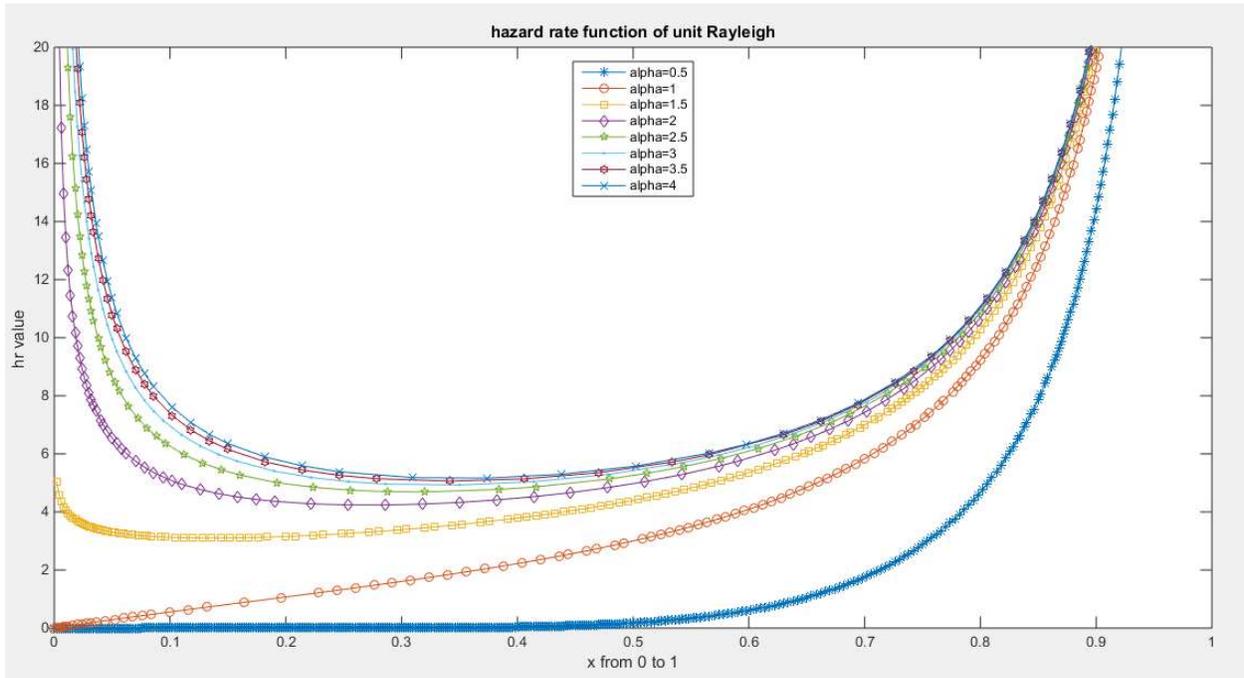

Fig. 8: hazard rate function of MBUR Distribution.

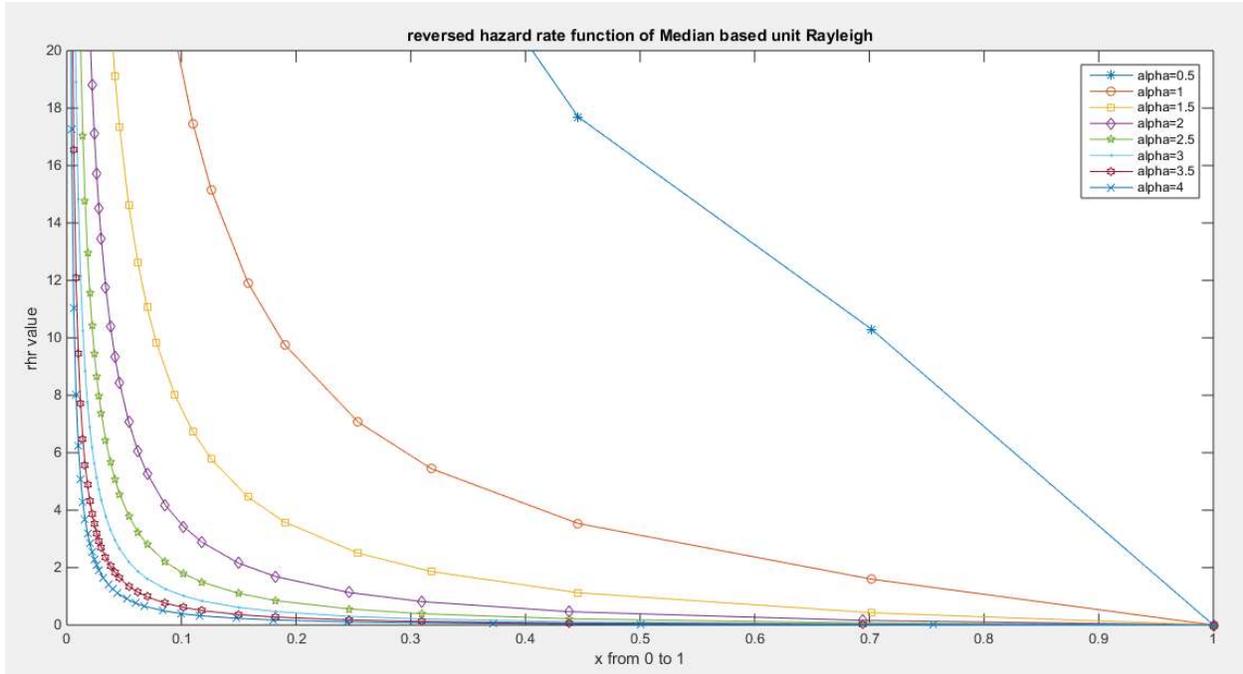

Fig. 9: reversed hazard rate function of MBUR Distribution.

To generate random variable distributed as MBUR:

1. Generate uniform random variable (0,1): $u \sim uniform(0,1)$.
2. Choose the parameter alpha.
3. Substitute the above values of $u \in (0,1)$ and the chosen alpha in the quantile function, to obtain x distributed as $x \sim MBUR(\alpha)$

## 2.3. Discussion and analysis of the above functions

See supplementary materials (section 1).

**2.4. r$^{th}$ Raw Moments** for a random variable y distributed as MBUR is shown in equation (11)

$$E(y^r) = \int_0^1 y^r \frac{6}{\alpha^2}\left[1 - y^{\frac{1}{\alpha^2}}\right] y^{\left(\frac{2}{\alpha^2}-1\right)} dy = \frac{6}{(2+r\alpha^2)(3+r\alpha^2)} \dots\dots\dots\dots\dots\dots\dots\dots (11)$$

$$E(y) = \frac{6}{(2+\alpha^2)(3+\alpha^2)}, \quad E(y^2) = \frac{6}{(2+2\alpha^2)(3+2\alpha^2)}$$

$$E(y^3) = \frac{6}{(2+3\alpha^2)(3+3\alpha^2)}, \quad E(y^4) = \frac{6}{(2+4\alpha^2)(3+4\alpha^2)}$$

$$\text{var}(y) = E(y^2) - [E(y)]^2 = \frac{3\alpha^4(13 + 10\alpha^2 + \alpha^4)}{(3 + 5\alpha^2 + 2\alpha^4)(6 + 5\alpha^2 + \alpha^4)^2}$$

**Coefficient of Skewness:** is shown in equation (12)

$$E\frac{(y-\mu)^3}{\sigma^3} = \frac{E(y^3) - 3\mu E(y^2) + 3\mu^2 E(y) - \mu^3}{\sigma^3}$$

$$= \frac{E(y^3) - 3\mu[E(y^2) - \mu E(y)] - \mu^3}{\sigma^3} = \frac{E(y^3) - 3\mu[E(y^2) - \mu\mu] - \mu^3}{\sigma^3}$$

$$\text{coefficient of skewness} = \frac{E(y^3) - 3\mu\sigma^2 - \mu^3}{\sigma^3} = \frac{E(y^3) - \mu(3\sigma^2 + \mu^2)}{\sigma^3} \ldots\ldots\ldots\ldots\ldots (12)$$

**Coefficient of Kurtosis:** is shown in equation (13)

$$E\frac{(y-\mu)^4}{\sigma^4} = \frac{E(y^4) - 4\mu E(y^3) + 6\mu^2 E(y^2) - 3\mu^4}{\sigma^4}$$

$$= \frac{E(y^4) - 4\mu E(y^3) + 6\mu^2[\sigma^2 + \mu^2] - 3\mu^4}{\sigma^4}$$

$$= \frac{E(y^4) - 4\mu E(y^3) + 6\mu^2\sigma^2 + 6\mu^4 - 3\mu^4}{\sigma^4}$$

$$= \frac{E(y^4) - 4\mu E(y^3) + 6\mu^2\sigma^2 + 3\mu^4}{\sigma^4}$$

$$\text{coefficient of Kurtosis} = \frac{E(y^4) - 4\mu E(y^3) + 3\mu^2[2\sigma^2 + \mu^2]}{\sigma^4} \ldots\ldots\ldots\ldots\ldots\ldots (13)$$

**Coefficient of Variation:** is shown in see equation (14)

$$CV = \frac{S}{\mu} \ldots\ldots\ldots\ldots\ldots\ldots\ldots\ldots\ldots\ldots\ldots\ldots\ldots\ldots\ldots\ldots (14)$$

where S is standard deviation and mu is the mean

Figure 10 illustrates the graph for the specified coefficients. In subplot (1), the variance increases as the alpha level rises, reaching a maximum value between 0.05 and 0.06 at a parameter level between 1 and 1.5. After that point, the variance begins to decrease with further increases in the parameter values.

In subplot (2), the Fisher coefficient of skewness shows negative skewness (left skewness) at low parameter levels, reaching a zero value when the parameter equals 1. At this point, the probability density function (PDF) of the distribution becomes symmetrical around 0.5. Beyond this parameter value of 1, the coefficient of skewness increases as the alpha level rises, indicating right skewness in the distribution.

Subplot (3) reveals that the coefficient of kurtosis starts at approximately 6 when the alpha level is 0.1, decreasing to around 2 at an alpha level of 1. Following this, the coefficient rises again as the alpha level increases, demonstrating a wide variety of kurtosis shapes. The distribution exhibits a mesokurtic shape (kurtosis equals 3) at approximately parameter level of 0.7 and 1.5. It displays a leptokurtic shape, characterized by fatter tails, when there is positive excess kurtosis (kurtosis greater than 3), which occurs at parameter levels below 0.7 and above 1.5. Conversely, it shows a platykurtic shape, with thinner tails, when there is negative excess kurtosis (kurtosis less than 3), which happens approximately at parameter levels between 0.7 and 1.5. These coefficients, which reflect the shape of the PDF, are fundamental to the flexibility of the new distribution, allowing it to accommodate a wide variety of data shapes. They play a crucial role in making this distribution outperform other unit distributions in certain data analyses.

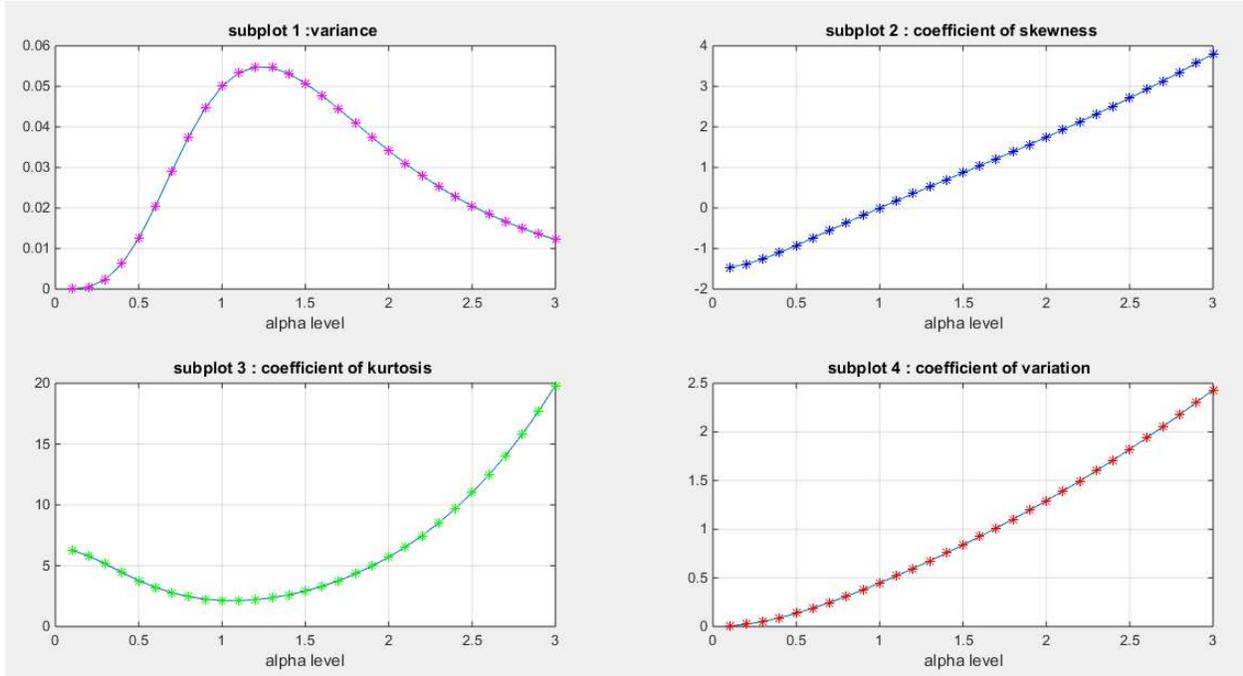

Fig. 10 shows that the maximum value of variance is attained between 0.05 and 0.06 when alpha values are between 1 and 1.5. At alpha level one, the coefficient of skewness is zero, coefficient of kurtosis is around 2 (2.1429) and the variance is 0.05. When alpha level is 0.668 the coefficient of kurtosis equals 2.9. When alpha level is 1.5, the coefficient of kurtosis is 2.9172.

**2.5. r$^{th}$ incomplete Moments:** for a random variable y distributed as MBUR is defined in equation (15): (See supplementary materials section1 for derivation)

$$E(y^r | y < t) = \int_0^t y^r \frac{6}{\alpha^2}\left[1 - y^{\frac{1}{\alpha^2}}\right] y^{\left(\frac{2}{\alpha^2}-1\right)} dy = \frac{6 y^{\frac{2}{\alpha^2}+r}}{(2+r\alpha^2)} - \frac{6 y^{\frac{3}{\alpha^2}+r}}{(3+r\alpha^2)} \ldots\ldots\ldots\ldots (15)$$

**2.6. Stress- strength reliability**

From a reliability prospective, if an element in a system has a random strength X that is strained with a random stress Y, this element will immediately break down if the stress overrides the strength. Conversely, it will adequately operate if the strength surpasses the stress. If X and Y are independent random variables denoting strength and stress respectively, and both follow MBUR distribution with parameters $\alpha_1$ and $\alpha_2$ respectively, then the reliability measure of this element can be deduced from appropriate equation (16) .(see suppl. Mat. Section1)

$$R = Pr(Y < X) = \int_0^1 Pr(Y < X | X = x) f_X(x) dy$$

$$R = Pr(Y < X) = \int_0^1 F(x; \alpha_2) f(x; \alpha_1) dy$$

$$R = \alpha_2^2 \left\{ \frac{13}{(\alpha_1^2 + \alpha_2^2)} - \frac{18}{(2\alpha_1^2 + 3\alpha_2^2)} - \frac{12}{(3\alpha_1^2 + 2\alpha_2^2)} \right\} \ldots\ldots\ldots\ldots\ldots\ldots\ldots\ldots\ldots. (16)$$

## 2.7. Lorenz, Bonferroni curves and Gini index

These indices have many applications in medicine, insurance, demography, and economics for studying wealth and poverty. They can be applied to variables defined as proportions where y is a random variable distributed as MBUR. Lorenz curve, Bonferroni curves and Gini index are defined in equation (17), (18), (19) respectively (see supplementary materials section 1 for derivation)

$$L(p) = \frac{\int_0^t y f(y) dy}{\int_0^1 y f(y) dy} = y^{\frac{2}{\alpha^2}+1}\left\{(3+\alpha^2) - y^{\frac{1}{\alpha^2}}(2+\alpha^2)\right\} \ldots \ldots \ldots (17)$$

$$B(p) = \frac{L(p)}{F(y)} = \frac{y^{\frac{2}{\alpha^2}+1}\left\{(3+\alpha^2) - y^{\frac{1}{\alpha^2}}(2+\alpha^2)\right\}}{3y^{\frac{2}{\alpha^2}} - 2y^{\frac{3}{\alpha^2}}}$$

$$B(p) = \frac{L(p)}{F(y)} = \frac{y\left\{(3+\alpha^2) - y^{\frac{1}{\alpha^2}}(2+\alpha^2)\right\}}{3 - 2y^{\frac{1}{\alpha^2}}} \ldots \ldots \ldots (18)$$

$$\text{Gini index} = 1 - 2\int_0^1 L(p) dp = 1 - \left(\frac{\alpha^2(5+3\alpha^2)}{(1+\alpha^2)(3+2\alpha^2)}\right) \ldots \ldots \ldots (19)$$

## 2.8. Renyi entropy

Entropy quantifies the variation in uncertainty within a random variable, in the paper context, y is distributed as MBUR. Renyi entropy $T_R(\gamma)$ is a well-known measure defined as follows in equation (20). For MBUR it is defined in equation (21):

$$T_R(\gamma) = \frac{1}{1-\gamma}\log\left\{\int_0^1 [f(y)]^\gamma dy\right\} \ldots \ldots \ldots (20)$$

$$T_R(\gamma) = \frac{1}{1-\gamma}\log\left\{\int_0^1 \left[\left(\frac{6}{\alpha^2}\right)\left(1-y^{\frac{1}{\alpha^2}}\right)y^{\left(\frac{2}{\alpha^2}-1\right)}\right]^\gamma dy\right\} \ldots \ldots \ldots (21)$$

Expanding the following term using the binomial expansion as in equation (22):

$$\left(1-y^{\frac{1}{\alpha^2}}\right)^\gamma = \sum_{m=0}^\infty (-1)^m \binom{\gamma}{m}\left(y^{\frac{1}{\alpha^2}}\right)^m \ldots \ldots \ldots (22)$$

substitute equation (22) into equation (21) gives equations (23), (24) & (25).

$$= \frac{1}{1-\gamma}\log\left\{\int_0^1 \left[\left(\frac{6}{\alpha^2}\right)^\gamma \sum_{m=0}^\infty (-1)^m \binom{\gamma}{m}\left(y^{\frac{1}{\alpha^2}}\right)^m y^{\left(\frac{2\gamma-\gamma\alpha^2}{\alpha^2}\right)}\right] dy\right\}$$

The integral will pass to the variable for integration:

$$= \frac{1}{1-\gamma}\log\left\{\left(\frac{6}{\alpha^2}\right)^\gamma \sum_{m=0}^\infty (-1)^m \binom{\gamma}{m}\int_0^1 \left[\left(y^{\frac{1}{\alpha^2}}\right)^m y^{\left(\frac{2\gamma-\gamma\alpha^2}{\alpha^2}\right)}\right] dy\right\} \ldots \ldots \ldots (23)$$

$$= \frac{1}{1-\gamma}\log\left\{\left(\frac{6}{\alpha^2}\right)^\gamma \sum_{m=0}^\infty (-1)^m \binom{\gamma}{m}\left(\frac{\alpha^2}{m+2\gamma-\gamma\alpha^2+\alpha^2}\right)\right\} \ldots \ldots \ldots (24)$$

$$= \frac{1}{1-\gamma}\log\left\{6^\gamma \alpha^{-2(\gamma-1)} \sum_{m=0}^\infty (-1)^m \binom{\gamma}{m}(m+2\gamma-\gamma\alpha^2+\alpha^2)^{-1}\right\} \ldots \ldots \ldots (25)$$

## 2.9. Mean residual life function: see Figure 11

It is defined for the MBUR random variable as shown in equation (26). Taking the limits at the ends of the unit interval is shown in equation (27)

$$\mu(y|\alpha) = E(Y - y|Y > y) = \frac{1}{S(y)} \int_y^1 S(y) dy$$

$$= \frac{\left(1 - \frac{3\alpha^2}{2+\alpha^2} + \frac{2\alpha^2}{3+\alpha^2}\right) - y\left\{1 - \frac{3\alpha^2 y^{\frac{2}{\alpha^2}}}{2+\alpha^2} + \frac{3\alpha^2 y^{\frac{2}{\alpha^2}}}{3+\alpha^2}\right\}}{1 - 3y^{\frac{2}{\alpha^2}} + 2y^{\frac{3}{\alpha^2}}} \quad \ldots (26)$$

$$\lim_{y \to 0} \mu(y|\alpha) = 1 - \frac{3\alpha^2}{2+\alpha^2} + \frac{2\alpha^2}{3+\alpha^2} \quad \text{and} \quad \lim_{y \to 1} \mu(y|\alpha) = 0 \quad \ldots (27)$$

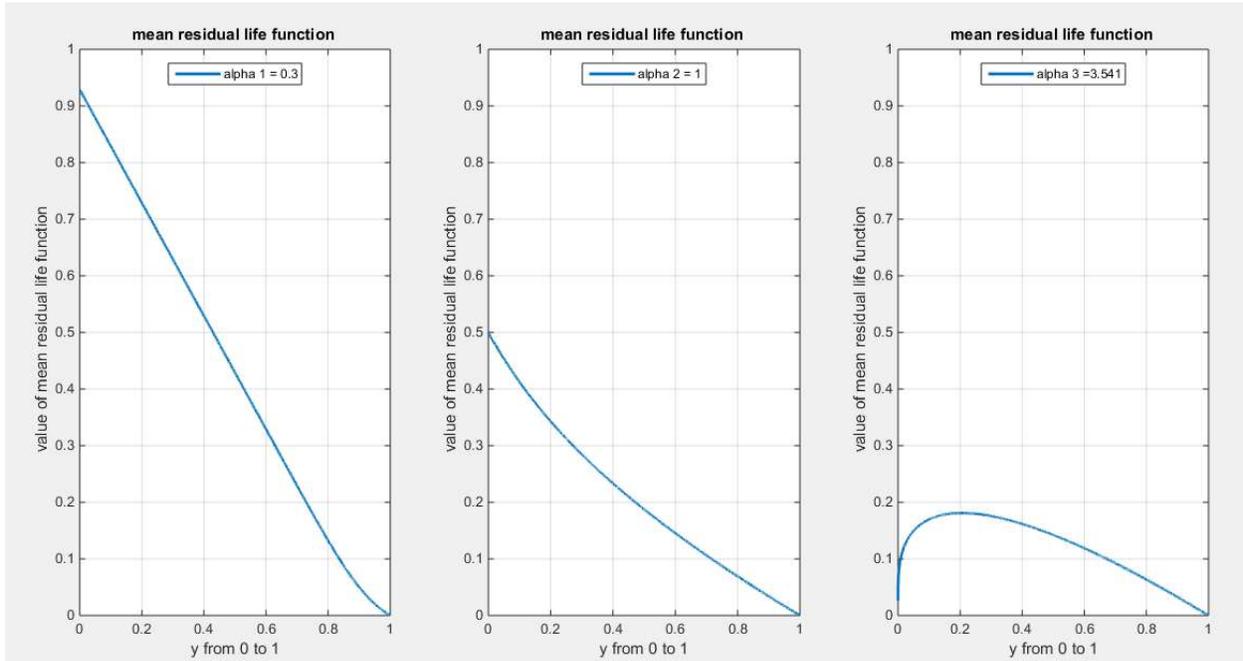

Fig. 11 shows mean residual life function at different levels of alpha.

### 2.10. Stochastic Ordering: see Figure 12

This ordering judges the comparative conduct of a variable. A random variable X is considered smaller than the random variable Y in the following orders:

1. Stochastic order $(X \leq_{st} Y)$ if $F_X(x) \geq F_Y(x)$ for all x.
2. Hazard rate order $(X \leq_{hr} Y)$ if $h_X(x) \geq h_Y(x)$ for all x.
3. Mean residual life order $(X \leq_{mrl} Y)$ if $m_X(x) \leq m_Y(x)$ for all x.
4. Likelihood ratio order $(X \leq_{lr} Y)$ if $\frac{f_X(x)}{f_Y(x)}$ decreases in x.

The following results are due to Shaked and Shanthikumar (23). They used the results to evaluate the stochastic ordering of a distribution:

$$X \leq_{lr} Y \implies X \leq_{hr} Y \implies X \leq_{mrl} Y \quad \text{hence} : X \leq_{st} Y$$

Theorem: let $X \sim MBUR(\alpha_1)$ and $Y \sim MBUR(\alpha_2)$ if $\alpha_1 < \alpha_2$, then

$X \leq_{lr} Y$, hence $X \leq_{hr} Y$, $X \leq_{mrl} Y$ and $X \leq_{st} Y$.

Proof: see equation (28)

$$\frac{f_X(y;\alpha_1)}{f_Y(y;\alpha_2)} = \frac{\frac{6}{\alpha_1^2}\left[1-y^{\frac{1}{\alpha_1^2}}\right]y^{\left(\frac{2}{\alpha_1^2}-1\right)}}{\frac{6}{\alpha_2^2}\left[1-y^{\frac{1}{\alpha_2^2}}\right]y^{\left(\frac{2}{\alpha_2^2}-1\right)}} \quad\ldots\ldots\ldots\ldots\ldots\ldots\ldots\ldots\ldots\ldots\ldots\ldots\ldots (28)$$

Taking the log of equation (28) gives equation (29)

$$\ln\left(\frac{6}{\alpha_1^2}\right) + \ln\left(1-y^{\frac{1}{\alpha_1^2}}\right) + \ln\left[y^{\left(\frac{2}{\alpha_1^2}-1\right)}\right] - \ln\left(\frac{6}{\alpha_2^2}\right) - \ln\left[1-y^{\frac{1}{\alpha_2^2}}\right] - \ln\left[y^{\left(\frac{2}{\alpha_2^2}-1\right)}\right] \ldots\ldots (29)$$

Taking the first derivative of the likelihood ratio order with respect to the variable y as shown in equation (30):

$$\frac{d\ LRO}{dy} = \frac{2}{y}\left\{\frac{1}{\alpha_1^2}-\frac{1}{\alpha_2^2}\right\} + \frac{1}{y}\left\{\frac{\frac{-1}{\alpha_1^2}y^{\frac{1}{\alpha_1^2}}}{1-y^{\frac{1}{\alpha_1^2}}} + \frac{\frac{1}{\alpha_2^2}y^{\frac{1}{\alpha_2^2}}}{1-y^{\frac{1}{\alpha_2^2}}}\right\} \quad\ldots\ldots\ldots\ldots\ldots\ldots\ldots\ldots\ldots (30)$$

Figure (12) describes the behavior of the previous function defined in equation (30). Taking the limit at the ends of the unit interval gives equation (31).

$$\lim_{y\to 1} LRO = 3\left\{\frac{1}{\alpha_1^2}-\frac{1}{\alpha_2^2}\right\} \text{ and } \lim_{y\to 0} LRO = +\infty \quad\ldots\ldots\ldots\ldots\ldots\ldots\ldots\ldots\ldots\ldots(31)$$

See supplementary materials (section1) for derivation of equation 31.( the Likelihood Ratio Order (LRO))

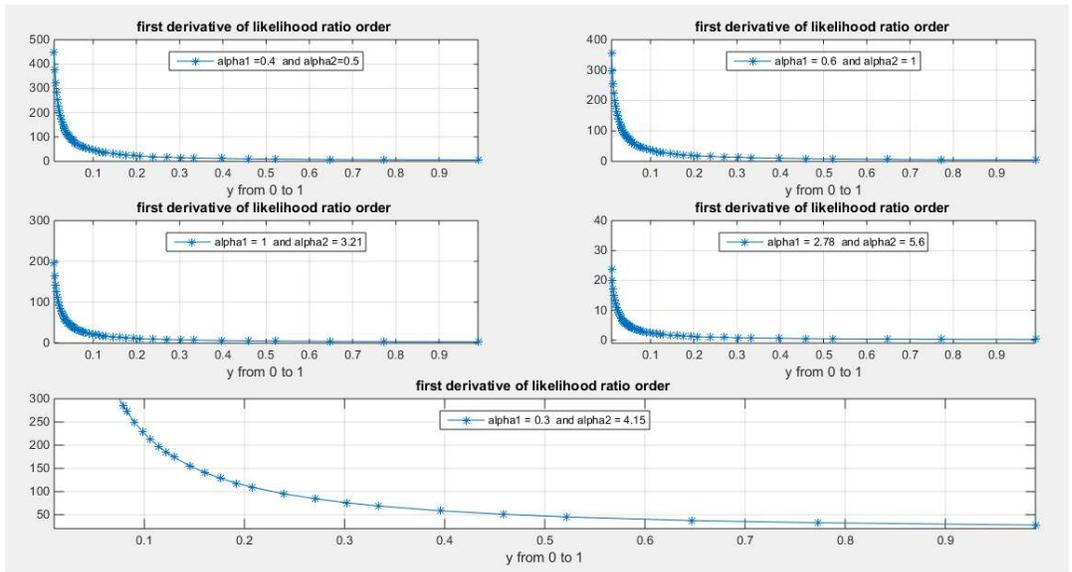

Fig.12 shows the first derivative of the likelihood ratio order with respect to random variable y for all possible values of the parameter alpha with $\alpha_1 < \alpha_2$. It is a decreasing function in y and hence all elements of stochastic ordering are true.

## 2.11. Probability Weighted Moments

Probability weighted moments are less vulnerable to extreme values and are tractable to obtain when ML estimators struggle for estimation. This can be deduced from the following equations (32) & (33):

$$E(y^r[F(y)]^s) = \int_0^1 y^r[F(y)]^s f(y) dy \quad\quad\quad (32)$$

$$= \int_0^1 y^r \left[3y^{\frac{2}{\alpha^2}} - 2y^{\frac{3}{\alpha^2}}\right]^s \frac{6}{\alpha^2}\left[1 - y^{\frac{1}{\alpha^2}}\right] y^{\left(\frac{2}{\alpha^2}-1\right)} dy$$

$$= \int_0^1 y^r \left(3y^{\frac{2}{\alpha^2}}\right)^s \left[1 - \frac{2}{3}y^{\frac{1}{\alpha^2}}\right]^s \frac{6}{\alpha^2}\left[1 - y^{\frac{1}{\alpha^2}}\right] y^{\left(\frac{2}{\alpha^2}-1\right)} dy$$

$$= \int_0^1 y^r \left(3^s y^{\frac{2s}{\alpha^2}}\right) \left[1 - \frac{2}{3}y^{\frac{1}{\alpha^2}}\right]^s \frac{6}{\alpha^2}\left[1 - y^{\frac{1}{\alpha^2}}\right] y^{\left(\frac{2}{\alpha^2}-1\right)} dy \quad\quad (33)$$

Binomial expansion of $\left[1 - \frac{2}{3}y^{\frac{1}{\alpha^2}}\right]^s$ is shown in equation (34):

$$\left[1 - \frac{2}{3}y^{\frac{1}{\alpha^2}}\right]^s = \sum_{m=0}^{\infty} (-1)^m \binom{s}{m} \left(\frac{2}{3}y^{\frac{1}{\alpha^2}}\right)^m \quad\quad\quad (34)$$

substitute equation (34) into equation (33).

$$= \frac{3^s 6}{\alpha^2} \int_0^1 y^r \left(y^{\frac{2s}{\alpha^2}}\right) \sum_{m=0}^{s} (-1)^m \binom{s}{m} \left(\frac{2}{3}y^{\frac{1}{\alpha^2}}\right)^m \left[1 - y^{\frac{1}{\alpha^2}}\right] y^{\left(\frac{2}{\alpha^2}-1\right)} dy$$

$$= \frac{3^s 6}{\alpha^2} \sum_{m=0}^{s} (-1)^m \binom{s}{m} \left(\frac{2}{3}\right)^m \int_0^1 y^r \left(y^{\frac{2s}{\alpha^2}}\right)\left(y^{\frac{1}{\alpha^2}}\right)^m \left[1 - y^{\frac{1}{\alpha^2}}\right] y^{\left(\frac{2}{\alpha^2}-1\right)} dy$$

$$= \frac{3^s 6}{\alpha^2} \sum_{m=0}^{s} (-1)^m \binom{s}{m} \left(\frac{2}{3}\right)^m \left\{\frac{\alpha^2}{(r\alpha^2 + 2s + m + 2)} - \frac{\alpha^2}{(r\alpha^2 + 2s + m + 3)}\right\}$$

$$= 3^s 6 \sum_{m=0}^{s} (-1)^m \binom{s}{m} \left(\frac{2}{3}\right)^m \left\{\frac{1}{(r\alpha^2 + 2s + m + 2)} - \frac{1}{(r\alpha^2 + 2s + m + 3)}\right\}$$

### 2.12. PDF and CDF of order statistics

Let $Y_1, Y_2, \ldots, Y_n$ be a sample randomly drawn from a MBUR distribution with sample size n and the corresponding order statistic distribution is $Y_{(1)}, Y_{(2)}, \ldots, Y_{(n)}$ so $Y_{(1)} \leq Y_{(2)} \leq \cdots, \leq Y_{(n)}$. The PDF of the j$^{th}$ order statistics $Y_{(j)}$ is defined as:

$$f_{Y_{(j)}}(y; \alpha) = \frac{n!}{(j-1)!(n-j)!} f(y;\alpha)[F(y;\alpha)]^{j-1}[1 - F(y;\alpha)]^{n-j}$$

Where $c = \frac{n!}{(j-1)!(n-j)!}$

$$f_{Y_{(j)}}(y;\alpha) = c\left[\frac{6}{\alpha^2}\left[1 - y^{\frac{1}{\alpha^2}}\right] y^{\left(\frac{2}{\alpha^2}-1\right)}\right]\left[3y^{\frac{2}{\alpha^2}} - 2y^{\frac{3}{\alpha^2}}\right]^{j-1}\left[1 - 3y^{\frac{2}{\alpha^2}} + 2y^{\frac{3}{\alpha^2}}\right]^{n-j}$$

PDF of the first or the smallest order statistics is given in equation (35):

$$f_{Y_{(1)}}(y;\alpha) = \frac{n!}{(n-1)!}\left[\frac{6}{\alpha^2}\left(1 - y^{\frac{1}{\alpha^2}}\right) y^{\left(\frac{2}{\alpha^2}-1\right)}\right]\left[1 - 3y^{\frac{2}{\alpha^2}} + 2y^{\frac{3}{\alpha^2}}\right]^{n-1} \quad\quad (35)$$

PDF of the largest order statistics is given in equation (36):

$$f_{Y_{(n)}}(y;\alpha) = \frac{n!}{(n-1)!} \left[\frac{6}{\alpha^2}\left(1-y^{\frac{1}{\alpha^2}}\right)y^{\left(\frac{2}{\alpha^2}-1\right)}\right]\left[3y^{\frac{2}{\alpha^2}} - 2y^{\frac{3}{\alpha^2}}\right]^{n-1} \ldots\ldots\ldots\ldots\ldots\ldots (36)$$

The CDF of the $j^{th}$ order statistics is defined as follows

$$F_{j:n}(y) = \sum_{j}^{n} C_{n:j}[F(y)]^j[1-F(y)]^{n-j}$$

So CDF of the $j^{th}$ order statistics from a MBUR random sample is defined in equation (37)

$$F_{j:n}(y) = \sum_{j}^{n} \binom{n}{j}\left[3y^{\frac{2}{\alpha^2}} - 2y^{\frac{3}{\alpha^2}}\right]^j \left[1 - 3y^{\frac{2}{\alpha^2}} + 2y^{\frac{3}{\alpha^2}}\right]^{n-j} \ldots\ldots\ldots\ldots\ldots\ldots (37)$$

## Results: Section 3:

### 3.1. Methods of Estimations:

### 3.1.1. Method of Moments (MOM):

Equating the sample's first moment equation (38), which is the mean, with the population's first moment equation (39) can provide an estimate for the parameter. This estimate can then be used as an initial guess in other methods that require numerical techniques to evaluate the parameter.

$$sample\ mean = \bar{y} = \frac{1}{n}\sum_{i=1}^{n} y_i \ \ldots\ldots\ldots\ldots\ldots\ldots\ldots\ldots\ldots\ldots\ldots\ldots\ldots (38)$$

$$population\ mean = E(y) = \frac{6}{(2+\alpha^2)(3+\alpha^2)} \ \ldots\ldots\ldots\ldots\ldots\ldots\ldots\ldots\ldots\ldots (39)$$

Equate the population's mean with the sample's mean to estimate the parameter equation (40).

$$E(y) = \frac{6}{(2+\alpha^2)(3+\alpha^2)} = \bar{y} \ \ldots\ldots\ldots\ldots\ldots\ldots\ldots\ldots\ldots\ldots\ldots\ldots\ldots (40)$$

To find the estimator for the alpha parameter, find the root of equation (41):

$$c = \frac{6}{\bar{y}} = (2+\alpha^2)(3+\alpha^2) = 6 + 5\alpha^2 + \alpha^4$$

$$0 = (\alpha^2)^2 + 5\alpha^2 + 6 - \frac{6}{\bar{y}} \ \ldots\ldots\ldots\ldots\ldots\ldots\ldots\ldots\ldots\ldots\ldots\ldots (41)$$

$$\alpha^2 = \frac{-5 + \sqrt{25 - 4(1)\left(6-\frac{6}{\bar{y}}\right)}}{2(1)}$$

$$\hat{\alpha} = \left\{\frac{-5 + \sqrt{25 - 24\left(\frac{\bar{y}-1}{\bar{y}}\right)}}{2}\right\}^{.5}, \quad 25 - 24\left(\frac{\bar{y}-1}{\bar{y}}\right) > 0, \alpha > 0$$

### 3.1.2. Maximum Likelihood Estimation (MLE):

Taking the first derivative of the log likelihood of the PDF of the MBUR distribution with respect to the alpha parameter as shown below:

The objective function to be maximized is the log-likelihood function as shown in equation (42) & (43).

$$L(\alpha) = \left(\frac{6}{\alpha^2}\right)^n \prod_{i=1}^{n}\left[1 - y_i^{\frac{1}{\alpha^2}}\right] \prod_{i=1}^{n} y_i^{\left(\frac{2}{\alpha^2}-1\right)}, \quad 0 < y < 1, \alpha > 0 \dots \dots \dots (42)$$

Maximize the log-likelihood function in equation (43) and this is equivalent to maximizing the likelihood function in equation (42) under certain regularity conditions. This is carried out by differentiating it as in equation (44) & (45) and using non-linear optimization algorithm like Newton-Raphson algorithm or quasi-Newton algorithm whichever is suitable.

$$l(\alpha) = n\ln(6) - 2n\ln(\alpha) + \sum_{i=1}^{n} \ln\left[1 - y_i^{\frac{1}{\alpha^2}}\right] + \left(\frac{2}{\alpha^2} - 1\right) \sum_{i=1}^{n} \ln(y_i) \dots \dots (43)$$

$$\frac{dl(\alpha)}{d\alpha} = \frac{-2n}{\alpha} + \sum_{i=1}^{n} \frac{-y_i^{\frac{1}{\alpha^2}}(\ln[y_i])(-2\alpha^{-3})}{1 - y_i^{\frac{1}{\alpha^2}}} - 4\alpha^{-3} \sum_{i=1}^{n} \ln(y_i) \dots \dots (44)$$

$$\frac{d^2 l(\alpha)}{d\alpha^2} = \frac{2n}{\alpha^2} - \frac{6}{\alpha^4} \sum_{i=1}^{n} \frac{y_i^{\frac{1}{\alpha^2}}(\ln[y_i])}{1 - y_i^{\frac{1}{\alpha^2}}} + \frac{12}{\alpha^4} \sum_{i=1}^{n} \ln(y_i) - \frac{4}{\alpha^6} \sum_{i=1}^{n} \frac{y_i^{\frac{1}{\alpha^2}}(\ln[y_i])^2}{\left(1 - y_i^{\frac{1}{\alpha^2}}\right)^2} \dots \dots (45)$$

Alpha can be estimated numerically using Newton Raphson algorithm.

$$\alpha_1 = \alpha_0 - \left[\frac{d^2 l(\alpha)}{d\alpha^2}\right]^{-1} \frac{dl(\alpha)}{d\alpha}$$

The expected information matrix $J_y$ is defined as the negative expected value of the second derivative of the log-likelihood, evaluated at the estimated parameter. Under certain regularity conditions and with a large sample size, the inverse of this information matrix, $\left[\frac{d^2 l(\alpha)}{d\alpha^2}\right]^{-1}$, represents the variance of the estimated parameter. Consequently, this estimator is approximately normally distributed.

$$\sqrt{n}\,(\alpha - \hat{\alpha}) \sim N(0, J_y^{-1})$$

This is used to construct an approximate confidence interval for the parameter in the form of $\hat{\alpha} \pm z_{1-\frac{\gamma}{2}} se(\hat{\alpha})$. SE is the square root of the inverse of the expected information matrix. $z_{1-\frac{\gamma}{2}}$ is the quantile of standard normal distribution.

### 3.1.3. Maximum Product of Spacing (MPS):

Maximize the following objective function in equation (46):

$$MPS = \frac{1}{n+1} \sum_{i=1}^{n} \log[F(y_i) - F(y_{i-1})] \dots \dots (46)$$

Take the derivatives of this function as in equations (47) & (48)

$$\frac{d\,MPS}{d\,\alpha} = \frac{1}{n+1} \sum_{i=1}^{n} \frac{[F'(y_i) - F'(y_{i-1})]}{[F(y_i) - F(y_{i-1})]} \quad \ldots \ldots \ldots \ldots \ldots \ldots \ldots \ldots \ldots \ldots (47)$$

$$\frac{d^2 MPS}{d\alpha^2} = \frac{1}{n+1} \sum_{i=1}^{n} \frac{[F''(y_i) F''(y_{i-1})]}{[F(y_i) - F(y_{i-1})]} - \frac{[F'(y_i) - F'(y_{i-1})]^2}{[F(y_i) - F(y_{i-1})]^2} \quad \ldots \ldots \ldots \ldots \ldots (48)$$

Alpha can be estimated numerically using Newton Raphson method.

### 3.1.4. Anderson Darling Estimator (AD)

Minimize the following objective function in equation (49):

$$AD = -n - \sum_{i=1}^{n} \left(\frac{2i-1}{n}\right) \{\log[F(x_i)] + \log[1 - F(x_{n-i+1})]\} \quad \ldots \ldots \ldots \ldots \ldots \ldots (49)$$

Take the derivatives of this function as in equations (50) & (51)

$$\frac{dAD}{d\alpha} = -\sum_{i=1}^{n} \left(\frac{2i-1}{n}\right) \left\{ \frac{F'(x_i)}{F(x_i)} + \frac{-F'(x_{n-i+1})}{1 - F(x_{n-i+1})} \right\} \quad \ldots \ldots \ldots \ldots \ldots \ldots \ldots \ldots (50)$$

$$\frac{d^2 AD}{d\alpha^2} = -\sum_{i=1}^{n} \left(\frac{2i-1}{n}\right) \left\{ \frac{F''(x_i)}{F(x_i)} - \frac{[F'(x_i)]^2}{[F(x_i)]^2} - \frac{F''(x_{n-i+1})}{1 - F(x_{n-i+1})} - \frac{[F'(x_{n-i+1})]^2}{[1 - F(x_{n-i+1})]^2} \right\} \quad \ldots \ldots (51)$$

Alpha can be estimated numerically using Newton Raphson method.

### 3.1.5. Percentile method (PERC):

Minimize the following objective function in equation (52):

$$Percentile = \sum_{i=1}^{n} \left\{ y_i - F^{-1}\left(\alpha, ecdf = \frac{i}{n+1}\right) \right\}^2 \quad \ldots \ldots \ldots \ldots \ldots \ldots \ldots \ldots (52)$$

Take the derivatives of this function as in equations (53) & (54)

$$\frac{d\,Percentile}{d\alpha} = 2 \sum_{i=1}^{n} \{y_i - F^{-1}\}\{-F^{-1}\}' \quad \ldots \ldots \ldots \ldots \ldots \ldots \ldots \ldots \ldots \ldots (53)$$

$$\frac{d^2 Percentile}{d\alpha^2} = 2 \sum_{i=1}^{n} \{(-F^{-1})'\,(-F^{-1})' + (y_i - F^{-1})(-F^{-1})''\} \quad \ldots \ldots \ldots \ldots \ldots \ldots (54)$$

Alpha can be estimated numerically using Newton Raphson method.

### 3.1.6. Cramer Von Mises (CVM)

Minimize the following objective function in equation (55):

$$CVM = \frac{1}{12n} + \sum_{i=1}^{n} \left\{ F(x_i) - \frac{2i-1}{2n} \right\}^2 \quad \ldots \ldots \ldots \ldots \ldots \ldots \ldots \ldots \ldots \ldots (55)$$

Take the derivatives of this function as in equations (56) & (57)

$$\frac{dCVM}{d\alpha} = 2\sum_{i=1}^{n}\left\{F(x_i) - \frac{2i-1}{2n}\right\}\{F'(x_i)\} \quad\quad\quad\quad\quad\quad\quad\quad\quad\quad\quad\quad\quad (56)$$

$$\frac{d^2CVM}{d\alpha^2} = 2\sum_{i=1}^{n}\left\{F'(x_i)F'(x_i) + \left(F(x_i) - \frac{2i-1}{2n}\right)(F''(x_i))\right\} \quad\quad\quad\quad\quad (57)$$

Alpha can be estimated numerically using Newton Raphson method.

### 3.1.7. Least Squares Method (LS):

Minimize the following objective function in equation (58):

$$LS = \sum_{i=1}^{n}\left\{F(x_i) - \frac{i}{n+1}\right\}^2 \quad\quad\quad\quad\quad\quad\quad\quad\quad\quad\quad\quad\quad\quad\quad (58)$$

Take the derivatives of this function as in equations (59) & (60)

$$\frac{\partial LS}{\partial \alpha} = 2\sum_{i=1}^{n}\left\{F(x_i) - \frac{i}{n+1}\right\}F'(x_i) \quad\quad\quad\quad\quad\quad\quad\quad\quad\quad\quad\quad (59)$$

$$\frac{\partial^2 LS}{\partial \alpha^2} = 2\sum_{i=1}^{n}\left\{F'(x_i)F'(x_i) + \left(F(x_i) - \frac{i}{n+1}\right)F''(x_i)\right\} \quad\quad\quad\quad\quad\quad (60)$$

Alpha can be estimated numerically using Newton Raphson method.

### 3.1.8. Weighted Least Squares Method (WLS):

Minimize the following objective function in equation (61):

$$WLS = \sum_{i=1}^{n}\left[\frac{(n+1)^2(n+2)}{i(n+1-i)}\right]\left\{F(x_i) - \frac{i}{n+1}\right\}^2 \quad\quad\quad\quad\quad\quad (61)$$

Take the derivatives of this function as in equations (62) & (63)

$$\frac{dLS}{d\alpha} = 2\sum_{i=1}^{n}\left[\frac{(n+1)^2(n+2)}{i(n+1-i)}\right]\left\{F(x_i) - \frac{i}{n+1}\right\}F'(x_i) \quad\quad\quad\quad\quad (62)$$

$$\frac{d^2LS}{d\alpha^2} = 2\sum_{i=1}^{n}\left[\frac{(n+1)^2(n+2)}{i(n+1-i)}\right]\left\{F'(x_i)F'(x_i) + \left(F(x_i) - \frac{i}{n+1}\right)F''(x_i)\right\} \quad (63)$$

Alpha can be estimated numerically using Newton Raphson method.

### 3.2. Simulation

A simulation study is conducted utilizing the following sample sizes $n = (20, 80, 160, 260, 500)$, and replicate N=1000 times. Various methods of estimation are utilized and compared with one another. The parameter alpha value chosen is $\alpha = (2.5)$. For alpha value $= (0.5)$, see the results in supplementary material section 2.

Steps:

1- Generate random variable from the MBUR Distribution with specified alpha
2- Choose the sample size n.

3- Replicate the method of estimation N times.
4- Calculate various metrics to compare methods and assess the impact of increasing sample size on estimators.

a) $Average\ Absolute\ Bias\ (AAB) = \frac{1}{N}\sum_{i=1}^{N}|\hat{\alpha} - \alpha|$
b) $Mean\ Square\ Error(MSE) = \frac{1}{N}\sum_{I=1}^{N}(\hat{\alpha} - \alpha)^2$
c) $Root\ of\ Mean\ Square\ Error(MSE) = \sqrt{MSE}$
d) $Mean\ Relative\ Error(MRE) = \frac{1}{N}\sum_{i=1}^{N}\frac{|\hat{\alpha}-\alpha|}{\alpha}$

Also the mean of estimated alpha from the 1000 replicates is evaluated with the standard error. For the chosen alpha level 2.5, the following results are obtained in the successive Tables 2-6, mean in Table 2, SE in Table 3, AAB in Table 4, MSE in Table 5, MRE in Table 6. For better visualization of the results each table is represented by a Heat-map graph as shown in Figures 13-17

Table (2): shows the mean from the 1000 replicates for each method

| mean | MOM | MLE | MPS | AD | PERC | CVM | LS | WLS |
|---|---|---|---|---|---|---|---|---|
| n=20 | 2.6001 | 2.4561 | 2.5321 | 2.4725 | 2.3617 | 2.4727 | 2.4755 | 2.4905 |
| n=80 | 2.52 | 2.486 | 2.5043 | 2.4896 | 2.4538 | 2.4896 | 2.4908 | 2.4943 |
| n=160 | 2.5069 | 2.4936 | 2.5039 | 2.495 | 2.4711 | 2.4953 | 2.496 | 2.4977 |
| n=260 | 2.5030 | 2.4972 | 2.5042 | 2.4991 | 2.4797 | 2.5004 | 2.5008 | 2.5008 |
| n=500 | 2.5028 | 2.4991 | 2.5032 | 2.4996 | 2.491 | 2.4997 | 2.5002 | 2.5004 |

Table 2 shows that as sample size increases, the estimated parameter approaches the true value. The percentile method is the least efficient to approach the true regardless the sample size. MOM and MPS have nearly comparable results at different sample sizes. AD and CVM have nearly similar results. And LS and WLS have similar results at large sample sizes. Figure 13 visually illustrates these results.

Table (3): shows the SE from the 1000 replicates for each method

| SE | MOM | MLE | MPS | AD | PERC | CVM | LS | WLS |
|---|---|---|---|---|---|---|---|---|
| n=20 | 0.013 | 0.0071 | 0.0065 | 0.0072 | 0.0123 | 0.0084 | 0.0083 | 0.0074 |
| n=80 | 0.0057 | 0.0033 | 0.0032 | 0.0034 | 0.0063 | 0.0037 | 0.0037 | 0.0035 |
| n=160 | 0.0041 | 0.0022 | 0.0022 | 0.0024 | 0.0046 | 0.0025 | 0.0025 | 0.0024 |
| n=260 | 0.0031 | 0.0018 | 0.0018 | 0.0019 | 0.0036 | 0.002 | 0.002 | 0.0019 |
| n=500 | 0.0023 | 0.0013 | 0.0013 | 0.0014 | 0.0027 | 0.0014 | 0.0014 | 0.0014 |

Table 3 shows that as sample size increases the standard error (SE) decreases. Percentile method has the highest SE followed by the MOM at all different sample sizes. The MLE & MPS have the lowest SE at n=500. MLE and MPS have nearly equal results at different sample sizes. This is also true as regards the pair of AD and WLS methods and the pair of CVM and LS methods. Figure 14 visually demonstrates these results.

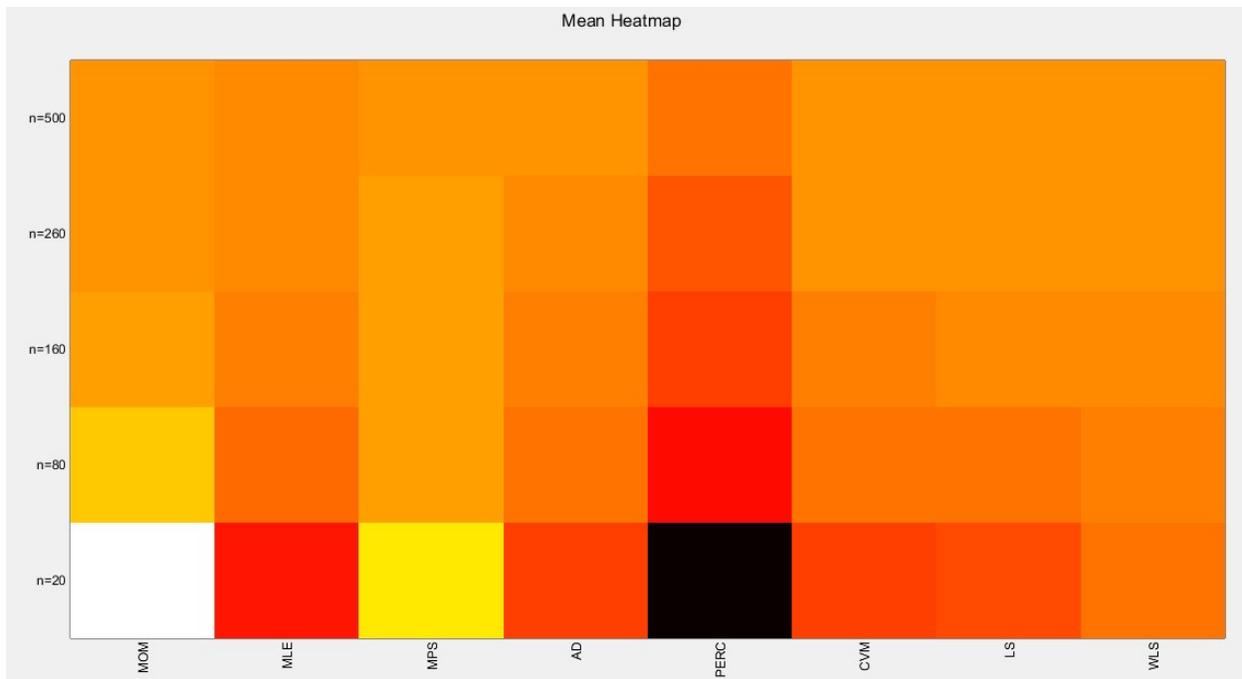

Fig 13 shows the Heat-map for the mean of the estimated alpha parameter from running the simulation using different methods for estimation with alpha value 2.5. As the sample size increases the estimated alpha approaches the true value of the parameter

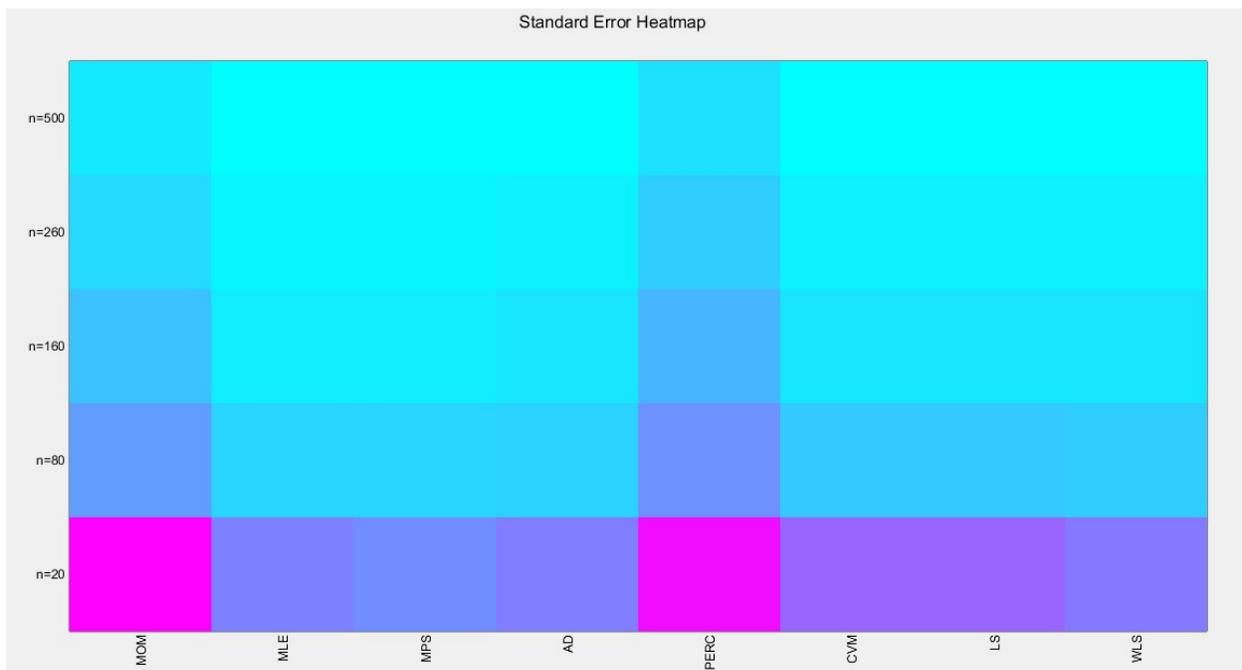

Fig 14 shows the Heat-map for the standard error (SE) of the estimated alpha parameter from running the simulation using different estimation methods with alpha value 2.5.

Table (4): shows the AAB from the 1000 replicates for each method

| AAB | MOM | MLE | MPS | AD | PERC | CVM | LS | WLS |
|---|---|---|---|---|---|---|---|---|
| n=20 | 0.3221 | 0.1673 | 0.1631 | 0.1706 | 0.3296 | 0.1912 | 0.1902 | 0.1776 |
| n=80 | 0.1444 | 0.0827 | 0.0809 | 0.085 | 0.1649 | 0.0902 | 0.0901 | 0.0854 |
| n=160 | 0.1037 | 0.0561 | 0.0552 | 0.0595 | 0.1195 | 0.0626 | 0.0625 | 0.0596 |
| n=260 | 0.0791 | 0.0457 | 0.0456 | 0.0481 | 0.0917 | 0.0506 | 0.0506 | 0.0481 |
| n=500 | 0.0579 | 0.0328 | 0.0327 | 0.0341 | 0.0667 | 0.0355 | 0.0354 | 0.0341 |

Table 4 shows that as the sample size increases, the average absolute bias (AAB) decreases. The percentile method has the highest value of AAB followed by MOM at all different sample sizes. MLE and MPS yield near identical results at different sample sizes. AD and WLS have approximately equal results. CVM and LS have nearly similar results. Figure 15 visually depicts these findings.

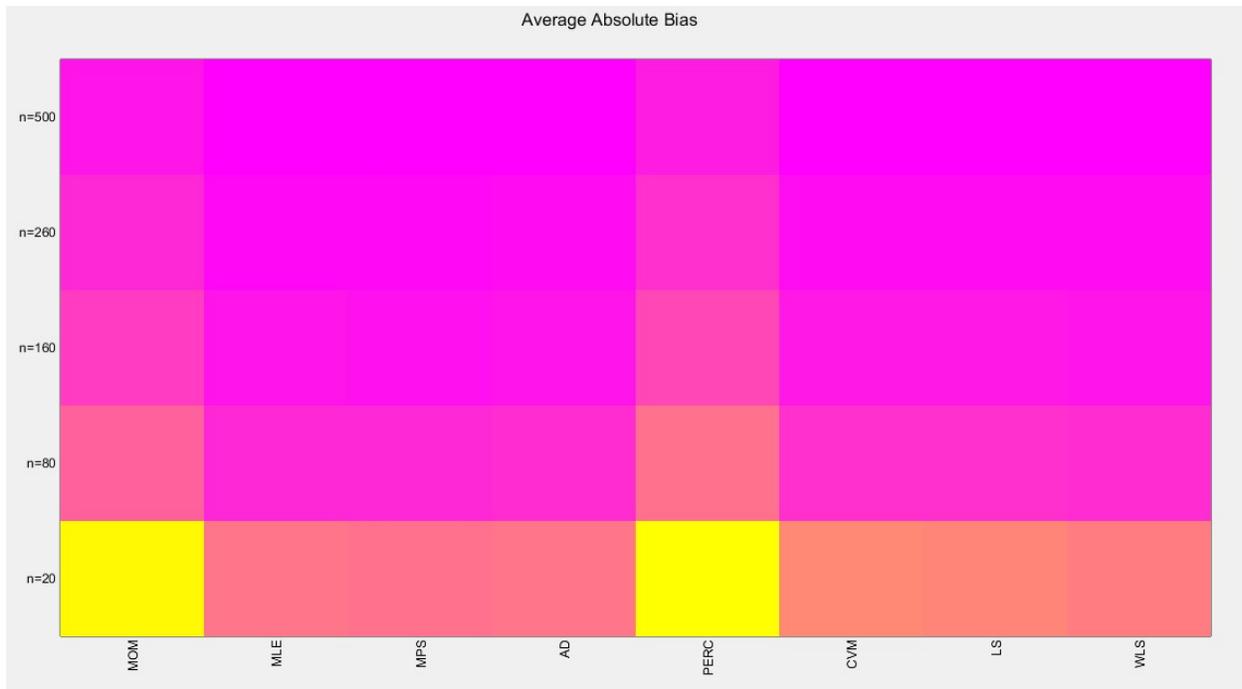

Fig 15 shows the Heat-map for the average absolute bias (AAB) of the estimated alpha parameter from running the simulation using different estimation methods with alpha value 2.5.

Table (5): shows the MSE from the 1000 replicates for each method

| MSE | MOM | MLE | MPS | AD | PERC | CVM | LS | WLS |
|---|---|---|---|---|---|---|---|---|
| n=20 | 0.1798 | 0.0519 | 0.0427 | 0.0521 | 0.1701 | 0.0708 | 0.0698 | 0.0553 |
| n=80 | 0.0333 | 0.0119 | 0.0102 | 0.012 | 0.0417 | 0.0137 | 0.0137 | 0.0120 |
| n=160 | 0.0166 | 0.0051 | 0.0048 | 0.0057 | 0.0224 | 0.0063 | 0.0063 | 0.0056 |
| n=260 | 0.0098 | 0.0032 | 0.0032 | 0.0036 | 0.013 | 0.004 | 0.004 | 0.0036 |
| n=500 | 0.0053 | 0.0017 | 0.0017 | 0.0018 | 0.0071 | 0.002 | 0.002 | 0.0018 |

Table (6): shows the MRE from the 1000 replicates for each method

| MRE | MOM | MLE | MPS | AD | PERC | CVM | LS | WLS |
|---|---|---|---|---|---|---|---|---|
| n=20 | 0.1288 | 0.0669 | 0.0652 | 0.0682 | 0.1318 | 0.0765 | 0.0761 | 0.0710 |
| n=80 | 0.0578 | 0.0331 | 0.0324 | 0.0340 | 0.066 | 0.0361 | 0.0361 | 0.0342 |
| n=160 | 0.0415 | 0.0224 | 0.0221 | 0.0238 | 0.0478 | 0.025 | 0.025 | 0.0238 |
| n=260 | 0.0317 | 0.0183 | 0.0182 | 0.0192 | 0.0367 | 0.0202 | 0.0202 | 0.0192 |
| n=500 | 0.0231 | 0.0131 | 0.0131 | 0.0136 | 0.0267 | 0.0142 | 0.0142 | 0.0137 |

The tables indicate that increasing the sample size leads to a decreases in the SE, AAB, MSE and MRE indices. For each method used, the indices decrease as the sample size increases. The values obtained from the MLE and MPS methods are nearly equal, especially with larger sample sizes (n=260 and n=500). Similarly, the AD and WLS methods yield comparable results. Additionally, the CVM and LS methods show approximately equal indices as the sample size increases. Overall, the methods demonstrate consistent results regarding the estimation values. Figure 16-17 display these findings

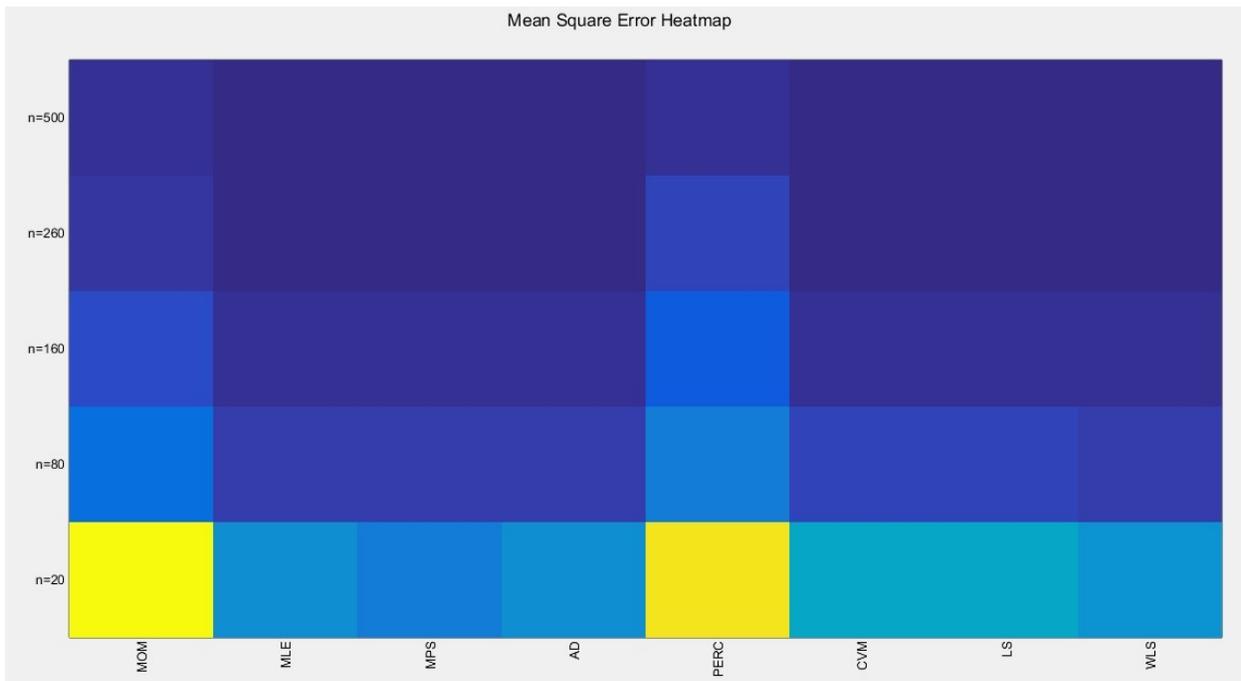

Fig 16 shows the Heat-map for the mean square error (MSE) of the estimated alpha parameter from running the simulation using different estimation methods with alpha value 2.5.

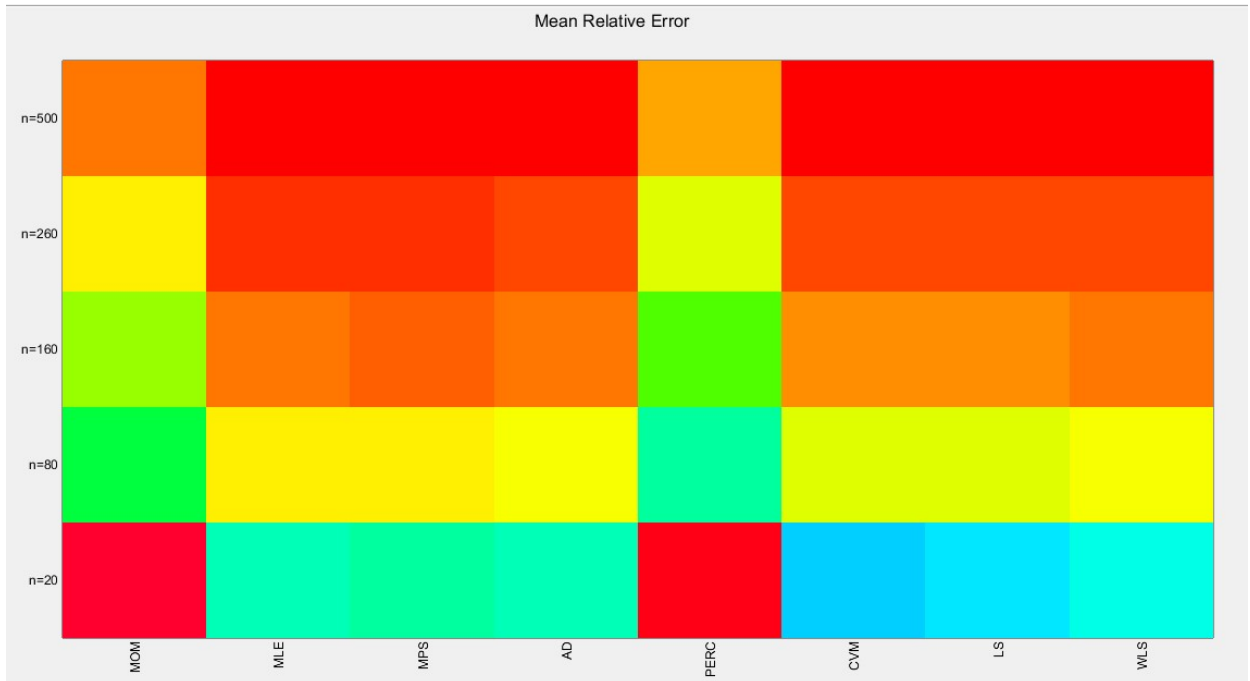

Fig 17 shows the Heat-map for the mean relative error (MRE) of the estimated alpha parameter from running the simulation using different estimation methods with alpha value 2.5.

## Discussion: Section 4: *Some Real Data Analysis:*

The data sets are sourced from the OECD, or Organization for Economic Co-operation and Development. It provides information on the economy, social events, education, health, labor, and the environment in the member countries. Matlab 2014 R was used for analysis where the mle function utilizes the derivative free Nelder-Mead algorithm for optimization. https://stats.oecd.org/index.aspx?DataSetCode=BLI .

***First data***: (Dwelling Without Basic Facilities), see table 7. These observations assess the percentage of homes in the affected countries that lack essential utilities such as indoor plumbing, central heating, and clean drinking water supplies.

Table (7): shows Dwelling without Basic facilities data set

| 0.008 | 0.007 | 0.002 | 0.094 | 0.123 | 0.023 | 0.005 | 0.005 | 0.057 | 0.004 |
|---|---|---|---|---|---|---|---|---|---|
| 0.005 | 0.001 | 0.004 | 0.035 | 0.002 | 0.006 | 0.064 | 0.025 | 0.112 | 0.118 |
| 0.001 | 0.259 | 0.001 | 0.023 | 0.009 | 0.015 | 0.002 | 0.003 | 0.049 | 0.005 |
| 0.001 | | | | | | | | | |

***Second data***: (Quality of Support Network), see table 8. This dataset examines the extent to which individuals can rely on sources of support, such as family, friends, or community members, during times of need and distress. It is presented as the percentage of individuals who have found social support in times of crisis.

Table (8): shows Quality of support Network data set

| 0.98 | 0.96 | 0.95 | 0.94 | 0.93 | 0.8 | 0.82 | 0.85 | 0.88 | 0.89 |
|---|---|---|---|---|---|---|---|---|---|
| 0.78 | 0.92 | 0.92 | 0.9 | 0.96 | 0.96 | 0.94 | 0.77 | 0.95 | 0.91 |

***Third data***: (Educational Attainment), see table 9. The observations measure the percentage of the OECD population that has completed their high-level education, such as high school or an equivalent degree.

Table (9): shows Educational attainment data set

| 0.84 | 0.86 | 0.8  | 0.92 | 0.67 | 0.59 | 0.43 | 0.94 | 0.82 | 0.91 |
|------|------|------|------|------|------|------|------|------|------|
| 0.91 | 0.81 | 0.86 | 0.76 | 0.86 | 0.76 | 0.85 | 0.88 | 0.63 | 0.89 |
| 0.89 | 0.94 | 0.74 | 0.42 | 0.81 | 0.81 | 0.93 | 0.55 | 0.92 | 0.9  |
| 0.63 | 0.84 | 0.89 | 0.42 | 0.82 | 0.92 |      |      |      |      |

***Fourth data***: (Flood Data), see table 10. These are 20 observations regarding the maximum flood level of the Susquehanna River at Harrisburg, Pennsylvania. (24).

Table (10): shows Flood Data set

| 0.26 | 0.27   | 0.3  | 0.32 | 0.32 | 0.34 | 0.38 | 0.38 | 0.39 | 0.4  |
|------|--------|------|------|------|------|------|------|------|------|
| 0.41 | 0.42   | 0.42 | 0.42 | 045  | 0.48 | 0.49 | 0.61 | 0.65 | 0.74 |

***Fifth data***: (Time between Failures of Secondary Reactor Pumps), (21) see table11.

Table (11): shows time between Failures data set

| 0.216  | 0.015  | 0.4082 | 0.0746 | 0.0358 | 0.0199 | 0.0402 | 0.0101 | 0.0605 |
|--------|--------|--------|--------|--------|--------|--------|--------|--------|
| 0.0954 | 0.1359 | 0.0273 | 0.0491 | 0.3465 | 0.007  | 0.656  | 0.106  | 0.0062 |
| 0.4992 | 0.0614 | 0.532  | 0.0347 | 0.1921 |        |        |        |        |

The analysis of the above data sets aims to determine how these sets align with the following distributions: Beta, Topp Leone, Unit Lindely, Kumaraswamy. The fitting of these data sets will be compared to the fitting of the new MBUR distribution. The tools used for this comparison include the following metrics: LL(log-likelihood), Akaike Information Criterion (AIC), corrected AIC (CAIC), Bayesian Information Criterion (BIC), and Hannan-Quinn Information Criterion (HQIC). Additionally, the Kolmogorov-Smirnov (K-S) test will be conducted. The test's results will include its value, along with the outcome of the null hypothesis (H0), which assumes that the data set follows the tested distribution; if this assumption is not met, the null hypothesis will be rejected. The P-value for the test will also be recorded. Furthermore, the Cramér-von Mises test and the Anderson-Darling test will be performed, with their respective values reported. Figures depicting the empirical cumulative distribution function (eCDF) and the theoretical cumulative distribution functions (CDF) of the five distributions will be illustrated, each in its place. Finally, the values of the estimated parameters, along with their estimated variances and standard errors, will be reported. The competitors' distributions are:

1- Beta Distribution:

$$f(y; \alpha, \beta) = \frac{\Gamma(\alpha + \beta)}{\Gamma(\alpha)\Gamma(\beta)} y^{\alpha-1}(1-y)^{\beta-1}, 0 < y < 1, \alpha > 0, \ \beta > 0$$

2- Kumaraswamy Distribution:

$$f(y; \alpha, \beta) = \alpha\beta y^{\alpha-1}(1-y^\alpha)^{\beta-1}, 0 < y < 1, \quad \alpha > 0, \ \beta > 0$$

3- Median Based Unit Rayleigh:

$$f(y; \alpha) = \frac{6}{\alpha^2}\left[1 - y^{\frac{1}{\alpha^2}}\right] y^{\left(\frac{2}{\alpha^2}-1\right)}, \ 0 < y < 1, \quad \alpha > 0$$

4- Topp-Leone Distribution:

$$f(y;\theta) = \theta(2-2y)(2y-y^2)^{\theta-1}, 0 < y < 1, \quad \theta > 0$$

5- Unit-Lindley:

$$f(y;\theta) = \frac{\theta^2}{1+\theta}(1-y)^3 \exp\left(\frac{-\theta y}{1-y}\right), \quad 0 < y < 1, \quad \theta > 0$$

Comparison tools are: (k) is the number of parameter, (n) is the number of observations.

$$AIC = -2MLL + 2k \quad , \quad CAIC = -2MLL + \frac{2kn}{n-k-1} \quad , \quad BIC = -2MLL + k\log(n)$$

$$HQIC = -2\log L + 2k * \ln[\ln(n)]$$

$$KS - test = Sup_n |F_n - F|, \quad F_n = \frac{1}{n}\sum_{i=1}^{n} I_{x_i < x}$$

$$Cramer - Von - Mises - test(CVM) = \frac{1}{12n} + \sum_{i=1}^{n}\left\{F(x_i) - \frac{2i-1}{2n}\right\}^2$$

$$Anderson - Darling - test(AD) = -n - \sum_{i=1}^{n}\left(\frac{2i-1}{n}\right)\{log[F(x_i)] + log[1 - F(x_{n-i+1})]\}$$

Total time on Test (TTT) can be calculated with the following approaches.

***First Approach (Empirical approach):***

1. Sort or order the data $X_1 \leq X_2 \leq \cdots \leq X_n$
2. Calculate the scaled TTT value by computing the following:

$$TTT(i) = \sum_{j=1}^{i}(N-j+1)(X_j - X_{j-1})$$

3. Normalize the cumulative scaled TTT plot values.

$$\frac{TTT(i)}{TTT(n)} \quad where \quad TTT(n) = \sum_{j=1}^{n}(N-j+1)(X_j - X_{j-1})$$

4. Plot the x-axis values (i/n) against the y-axis values which are the normalized cumulative scaled TTT values.

***Second Approach (theoretical approach):*** scaled TTT transform curve using survival function and the theoretical quantile.

$$T_s(u) = \frac{T(u)}{T(1)} = \frac{\int_0^{F^{-1}(u)}[1-F(x)]dx}{\int_0^{F^{-1}(1)}[1-F(x)]dx}$$

Where the theoretical quantile function is:

$$F^{-1}(u) = \inf\{x|F(x) \geq u\} \quad u \in [0,1]$$

Both graphs are provided for each data set.

The rationale for selecting these datasets is the characteristics they exhibit. Descriptive statistics definitively reveal empirical skewness and kurtosis from the data, compelling the author to choose the most appropriate competitor distributions to accommodate these findings.

**4.1. Analysis of the first data set.** See supplementary materials (section 3)

**4.2. Analysis of the second data set.** See Table 12 & 13. See Figures 18-25

Table (12): Descriptive statistics of the second data set

| min | mean | std | skewness | kurtosis | 25$_{percentile}$ | 50$_{perc}$ | 75$_{perc}$ | max |
|---|---|---|---|---|---|---|---|---|
| 0.77 | 0.9005 | 0.064 | -0.9147 | 2.6716 | 0.865 | 0.92 | 0.95 | 0.98 |

The data demonstrates a left skewness and a negative excess kurtosis, indicating a platykurtic shape. This is supported by the histogram and box plot shown in Figure 18. Plotting the empirical survival function reveals a rapid decay, suggesting a light tail, as illustrated in Figure 19. This observation is further reinforced by the log-log plot in Figure 20. When the author plots the logarithm of the observations against the logarithm of the empirical survival function, the author sees a concave curve, which indicates faster decay and supports the concept of a light tail. Additionally, a quantile analysis was conducted by comparing the empirical 1st, 5th, and 10th quantiles with the corresponding theoretical quantiles from the standard uniform distribution. The findings show a light tail: specifically, (0.7700 < 0.7721), (0.7750 < 0.7805), and (0.7900 < 0.7910). Fitting the MBUR model to the data produced an estimated alpha value of 0.3591, which is less than 1. This is consistent with the empirical survival function depicted in Figure 19, which bears similarity to the survival function in Figure 6, where alpha is also less than 1.

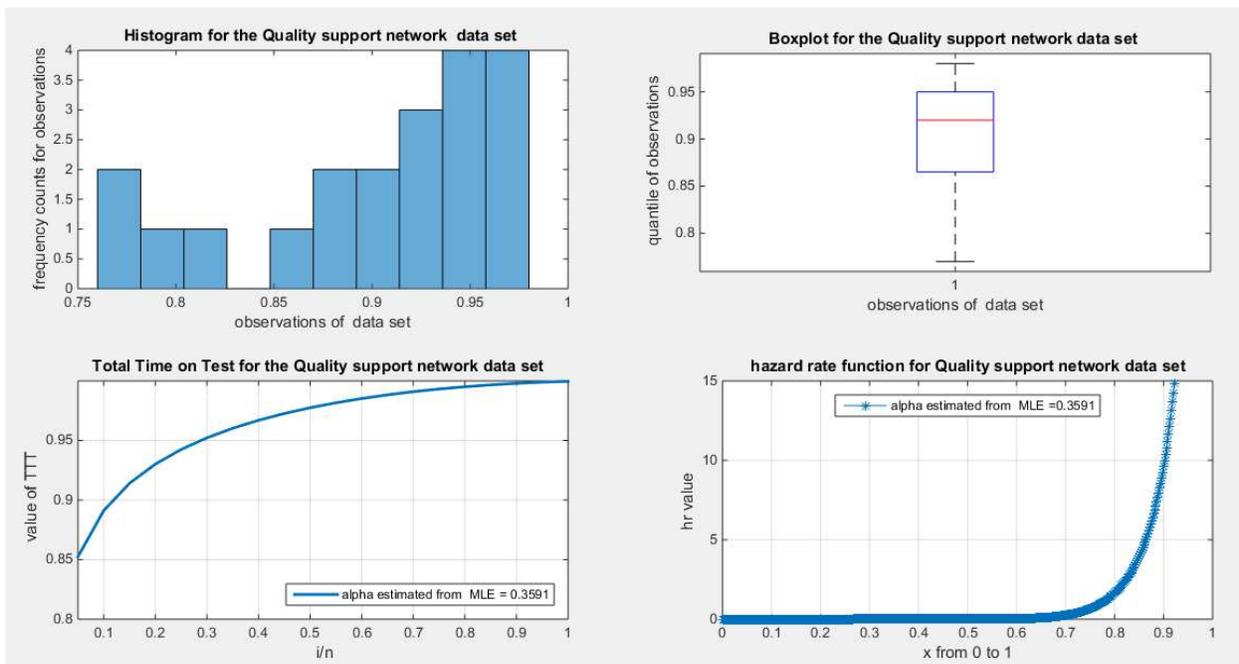

Fig.18 shows the histogram with left skewness and associated boxplot with no outliers or extreme values. The TTT plot shows concave shape which supports increased failure rate that is obvious in the shape of the hazard function on the right lower graph.

After fitting the MBUR distribution to the data, the scaled TTT plot reveals a concave shape, indicating an increased failure rate. This pattern is evident in the shape of the hazard rate function. Figure 18 illustrates this concavity through the theoretical scaled TTT plot, which supports the increase in the hazard rate. Similarly, Figure 21 displays the empirical scaled TTT plot, also confirming this increase in the hazard rate.

Out of the five distributions analyzed, all successfully fitted the data. However, the MBUR distribution emerged as the most effective model. This was evidenced by its superior validation indices compared to the other distributions, as illustrated in in Table 13.

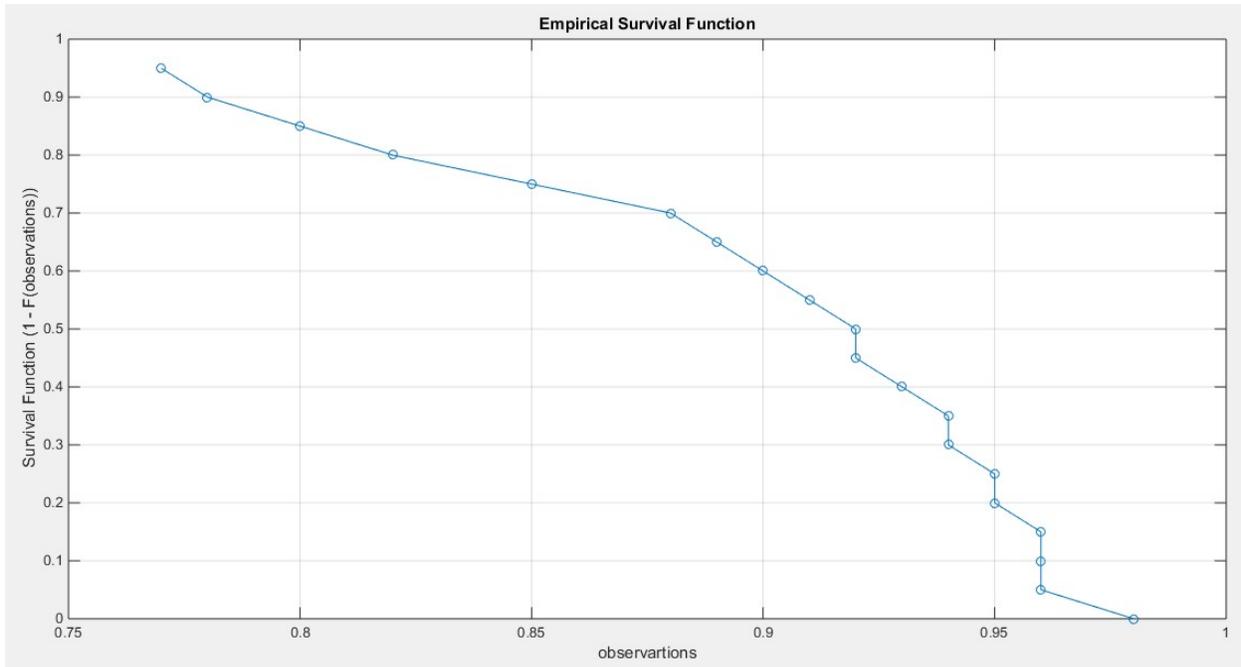

Fig. 19 shows the empirical survival function concave graph curved upward like the one in fig. (6) where the alpha is <1.

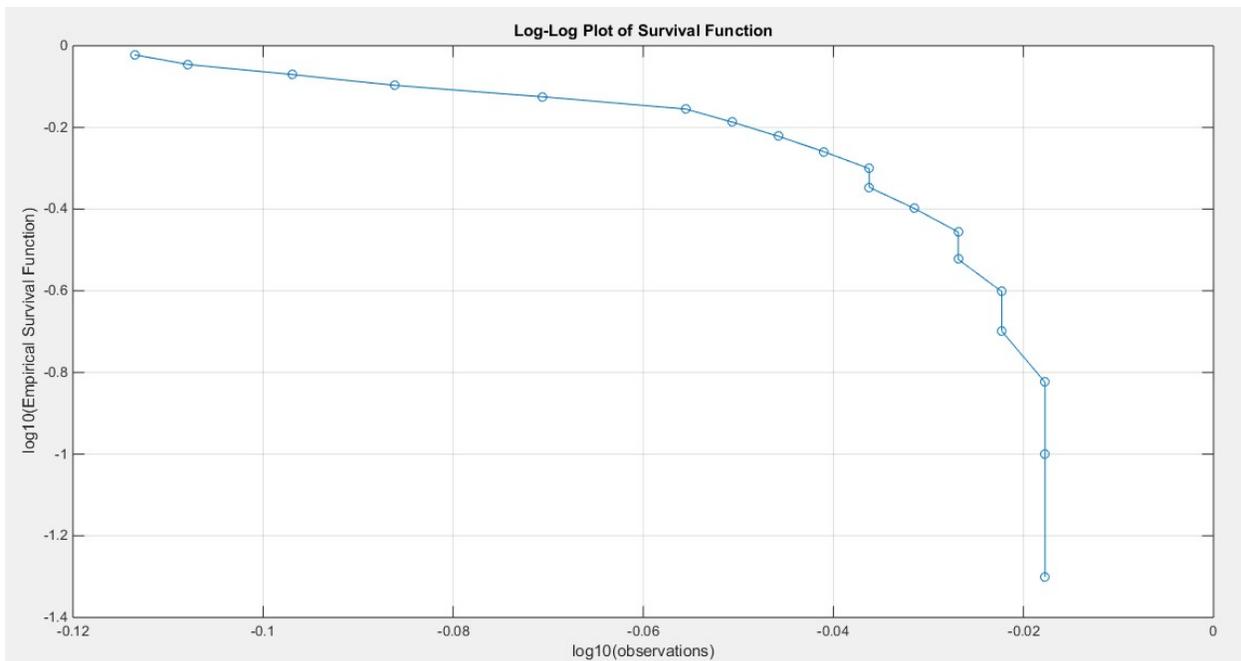

Fig 20 shows the log-log plot of the log observations against the log empirical survival function with concave graph denoting fast decay and hence light tail

The P-values for the estimators of alpha and beta parameters of the Beta distribution and Kumaraswamy distributions are significant $p < 0.001$. P-values for the estimators of alpha of the MBUR distribution is significant $p < 0.001$, for the estimators of theta of the Topp-Leone distribution is significant $p < 0.001$, and for the estimators of theta of the Unit Lindley distribution is significant $p < 0.001$.

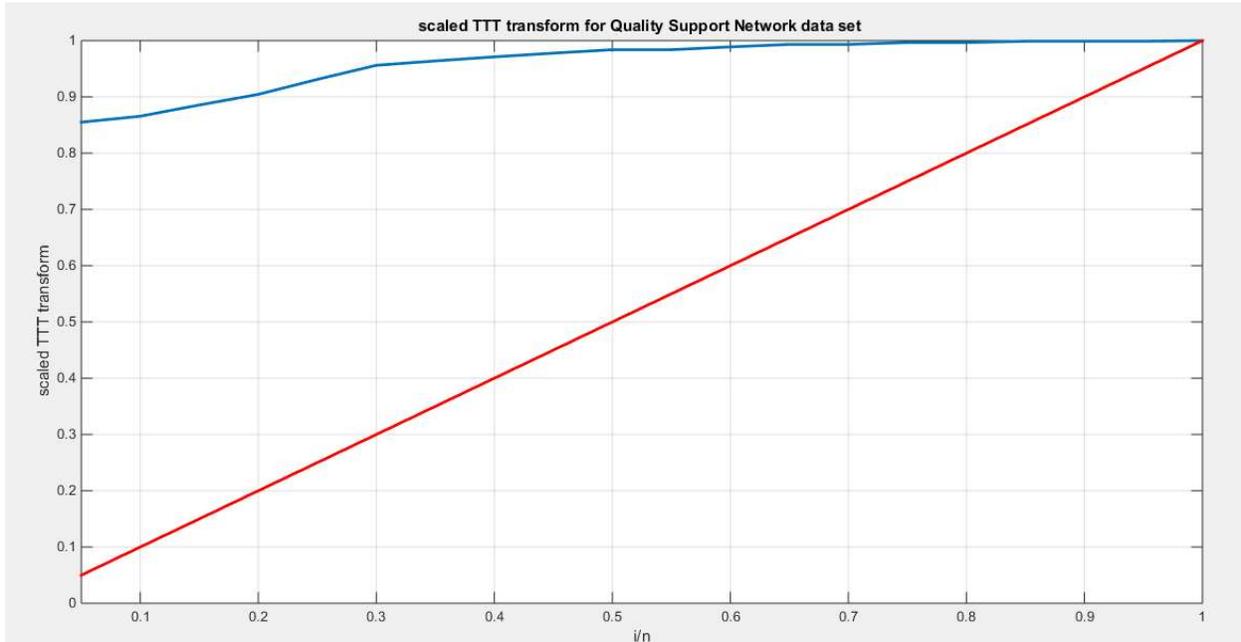

Fig. 21 shows the scaled TTT plot for the Quality support network data set with a concave shape supporting the increased hazard rate as reflected in the shape of the hazard function.

The KS test effectively measures the maximum distance between the empirical cumulative distribution function (eCDF) and the theoretical cumulative distribution function (CDF). Its straightforward nature and broad applicability make it a valuable tool, as it imposes no assumptions on the distribution parameters. However, it is less sensitive to deviations in the tails of the distribution, as it primarily focuses on the center. In contrast, the AD test excels in detecting deviations in the tails and is particularly suited for distributions with extreme values. Despite this advantage, the necessity for calculating critical values for newly emerging distributions can hinder its application. The CVM test takes a different approach by measuring the overall distance between the eCDF and the theoretical CDF, treating all parts of the distribution equally. This means it effectively balances sensitivity to deviations across the tail and the center, making it a compelling choice in many scenarios. Given the complexity of skewed data, it is crucial to utilize more than one test. Each test highlights specific characteristics of the data, offering a more comprehensive understanding of the fitting distribution. When combined with visual aids, such as QQ plots and PP plots, this methodology significantly enhances the analysis, driving more informed decisions. Therefore, when assessing the goodness of fit of a distribution, it is important to consider the results of the three tests mentioned above, along with the information obtained from the QQ plot and PP plot. Key aspects to observe include how closely the points align with the diagonal, the degree of deviation from the diagonal, and the percentage of observations that deviate from it.

The MBUR model fits the second dataset, as evidenced by its failure to reject the Kolmogorov-Smirnov (KS) test. The QQ plot shows almost perfect alignment with the diagonal, with only slight deviations at the lower end of the distribution. Since the KS test is less sensitive to deviations in the tail of the distribution, the author also conducted the Anderson-Darling (AD) test and the Crámer-von Mises (CVM) test using Monte Carlo simulations.

The observed value of the AD test statistic was 0.3184, while the critical values obtained from the simulations were 2.4433 (95th quantile) and 3.0146 (97.5th quantile). Since the observed AD value is less than the critical values from the simulation, the author fails to reject the null hypothesis, indicating that MBUR could be a generating process for the data. The approximate p-value for this test was 0.929, which is greater than 0.025, further confirming that MBUR fits the second dataset. Additionally, the CVM test from the observed data revealed a value of 0.0407. The CVM test conducted using Monte Carlo simulations yielded critical values of 0.4578 (95th quantile) and 0.5781 (97.5th quantile). Again, since the observed CVM value is less than the critical values, the author fails to reject the null hypothesis that the data was generated by MBUR. The approximate p-value for this test is 0.936, which is also greater than 0.025, supporting the conclusion that MBUR fits the second dataset. Overall, combining various goodness-of-fit statistics with visualizations enhances the results of the analysis. This is shown in the Table13 and Figures 22-25

Table (13): Estimators and validation indices for the Second data set

| | Beta | | Kumaraswamy | | MBUR | Topp-Leone | Unit-Lindley |
|---|---|---|---|---|---|---|---|
| theta | $\alpha = 21.7353$ | | $\alpha = 16.5447$ | | 0.3591 | 71.2975 | 0.1334 |
| | $\beta = 2.4061$ | | $\beta = 2.772$ | | | | |
| Var | 86.461 | 9.0379 | 15.7459 | 3.2005 | 0.000837 | 254.1667 | 0.00045 |
| | 9.0379 | 1.0646 | 3.2005 | 1.0347 | | | |
| SE | 2.079 | | 0.8873 | | 0.0063 | 3.565 | 0.0047 |
| | 0.231 | | 0.2275 | | | | |
| AIC | -56.5056 | | -56.7274 | | -58.079 | -56.6796 | -57.3746 |
| CAIC | -55.7997 | | -56.0215 | | -57.8567 | -56.4574 | -57.1523 |
| BIC | -54.5141 | | -54.7359 | | -57.0832 | -55.6839 | -56.3788 |
| HQIC | -56.1168 | | -56.3386 | | -57.8846 | -56.4852 | -57.1802 |
| LL | 30.2528 | | 30.3637 | | 30.0395 | 29.3398 | 29.6873 |
| K-S | 0.0974 | | 0.0995 | | 0.1309 | 0.1327 | 0.1057 |
| H$_0$ | Fail to reject | | Fail to reject | | Fail to reject | Fail to reject | Reject to reject |
| P-value | 0.9416 | | 0.9513 | | 0.8399 | 0.4627 | 0.954 |
| AD | 0.3828 | | 0.3527 | | 0.3184 | 0.9751 | 0.2749 |
| CVM | 0.0566 | | 0.0498 | | 0.0407 | 0.1719 | 0.0261 |

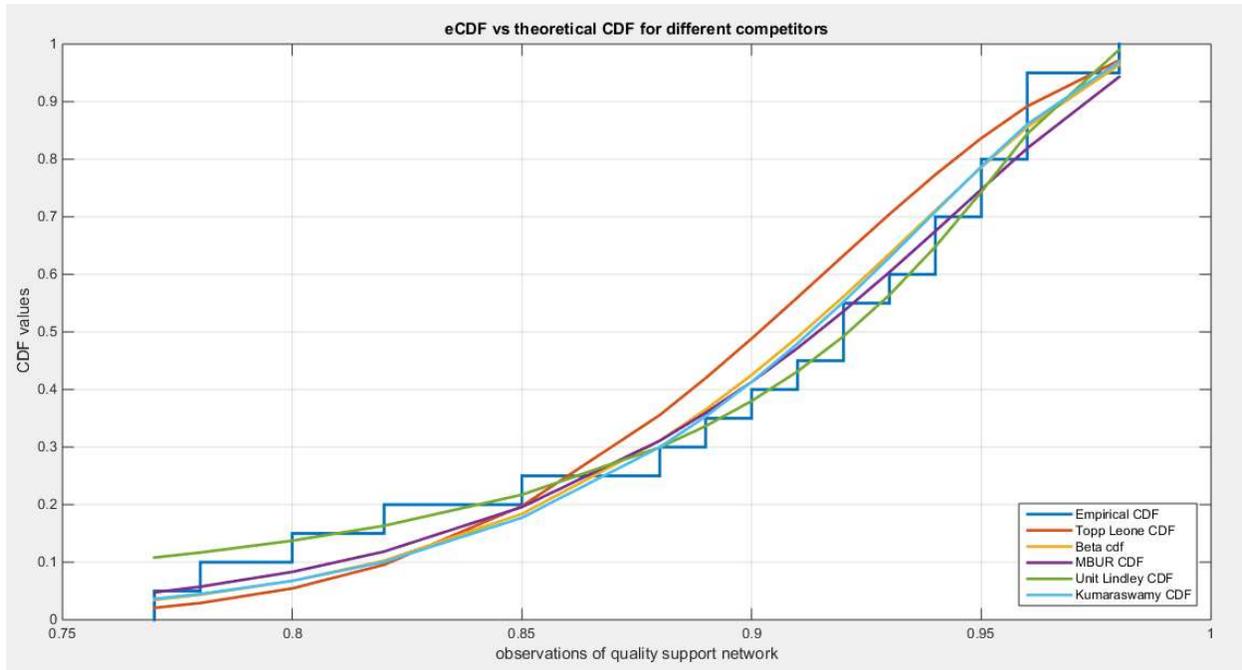

Fig. 22 shows the eCDF vs. theoretical CDF of the 5 distributions for the 2nd data set (Quality of support network).

According to the AIC, AIC corrected, BIC, and Hannan–Quinn Information Criterion (HQIC), the MBUR distribution is the best fit for the data, followed by the Unit Lindley, Topp-Leone, Kumaraswamy, and finally the Beta distribution. This conclusion is based on the MBUR having the lowest values of these indices (or the largest negative values). However, it is worth noting that the MBUR has the second lowest value for the Anderson-Darling (AD) test, Cramer-von Mises (CVM) test, and Kolmogorov-Smirnov (KS) test, coming in just after the Unit Lindley. Figure 22 illustrates that the theoretical cumulative distribution functions (CDFs) for the various distributions closely follow the empirical CDF. Meanwhile, Figure 23 presents the fitted probability density functions (PDFs). An important observation from this analysis is that the metric values for the MBUR distribution are comparable to those of the Topp-Leone and Unit Lindley distributions, indicating that the new MBUR distribution has performed well in fitting the data. Figure 24 shows the quantile-quantile (QQ) plot for the fitted MBUR distribution, which exhibits nearly perfect alignment along the diagonal, with only slight deviations at the lower tail. The log-likelihood function is maximized at an alpha level of 0.3519. Finally, Figure 25 provides the QQ plot and probability-probability (PP) plot for the other distributions, which also demonstrate near-perfect alignment along the diagonal.

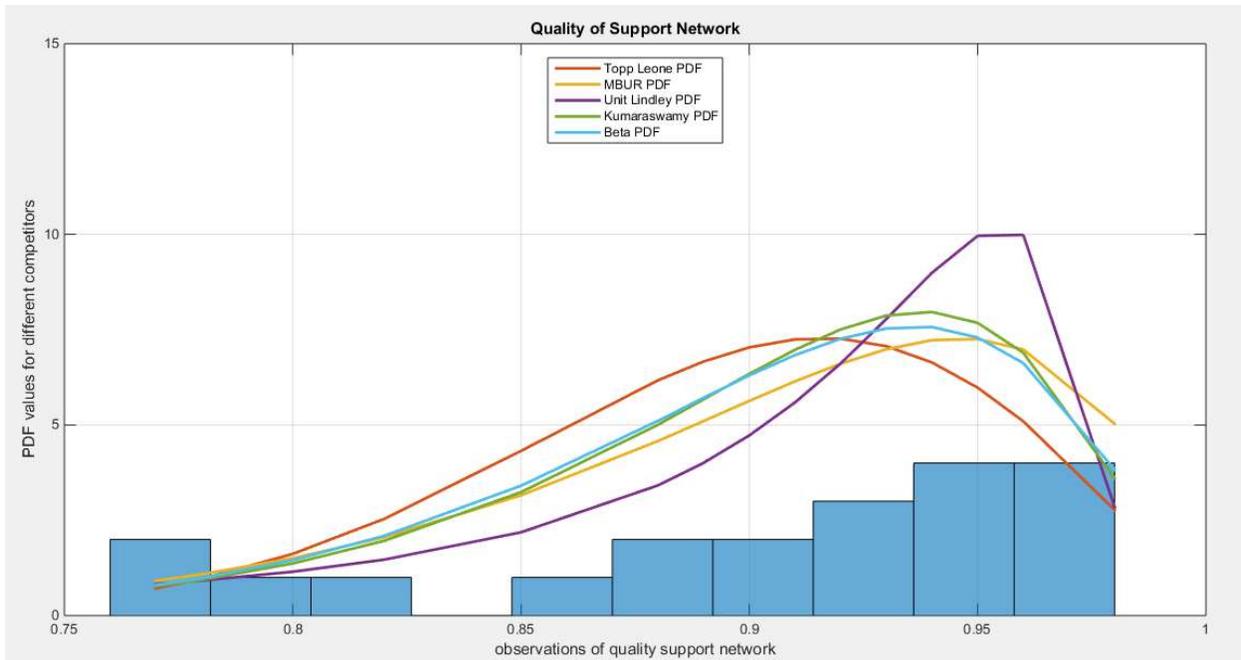

Fig. 23 shows the fitted PDFs for the different competitors

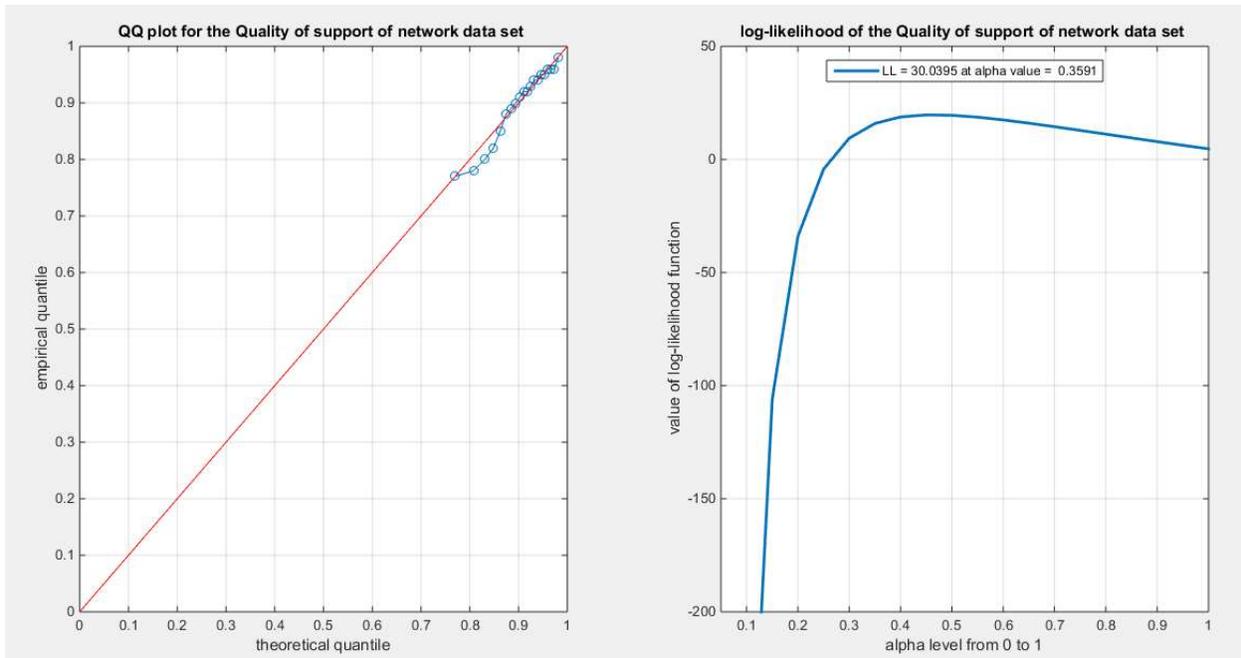

Fig. 24 shows the QQ plot for quality of support of network data set, on the left hand side of the graph and the log-likelihood on the right after fitting BMUR distribution.

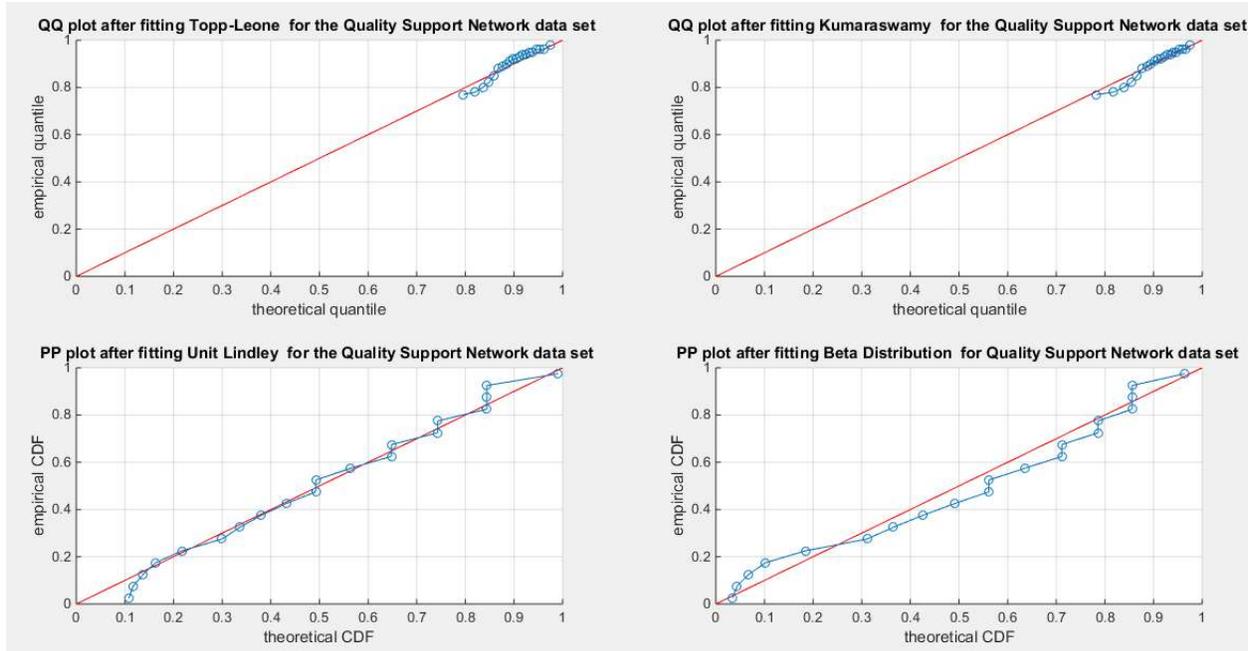

Fig. 25 shows the QQ plot after fitting both Topp-Leone and Kumaraswamy distributions. The PP plot after fitting both Beta and Unit Lindley distributions are also seen.

**4.3. Analysis of the third data set.** See supplementary materials (section 3)

**4.4. Analysis of the fourth data set**. See supplementary materials (section 3)

**4.5. Analysis of the fifth data set**. See Table 14 & 15. See Figures 26-33

Table (14): Descriptive statistics of the fifth data set

| min | mean | std | skewness | kurtosis | $25_{perc}$ | $50_{perc}$ | $75_{perc}$ | max |
|---|---|---|---|---|---|---|---|---|
| 0.0062 | 0.1578 | 0.1931 | 1.4614 | 3.9988 | 0.0292 | 0.0614 | 0.21 | 0.656 |

The data shows a right skewness and a positive excess kurtosis (leptokurtic shape), which is supported by the histogram and box plot in Figure 26. When plotting the empirical survival function, the author observes a slower decay, indicating heavier tails, as illustrated in Figure 27. This observation is further supported by the log-log plot in Figure 28. When the author plots the logarithm of the observations against the logarithm of the empirical survival function, the resulting straight line indicates a slower decay, which is characteristic of heavier tails. In quantile analysis, the author compared the empirical 99th quantile (0.6560) with the 99th theoretical quantile of a standard uniform distribution (0.6494). This comparison reveals that the empirical value is greater than the theoretical one (0.6560 > 0.6494), suggesting a heavier tail. Additionally, when fitting the MBUR model to the data, the author obtained an estimated alpha value of 1.7886, which is greater than 1. This is consistent with the empirical survival function shown in Figure 27, which is similar to the survival function depicted in Figure 6, where alpha is also greater than 1.

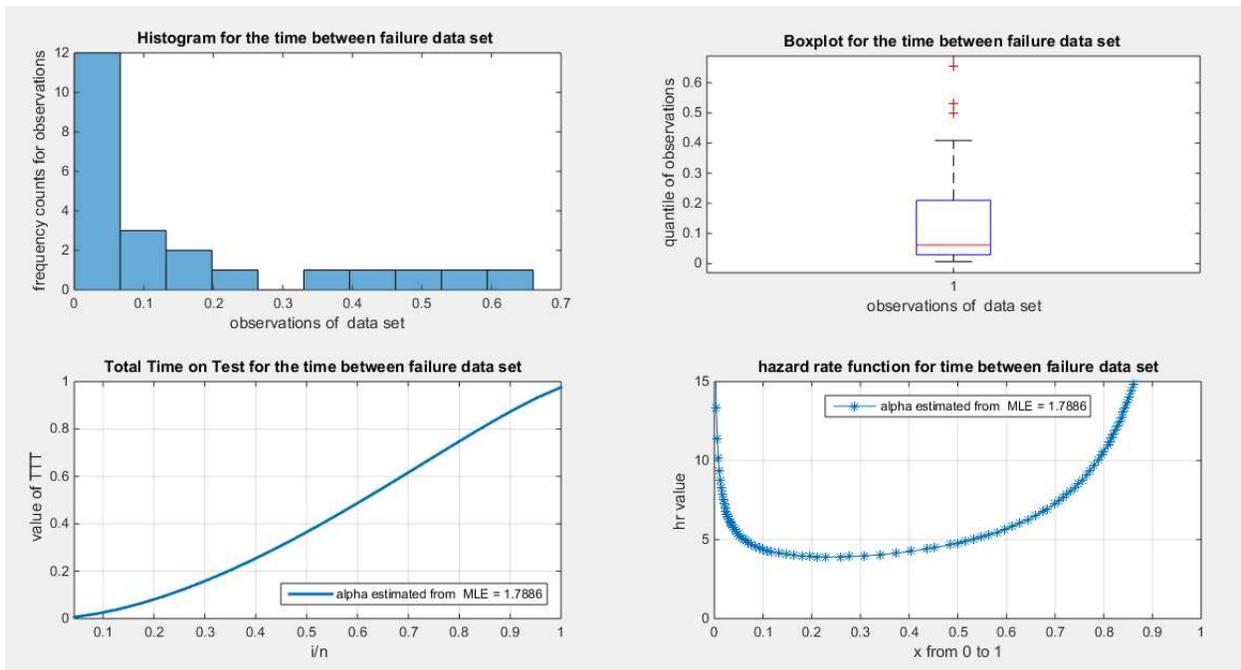

Fig. 26 shows the histogram with right skewness and associated boxplot with 3 outliers or extreme values on the upper tail of the distribution. The TTT plot shows convex shape which supports initial decreased failure rate that is obvious in the shape of the hazard function on the right lower graph.

The second approach illustrated in Figure 26 for calculating and graphing the TTT plot does not exhibit the typical convexity followed by concavity that characterizes the bathtub shape seen in the hazard rate function. In contrast, Figure 29, which employs the first approach for calculation and graphing, more accurately represents this relationship.

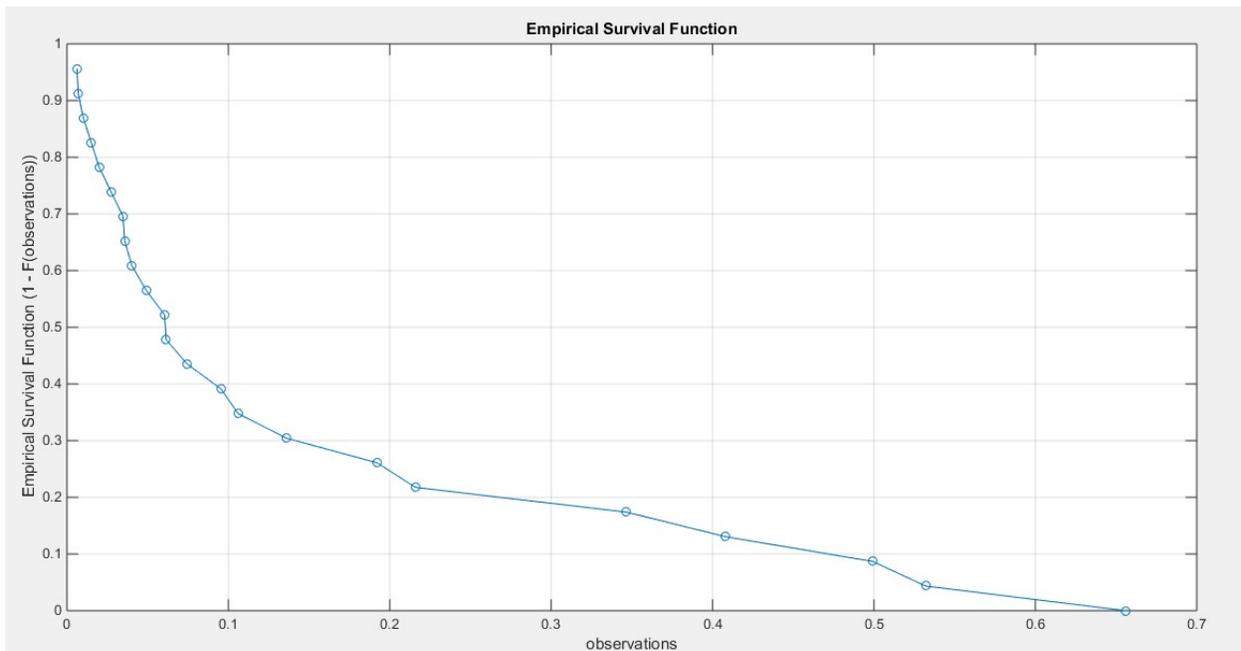

Fig. 27 shows the empirical survival function convex graph curved downward like the one in Fig. 6 where the alpha is >1.

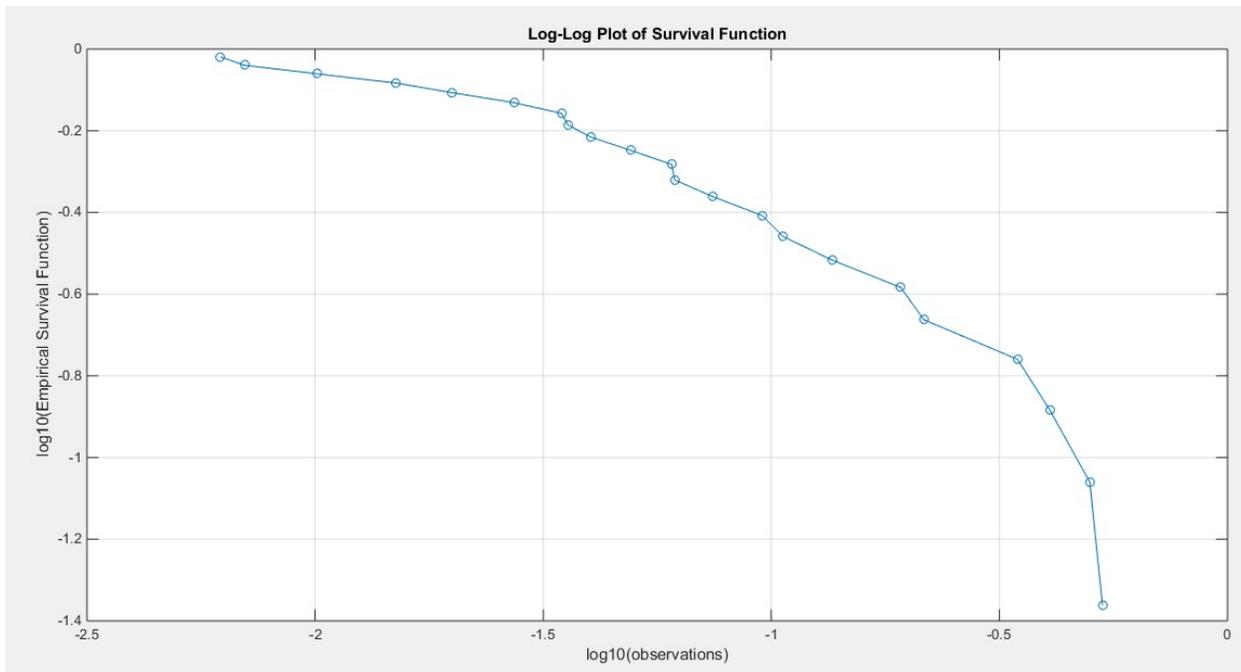

Fig 28 shows the log-log plot of the log observations against the log empirical survival function with straight graph at higher values of observations denoting slow decay and hence heavier tail

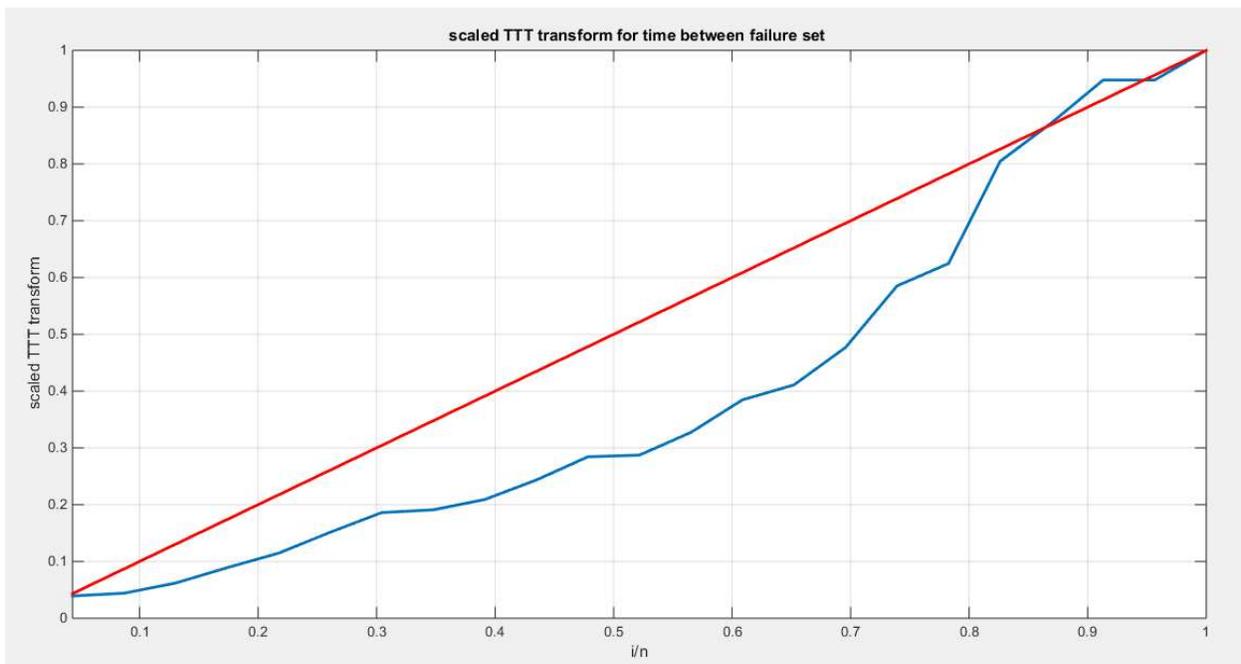

Fig.29 shows the scaled TTT plot for the time between failure dataset with a convex shape followed by a concave shape more obvious on the upper part of the graph, supporting the decreased hazard rate followed by the increased hazard rate as reflected by the bathtub shape of the hazard function.

The MBUR distribution is the best fit for the time between failures data among the five distributions evaluated, followed by Kumaraswamy, Beta, and Topp-Leone. The Unit Lindley distribution, however, did not fit the data well. The MBUR has the most significant negative values for AIC, AIC corrected, BIC, and HQIC. Despite this, it is the second distribution to have the smallest values for the AD test and the CVM test. Figure 30 illustrates that the eCDF closely follows the theoretical CDF for the fitted distributions, particularly at the tails, though there is a slight deviation at the center. Figure 31 displays the PDFs for the various competing distributions. In Figure 32, the QQ plot demonstrates good alignment with the diagonal after fitting the MBUR, with the maximum likelihood estimate achieved at an alpha level of 1.7886. Figure 33 shows a generally close alignment with the diagonal line for the other fitted distributions, especially at the tails, with slight deviations at the center, indicating that these distributions capture the characteristics of the data well. The PP plot further illustrates that the Unit Lindley distribution does not align closely with the diagonal. The P-values for the estimators of alpha and beta parameters of the Beta distribution and Kumaraswamy distributions are significant $p < 0.001$. P-value for the estimators of alpha of the MBUR distribution is significant $p < 0.001$. P-value for the estimators of theta of the Topp-Leone distribution is significant $p < 0.001$.

Table (15): Estimators and validation indices for the Fifth data set

| | Beta | | Kumaraswamy | | MBUR | Topp-Leone | Unit-Lindley |
|---|---|---|---|---|---|---|---|
| theta | $\alpha = 0.6307$ | | $\alpha = 0.6766$ | | 1.7886 | 0.4891 | 4.1495 |
| | $\beta = 3.2318$ | | $\beta = 2.936$ | | | | |
| Var | 0.071 | 0.2801 | 0.0198 | 0.1033 | 0.018 | 0.0104 | 0.5543 |
| | 0.2801 | 1.647 | 0.1033 | 0.9135 | | | |
| SE | 0.0555 | | 0.0293 | | 0.0279 | 0.0213 | 0.1552 |
| | 0.2676 | | 0.1993 | | | | |
| AIC | -36.0571 | | -36.6592 | | -37.862 | -35.5653 | -27.007 |
| CAIC | -35.4571 | | -36.0592 | | -37.6712 | -35.3749 | -26.8165 |
| BIC | -33.7861 | | -34.3882 | | -36.7262 | -34.4298 | -25.8715 |
| HQIC | -35.4859 | | -36.0881 | | -37.5764 | -35.2798 | -26.7214 |
| LL | 20.0285 | | 20.3296 | | 19.9310 | 18.7827 | 14.5035 |
| K-S | 0.1541 | | 0.1393 | | 0.1584 | 0.1962 | 0.3274 |
| H$_0$ | Fail to reject | | Fail to reject | | Fail to reject | Fail to Reject | Reject |
| P-value | 0.5918 | | 0.7123 | | 0.5575 | 0.2982 | 0.0107 |
| AD | 0.6886 | | 0.5755 | | 0.6703 | 1.1022 | 4.7907 |
| CVM | 0.1264 | | 0.0989 | | 0.1253 | 0.2149 | 0.8115 |

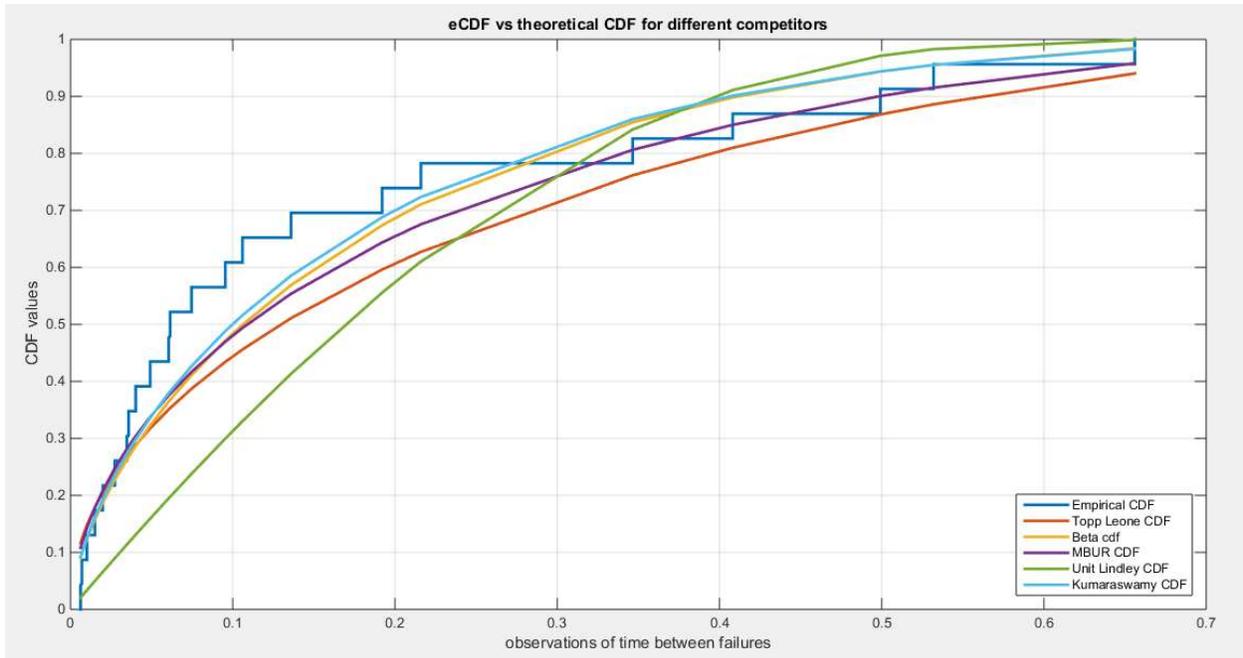

Fig. 30 shows the eCDF vs. theoretical CDF of the 5 distributions for the 5th data set (Time between failures of Secondary Reactor Pumps).

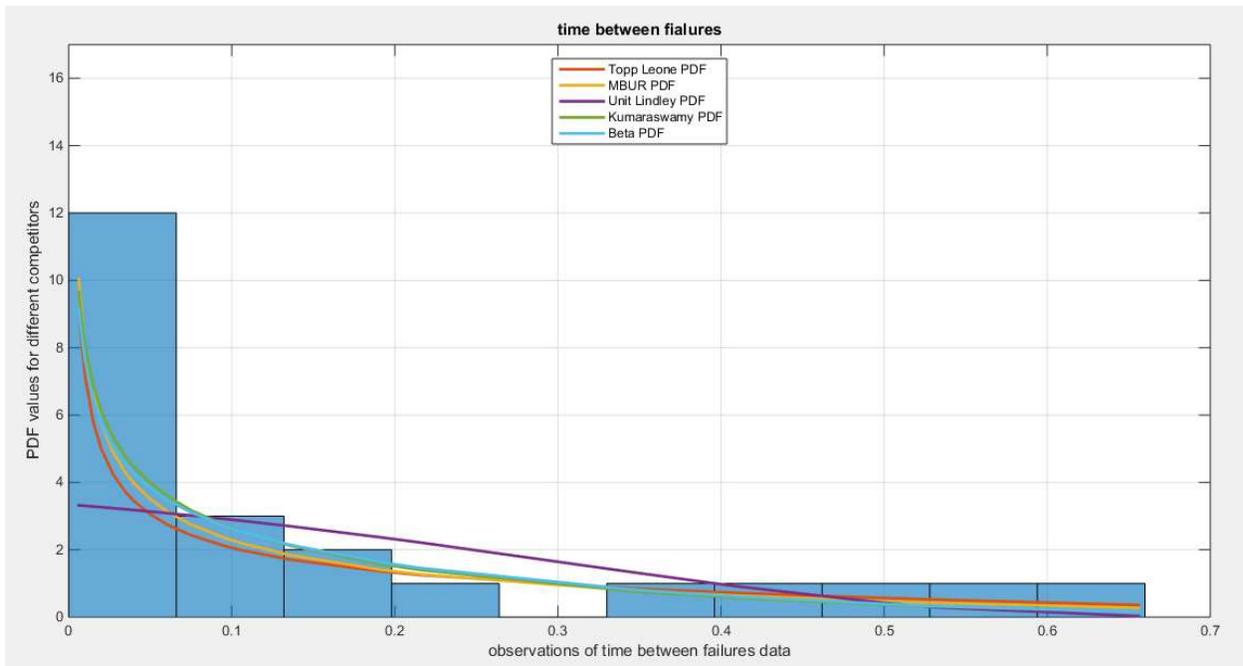

Fig. 31 shows the fitted PDFs for the different competitors.

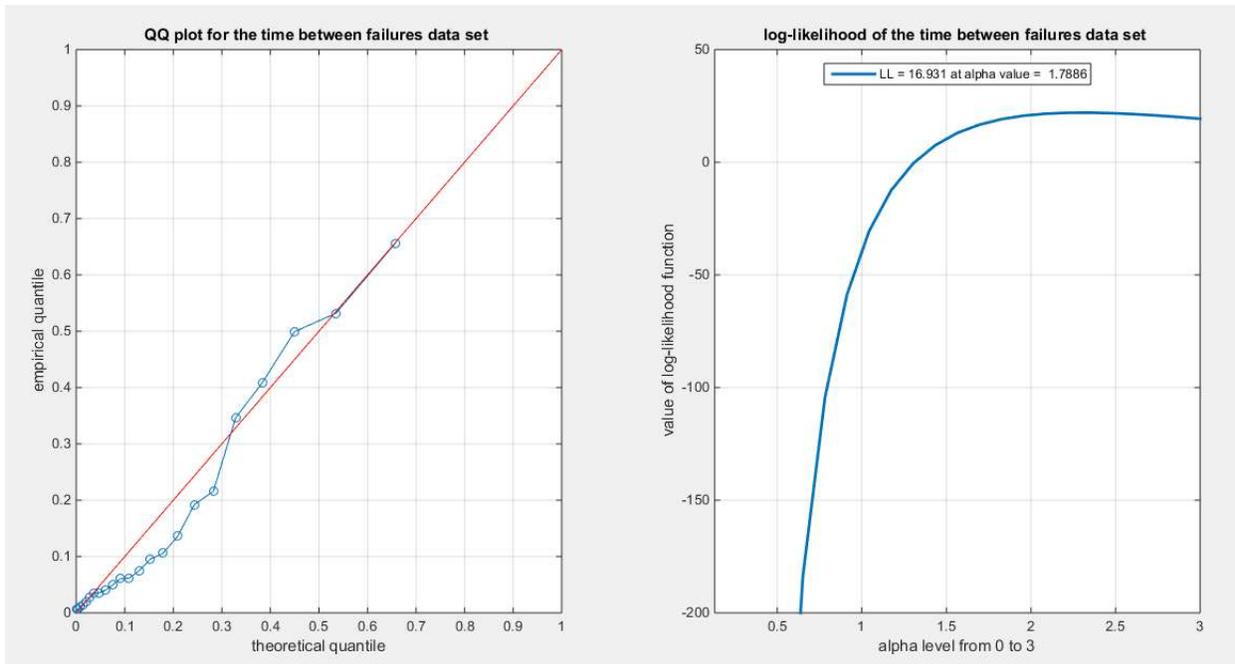

Fig. 32 shows the QQ plot for time between failures data set, on the left hand side of the graph and the log-likelihood on the right, after fitting the MBUR distribution.

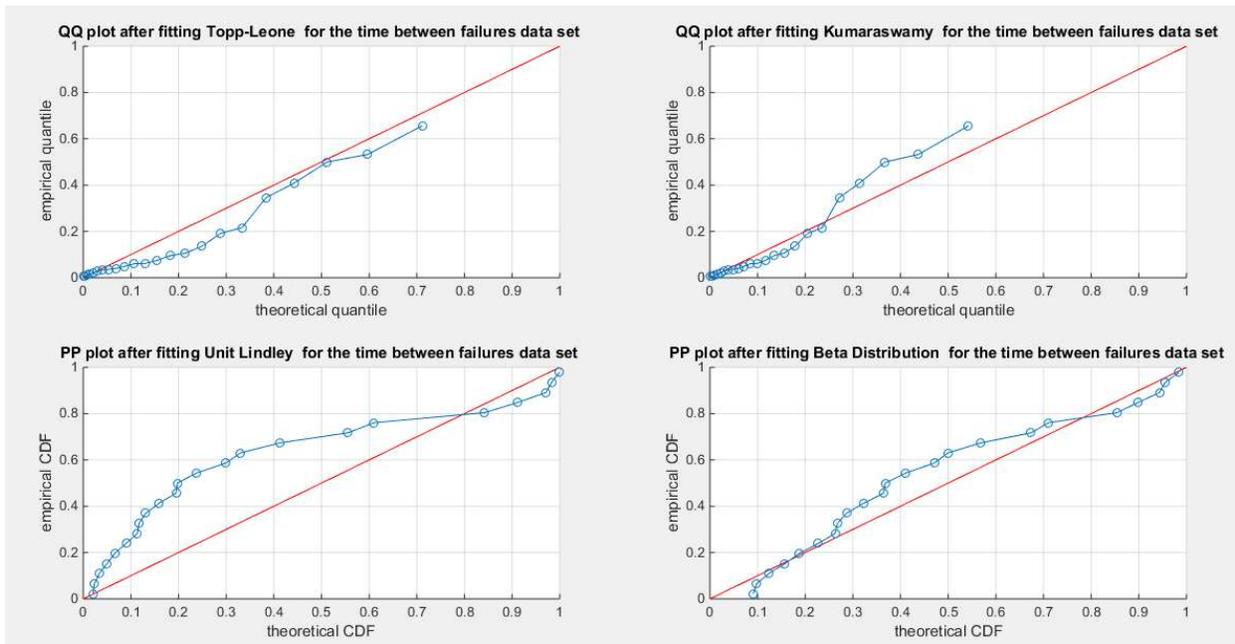

Fig. 33 shows the QQ plot after fitting both Topp-Leone and Kumaraswamy distributions. The PP plot after fitting both Beta and Unit Lindley distribution are also seen.

The MBUR model unequivocally fits the fifth dataset, as demonstrated by the results of the Kolmogorov-Smirnov (KS) test, which successfully failed to reject the null hypothesis that the data adheres to the MBUR distribution. This conclusion is visually substantiated by the QQ plot, which shows a strong alignment along the diagonal, indicating a robust correspondence between the theoretical distribution and the empirical data, though minor deviations can be observed at the lower end. To address the potential limitations of the KS test—particularly its insensitivity to deviations in the distribution tails—the author conducted additional analyses using the AD test and the CVM test, both of which utilized Monte Carlo simulations for enhanced accuracy. The AD test produced a statistic of 0.6703. The critical values from the simulations for this test were clear: 2.6428 for the 95th quantile and 3.3935 for the 97.5th quantile. The value for 2.5th quantile is 0.2309. The observed AD statistic is significantly less than these critical values, compellingly leading the author to fail to reject the null hypothesis. This strongly indicates that the MBUR model can indeed act as a generating process for the observed data. The p-value corresponding to this test was a definitive 0.594, exceeding the conventional significance threshold of 0.025, thus firmly establishing MBUR as an appropriate fit for the fifth dataset. Likewise, the CVM test reinforced this conclusion with an observed statistic of 0.1253. Critical values derived from Monte Carlo simulations were found to be 0.4858 (95th quantile) and 0.6099 (97.5th quantile). The value for the 2.5th quantile is 0.03. As with the AD test, the observed CVM statistic fell below these critical thresholds, leading to a resolute failure to reject the null hypothesis that the data originated from the MBUR model. The approximate p-value for this test, calculated at 0.485, also exceeds the 0.025 significance level, decisively affirming that MBUR is an excellent fit for the dataset in question.

Integrating various goodness-of-fit statistics with effective visualizations significantly enhances the analysis results, leading to clearer insights and more informed decisions. This is shown in the Table15 and Figures 30-33

When using AIC and BIC to compare distributions that fit specific data, both metrics aim to balance maximizing model fit, reflected in the highest negative values of log-likelihood, with minimizing model complexity, which is represented by the number of parameters in the model. This balance helps avoid overfitting, particularly in cases where a model may be too complex and have too many parameters. Such complex models can capture not only the true underlying structure of the data but also random noise, leading to poor generalization when new data is introduced.

The log-likelihood (LL) measures how well a model fits the data; higher LL values indicate a better fit. However, simply adding more parameters (k) tends to increase LL, even if those additional parameters are not meaningful. Therefore, AIC and BIC serve as trade-offs between model fit and complexity, addressing the challenge of balancing overfitting (too complex a model) and underfitting (too simple a model). To mitigate overfitting, AIC and BIC introduce penalties for complexity that are proportional to the number of parameters (k). The AIC penalty is a linear penalty of 2k, whereas the BIC penalty is k * ln(n), which increases with sample size.

AIC and BIC are used to select the best-fitting distribution among candidates. They depend on LL, meaning that the model which better captures the structure of the data will display more negative values for AIC and BIC. In cases where the data exhibits complex features such as skewness and heavy tails, a model with more parameters may yield more negative AIC and BIC values. Conversely, if the data is simpler (e.g., symmetric with a small sample size), more straightforward models are often preferred.

The more negative the values of AIC and BIC, the better the model. By themselves, these values are meaningless, but they are useful for comparing models. A difference greater than 10 between two models suggests that the model with the more negative value is significantly better. AIC is typically used when the goal is prediction and the dataset is small, while BIC is preferred when identifying the true model is critical and the dataset is large, due to differing penalty structures.

Regarding the datasets discussed in this paper, they exhibit skewness and kurtosis, indicating their complexity. The sizes of these datasets range between 20 and 36, making them small to moderate in size. The new MBUR distribution can effectively fit all the data using just one parameter, resulting in a relatively small penalty from AIC and BIC. This represents an advantage of the MBUR distribution over other distributions, such as the beta and Kumaraswamy distributions, which require multiple parameters.

## Section 5: Conclusion:

The discovery of new distributions to fit data across various fields is crucial for scientists to better understand emerging phenomena in our rapidly changing world and environment. The new MBUR distribution is characterized by a single parameter that needs to be estimated. It features a well-defined (CDF) and a well-defined quantile function. This distribution can accommodate a wide variety of highly skewed data, whether exhibiting right or left skewness.

The shape of the hazard rate function is highly dependent on this parameter, allowing it to be increasing, decreasing, or resembling a bathtub shape. The MBUR distribution is comparable to other distributions, such as the beta distribution and the Kumaraswamy distribution, which also have hazard functions that exhibit similar behaviors, though these competitors have two parameters. As a result, the estimation of the parameter for the MBUR distribution is less cumbersome.

Additionally, it can be tractably estimated using numerical methods. Moreover, due to the closed form of the quantile function, the MBUR distribution is suitable for use in quantile regression analysis of proportions and ratios. The limitation of the distribution is that it is unable to accommodate multimodal-shaped data or multi-antimodal data shapes.

## Future work:

The quality of data fitting directly affects the analysis outcomes in various fields, including regression, survival data analysis, reliability analysis, and time series analysis. This new distribution can be applied in these areas. Additionally, Bayesian estimation of the parameters can also be explored. Generalization of the distribution can be accomplished by adding new parameters that may extend its ability to accommodate more data shapes.

**Declarations:**
**Ethics approval and consent to participate**
Not applicable.
**Consent for publication**
Not applicable
**Availability of data and material**
Not applicable. Data sharing not applicable to this article as no datasets were generated or analyzed during the current study.
**Competing interests**
The author declares no competing interests of any type.
**Funding**
No funding resource. No funding roles in the design of the study and collection, analysis, and interpretation of data and in writing the manuscript are declared
**Authors' contribution**
AI carried the conceptualization by formulating the goals, aims of the research article, formal analysis by applying the statistical, mathematical and computational techniques to synthesize and analyze the hypothetical data, carried the methodology by creating the model, software programming and implementation, supervision, writing, drafting, editing, preparation, and creation of the presenting work.
**Acknowledgement**
Not applicable

# Supplementary material (section 1)

### 2.3.Discussion and analysis of the above functions

### 2.3.1.Analysis of CDF behavior :

The CDF is a function of the parameter as shown in figure (10) below:

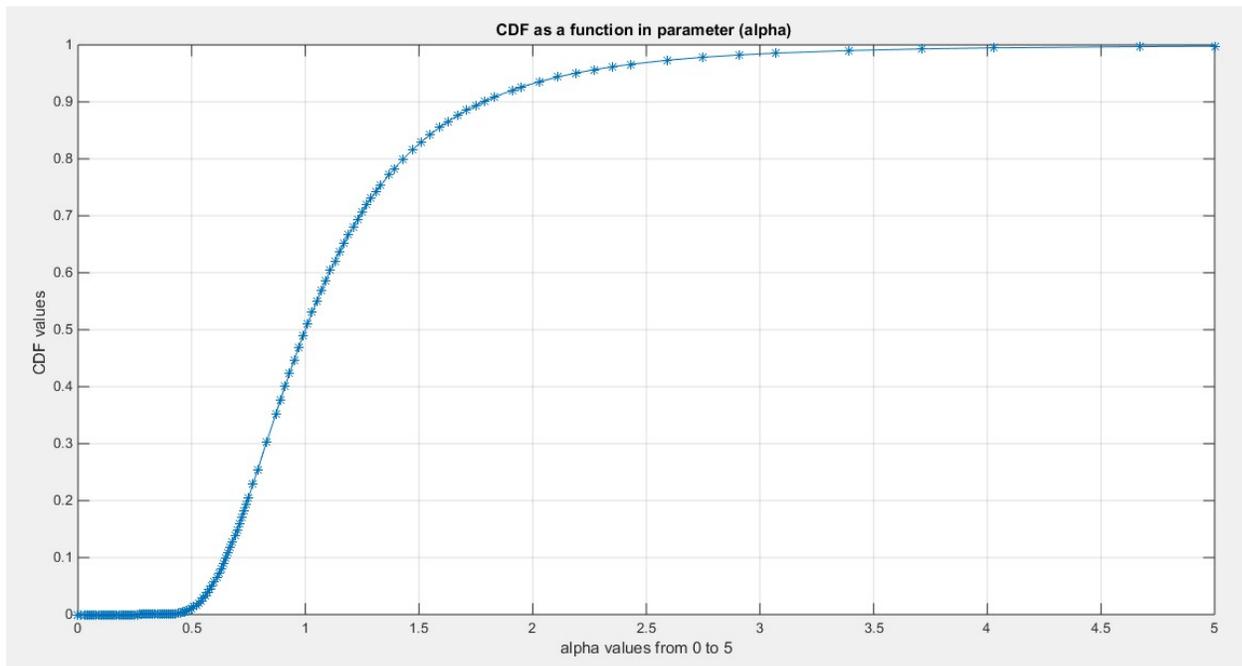

Fig.10 shows that the CDF function is an increasing function in parameter.

By setting the CDF fixed at specific y value

$F(y; \alpha) = F(y) \ for \ any, \alpha_2 \geq \alpha_1 \ then \ F(y; \alpha_2) \geq F(y; \alpha_1)$

The first derivative of the CDF with respect to $\alpha$ is

$$\frac{\partial F(y; \alpha)}{\partial \alpha} = \left(\frac{-12}{\alpha^3}\right) y^{2\alpha^{-2}} \ln y + \left(\frac{12}{\alpha^3}\right) y^{3\alpha^{-2}} \ln y = \left(\frac{12}{\alpha^3}\right) \ln y \ \{y^{3\alpha^{-2}} - y^{2\alpha^{-2}}\}$$

Figure (11) shows the graph of this function:

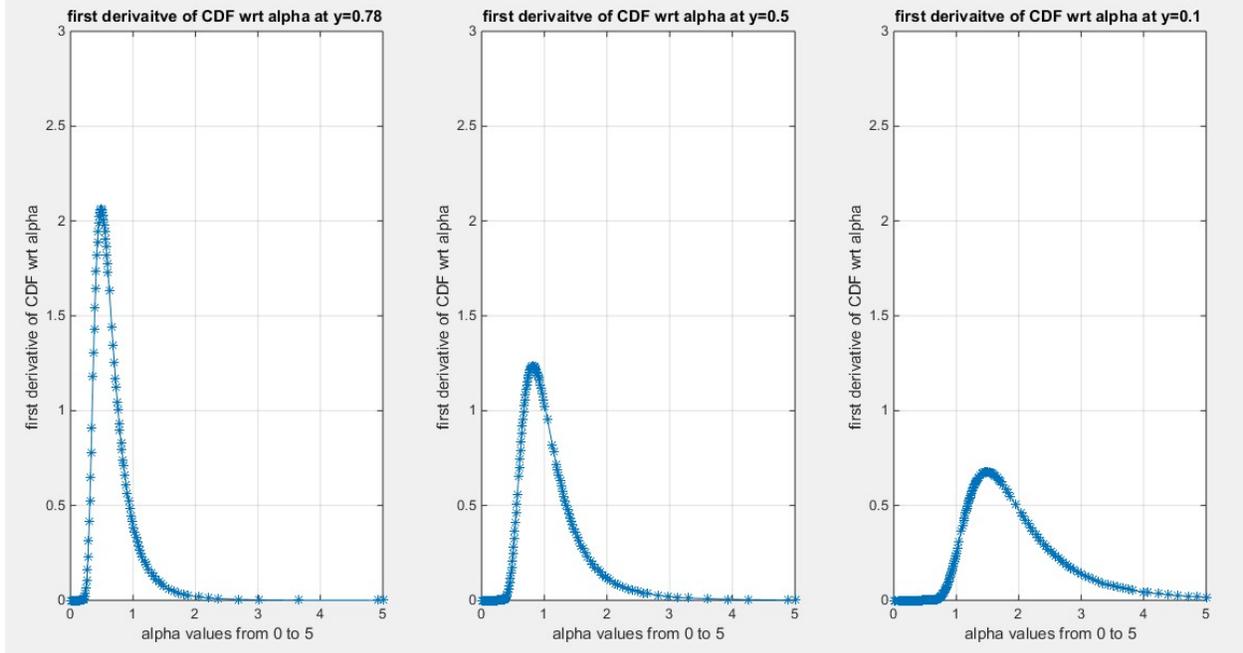

Fig. 11 shows that the function is always positive and concave for $\alpha \in (0,5)$ and $\forall\, y \in (0,1)$. This is applicable to all values of y and alphas.

The second derivative of the CDF with respect to alpha is:

$$\frac{\partial^2 F(x;\alpha)}{\partial \alpha^2} = \left(\frac{36}{\alpha^4}\right)\ln y \;\{y^{2\alpha^{-2}} - y^{3\alpha^{-2}}\} + \left(\frac{24}{\alpha^6}\right)(\ln y)^2\{2y^{2\alpha^{-2}} - 3y^{3\alpha^{-2}}\}$$

This function changes the sign for $\forall\, y \in (0,1)$ as shown in the figure (12):

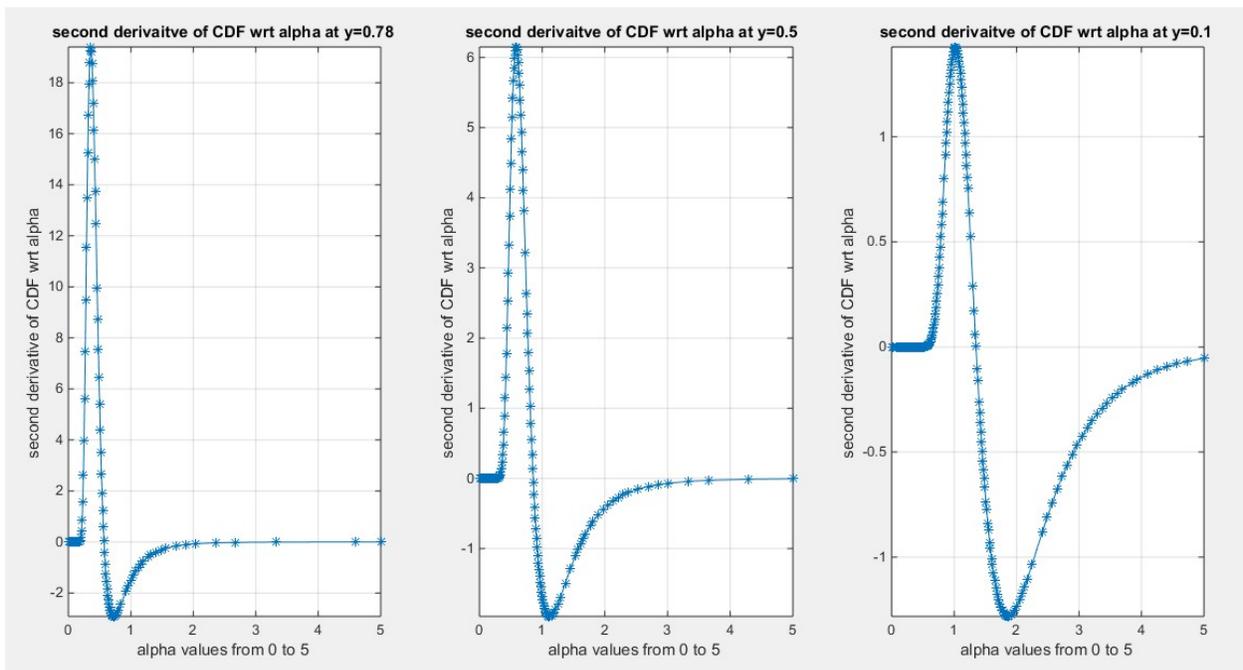

Fig.12 shows that for $\forall\, y \in (0,1)$ and $\alpha \in (0,5)$ the function changes the sing. This is applicable to all values of y and alphas

Figure (12) shows the same changes, which is true for all values of y and for all values of the parameter alpha larger than zero. As y increases, the level of alpha which the function changes its sign at is decreased. The larger the y, the lesser the value of alpha at which the function changes its sign at. The direction of the change is from positive to negative and then it never exceeds the zero, the function is always negative after it changes its sign. This is applicable for all values of y and alphas.

The first derivative of the log CDF with respect to alpha is

$$\frac{\partial \log F(y;\alpha)}{\partial \alpha} = \frac{F'(y;\alpha)}{F(y;\alpha)}$$

This function is shown in figure (13)

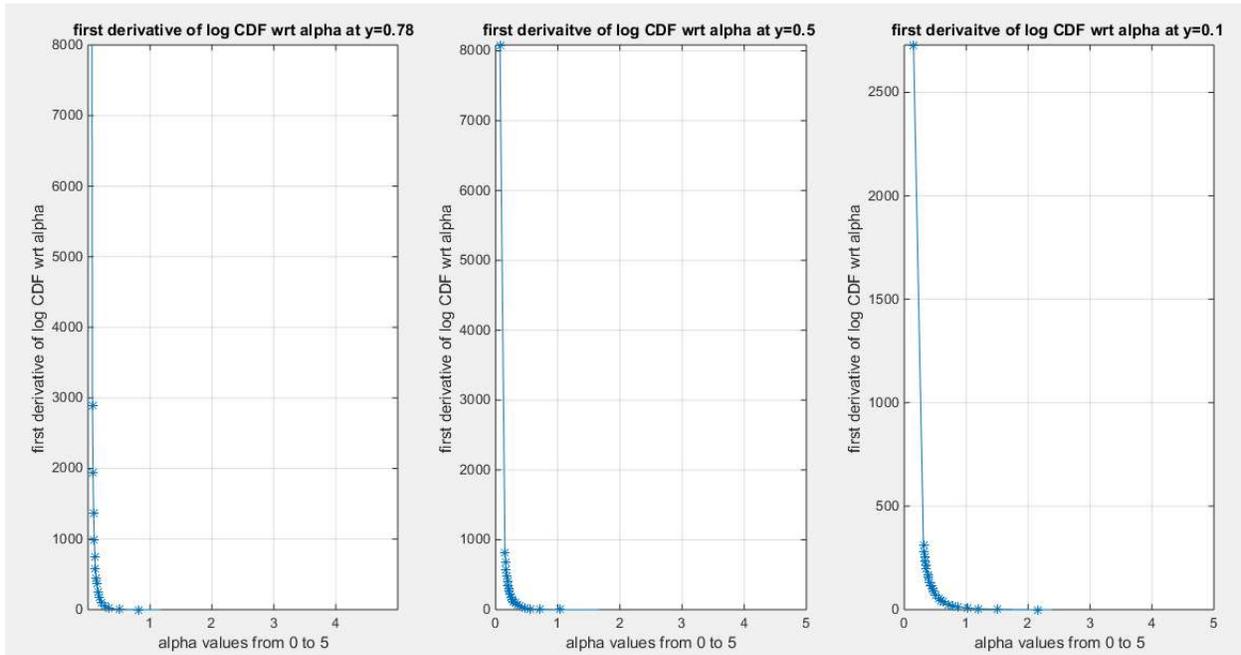

Fig.13 shows that the first derivative of log CDF with respect to alpha is a decreasing function and always positive for $\forall y \in (0,1)$ and all $\in (0,5)$. This is applicable to all values of y and alphas.

The second derivative of log CDF with respect to alpha is

$$\frac{\partial \log F(y;\alpha)}{\partial \alpha} = \frac{F''(y;\alpha)}{F(y;\alpha)} - \left(\frac{F'(y;\alpha)}{F(y;\alpha)}\right)^2$$

This function is shown in figure (14).

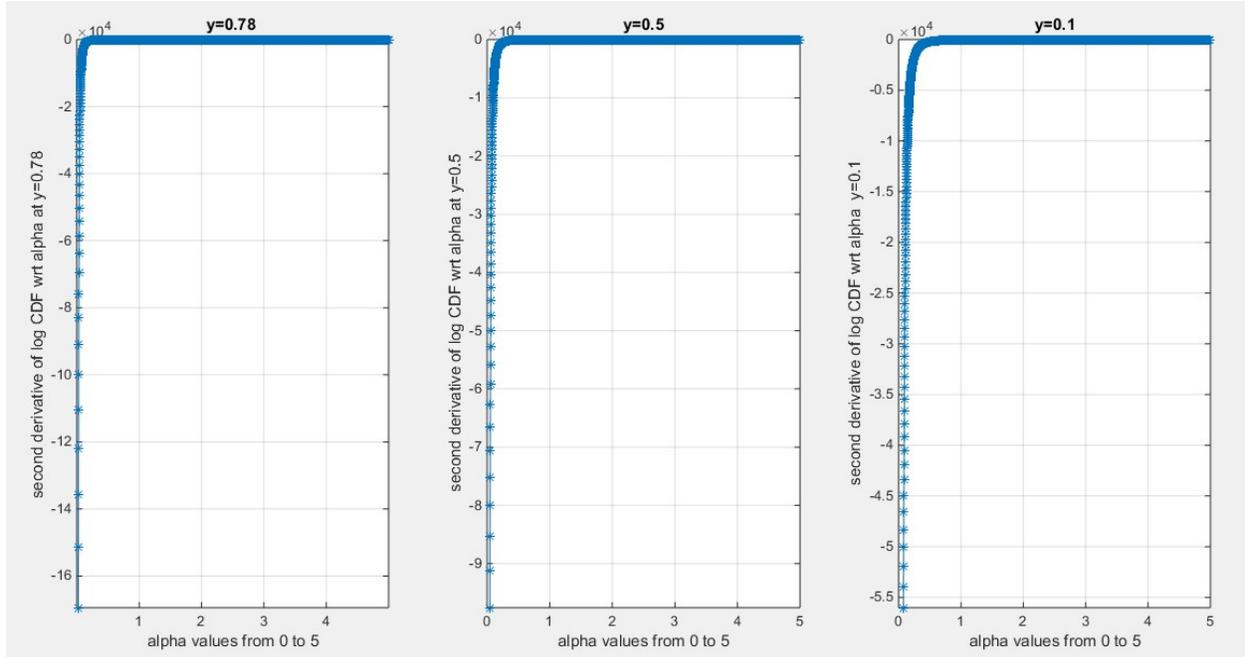

Fig.14 shows that the second derivative of log CDF with respect to alpha is an increasing function and always negative for $\forall y \in (0,1)$ and all $\alpha \in (0,5)$. This is applicable to all values of y and alphas.

Figure (15) shows the log CDF of MBUR defined as $\log F(y; \alpha) = \log\left\{3y^{\frac{2}{\alpha^2}} - 2y^{\frac{3}{\alpha^2}}\right\}$

The first derivative of the log CDF with respect to y variable is shown in figure (16) and is defined as

$$\frac{\partial \log F(y;\alpha)}{\partial y} = \frac{6\,\alpha^{-2}y^{-1}\left\{y^{2\alpha^{-2}} - y^{3\alpha^{-2}}\right\}}{3y^{2\alpha^{-2}} - 2y^{3\alpha^{-2}}}$$

The second derivative of the log CDF with respect to y variable is defined as:

$$\frac{\partial^2 \log F(y;\alpha)}{\partial y^2} =$$

$$\frac{6\,\alpha^{-2}(2\,\alpha^{-2} - 1)y^{2\alpha^{-2}-2} - 6\,\alpha^{-2}(3\,\alpha^{-2} - 1)y^{3\alpha^{-2}-2}}{3y^{2\alpha^{-2}} - 2y^{3\alpha^{-2}}} - \frac{\left(6\,\alpha^{-2}y^{2\alpha^{-2}-1} - 6\,\alpha^{-2}y^{3\alpha^{-2}-1}\right)^2}{\left(3y^{2\alpha^{-2}} - 2y^{3\alpha^{-2}}\right)^2}$$

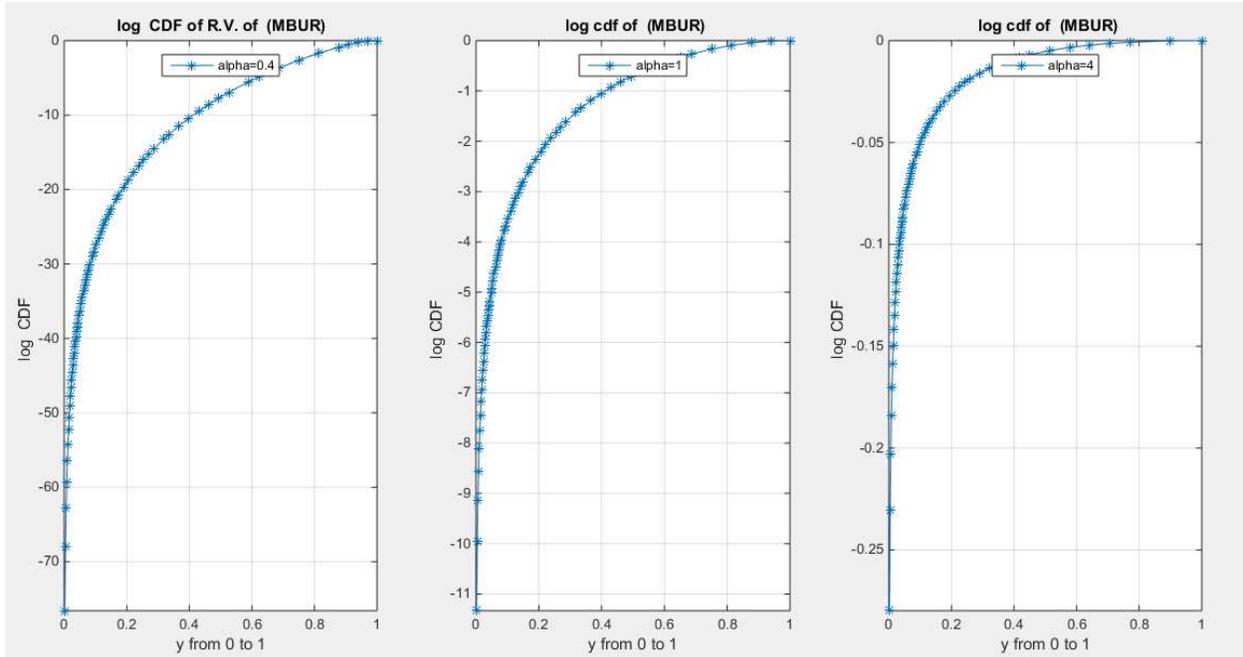

Fig.15 shows that the log CDF is an increasing function and always negative for $\forall\, y \in (0,1)$ and for $\alpha > 1$, $\alpha = 1$, $\alpha < 1$.

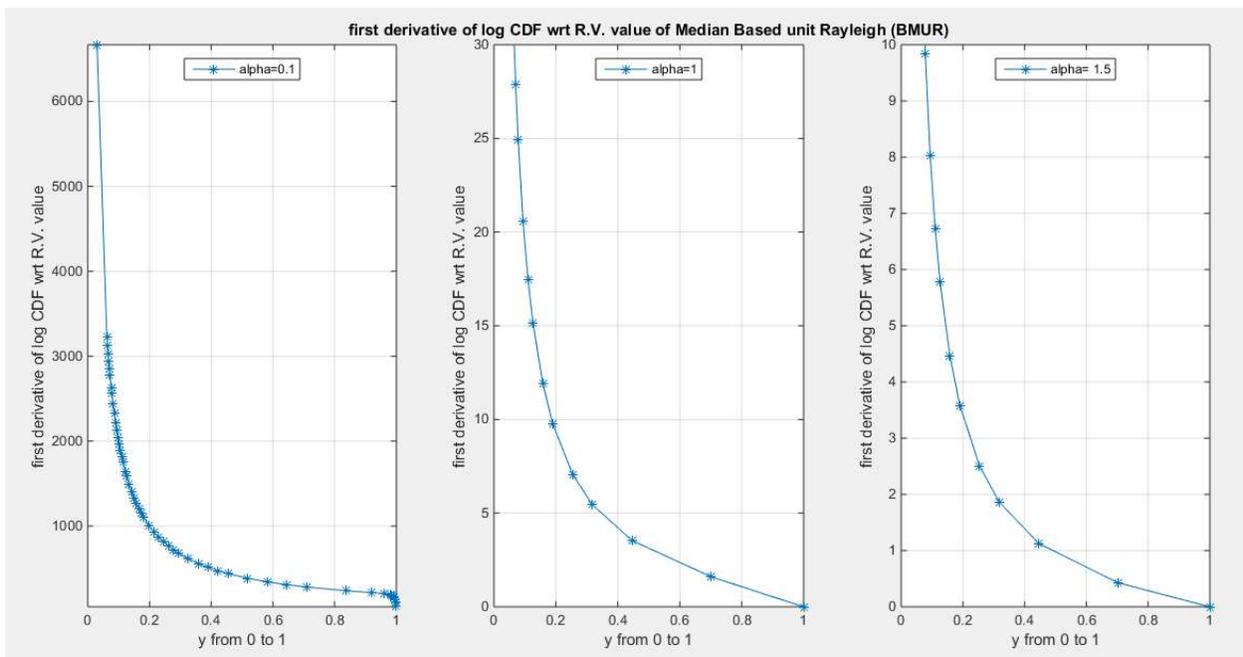

Fig.16 shows that first derivative of log CDF is a positive decreasing function for $\forall\, y \in (0,1)$ and $\forall \alpha \in (0, \infty)$.

The behavior of the second derivative of the log CDF depends on the value of alpha. If $\alpha > 1$, the second derivative is always negative. If $\alpha = 1$, the second derivative is positive then negative. If $\alpha < 1$, It is initially negative then positive then negative again. Figure (17) shows this behavior.

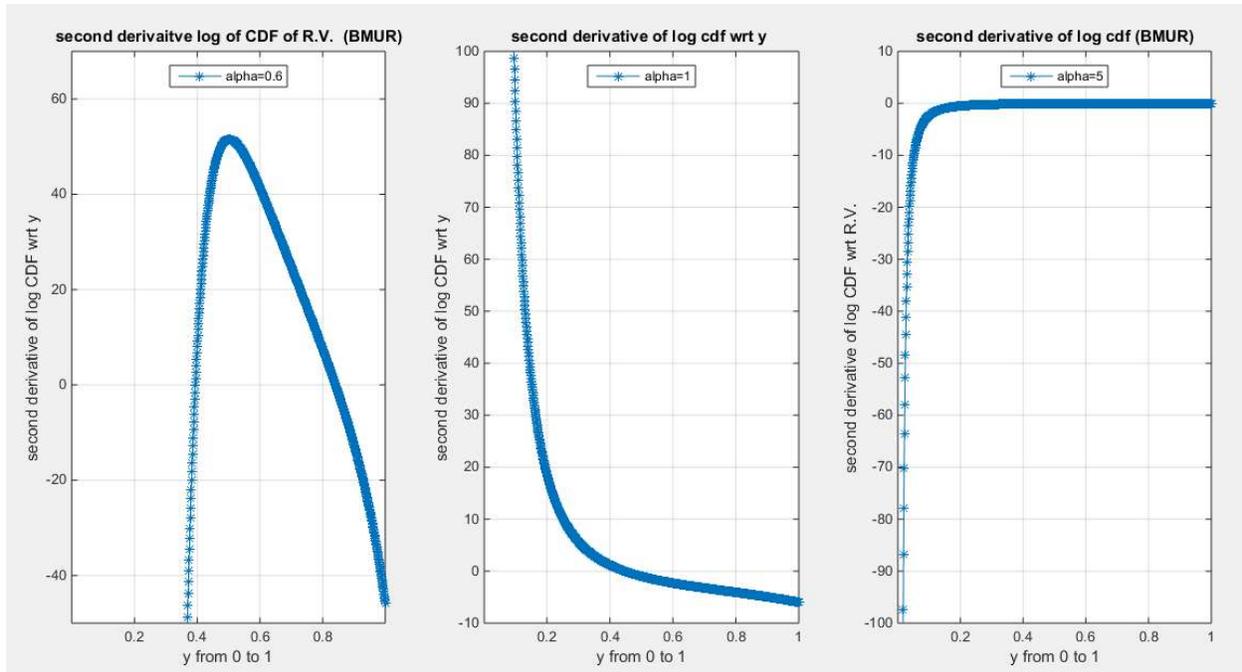

Fig. 17 shows that the second derivative of log CDF with respect to y variable largely depends on the alpha levels.

### 2.3.2. Analysis of the pdf :

***Special case of the MBUR:***

if $\alpha = 1$ then $f(y;\alpha) = 6[1-y]y$, this is a beta distribution with $\alpha = \beta = 2$ which is a special case of the new distribution (MBUR).

The Generalized Beta type I distribution with parameters $\theta = 2$, $\beta = 2$, $\gamma = 1/\alpha^2$ is a special case of the new distribution (MBUR) with $\gamma = 1/\alpha^2$ as illustrated below:

if $\alpha^2 \in (0,1)$ then $f(y;\alpha)$ is a generalized beta type I distribution which has pdf:

$f(y; \theta,\beta,\gamma) = \gamma\, B(\theta,\beta)^{-1}\, y^{\theta\gamma-1}(1-y^\gamma)^{\beta-1}$ where

$\theta = 2$, $\beta = 2$, $\gamma = 1/\alpha^2$

As an example: if $\alpha^2 = 0.1$

Replacing this $\alpha^2$ in MBUR lead to the following pdf:

$f(y;\alpha^2) = \dfrac{6}{0.1}\left[1 - y^{\frac{1}{0.1}}\right] y^{\left(\frac{2}{0.1}-1\right)} = 60\,[1-y^{10}]y^{20-1}$

This is equivalent to $f(y;\theta,\beta,\gamma) = \gamma\, B(\theta,\beta)^{-1}\, y^{\theta\gamma-1}(1-y^\gamma)^{\beta-1}$

$f(y;\,2,2,10) = (10)\,(6)\,y^{(2*10)-1}(1-y^{10})^{2-1}$ which is the Generalized Beta type I distribution.

***Mode of MBUR distribution***:

The mode of the distribution is defined in the following equation. It is a decreasing function in the parameter alpha, see figure (18):

$$y = \left\{\frac{2\alpha^{-2} - 1}{3\alpha^{-2} - 1}\right\}^{\alpha^2}$$

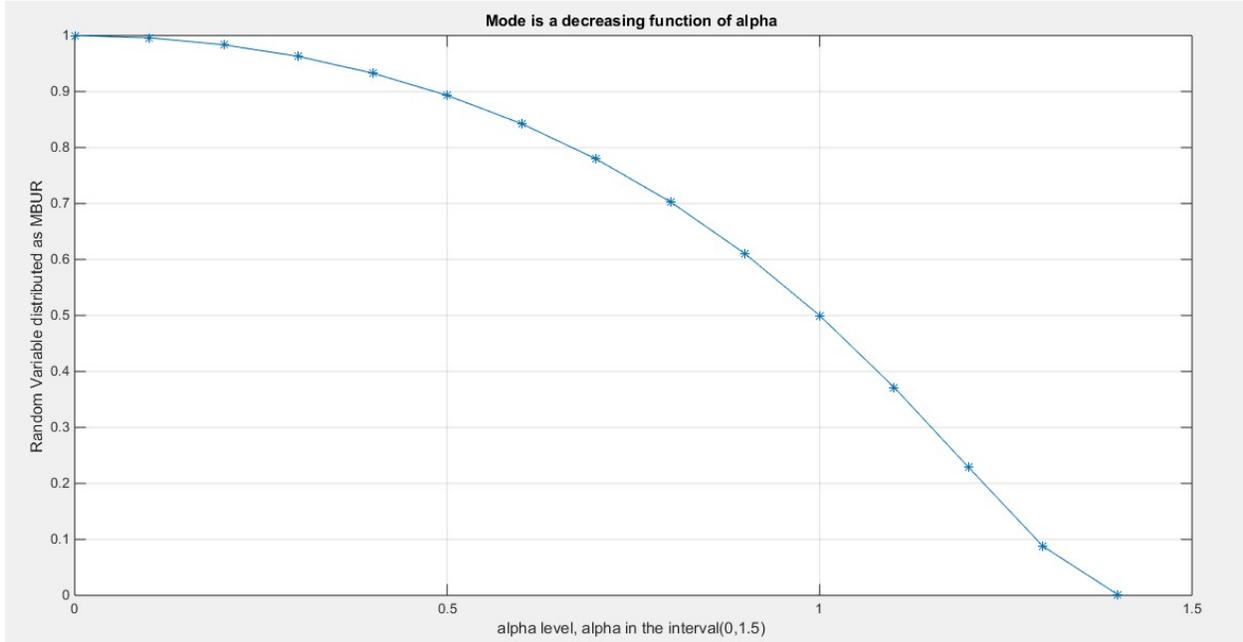

Fig.18 shows the mode function. It is a decreasing function in the parameter, alpha.

As obvious from the definition of the mode, when:

$if\ \alpha \in (0, 1.5)\ then\ y \in (0,1)$ , the mode exists.

$if\ \alpha \in [\ 1.5, 1.8\ )\ then\ y \in complex\ plane$ ,

That is to mean mode does not exist in real plane.

$if\ \alpha \in [\ 1.8, \infty\ )\ then\ y > 1$ , the mode does not exist, as shown in the figure (19):

Differentiating the mode with respect to alpha:

$$\frac{dy}{d\alpha} = (2\alpha)\left\{\frac{\frac{2}{\alpha^2} - 1}{\frac{3}{\alpha^2} - 1}\right\}^{\alpha^2} \ln\left\{\frac{\frac{2}{\alpha^2} - 1}{\frac{3}{\alpha^2} - 1}\right\}\left\{\frac{\left(\frac{-4}{\alpha^3}\right)\left(\frac{3}{\alpha^2} - 1\right) - \left(\frac{-6}{\alpha^3}\right)\left(\frac{2}{\alpha^2} - 1\right)}{\left(\frac{3}{\alpha^2} - 1\right)^2}\right\}$$

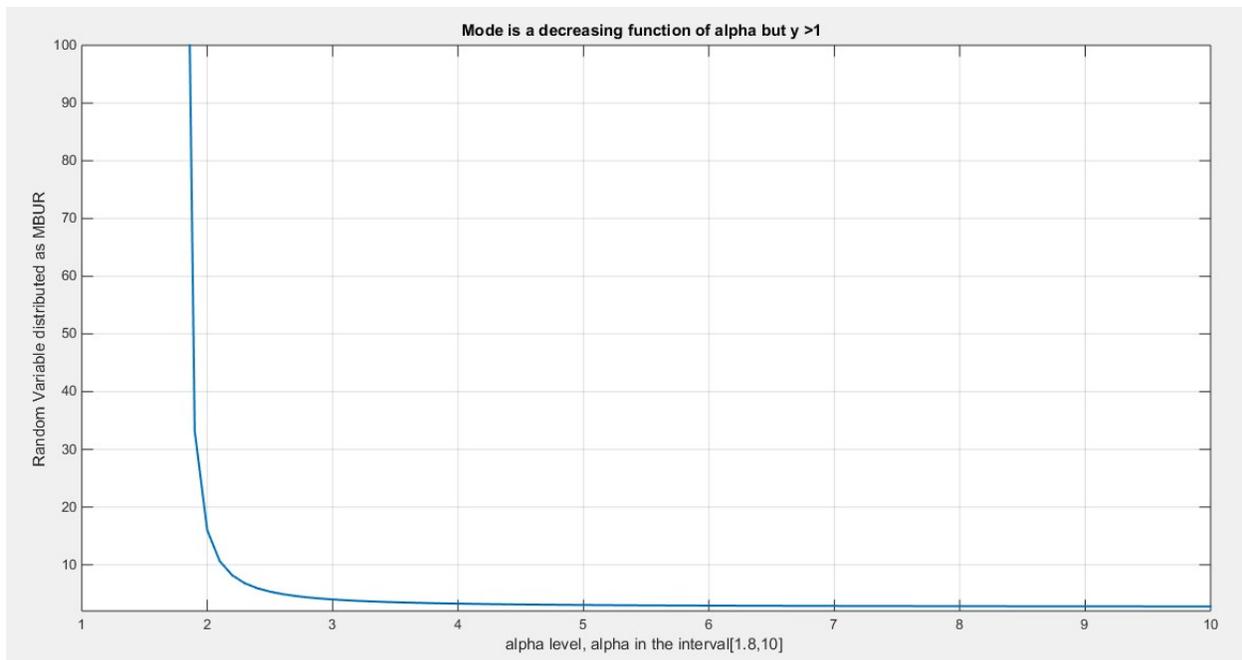

Fig.19 shows that $if\ \alpha \in [\,1.8, \infty\,)\ then\ y > 1$. Mode does not exist.

Graphical analysis of this derivative reveals that this function is not defined to give real number for $\forall\ \alpha \in (0, \infty)$ as shown in the figure (20):

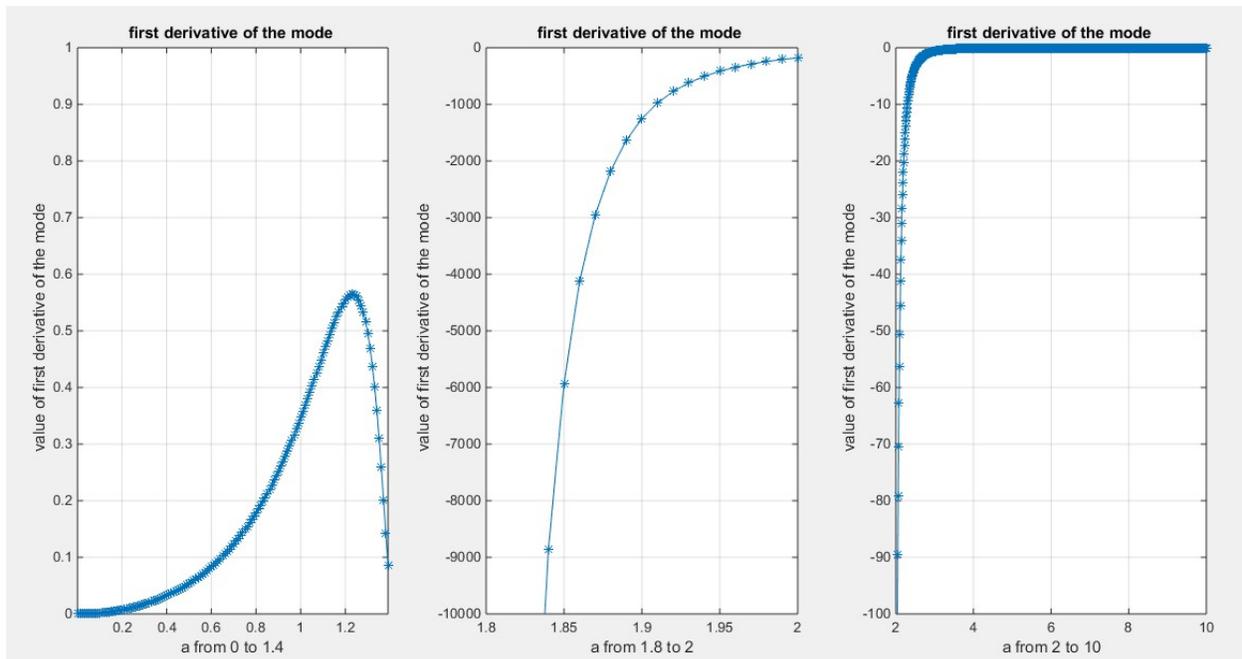

Fig. 20 shows that the function yielded positive real numbers on the interval $\in (0,1.4]$, on the interval $\alpha \in [1.42,1.8)$, the results are complex number. On the interval $\in [1.8, \infty)$, the results are always negative and the variable is larger than one.

***Concavity of MBUR***: The concavity of the PDF largely depends on the alpha level. For alpha less than or equal to one the PDF is concave and the mode exists. If the alpha levels are larger than one, the concavity may not be obvious. If $\alpha = 1$, the PDF is symmetric. If the $\alpha < 1$, the PDF is left skewed. If $\alpha > 1$, the PDF is right skewed. Figures (21-22) highlight these issues.

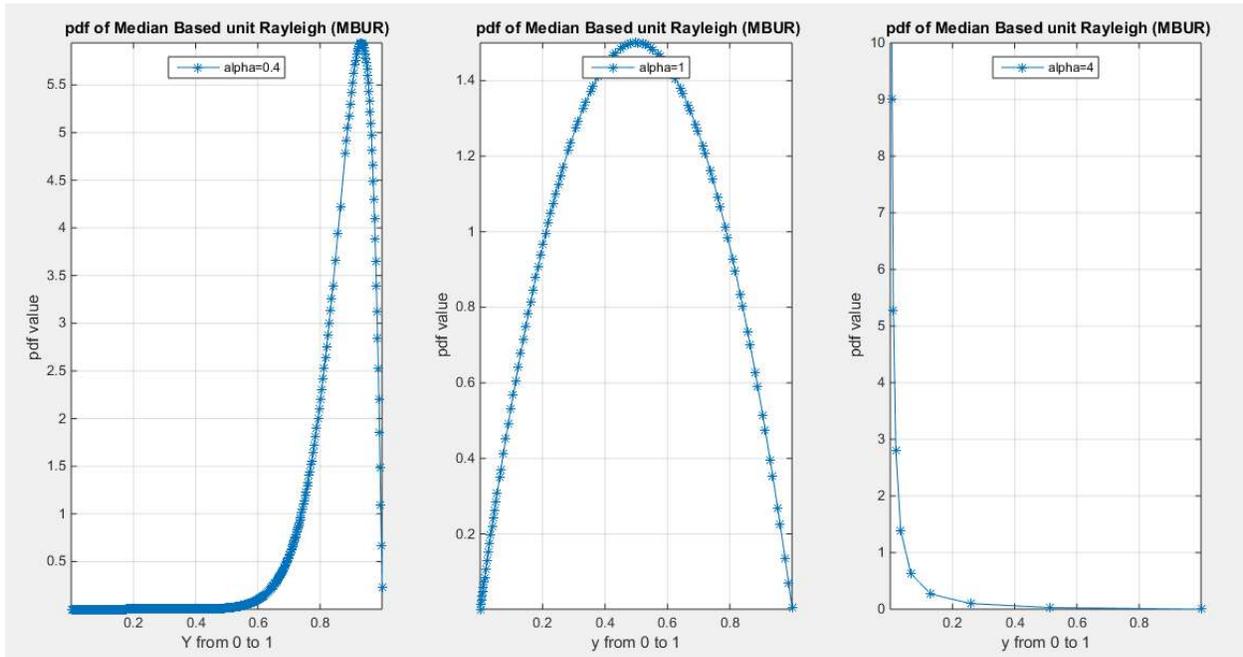

Fig.21 shows that the pdf of MBUR is concave if $\alpha \leq 1$. And the modes exist.

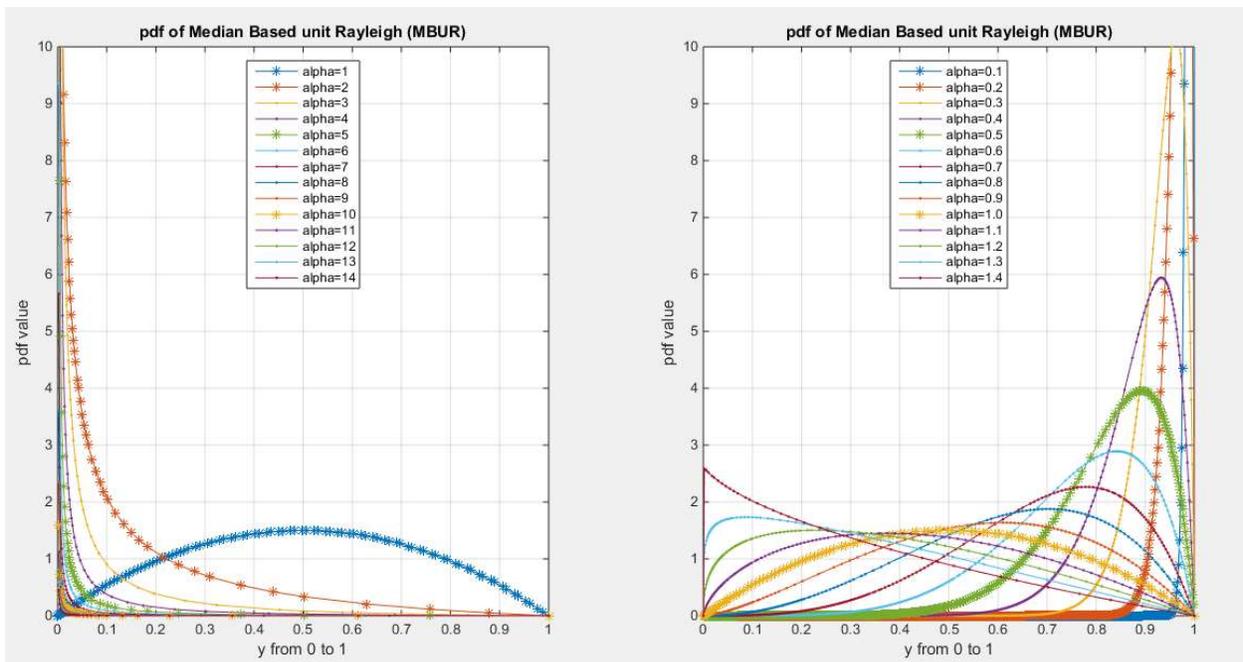

Fig.22 shows that if $\alpha = 1$, the pdf is symmetric. If $1 < \alpha < 1.5$, the mode also exists. If $\alpha > 1.5$; the distribution is right skewed and the mode does not exist (anti-unimodal).

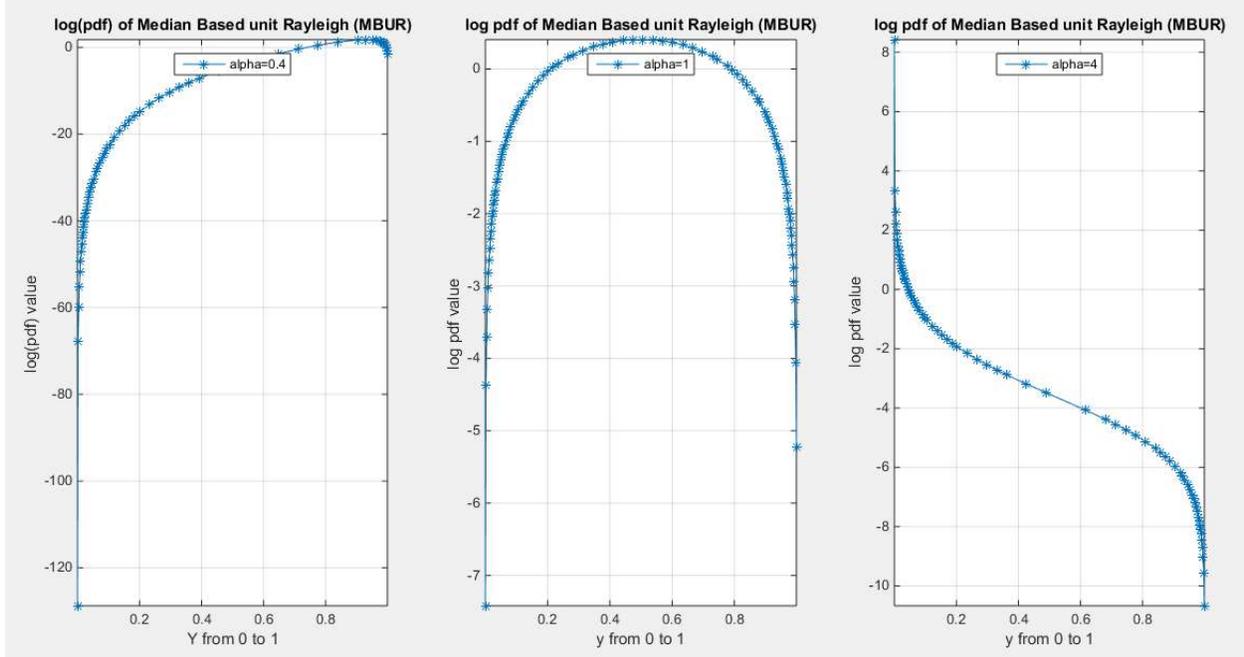

Fig.23 shows log pdf of MBUR for $\forall y \in (0,1)$ with $\alpha < 1$, $\alpha = 1$ & $\alpha > 1$

$$\log f(y;\alpha) = \ln 6 - \ln \alpha^2 + \ln(1 - y^{\alpha^{-2}}) + (2\alpha^{-2} - 1)\ln y$$

if $\alpha \leq 1$, then log pdf is initally negative, then positive and finally negative. If $\alpha > 1$, so the log pdf is initially positive then lastly negative.

if $\alpha \leq 1$, $\lim_{y \to 0} \log f(y;\alpha) = \lim_{y \to 1} \log f(y;\alpha) = -\infty$

f $\alpha > 1$, $\lim_{y \to 0} \log f(y;\alpha) = +\infty$ , $\lim_{y \to 1} \log f(y;\alpha) = -\infty$

The following equations show the first and second derivatives of log PDF with respect to y, see figure (24-25) for their behaviors :

$$\frac{d\log f(y;\alpha)}{dy} = \frac{(-\alpha^{-2})(y)^{\alpha^{-2}-1}}{1 - (y)^{\alpha^{-2}}} + (2\alpha^{-2} - 1)\left(\frac{1}{y}\right)$$

$$\frac{d^2 \log f(y;\alpha)}{dy^2} = \frac{(-\alpha^{-2})(\alpha^{-2} - 1)(y)^{\alpha^{-2}-1}}{1 - (y)^{\alpha^{-2}}} - \left(\frac{\alpha^{-2}(y)^{\alpha^{-2}-1}}{1 - (y)^{\alpha^{-2}}}\right)^2 - \left(\frac{2\alpha^{-2} - 1}{y^2}\right)$$

if $\alpha \leq 1$, then first derivative of log PDF is decreasing being initially positive then finally negative. If $\alpha > 1$, then the log PDF is always negative (initially increasing then lastly decreasing) and concave as shown in the figure (24).

if $\alpha \leq 1$, $\lim_{y \to 0} \frac{\partial lo\ f(y;\alpha)}{\partial y} = +\infty$ , $\lim_{y \to 1} \frac{\partial \log f(y;\alpha)}{\partial y} = -\infty$

f $\alpha > 1$, $\lim_{y \to 0} \frac{\partial \log f(y;\alpha)}{\partial y} = \lim_{y \to 1} \frac{\partial \log f(y;\alpha)}{\partial y} = -\infty$

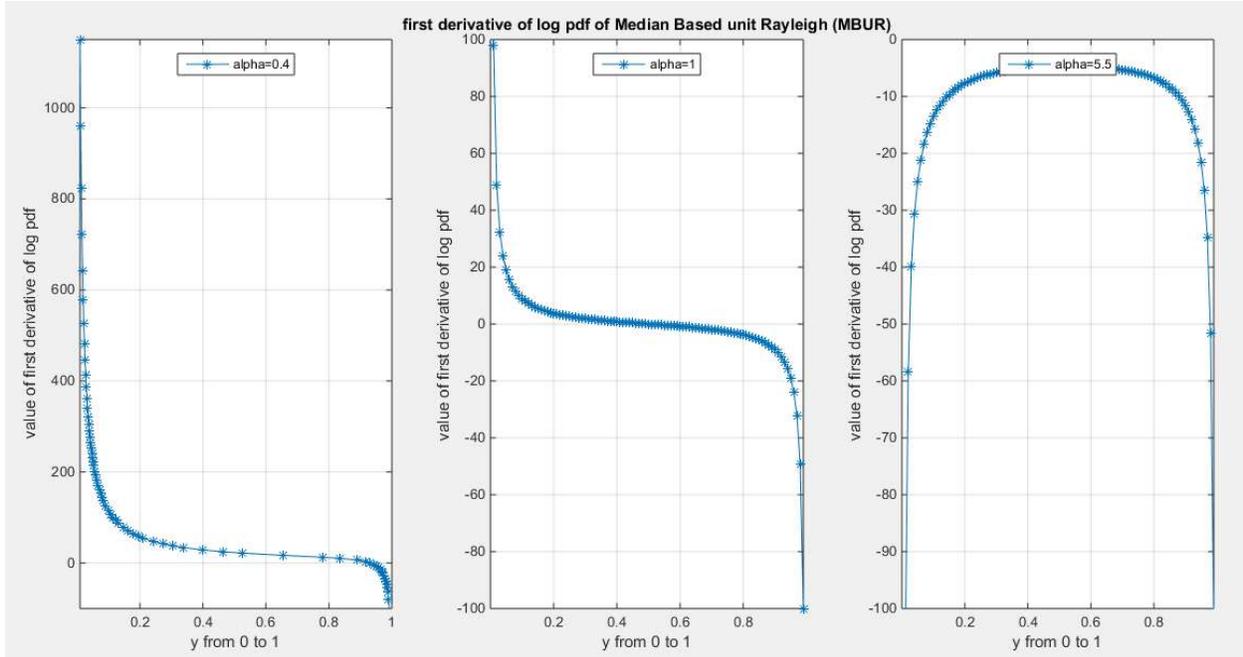

Fig.24 shows the first derivative of log pdf of MBUR for $\forall \, y \in (0,1)$ and $\alpha < 1 \quad \alpha = 1 \, \& \, \alpha > 1$.

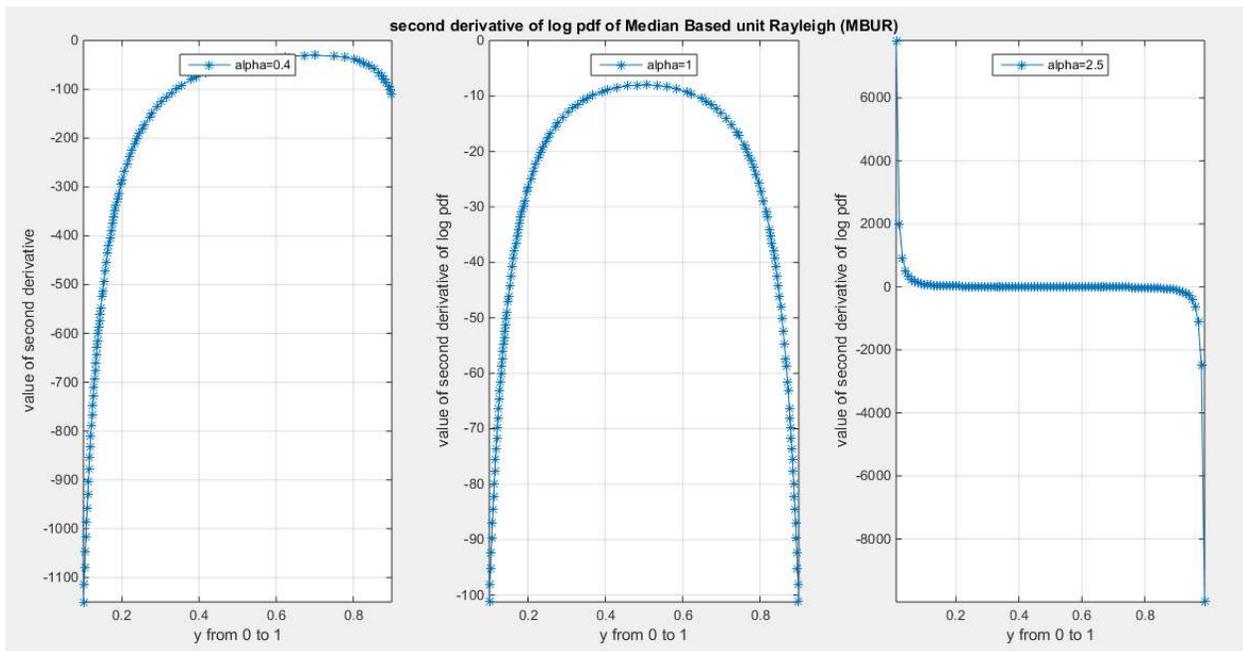

Fig.25 shows the second derivative of log PDF of MBUR for $\forall \, y \in (0,1)$ and $\alpha < 1, \, \alpha = 1 \, \& \, \alpha > 1$

$if \, \alpha \leq 1$, then second derivative of log pdf is always negative and concave. If $\alpha > 1$, then second derivative of the log PDF is initially positive then lastly negative.

$if \, \alpha \leq 1 \, , \, \lim_{y \to 0} \frac{\partial^2 \log f(y;\alpha)}{\partial y^2} = \lim_{y \to 1} \frac{\partial^2 \log f(y;\alpha)}{\partial y^2} = -\infty$

$f\ \alpha > 1$, $\lim\limits_{y \to 0} \dfrac{\partial^2 \log f(y;\alpha)}{\partial y^2} = +\infty$, $\lim\limits_{y \to 1} \dfrac{\partial^2 \log f(y;\alpha)}{\partial y^2} = -\infty$

### 2.3.3. Analysis of log-likelihood:

Discussing the behavior of PDF with respect to the parameter is crucial for estimation. Figures (26-29) depict this behavior of the PDF, log PDF, the first and, the second derivative of this log PDF. The following eqautions are the log-likelihhod and its first and second derivative with respect to the parameter.

$$l(\alpha) = \ln(6) - \ln(\alpha^2) + \ln\left[1 - y^{\frac{1}{\alpha^2}}\right] + \left(\frac{2}{\alpha^2} - 1\right)\ln(y)$$

$$\frac{dl(\alpha)}{d\alpha} = \frac{-2}{\alpha} + \frac{-y^{\frac{1}{\alpha^2}}(\ln[y])(-2\alpha^{-3})}{1 - y^{\frac{1}{\alpha^2}}} - 4\alpha^{-3}\ln(y)$$

$$\frac{d^2 l(\alpha)}{d\alpha^2} = \frac{2}{\alpha^2} - \left(\frac{6}{\alpha^4}\right)\left(\frac{y^{\frac{1}{\alpha^2}}(\ln[y])}{1 - y^{\frac{1}{\alpha^2}}}\right) + \left(\frac{12}{\alpha^4}\right)(\ln(y)) - \left(\frac{4}{\alpha^6}\right)\left(\frac{y^{\frac{1}{\alpha^2}}(\ln[y])^2}{\left(1 - y^{\frac{1}{\alpha^2}}\right)^2}\right)$$

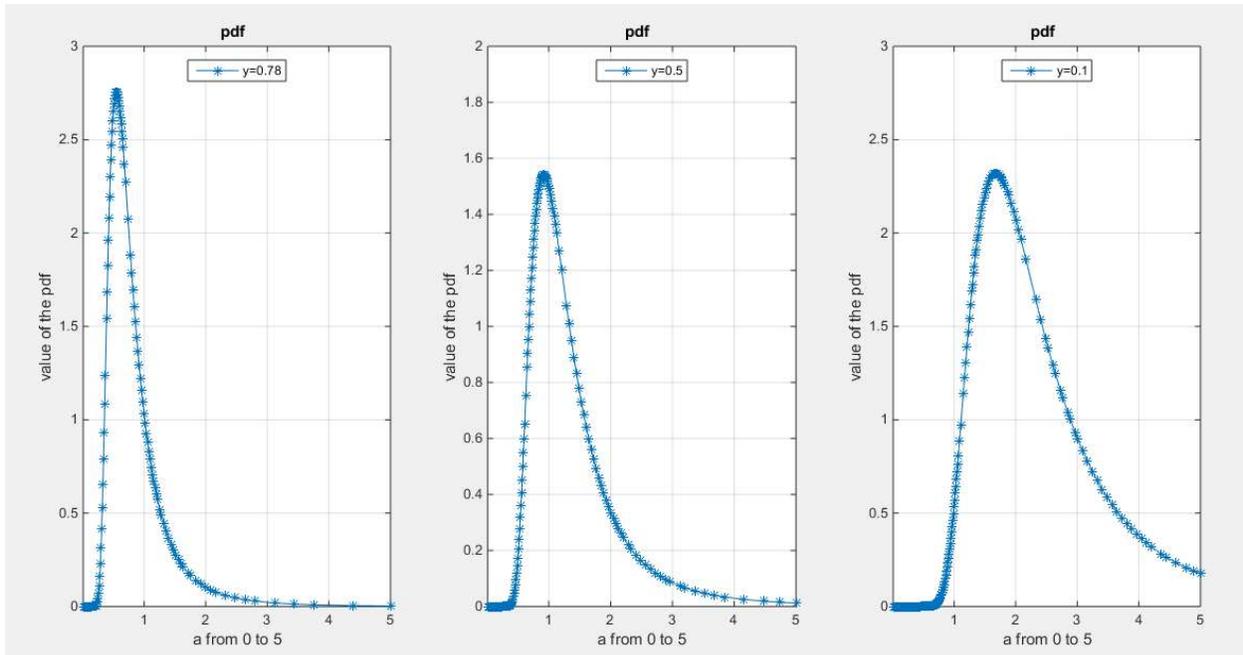

Fig.26 shows the pdf as a function in alpha $\alpha \in (0,5)$ for $y = 0.78$  $y = 0.5$ and $y = 0.1$. It is always positive and concave. This is applicable to all level of alpha.

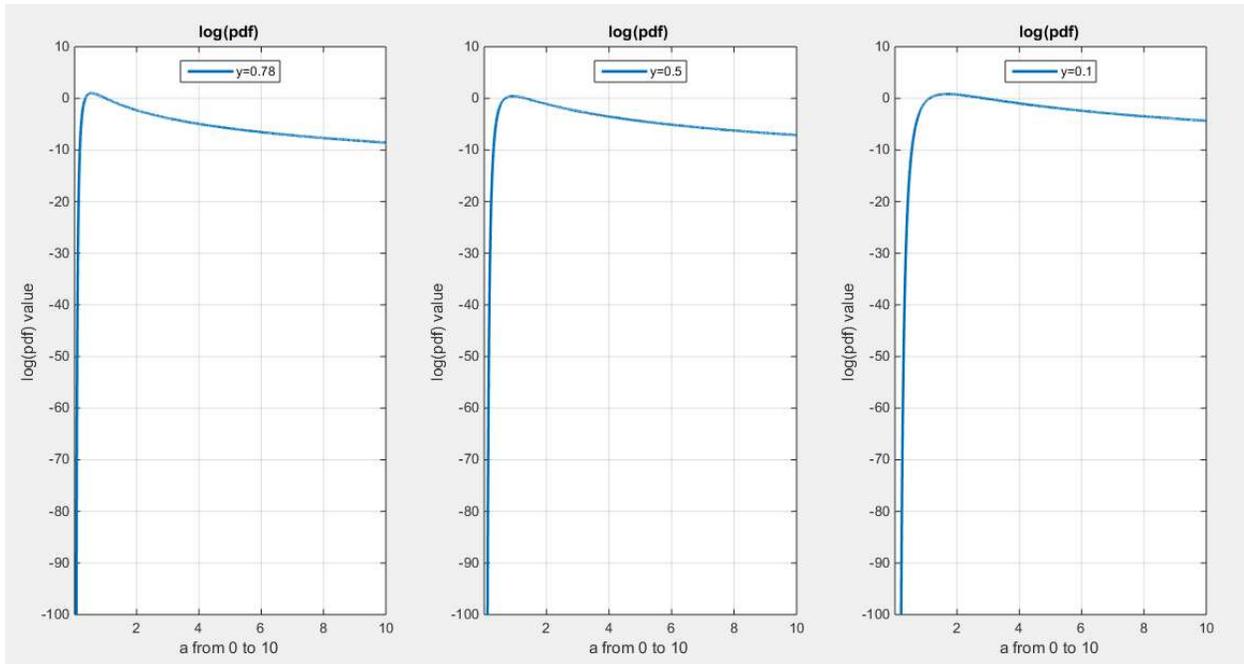

Fig.27 shows the log PDF as a function of alpha $\alpha \in (0,10)$ for $= 0.78$, $y = 0.5$ and $y = 0.1$. It is initially negative, then positive then again negative. And this is applicable to all levels of alpha.

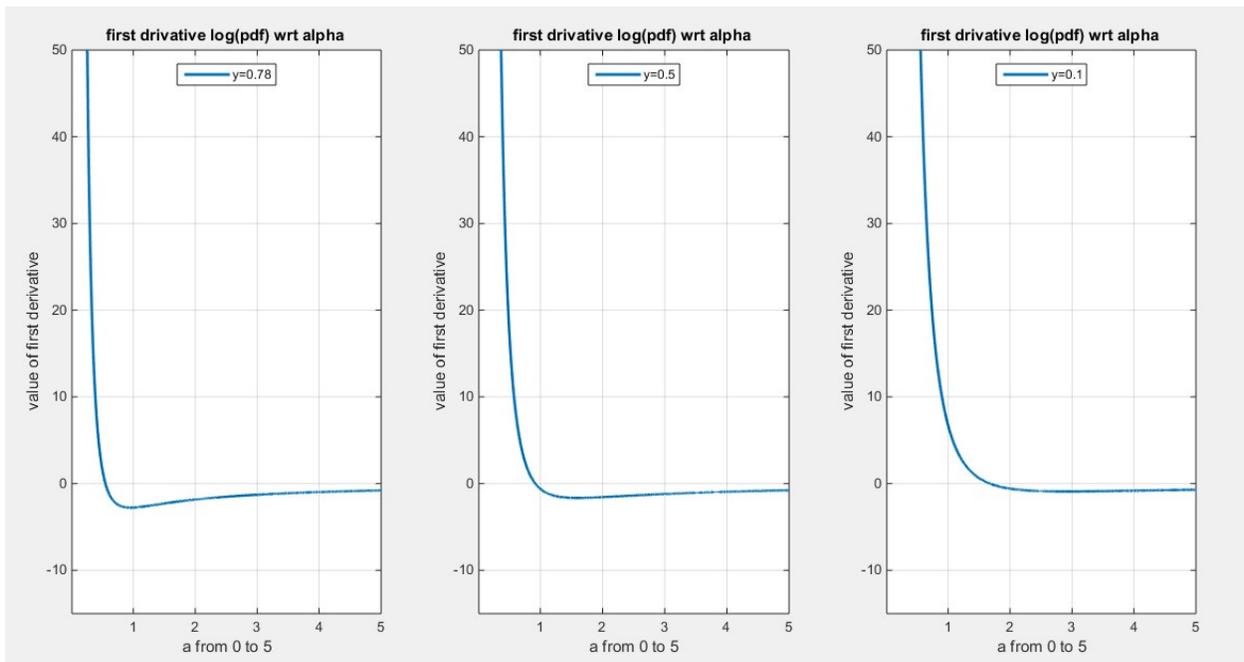

Fig.28 shows the first derivative of log PDF as a function in alpha $\alpha \in (0,5)$ for $= 0.78$, $y = 0.5$ and $y = 0.1$. It is initially positive, then persistently becomes negative.

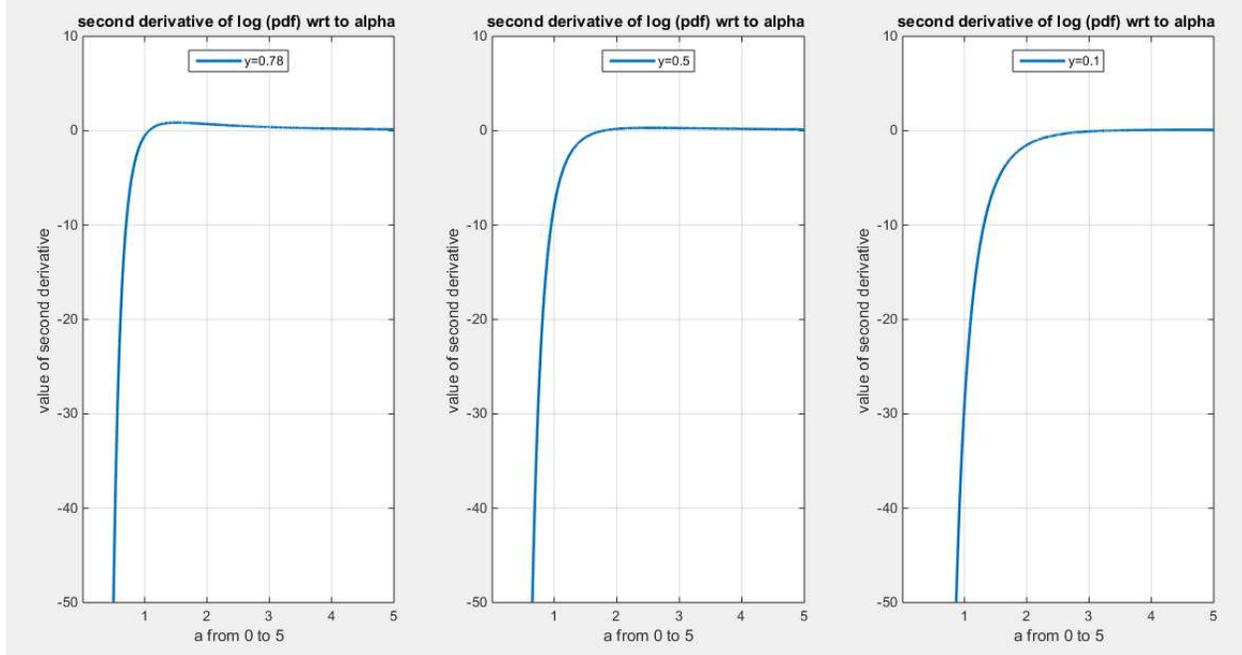

Fig.29 shows the second derivative of log PDF as a function in alpha $\alpha \in (0,5)$ for $= 0.78$, $y = 0.5$ and $y = 0.1$. It is initially negative, then persistently becomes posistive.

### 2.3.4. Analysis of hazard function:

Differentiating the hazard function with respect to the variable y is essential to understand the behavior of the hazard function. Figures (30-31) demonstrate these concepts.

$$\frac{dh(y)}{d(y)} = \frac{d}{d(y)}\left[\frac{f(y)}{S(y)}\right] = \frac{f'(y)S(y) - f(y)S'(y)}{[S(y)]^2}$$

$$f'(y) = \frac{6}{\alpha^2}(y)^{2\alpha^{-2}-2}\left\{\left[(2\alpha^{-2}-1) - (3\alpha^{-2}-1)(y)^{\alpha^{-2}}\right]\right\}$$

$$S'(y) = \frac{\partial S(y)}{\partial y} = \frac{6}{\alpha^2}y^{\left(\frac{3}{\alpha^2}-1\right)} - \frac{6}{\alpha^2}y^{\left(\frac{2}{\alpha^2}-1\right)} = \frac{6}{\alpha^2}y^{\left(\frac{2}{\alpha^2}-1\right)}\left\{y^{\left(\frac{1}{\alpha^2}\right)} - 1\right\}$$

if $\alpha < 1$, $\lim_{y \to 0} h'(y) = 0$ and $\lim_{y \to 1} h'(y) = +\infty$

if $\alpha = 1$, $\lim_{y \to 1} h'(y) = +\infty$ and $\lim_{y \to 0} h'(y) \cong 5.8853$ and has minimum value of (5.1298) at (y=0.162).

if $\alpha > 1$, $\lim_{y \to 0} h'(y) = -\infty$ and $\lim_{y \to 1} h'(y) = +\infty$

This can explain the behavior of the hazard function

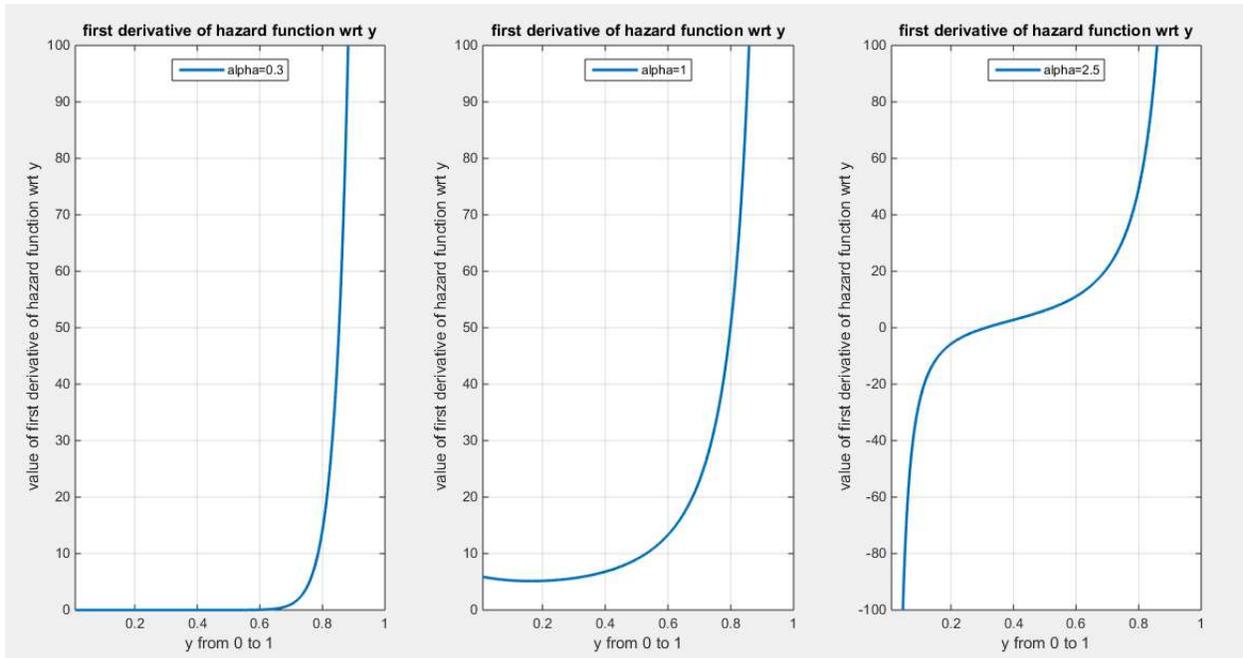

Fig.30 shows that the first derivative of the hazard function with respect to $\forall\, y \in (0,1)$ is increasing function with respect to variable y. The shape depends on the alpha level.

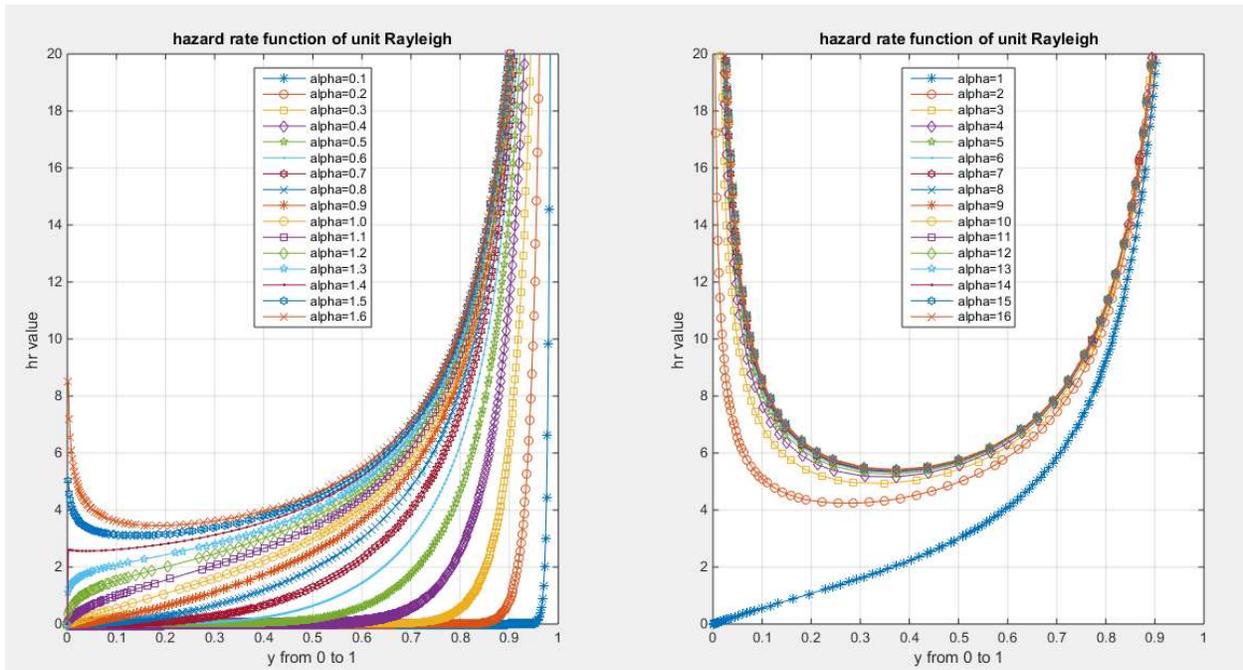

Fig.31 shows that the hazard function can attain many different shapes depending on level of the alpha.

From the graph on the left side:

$if\ \alpha \leq 1$ , $hr$ is increasing function with convex properties.

$if\ \alpha \in (1,1.4)$ , $hr$ is increasing function with concavity properties.

if $\alpha = 1.4$, $hr$ is increasing function.

if $\alpha \in [1.5, 1.6]$, $hr$ has bathtub shape.

From the graph on the right side where $\alpha \in [1.5, \infty)$, $hr$ has bathtub shape.

## 2.5. $r^{th}$ incomplete Moments: (see the text)

for a random variable y distributed as MBUR is defined in equation (15): $E(y^r | y < t) = \int_0^{y=t} y^r \frac{6}{\alpha^2} \Big[ 1 -$

$y^{\frac{1}{\alpha^2}} \Big] y^{\left(\frac{2}{\alpha^2}-1\right)} dy = \frac{6 y^{\frac{2}{\alpha^2}+r}}{(2+r\alpha^2)} - \frac{6 y^{\frac{3}{\alpha^2}+r}}{(3+r\alpha^2)} \ldots \ldots (15)$

$\int_0^{y=t} \frac{6}{\alpha^2} y^{\left(\frac{2}{\alpha^2}-1+r\right)} dy - \int_0^{y=t} \frac{6}{\alpha^2} y^{\left(\frac{3}{\alpha^2}-1+r\right)} dy =$

$\frac{6}{\alpha^2} \left( \frac{y^{\left(\frac{2}{\alpha^2}+r\right)}}{\frac{2}{\alpha^2}+r} \right) \Big|_0^t - \frac{6}{\alpha^2} \left( \frac{y^{\left(\frac{3}{\alpha^2}+r\right)}}{\frac{3}{\alpha^2}+r} \right) \Big|_0^t = \frac{6 t^{\frac{2}{\alpha^2}+r}}{(2+r\alpha^2)} - \frac{6 t^{\frac{3}{\alpha^2}+r}}{(3+r\alpha^2)}$

But the t is only a value of the random variable y so to generalize the equation use y instead of the t value as in equation 15 (see text).

## 2.6. Stress- strength reliability (see text)

From a reliability prospective, if an element in a system has a random strength X that is strained with a random stress Y, this element will immediately break down if the stress overrides the strength. Conversely, it will adequately operate if the strength surpasses the stress. If X and Y are independent random variables denoting strength and stress respectively, and both follow MBUR distribution with parameters $\alpha_1$ and $\alpha_2$ respectively, then the reliability measure of this element can be deduced from appropriate equation (16)

$R = Pr(Y < X) = \int_0^1 Pr(Y < X | X = x) f_X(x) dy$

$R = Pr(Y < X) = \int_0^1 F(x; \alpha_2) f(x; \alpha_1) dy$

$= \int_0^1 \left\{ 3 y^{\frac{2}{\alpha_2^2}} - 2 y^{\frac{3}{\alpha_2^2}} \right\} \left\{ \frac{6}{\alpha_1^2} \left[ 1 - y^{\frac{1}{\alpha_1^2}} \right] y^{\left(\frac{2}{\alpha_1^2}-1\right)} \right\} dy$

$= \int_0^1 \left\{ 3 y^{\frac{2}{\alpha_2^2}} - 2 y^{\frac{3}{\alpha_2^2}} \right\} \left\{ \frac{6}{\alpha_1^2} y^{\left(\frac{2}{\alpha_1^2}-1\right)} - \frac{6}{\alpha_1^2} y^{\left(\frac{3}{\alpha_1^2}-1\right)} \right\} dy$

$= \int_0^1 \left\{ 3 y^{\frac{2}{\alpha_2^2}} - 2 y^{\frac{3}{\alpha_2^2}} \right\} \left\{ \frac{6}{\alpha_1^2} y^{\left(\frac{2}{\alpha_1^2}-1\right)} - \frac{6}{\alpha_1^2} y^{\left(\frac{3}{\alpha_1^2}-1\right)} \right\} dy$

$= \int_0^1 \frac{18}{\alpha_1^2} y^{\frac{2}{\alpha_2^2}+\frac{2}{\alpha_1^2}-1} - \frac{18}{\alpha_1^2} y^{\frac{2}{\alpha_2^2}+\frac{3}{\alpha_1^2}-1} - \frac{12}{\alpha_1^2} y^{\frac{3}{\alpha_2^2}+\frac{2}{\alpha_1^2}-1} + \frac{12}{\alpha_1^2} y^{\frac{3}{\alpha_2^2}+\frac{3}{\alpha_1^2}-1} dy$

$$= \frac{18}{\alpha_1^2}\left(\frac{y^{\frac{2}{\alpha_2^2}+\frac{2}{\alpha_1^2}}}{\frac{2}{\alpha_2^2}+\frac{2}{\alpha_1^2}}\right)\Bigg|_0^1 - \frac{18}{\alpha_1^2}\left(\frac{y^{\frac{2}{\alpha_2^2}+\frac{3}{\alpha_1^2}}}{\frac{2}{\alpha_2^2}+\frac{3}{\alpha_1^2}}\right)\Bigg|_0^1 - \frac{12}{\alpha_1^2}\left(\frac{y^{\frac{3}{\alpha_2^2}+\frac{2}{\alpha_1^2}-1}}{\frac{3}{\alpha_2^2}+\frac{2}{\alpha_1^2}}\right)\Bigg|_0^1 + \frac{12}{\alpha_1^2}\left(\frac{y^{\frac{3}{\alpha_2^2}+\frac{3}{\alpha_1^2}-1}}{\frac{3}{\alpha_2^2}+\frac{3}{\alpha_1^2}}\right)\Bigg|_0^1$$

$$= \frac{18}{\alpha_1^2}\left(\frac{1}{\frac{2}{\alpha_2^2}+\frac{2}{\alpha_1^2}}\right) - \frac{18}{\alpha_1^2}\left(\frac{1}{\frac{2}{\alpha_2^2}+\frac{3}{\alpha_1^2}}\right) - \frac{12}{\alpha_1^2}\left(\frac{1}{\frac{3}{\alpha_2^2}+\frac{2}{\alpha_1^2}}\right) + \frac{12}{\alpha_1^2}\left(\frac{1}{\frac{3}{\alpha_2^2}+\frac{3}{\alpha_1^2}}\right)$$

$$= \left(\frac{18}{\frac{2\alpha_1^2}{\alpha_2^2}+2}\right) - \left(\frac{18}{\frac{2\alpha_1^2}{\alpha_2^2}+3}\right) - \left(\frac{12}{\frac{3\alpha_1^2}{\alpha_2^2}+2}\right) + \left(\frac{12}{\frac{3\alpha_1^2}{\alpha_2^2}+3}\right)$$

$$= \left(\frac{18}{\frac{2\alpha_1^2}{\alpha_2^2}+\frac{2\alpha_2^2}{\alpha_2^2}}\right) - \left(\frac{18}{\frac{2\alpha_1^2}{\alpha_2^2}+\frac{3\alpha_2^2}{\alpha_2^2}}\right) - \left(\frac{12}{\frac{3\alpha_1^2}{\alpha_2^2}+\frac{2\alpha_2^2}{\alpha_2^2}}\right) + \left(\frac{12}{\frac{3\alpha_1^2}{\alpha_2^2}+\frac{3\alpha_2^2}{\alpha_2^2}}\right)$$

$$= \left(\frac{18\,\alpha_2^2}{2\alpha_1^2+2\alpha_2^2}\right) - \left(\frac{18\alpha_2^2}{2\alpha_1^2+3\alpha_2^2}\right) - \left(\frac{12\alpha_2^2}{3\alpha_1^2+2\alpha_2^2}\right) + \left(\frac{12\alpha_2^2}{3\alpha_1^2+3\alpha_2^2}\right)$$

$$= \frac{18}{2}\left(\frac{\alpha_2^2}{\alpha_1^2+\alpha_2^2}\right) - \left(\frac{18\alpha_2^2}{2\alpha_1^2+3\alpha_2^2}\right) - \left(\frac{12\alpha_2^2}{3\alpha_1^2+2\alpha_2^2}\right) + \frac{12}{3}\left(\frac{\alpha_2^2}{\alpha_1^2+\alpha_2^2}\right)$$

$$= 13\left(\frac{\alpha_2^2}{\alpha_1^2+\alpha_2^2}\right) - \left(\frac{18\alpha_2^2}{2\alpha_1^2+3\alpha_2^2}\right) - \left(\frac{12\alpha_2^2}{3\alpha_1^2+2\alpha_2^2}\right)$$

$$R = \alpha_2^2\left\{\frac{13}{(\alpha_1^2+\alpha_2^2)} - \frac{18}{(2\alpha_1^2+3\alpha_2^2)} - \frac{12}{(3\alpha_1^2+2\alpha_2^2)}\right\} \ldots\ldots(16)$$

**2.7. Lorenz, Bonferroni curves and Gini index (see text)**

These indices have many applications in medicine, insurance, demography, and economics for studying wealth and poverty. They can be applied to variables defined as proportions where y is a random variable distributed as MBUR. Lorenz curve, Bonferroni curves and Gini index are defined in equation (17), (18), (19) respectively (see supplementary materials section 1 for derivation)

$$L(p) = \frac{\int_0^t y\,f(y)dy}{\int_0^1 yf(y)dy} = \left\{\frac{6y^{\frac{2}{\alpha^2}+1}}{(2+\alpha^2)} - \frac{6y^{\frac{3}{\alpha^2}+1}}{(3+\alpha^2)}\right\} \div \left\{\frac{6}{(2+\alpha^2)(3+\alpha^2)}\right\}$$

$$= \left\{\frac{6y^{\frac{2}{\alpha^2}+1}(3+\alpha^2) - 6y^{\frac{3}{\alpha^2}+1}(2+\alpha^2)}{(2+\alpha^2)(3+\alpha^2)}\right\} \times \left\{\frac{(2+\alpha^2)(3+\alpha^2)}{6}\right\}$$

$$= \left\{\frac{6y^{\frac{2}{\alpha^2}+1}(3+\alpha^2) - 6y^{\frac{3}{\alpha^2}+1}(2+\alpha^2)}{6}\right\} = \left\{y^{\frac{2}{\alpha^2}+1}(3+\alpha^2) - y^{\frac{3}{\alpha^2}+1}(2+\alpha^2)\right\}$$

$$= y^{\frac{2}{\alpha^2}+1}\left\{(3+\alpha^2) - y^{\frac{1}{\alpha^2}}(2+\alpha^2)\right\} \ldots\ldots\ldots\ldots\ldots\ldots\ldots\ldots\ldots\ldots\ldots\ldots(17)$$

$$B(p) = \frac{L(p)}{F(y)} = \frac{y^{\frac{2}{\alpha^2}+1}\left\{(3+\alpha^2) - y^{\frac{1}{\alpha^2}}(2+\alpha^2)\right\}}{3y^{\frac{2}{\alpha^2}} - 2y^{\frac{3}{\alpha^2}}}$$

$$B(p) = \frac{L(p)}{F(y)} = \frac{y^{\frac{2}{\alpha^2}}\left\{y(3+\alpha^2) - y^{\frac{1}{\alpha^2}+1}(2+\alpha^2)\right\}}{y^{\frac{2}{\alpha^2}}\left(3 - 2y^{\frac{1}{\alpha^2}}\right)}$$

$$B(p) = \frac{L(p)}{F(y)} = \frac{y\left\{(3+\alpha^2) - y^{\frac{1}{\alpha^2}}(2+\alpha^2)\right\}}{3 - 2y^{\frac{1}{\alpha^2}}} \quad \ldots\ldots\ldots\ldots\ldots\ldots\ldots\ldots (18)$$

$$\text{Gini index} = 1 - 2\int_0^1 L(p)\,dp$$

$$= 1 - 2\int_0^1 L(p)\,dp = 1 - 2\int_0^1 y^{\frac{2}{\alpha^2}+1}\left\{(3+\alpha^2) - y^{\frac{1}{\alpha^2}}(2+\alpha^2)\right\}dy$$

$$= 1 - 2\int_0^1 \left\{y^{\frac{2}{\alpha^2}+1}(3+\alpha^2) - y^{\frac{3}{\alpha^2}+1}(2+\alpha^2)\right\}dy$$

$$= 1 - 2\left\{(3+\alpha^2)\left(\frac{y^{\frac{2}{\alpha^2}+1+1}}{\frac{2}{\alpha^2}+2}\right)\Big|_0^1 - (2+\alpha^2)\left(\frac{y^{\frac{3}{\alpha^2}+1+1}}{\frac{3}{\alpha^2}+2}\right)\Big|_0^1\right\}$$

$$= 1 - 2\left\{\frac{(3+\alpha^2)}{\frac{2}{\alpha^2}+2} - \frac{(2+\alpha^2)}{\frac{3}{\alpha^2}+2}\right\} = 1 - 2\left\{\frac{\alpha^2(3+\alpha^2)}{2+2\alpha^2} - \frac{\alpha^2(2+\alpha^2)}{3+2\alpha^2}\right\}$$

$$= 1 - 2\left\{\frac{\alpha^2(3+\alpha^2)3+2\alpha^2}{2+2\alpha^2} - \frac{\alpha^2(2+\alpha^2)2+2\alpha^2}{3+2\alpha^2}\right\}$$

$$= 1 - 2\left\{\frac{\alpha^2(3+\alpha^2)(3+2\alpha^2) - \alpha^2(2+\alpha^2)(2+2\alpha^2)}{(2+2\alpha^2)(3+2\alpha^2)}\right\}$$

$$= 1 - 2\left\{\frac{9\alpha^2 + 2\alpha^6 + 9\alpha^4 - 4\alpha^2 - 2\alpha^6 - 6\alpha^4}{(2+2\alpha^2)(3+2\alpha^2)}\right\}$$

$$= 1 - 2\left\{\frac{5\alpha^2 + 3\alpha^4}{2(1+\alpha^2)(3+2\alpha^2)}\right\}$$

$$\text{Gini index} = 1 - 2\int_0^1 L(p)\,dp = 1 - \left(\frac{\alpha^2(5+3\alpha^2)}{(1+\alpha^2)(3+2\alpha^2)}\right) \quad \ldots\ldots\ldots\ldots\ldots (19)$$

### 2.8. Renyi entropy (see the text)

Entropy quantifies the variation in uncertainty within a random variable, in the paper context, y is distributed as MBUR. Renyi entropy $T_R(\gamma)$ is a well-known measure defined as follows in equation (20). For MBUR it is defined in equation (21):

$$T_R(\gamma) = \frac{1}{1-\gamma}\log\left\{\int_0^1 [f(y)]^\gamma \, dy\right\} \quad \ldots\ldots (20)$$

$$T_R(\gamma) = \frac{1}{1-\gamma}\log\left\{\int_0^1 \left[\left(\frac{6}{\alpha^2}\right)\left(1 - y^{\frac{1}{\alpha^2}}\right)y^{\left(\frac{2}{\alpha^2}-1\right)}\right]^\gamma dy\right\}\ldots\ldots(21)$$

Expanding the following term using the binomial expansion as in equation (22):

$$\left(1 - y^{\frac{1}{\alpha^2}}\right)^\gamma = \sum_{m=0}^{\infty}(-1)^m \binom{\gamma}{m}\left(y^{\frac{1}{\alpha^2}}\right)^m \ldots\ldots(22)$$

substitute equation (22) into equation (21) gives equations (23), (24) & (25).

$$= \frac{1}{1-\gamma}\log\left\{\int_0^1 \left[\left(\frac{6}{\alpha^2}\right)^\gamma \sum_{m=0}^{\infty}(-1)^m \binom{\gamma}{m}\left(y^{\frac{1}{\alpha^2}}\right)^m y^{\left(\frac{2\gamma-\gamma\alpha^2}{\alpha^2}\right)}\right] dy\right\}$$

The integral will pass to the variable for integration:

$$= \frac{1}{1-\gamma}\log\left\{\left(\frac{6}{\alpha^2}\right)^\gamma \sum_{m=0}^{\infty}(-1)^m \binom{\gamma}{m}\int_0^1 \left[\left(y^{\frac{1}{\alpha^2}}\right)^m y^{\left(\frac{2\gamma-\gamma\alpha^2}{\alpha^2}\right)}\right] dy\right\}\ldots\ldots(23)$$

$$= \frac{1}{1-\gamma}\log\left\{\left(\frac{6}{\alpha^2}\right)^\gamma \sum_{m=0}^{\infty}(-1)^m \binom{\gamma}{m}\left(\frac{y^{\left(\frac{m+2\gamma-\gamma\alpha^2}{\alpha^2}\right)+1}}{\frac{m+2\gamma-\gamma\alpha^2}{\alpha^2}+1}\right)\Bigg|_0^1\right\}$$

$$= \frac{1}{1-\gamma}\log\left\{\left(\frac{6}{\alpha^2}\right)^\gamma \sum_{m=0}^{\infty}(-1)^m \binom{\gamma}{m}\left(\frac{\alpha^2}{m+2\gamma-\gamma\alpha^2+\alpha^2}\right)\right\}\ldots(24)$$

$$= \frac{1}{1-\gamma}\log\left\{6^\gamma \alpha^{-2(\gamma-1)}\sum_{m=0}^{\infty}(-1)^m \binom{\gamma}{m}(m+2\gamma-\gamma\alpha^2+\alpha^2)^{-1}\right\}\ldots(25)$$

**2.9. Mean residual life function:** see text

It is defined for the MBUR random variable as shown in equation (26). Taking the limits at the ends of the unit interval is shown in equation (27)

$$\mu(y|\alpha) = E(Y - y|Y > y) = \frac{1}{S(y)}\int_y^1 S(y)dy$$

$$= \frac{1}{S(y)}\int_y^1 1 - 3y^{\frac{2}{\alpha^2}} + 2y^{\frac{3}{\alpha^2}}dy = \frac{1}{S(y)}\left\{y - 3\left(\frac{y^{\frac{2}{\alpha^2}+1}}{\frac{2}{\alpha^2}+1}\right) + 2\left(\frac{y^{\frac{3}{\alpha^2}+1}}{\frac{3}{\alpha^2}+1}\right)\right\}\Bigg|_y^1$$

$$= \frac{1}{S(y)}\left[y - 3\left(\frac{y^{\frac{2}{\alpha^2}+1}}{\frac{2}{\alpha^2}+1}\right) + 2\left(\frac{y^{\frac{3}{\alpha^2}+1}}{\frac{3}{\alpha^2}+1}\right)\right]_y^1$$

$$= \frac{1}{S(y)}\left\{(1-y) - 3\left[\frac{1}{\frac{2}{\alpha^2}+1} - \frac{y^{\frac{2}{\alpha^2}+1}}{\frac{2}{\alpha^2}+1}\right] + 2\left[\frac{1}{\frac{3}{\alpha^2}+1} - \frac{y^{\frac{3}{\alpha^2}+1}}{\frac{3}{\alpha^2}+1}\right]\right\}$$

Rearrange the above equation:

$$= \frac{1}{S(y)}\left\{\left(1 - \frac{3}{\frac{2}{\alpha^2}+1} + \frac{2}{\frac{3}{\alpha^2}+1}\right) - y + 3\left(\frac{y^{\frac{2}{\alpha^2}+1}}{\frac{2}{\alpha^2}+1}\right) - 2\left(\frac{y^{\frac{3}{\alpha^2}+1}}{\frac{3}{\alpha^2}+1}\right)\right\}$$

$$= \frac{1}{S(y)} \left\{ \left(1 - \frac{3\alpha^2}{2+\alpha^2} + \frac{2\alpha^2}{3+\alpha^2}\right) - y + 3\left(\frac{\alpha^2 y^{\frac{2}{\alpha^2}+1}}{2+\alpha^2}\right) - 2\left(\frac{\alpha^2 y^{\frac{3}{\alpha^2}+1}}{3+\alpha^2}\right)\right\}$$

$$= \frac{1}{S(y)} \left\{ \left(1 - \frac{3\alpha^2}{2+\alpha^2} + \frac{2\alpha^2}{3+\alpha^2}\right) - y\left(1 - \frac{3\alpha^2 y^{\frac{2}{\alpha^2}}}{2+\alpha^2} + \frac{2\alpha^2 y^{\frac{3}{\alpha^2}}}{3+\alpha^2}\right)\right\}$$

$$= \frac{\left(1 - \frac{3\alpha^2}{2+\alpha^2} + \frac{2\alpha^2}{3+\alpha^2}\right) - y\left\{1 - \frac{3\alpha^2 y^{\frac{2}{\alpha^2}}}{2+\alpha^2} + \frac{3\alpha^2 y^{\frac{2}{\alpha^2}}}{3+\alpha^2}\right\}}{1 - 3y^{\frac{2}{\alpha^2}} + 2y^{\frac{3}{\alpha^2}}} \quad \ldots\ldots (26)$$

$$\lim_{y \to 0} \mu(y|\alpha) = 1 - \frac{3\alpha^2}{2+\alpha^2} + \frac{2\alpha^2}{3+\alpha^2} \quad \text{and} \quad \lim_{y \to 1} \mu(y|\alpha) = 0 \quad \ldots\ldots(27)$$

To derive the $\lim_{y \to 1} \mu(y|\alpha) = 0$, use L'Hopital rule

$$\frac{y + 3\left(\frac{y^{\frac{2}{\alpha^2}+1}}{\frac{2}{\alpha^2}+1}\right) - 2\left(\frac{y^{\frac{3}{\alpha^2}+1}}{\frac{3}{\alpha^2}+1}\right)}{1 - 3y^{\frac{2}{\alpha^2}} + 2y^{\frac{3}{\alpha^2}}}$$

$$\lim_{y \to 1} \frac{-1 + \left(3\left(\frac{2}{\alpha^2}+1\right)\frac{y^{\frac{2}{\alpha^2}}}{\frac{2}{\alpha^2}+1}\right) - \left(2\left(\frac{3}{\alpha^2}+1\right)\frac{y^{\frac{3}{\alpha^2}+1}}{\frac{3}{\alpha^2}+1}\right)}{-3\frac{2}{\alpha^2} y^{\frac{2}{\alpha^2}-1} + 2\frac{3}{\alpha^2} y^{\frac{3}{\alpha^2}-1}}$$

$$\lim_{y \to 1} \frac{-1 + (3) - (2)}{-3\frac{2}{\alpha^2} y^{\frac{2}{\alpha^2}-1} + 2\frac{3}{\alpha^2} y^{\frac{3}{\alpha^2}-1}} = \frac{0}{-3\frac{2}{\alpha^2} + 2\frac{3}{\alpha^2}} = \frac{0}{0}$$

Use L'Hopital for the second time on the following equation:

$$\lim_{y \to 1} \frac{-1 + \left(3y^{\frac{2}{\alpha^2}}\right) - \left(2y^{\frac{3}{\alpha^2}+1}\right)}{-3\frac{2}{\alpha^2} y^{\frac{2}{\alpha^2}-1} + 2\frac{3}{\alpha^2} y^{\frac{3}{\alpha^2}-1}}$$

$$\lim_{y \to 1} \frac{+\left(3\left(\frac{2}{\alpha^2}\right) y^{\frac{2}{\alpha^2}-1}\right) - \left(2\left(\frac{3}{\alpha^2}\right) y^{\frac{3}{\alpha^2}-1}\right)}{\frac{-6}{\alpha^2}\left(\frac{2}{\alpha^2}-1\right) y^{\frac{2}{\alpha^2}-1} + \frac{6}{\alpha^2}\left(\frac{3}{\alpha^2}-1\right) y^{\frac{3}{\alpha^2}-1}} = \frac{0}{\frac{-6}{\alpha^2}\left\{\frac{2}{\alpha^2}-1-\frac{3}{\alpha^2}+1\right\}}$$

$$\lim_{y \to 1} \frac{+\left(3\left(\frac{2}{\alpha^2}\right) y^{\frac{2}{\alpha^2}-1}\right) - \left(2\left(\frac{3}{\alpha^2}\right) y^{\frac{3}{\alpha^2}-1}\right)}{\frac{-6}{\alpha^2}\left(\frac{2}{\alpha^2}-1\right) y^{\frac{2}{\alpha^2}-1} + \frac{6}{\alpha^2}\left(\frac{3}{\alpha^2}-1\right) y^{\frac{3}{\alpha^2}-1}} = \frac{0}{\frac{-6}{\alpha^2}\left\{\frac{2}{\alpha^2}-\frac{3}{\alpha^2}\right\}} = \frac{0}{\frac{6}{\alpha^4}} = 0$$

## 2.10. Stochastic Ordering: (see the text)

This ordering judges the comparative conduct of a variable. A random variable X is considered smaller than the random variable Y in the following orders:

1. Stochastic order ($X \leq_{st} Y$) if $F_X(x) \geq F_Y(x)$ for all x.
2. Hazard rate order ($X \leq_{hr} Y$) if $h_X(x) \geq h_Y(x)$ for all x.
3. Mean residual life order ($X \leq_{mrl} Y$) if $m_X(x) \leq m_Y(x)$ for all x.
4. Likelihood ratio order ($X \leq_{lr} Y$) if $\frac{f_X(x)}{f_Y(x)}$ decreases in x.

The following results are due to Shaked and Shanthikumar. They used the results to evaluate the stochastic ordering of a distribution:

$$X \leq_{lr} Y \Rightarrow X \leq_{hr} Y \Rightarrow X \leq_{mrl} Y \quad hence: X \leq_{st} Y$$

Theorem: let $X \sim MBUR(\alpha_1)$ and $Y \sim MBUR(\alpha_2)$ if $\alpha_1 < \alpha_2$, then

$X \leq_{lr} Y$, hence $X \leq_{hr} Y$, $X \leq_{mrl} Y$ and $X \leq_{st} Y$.

Proof: see equation (28)

$$\frac{f_X(y; \alpha_1)}{f_Y(y; \alpha_2)} = \frac{\frac{6}{\alpha_1^2}\left[1 - y^{\frac{1}{\alpha_1^2}}\right] y^{\left(\frac{2}{\alpha_1^2}-1\right)}}{\frac{6}{\alpha_2^2}\left[1 - y^{\frac{1}{\alpha_2^2}}\right] y^{\left(\frac{2}{\alpha_2^2}-1\right)}} \quad \ldots (28)$$

Hint: X and Y both have the same PDF of MBUR but with different alpha, the pdfs are evaluated at the same value let's call it y and so the differentiation of the pdf will yield the same function but with different parameter and substitute the same value. So the differentiation is taken with respect to say Y variable is the same as the differentiation with the X variable, the only difference is the alpha and then substitute with y value in both function because the evaluation is at the same value.

Taking the log of equation (28) gives equation (29)

$$\ln\left(\frac{6}{\alpha_1^2}\right) + \ln\left(1 - y^{\frac{1}{\alpha_1^2}}\right) + \ln\left[y^{\left(\frac{2}{\alpha_1^2}-1\right)}\right] - \ln\left(\frac{6}{\alpha_2^2}\right) - \ln\left[1 - y^{\frac{1}{\alpha_2^2}}\right] - \ln\left[y^{\left(\frac{2}{\alpha_2^2}-1\right)}\right] \ldots (29)$$

Taking the first derivative of the likelihood ratio order with respect to the variable y as shown in equation (30):

$$\frac{d\,LR}{dy} = \frac{2}{y}\left\{\frac{1}{\alpha_1^2} - \frac{1}{\alpha_2^2}\right\} + \frac{1}{y}\left\{\frac{\frac{-1}{\alpha_1^2} y^{\frac{1}{\alpha_1^2}}}{1 - y^{\frac{1}{\alpha_1^2}}} + \frac{\frac{1}{\alpha_2^2} y^{\frac{1}{\alpha_2^2}}}{1 - y^{\frac{1}{\alpha_2^2}}}\right\} \quad \ldots\ldots\ldots (30)$$

How to derive equation 30: first take the derivative of equation 29

$$\frac{d}{dy} = \frac{-\alpha_1^{-2} y^{\alpha_1^{-2}-1}}{1 - y^{\alpha_1^{-2}}} + \frac{(2\alpha_1^{-2} - 1)}{y} - \frac{-\alpha_2^{-2} y^{\alpha_2^{-2}-1}}{1 - y^{\alpha_2^{-2}}} - \frac{(2\alpha_2^{-2} - 1)}{y} \quad \ldots (30.A)$$

Second rearrange equation 30.A

$$\frac{d}{dy} = y^{-1}(2\alpha_1^{-2} - 1 - 2\alpha_2^{-2} + 1) + y^{-1}\left(\frac{-\alpha_1^{-2} y^{\alpha_1^{-2}}}{1 - y^{\alpha_1^{-2}}} + \frac{\alpha_2^{-2} y^{\alpha_2^{-2}}}{1 - y^{\alpha_2^{-2}}}\right) \ldots (30.B)$$

$$\frac{d}{dy} = \frac{2}{y}(\alpha_1^{-2} - \alpha_2^{-2}) + \frac{1}{y}\left(\frac{-\alpha_1^{-2}y^{\alpha_1^{-2}}}{1-y^{\alpha_1^{-2}}} + \frac{\alpha_2^{-2}y^{\alpha_2^{-2}}}{1-y^{\alpha_2^{-2}}}\right) \ldots\ldots (30.C)$$

Taking L'Hohpital's rule on equation 30.C:

$$= \frac{2}{y}(\alpha_1^{-2} - \alpha_2^{-2}) + \frac{1}{y}\left(\frac{-\alpha_1^{-2}\alpha_1^{-2}y^{\alpha_1^{-2}-1}}{-\alpha_1^{-2}y^{\alpha_1^{-2}-1}} + \frac{\alpha_2^{-2}\alpha_2^{-2}y^{\alpha_2^{-2}-1}}{-\alpha_2^{-2}y^{\alpha_2^{-2}-1}}\right) \ldots . (30.D)$$

Taking the limit of equation 30.D:

$$= \lim_{y \to 1}\left\{\frac{2}{y}(\alpha_1^{-2} - \alpha_2^{-2}) + \frac{1}{y}\left(\frac{-\alpha_1^{-2}\alpha_1^{-2}y^{\alpha_1^{-2}-1}}{-\alpha_1^{-2}y^{\alpha_1^{-2}-1}} + \frac{\alpha_2^{-2}\alpha_2^{-2}y^{\alpha_2^{-2}-1}}{-\alpha_2^{-2}y^{\alpha_2^{-2}-1}}\right)\right\} .. (30.E)$$

$$= \lim_{y \to 1}\left\{\frac{2}{y}(\alpha_1^{-2} - \alpha_2^{-2}) + \frac{1}{y}\left(\frac{\alpha_1^{-2}}{1} + \frac{\alpha_2^{-2}}{-1}\right)\right\}$$

$$= \lim_{y \to 1}\left\{\frac{2}{y}(\alpha_1^{-2} - \alpha_2^{-2}) + \frac{1}{y}(\alpha_1^{-2} - \alpha_2^{-2})\right\} = \lim_{y \to 1}\left\{\frac{3}{y}(\alpha_1^{-2} - \alpha_2^{-2})\right\}$$

$$\lim_{y \to 1 \text{ or } 0}\left\{\frac{3}{y}(\alpha_1^{-2} - \alpha_2^{-2})\right\} = \lim_{y \to 1 \text{ or } 0} LRO = \ldots\ldots .. (31)$$

$$\lim_{y \to 1}\left\{\frac{3}{y}(\alpha_1^{-2} - \alpha_2^{-2})\right\} = 3\left\{\frac{1}{\alpha_1^2} - \frac{1}{\alpha_2^2}\right\} \text{ and } \lim_{y \to 0}\left\{\frac{3}{y}(\alpha_1^{-2} - \alpha_2^{-2})\right\} = +\infty$$

Where LRO is the likelihood ratio order

# Supplementary materials (section 2)

# ALPHA LEVEL =2.5

Table (1): characteristics of empirical distribution of estimated alpha using MOM

| MOM | n=20 | n=80 | n=160 | n=260 | n=500 |
|---|---|---|---|---|---|
| 2.5 Q | 1.9635 | 2.2062 | 2.2705 | 2.3197 | 2.3647 |
| 97.5 Q | 3.5095 | 2.9080 | 2.7614 | 2.7171 | 2.6581 |
| Skewness | 0.8381 | 0.4247 | 0.1649 | 0.2798 | 0.2249 |
| Kurtosis | 4.3346 | 3.2498 | 2.9082 | 3.0278 | 3.1826 |
| Fit N (5%) | $H_0$=1 (0.001) | $H_0$=1 (0.0053) | $H_0$=0 (0.4899) | $H_0$=0 (0.1637) | $H_0$=0 (0.2444) |
| Fit N (1%) | $H_0$=1 (0.001) | $H_0$=1 (0.0053 | $H_0$=0 (0.4899) | $H_0$=0 (0.1637) | $H_0$=0 (0.2444) |

The empirical distribution of the estimated parameter alpha using MOM is shown in table 1. Each column represents a specific sample size with 1000 replicates in each size. The 2.5 th quantile and the 97.5 th quantile of the 1000 values of the estimated parameter in each sample shows that as the sample size increases the 2.5 quantile rises while the 97.5 quantile decreases. In other words, the distance between the two quantiles decreases as the sample size increases and this is reflected on the confidence interval (CI). As the sample size increases the CI becomes narrower. The distribution exhibits a mild right skewness and a high positive excess kurtosis (leptokurtic shape) at small sample size. As sample size increases the skewness decreases trying to approach the zero level (skewness of standard normal) and kurtosis decreases trying to approach the kurtosis of standard normal. The empirical distribution fits standard normal starting at size 160 and larger than this at significance level 5% and 1% with associated P-value as shown in the table. $H_0$=1 means reject the null hypothesis that states the parameter distribution follows the standard normal distribution. While $H_0$=0 means fail to reject the null hypothesis. See the following Figures (1-4)

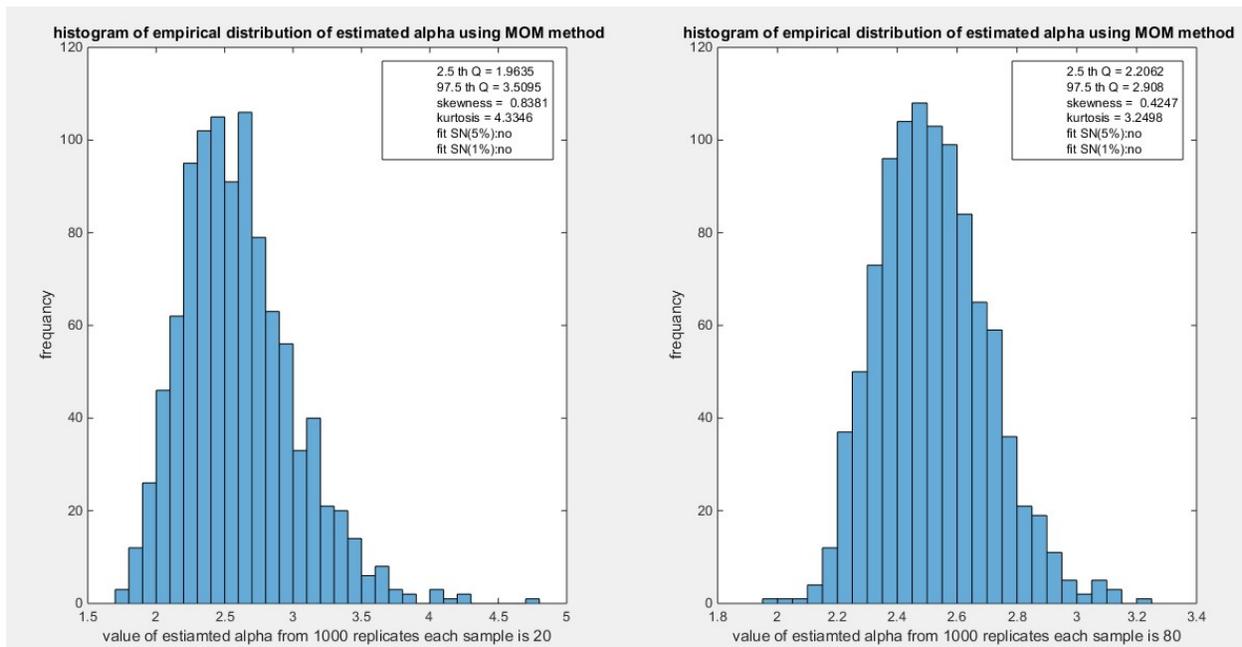

Fig.1 shows the histogram of the empirical distribution of the estimated alpha from the 1000 replicates with sample size (n=20) on the left and (n=80) on the right using MOM method

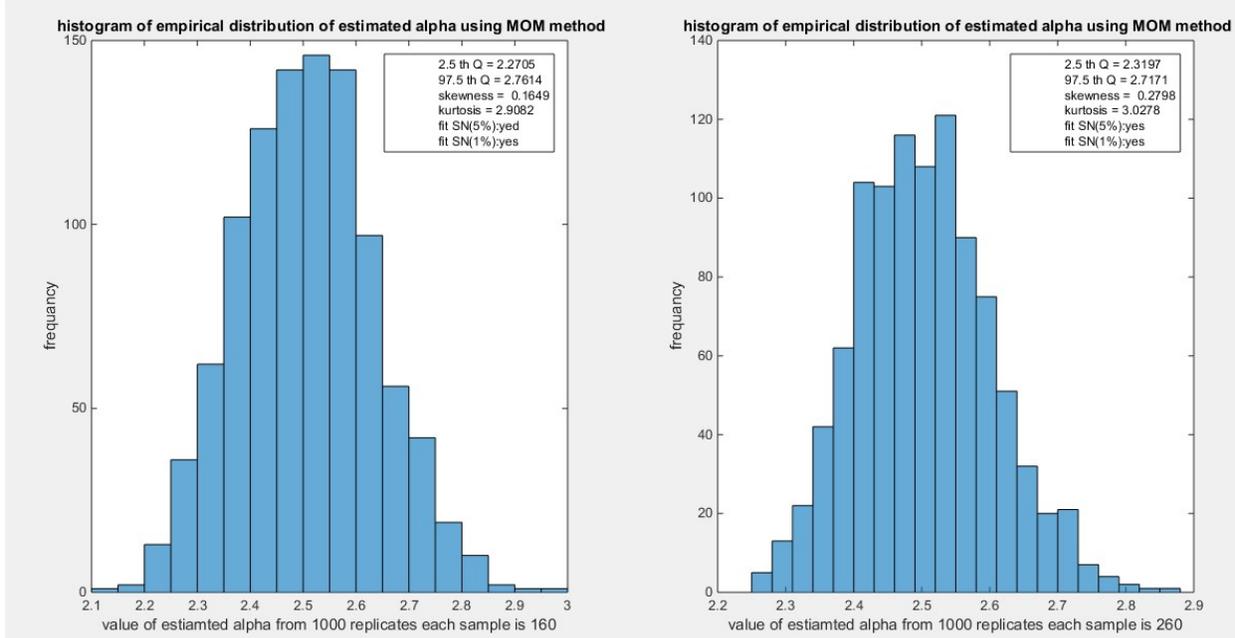

Fig.2 shows the histogram of the empirical distribution of the estimated alpha from the 1000 replicates with sample size (n=160) on the left and (n=260) on the right using MOM method

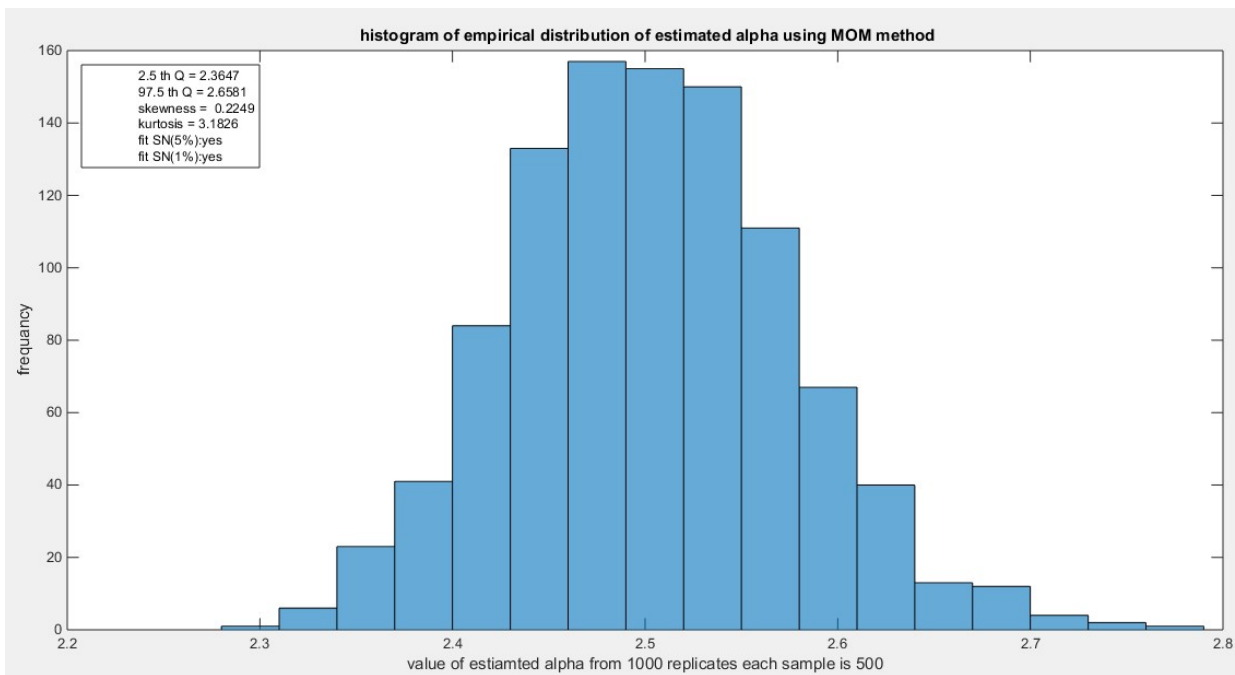

Fig. 3 shows the histogram of the empirical distribution of the estimated alpha from the 1000 replicates with sample size (n=500) using MOM method

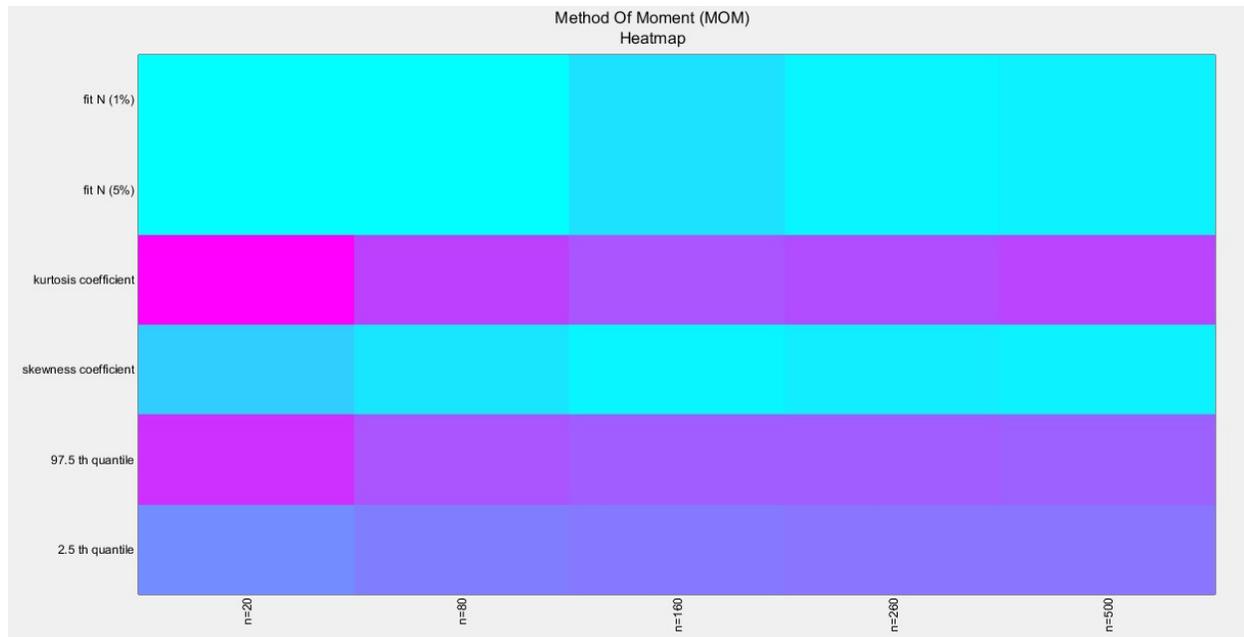

Fig.4 shows the heat map of the indices of the empirical distribution of the estimated alpha using (MOM) method and how these indices change with changing the sample size from 20 to 500. (p values are shown)

Table (2): characteristics of empirical distribution of estimated alpha using MLE

| MLE | n=20 | n=80 | n=160 | n=260 | n=500 |
|---|---|---|---|---|---|
| 2.5 Q | 1.8751 | 2.252 | 2.3332 | 2.3707 | 2.4151 |
| 97.5 Q | 2.7839 | 2.6506 | 2.6181 | 2.6006 | 2.5779 |
| Skewness | -1.2026 | -0.5556 | -0.4021 | -0.3011 | -0.1154 |
| Kurtosis | 5.6182 | 3.1668 | 3.3037 | 2.9514 | 3.0945 |
| Fit N (5%) | $H_0=1$ (0.001) | $H_0=1$ (0.001) | $H_0=1$ (0.001) | $H_0=0$ (0.1318) | $H_0=0$ (0.3172) |
| Fit N (1%) | $H_0=1$ (0.001) | $H_0=1$ (0.001) | $H_0=1$ (0.001) | $H_0=0$ (0.1318) | $H_0=0$ (0.3172) |

The empirical distribution of the estimated parameter alpha using MLE is shown in Table 2. Each column represents a specific sample size with 1000 replicates in each size. The 2.5 th quantile and the 97.5 th quantile of the 1000 values of the estimated parameter in each sample shows that as the sample size increases the 2.5 quantile rises while the 97.5 quantile decreases. In other words, the distance between the two quantiles decreases as the sample size increases and this is reflected on the confidence interval (CI). As the sample size increases the CI becomes narrower. The distribution exhibits a moderate left skewness and a high positive excess kurtosis (leptokurtic shape) at small sample size. As sample size increases the skewness decreases trying to approach the zero level (skewness of standard normal) and kurtosis decreases trying to approach the kurtosis of standard normal. The empirical distribution fits standard normal starting at size 260 and larger than this at significance level 5% and 1% with associated P-value as shown in the table. $H_0=1$ means reject the null hypothesis that states the parameter distribution follows the standard normal distribution. While $H_0=0$ means fail to reject the null hypothesis. The test used is the Lillietest in all the following tables. See following Figures (5-8)

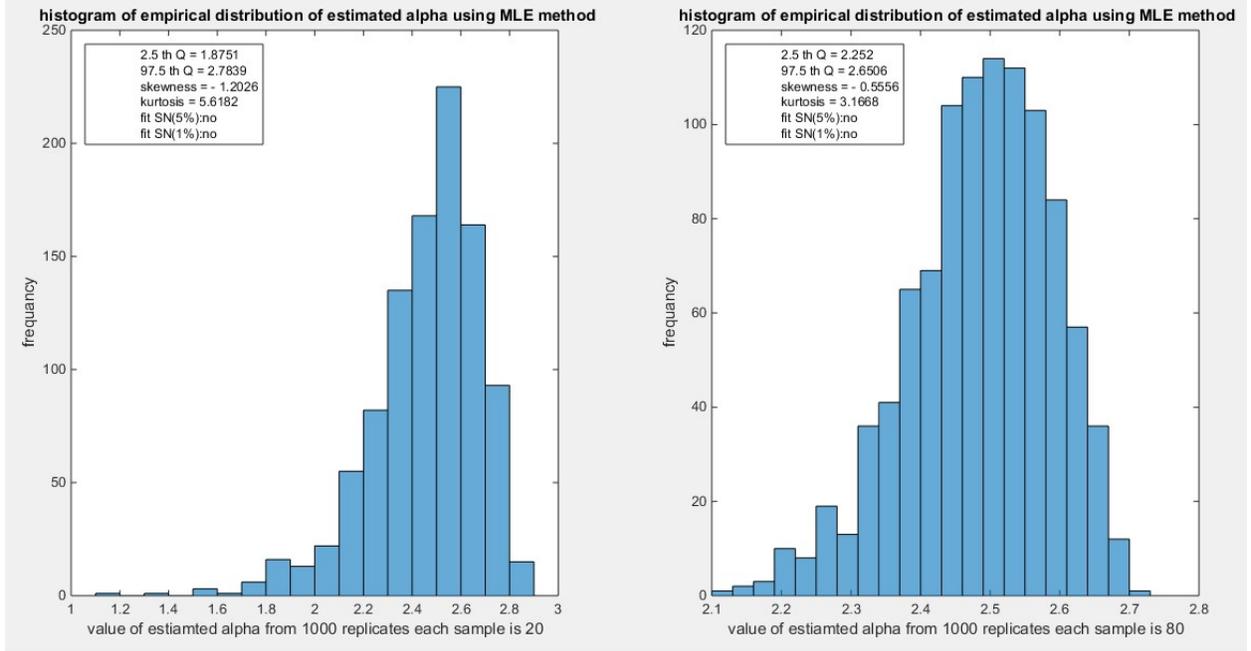

Fig. 5 shows the histogram of the empirical distribution of the estimated alpha from the 1000 replicates with sample size (n=20) on the left and (n=80) on the right using MLE method

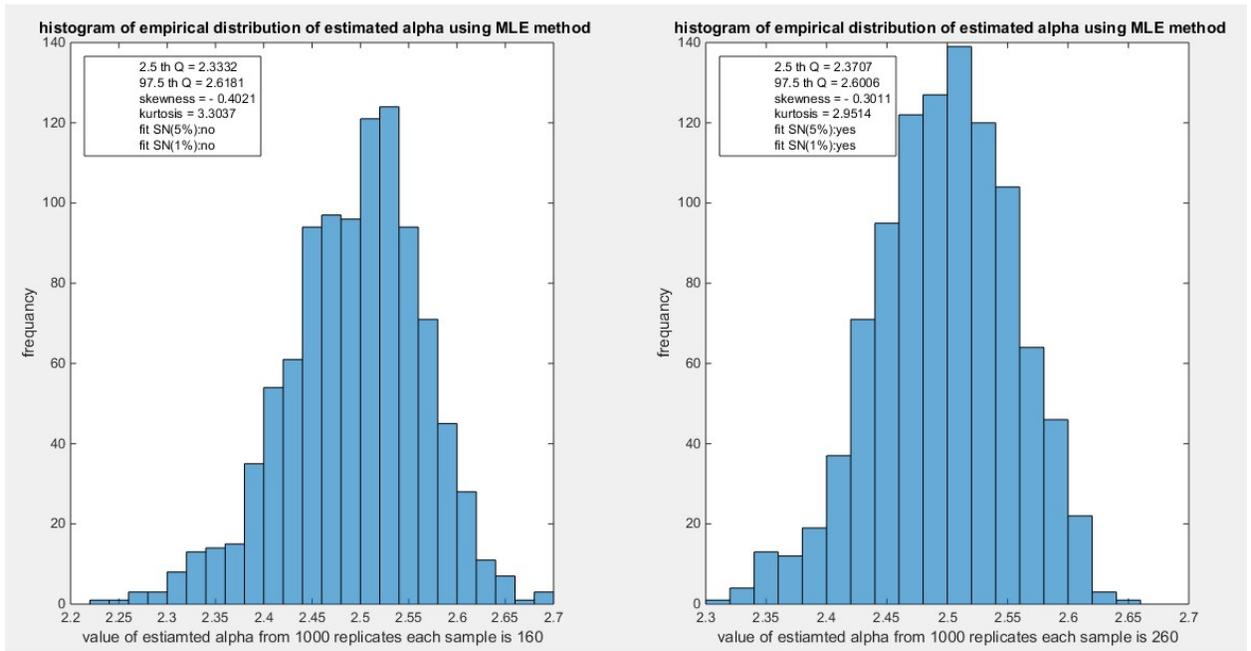

Fig.6 shows the histogram of the empirical distribution of the estimated alpha from the 1000 replicates with sample size (n=160) on the left and (n=260) on the right using MLE method

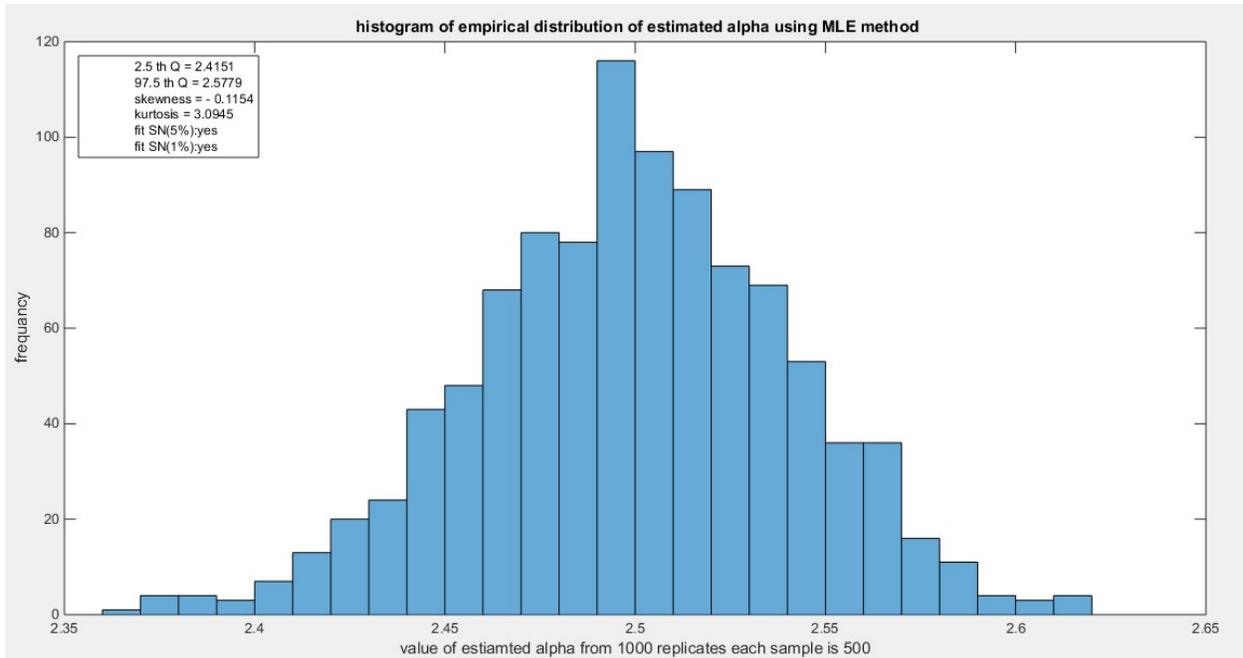

Fig.7 shows the histogram of the empirical distribution of the estimated alpha from the 1000 replicates with sample size (n=500) using MLE method

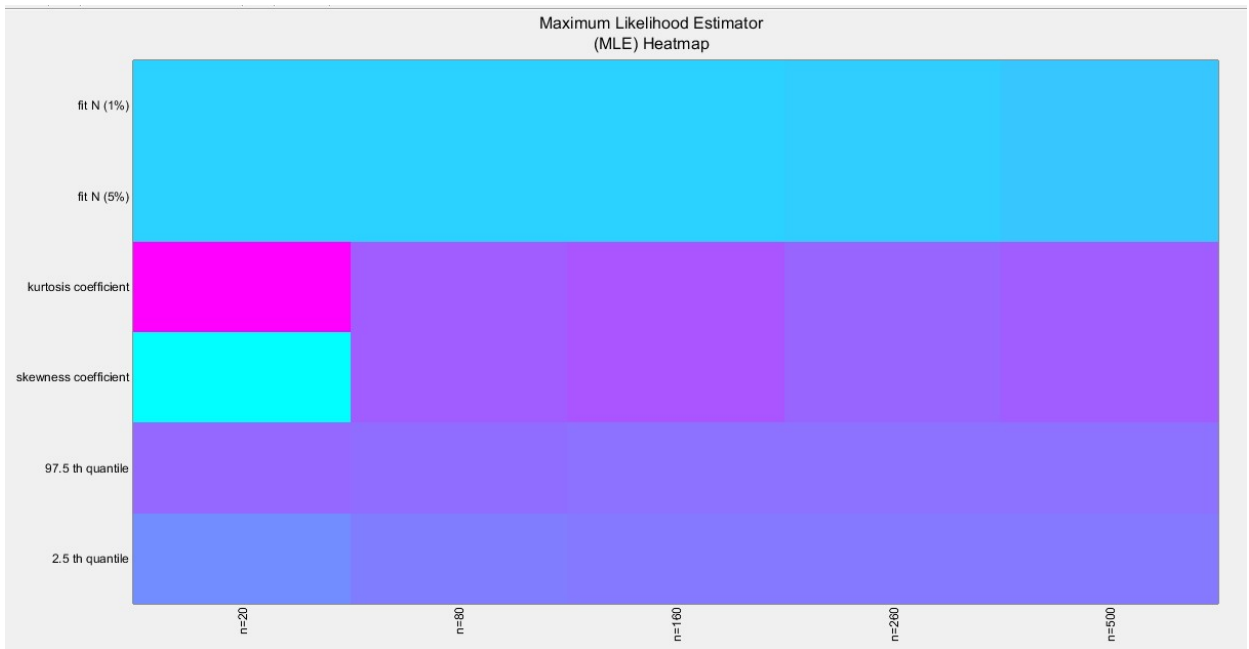

Fig.8 shows the heat map of the indices of the empirical distribution of the estimated alpha using (MLE) method and how these indices change with changing the sample size from 20 to 500. (p values are shown)

Table (3): characteristics of empirical distribution of estimated alpha using MPS

| MPS | n=20 | n=80 | n=160 | n=260 | n=500 |
|---|---|---|---|---|---|
| 2.5 Q | 2.1020 | 2.278 | 2.3455 | 2.3797 | 2.4192 |
| 97.5 Q | 2.9075 | 2.6649 | 2.6247 | 2.6056 | 2.5818 |
| Skewness | -0.4654 | -0.5367 | -0.3972 | -0.2999 | -0.1177 |
| Kurtosis | 3.8161 | 3.1241 | 3.2465 | 2.9476 | 3.1013 |
| Fit N (5%) | $H_0=1$ (0.0066) | $H_0=1$ (0.001) | $H_0=1$ (0.001) | $H_0=0$ (0.0663) | $H_0=0$ (0.3267) |
| Fit N (1%) | $H_0=1$ (0.0066) | $H_0=1$ (0.001) | $H_0=1$ (0.001) | $H_0=0$ (0.0663) | $H_0=0$ (0.3267) |

The empirical distribution of the estimated parameter alpha using MPS is shown in Table 3. Each column represents a specific sample size with 1000 replicates in each size. The 2.5 th quantile and the 97.5 th quantile of the 1000 values of the estimated parameter in each sample shows that as the sample size increases the 2.5 quantile rises while the 97.5 quantile decreases. In other words, the distance between the two quantiles decreases as the sample size increases and this is reflected on the confidence interval (CI). As the sample size increases the CI becomes narrower. The distribution exhibits a mild left skewness and a mild positive excess kurtosis (leptokurtic shape) at small sample size. As sample size increases the skewness decreases trying to approach the zero level (skewness of standard normal) and kurtosis decreases trying to approach the kurtosis of standard normal. The empirical distribution fits standard normal starting at size 260 and larger than this at significance level 5% and 1% with associated P-value as shown in the table. $H_0=1$ means reject the null hypothesis that states the parameter distribution follows the standard normal distribution. While $H_0=0$ means fail to reject the null hypothesis. See following Figures (9-12).

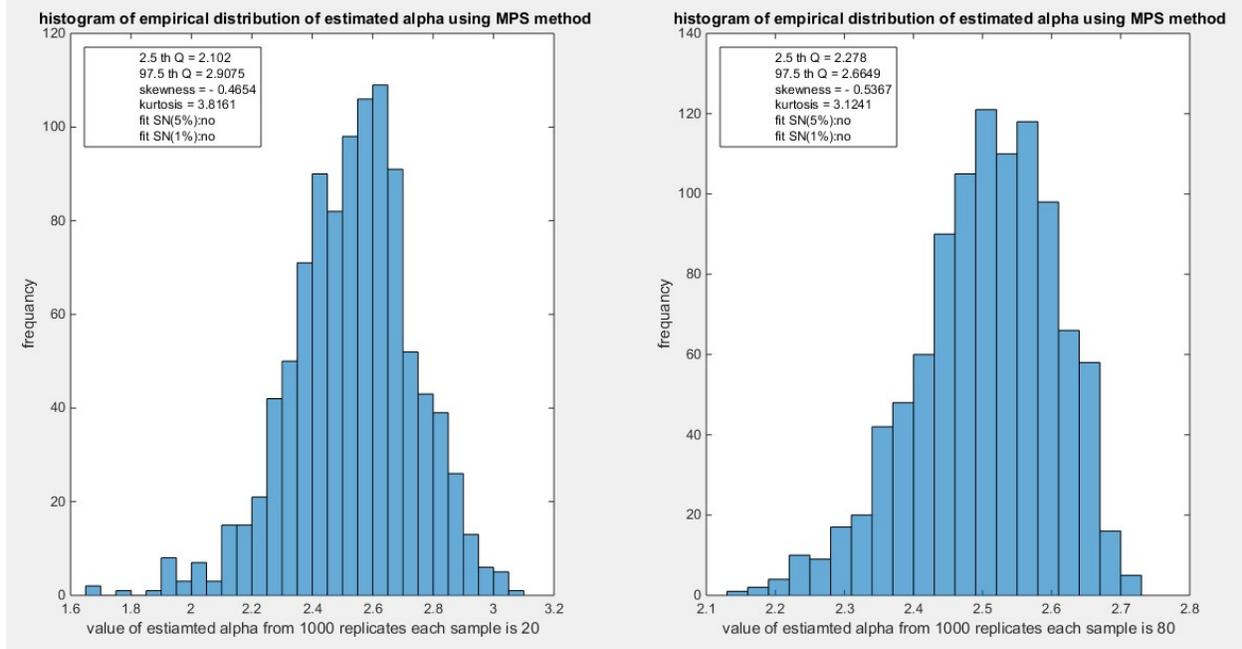

Fig.9 shows the histogram of the empirical distribution of the estimated alpha from the 1000 replicates with sample size (n=20) on the left and (n=80) on the right using MPS method

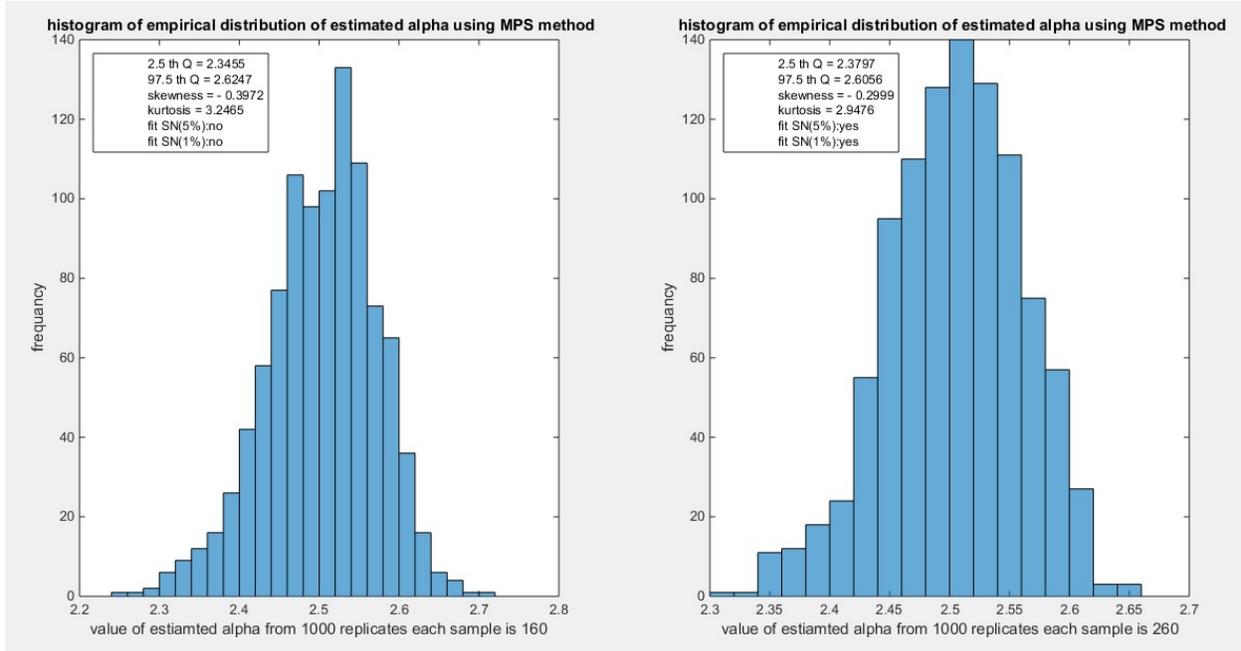

Fig.10 shows the histogram of the empirical distribution of the estimated alpha from the 1000 replicates with sample size (n=160) on the left and (n=260) on the right using MPS method

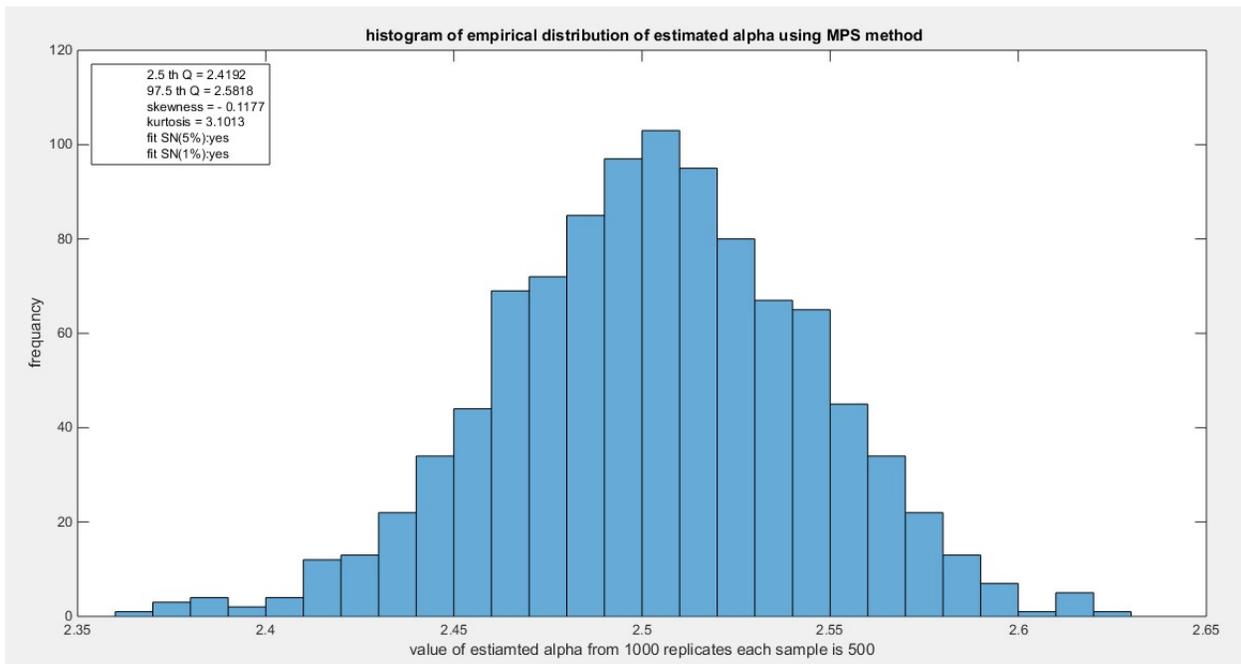

Fig. 11 shows the histogram of the empirical distribution of the estimated alpha from the 1000 replicates with sample size (n=500) using MPS method

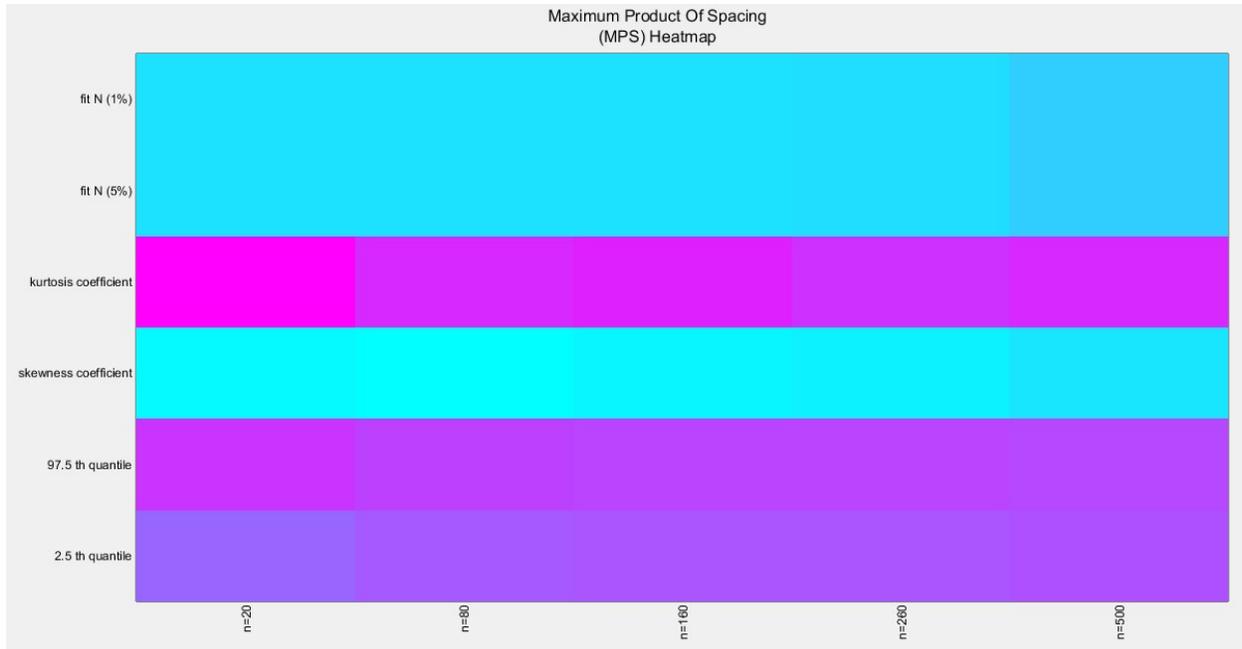

Fig.12 shows the heat map of the indices of the empirical distribution of the estimated alpha using (MPS) method and how these indices change with changing the sample size from 20 to 500.(p values are shown)

Table (4): characteristics of empirical distribution of estimated alpha using AD

| AD | n=20 | n=80 | n=160 | n=260 | n=500 |
|---|---|---|---|---|---|
| 2.5 Q | 1.9191 | 2.245 | 2.3306 | 2.3754 | 2.4105 |
| 97.5 Q | 2.8090 | 2.6669 | 2.6243 | 2.6072 | 2.5793 |
| Skewness | -1.0612 | -0.6927 | -0.3771 | -0.2281 | -0.0795 |
| Kurtosis | 4.8814 | 3.7495 | 3.3002 | 2.9371 | 2.9561 |
| Fit N (5%) | $H_0=1$ (0.001) | $H_0=1$ (0.001) | $H_0=1$ (0.0107) | $H_0=0$ (0.2142) | $H_0=0$ (0.5) |
| Fit N (1%) | $H_0=1$ (0.001) | $H_0=1$ (0.001) | $H_0=1$ (0.0107) | $H_0=0$ (0.2142) | $H_0=0$ (0.5) |

The empirical distribution of the estimated parameter alpha using AD is shown in Table 4. Each column represents a specific sample size with 1000 replicates in each size. The 2.5 th quantile and the 97.5 th quantile of the 1000 values of the estimated parameter in each sample shows that as the sample size increases the 2.5 quantile rises while the 97.5 quantile decreases. In other words, the distance between the two quantiles decreases as the sample size increases and this is reflected on the confidence interval (CI). As the sample size increases the CI becomes narrower. The distribution exhibits a moderate left skewness and a moderate positive excess kurtosis (leptokurtic shape) at small sample size. As sample size increases the skewness decreases trying to approach the zero level (skewness of standard normal) and kurtosis decreases trying to approach the kurtosis of standard normal. The empirical distribution fits standard normal starting at size 260 and larger than this at significance level 5% and 1% with associated P-value as shown in the table. $H_0=1$ means reject the null hypothesis that states the parameter distribution follows the standard normal distribution. While $H_0=0$ means fail to reject the null hypothesis. See the following Figures (13-16).

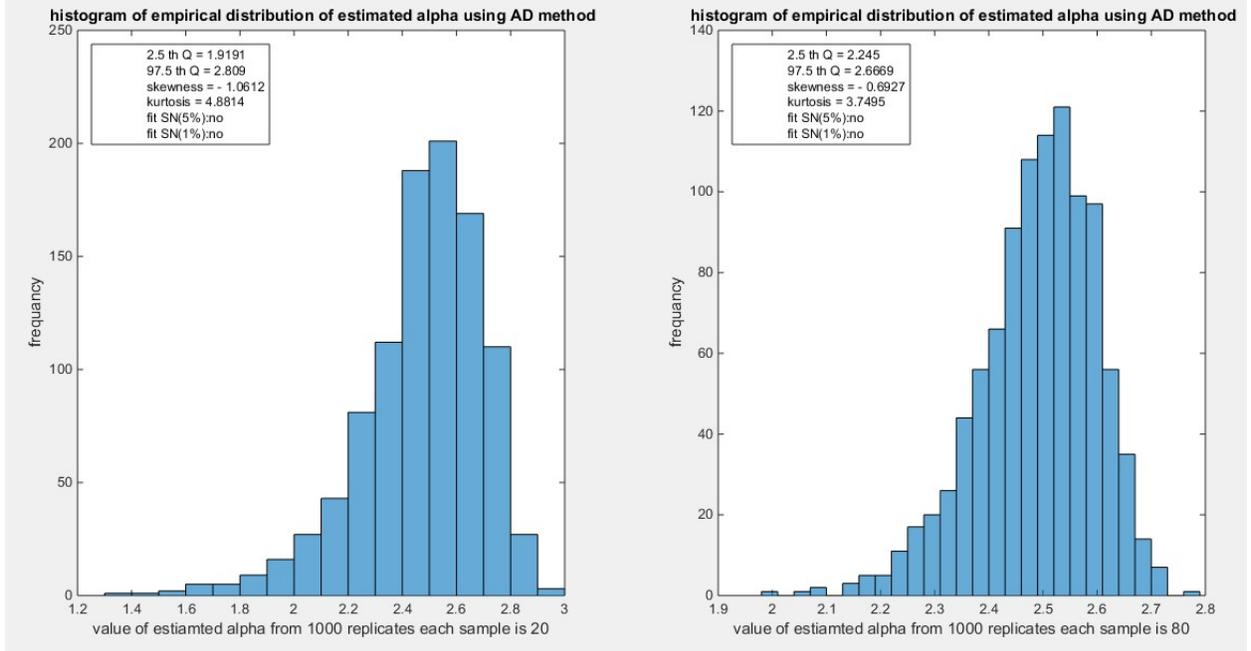

Fig. 13 shows the histogram of the empirical distribution of the estimated alpha from the 1000 replicates with sample size (n=20) on the left and (n=80) on the right using AD method

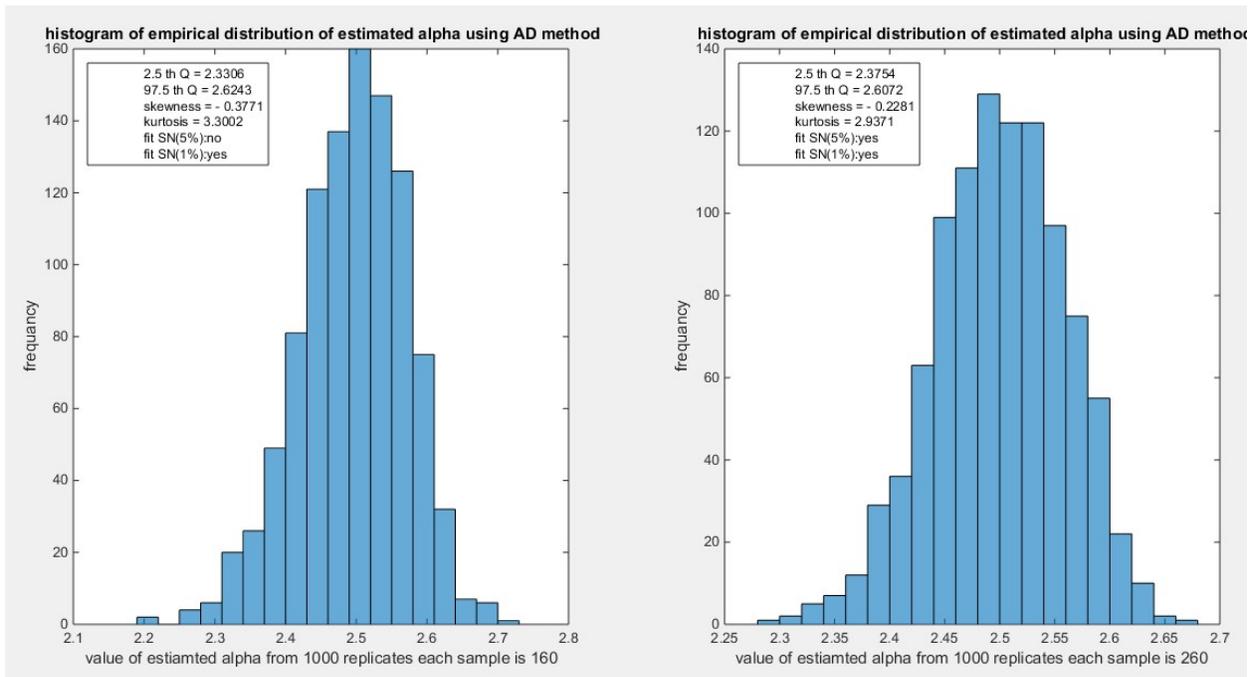

Fig. 14 shows the histogram of the empirical distribution of the estimated alpha from the 1000 replicates with sample size (n=160) on the left and (n=260) on the right using AD method

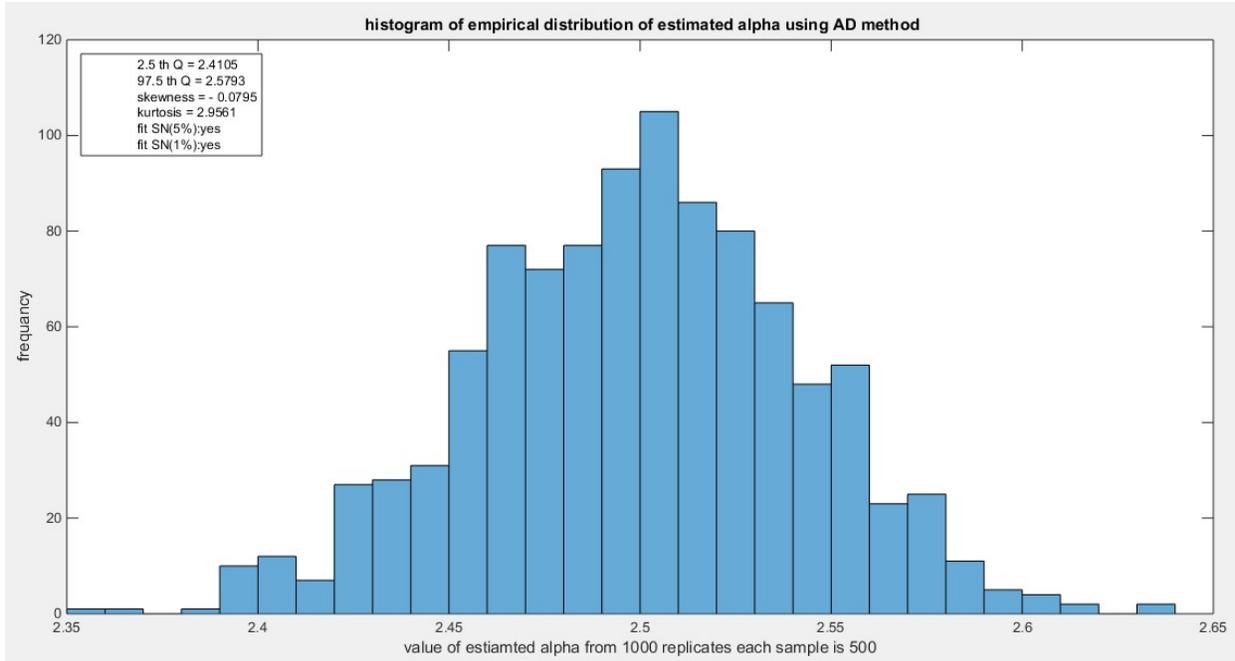

Fig. 15 shows the histogram of the empirical distribution of the estimated alpha from the 1000 replicates with sample size (n=500) using AD method

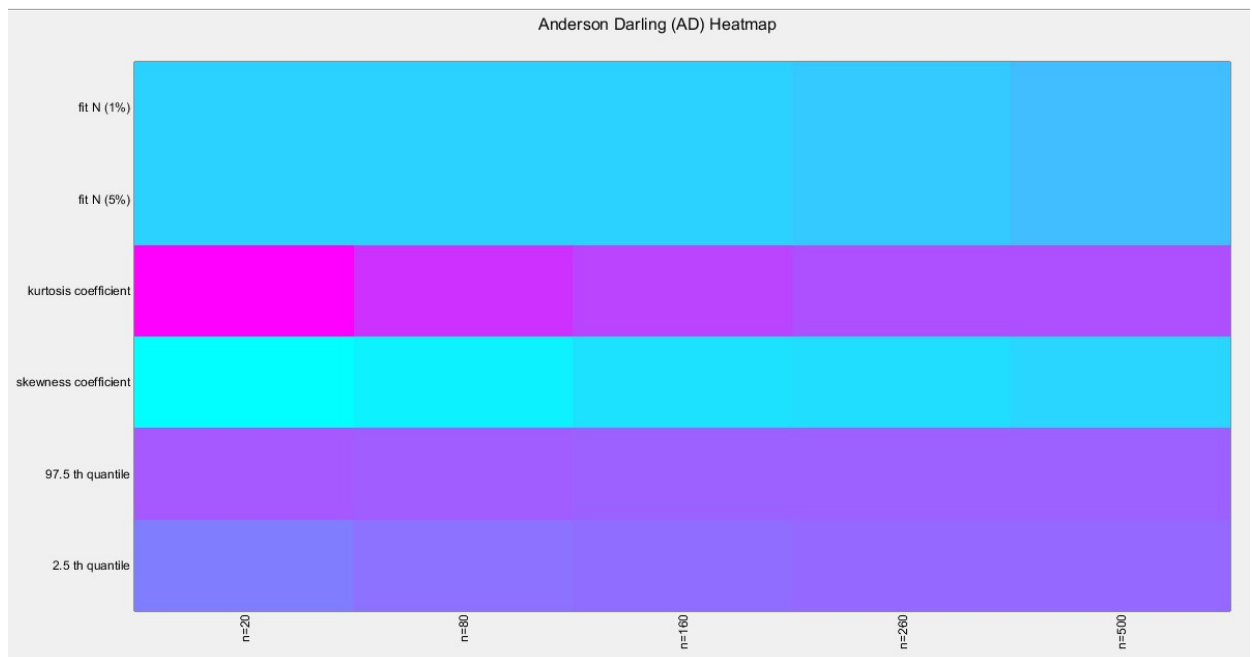

Fig. 16 shows the heat map of the indices of the empirical distribution of the estimated alpha using (AD) method and how these indices change with changing the sample size from 20 to 500. (p=values are shown)

Table (5): characteristics of empirical distribution of estimated alpha using CVM

| CVM | n=20 | n=80 | n=160 | n=260 | n=500 |
|---|---|---|---|---|---|
| 2.5 Q | 1.8245 | 2.2331 | 2.3265 | 2.3754 | 2.4088 |
| 97.5 Q | 2.896 | 2.6826 | 2.6384 | 2.6137 | 2.5853 |
| Skewness | -2.1132 | -0.7597 | -0.2994 | -0.1571 | -0.0392 |
| Kurtosis | 16.6523 | 4.7673 | 3.3267 | 2.9688 | 2.9101 |
| Fit N (5%) | $H_0$=1 (0.001) | $H_0$=1 (0.001) | $H_0$=0 (0.1544) | $H_0$=0 (0.326) | $H_0$=0 (0.5) |
| Fit N (1%) | $H_0$=1 (0.001) | $H_0$=1 (0.001) | $H_0$=0 (0.1544) | $H_0$=0 (0.326) | $H_0$=0 (0.5) |

The empirical distribution of the estimated parameter alpha using CVM is shown in Table 5. Each column represents a specific sample size with 1000 replicates in each size. Each column depicts the characteristics of the empirical distribution of the estimated alpha. The 2.5 th quantile and the 97.5 th quantile of the 1000 values of the estimated parameter in each sample shows that as the sample size increases the 2.5 th quantile rises while the 97.5 th quantile decreases. In other words, the distance between the two quantiles decreases as the sample size increases and this is reflected on the confidence interval (CI). As the sample size increases the CI becomes narrower. The distribution exhibits a moderate left skewness and a high positive excess kurtosis (leptokurtic shape) at small sample size. As sample size increases the skewness decreases trying to approach the zero level (skewness of standard normal) and kurtosis decreases trying to approach the kurtosis of standard normal. The empirical distribution fits standard normal starting at size 160 and larger than this at significance level 5% and 1% with associated P-value as shown in the table. $H_0$=1 means reject the null hypothesis that states the parameter distribution follows the standard normal distribution. While $H_0$=0 means fail to reject the null hypothesis. See the following Figures (17-20).

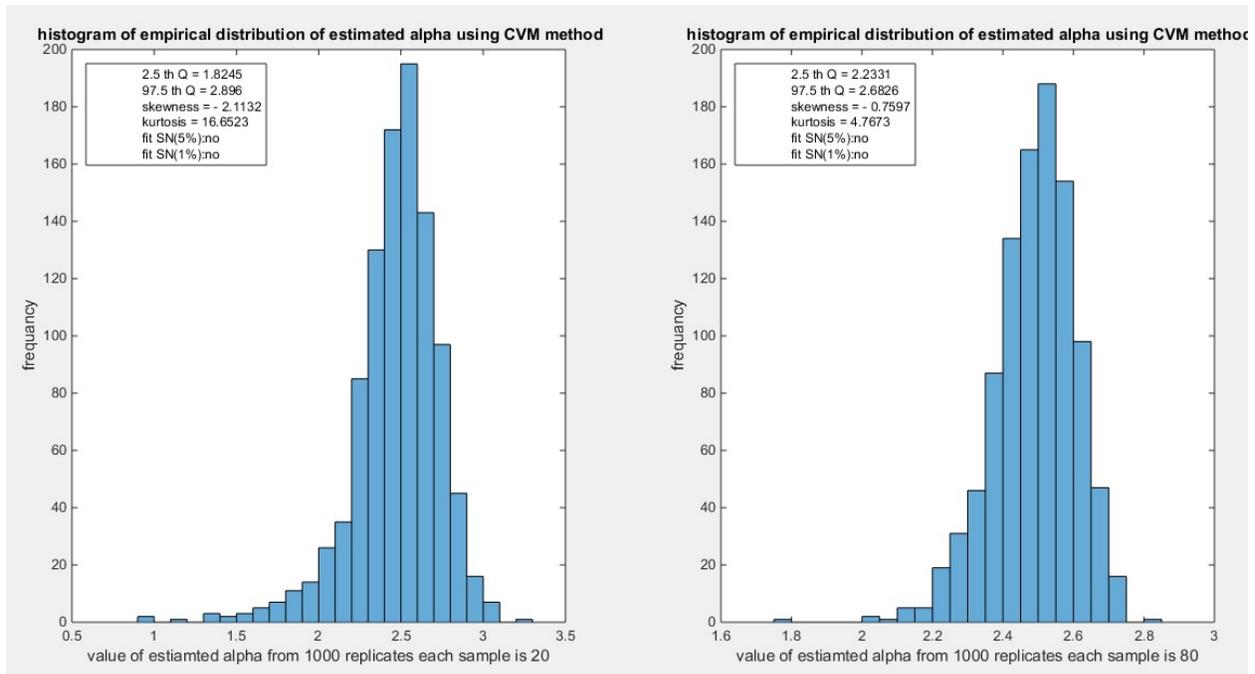

Fig. 17 shows the histogram of the empirical distribution of the estimated alpha from the 1000 replicates with sample size (n=20) on the left and (n=80) on the right using CVM method

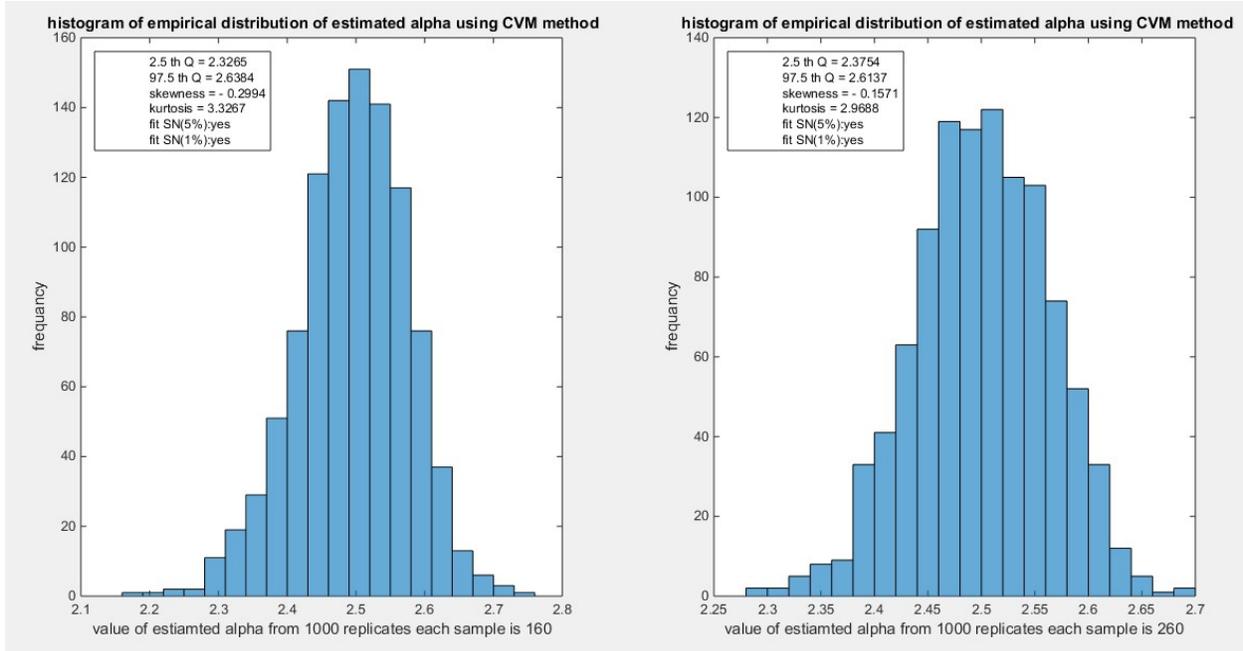

Fig. 18 shows the histogram of the empirical distribution of the estimated alpha from the 1000 replicates with sample size (n=160) on the left and (n=260) on the right using CVM method

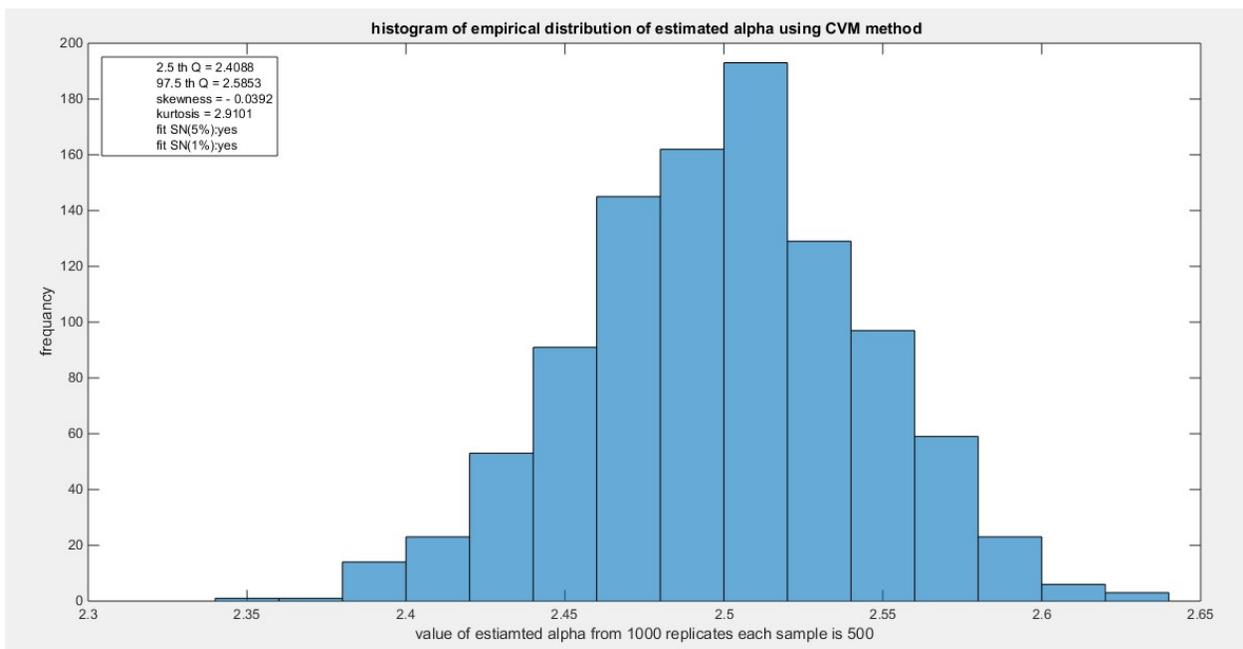

Fig.19 shows the histogram of the empirical distribution of the estimated alpha from the 1000 replicates with sample size (n =500) using CVM method

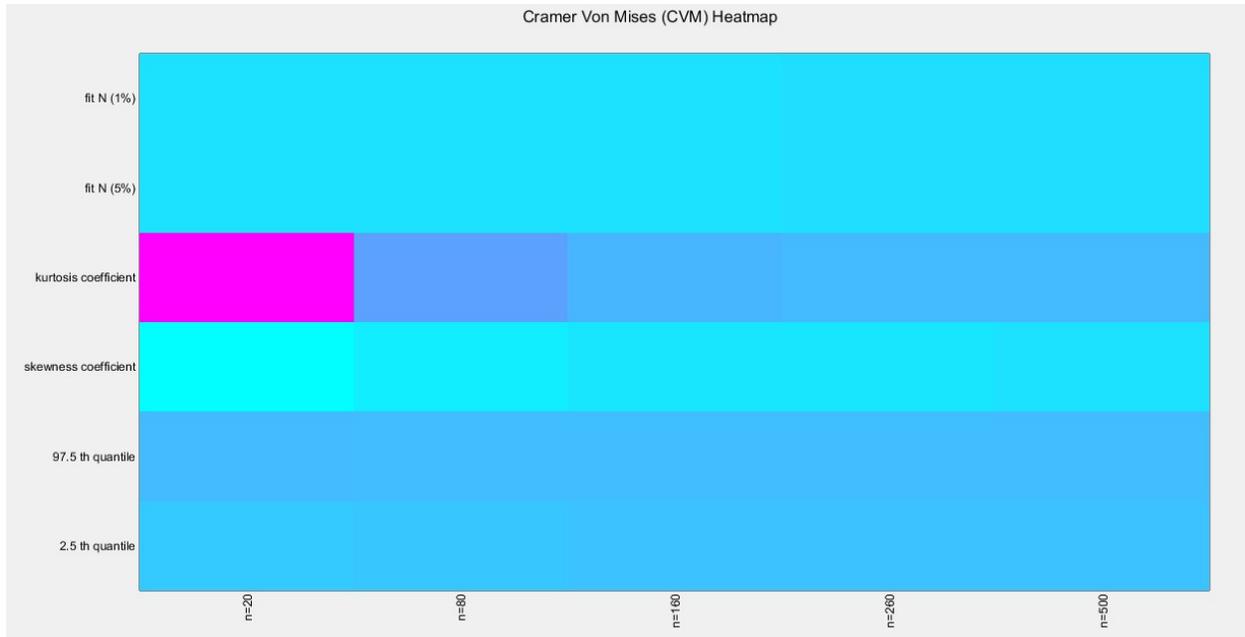

Fig. 20 shows the heat map of the indices of the empirical distribution of the estimated alpha using (CVM) method and how these indices change with changing the sample size from 20 to 500. (p value shown)

Table (6): characteristics of empirical distribution of estimated alpha using LS

| LS | n=20 | n=80 | n=160 | n=260 | n=500 |
|---|---|---|---|---|---|
| 2.5 Q | 1.8885 | 2.2345 | 2.3275 | 2.3758 | 2.4112 |
| 97.5 Q | 2.9203 | 2.6842 | 2.6390 | 2.6142 | 2.5855 |
| Skewness | -1.1974 | -0.7647 | -0.2988 | -0.1567 | -0.0182 |
| Kurtosis | 7.1553 | 4.7783 | 3.3263 | 2.9686 | 2.8689 |
| Fit N (5%) | $H_0=1$ (0.001) | $H_0=1$ (0.001) | $H_0=0$ (0.1454) | $H_0=0$ (0.3256) | $H_0=0$ (0.5) |
| Fit N (1%) | $H_0=1$ (0.001) | $H_0=1$ (0.001) | $H_0=0$ (0.1454) | $H_0=0$ (0.3256) | $H_0=0$ (0.5) |

The empirical distribution of the estimated parameter alpha using CVM is shown in Table 6. Each column represents a specific sample size with 1000 replicates in each size. Each column depicts the characteristics of the empirical distribution of the estimated alpha. The 2.5 th quantile and the 97.5 th quantile of the 1000 values of the estimated parameter in each sample shows that as the sample size increases the 2.5 th quantile rises while the 97.5 th quantile decreases. In other words, the distance between the two quantiles decreases as the sample size increases and this is reflected on the confidence interval (CI). As the sample size increases the CI becomes narrower. The distribution exhibits a moderate left skewness and a high positive excess kurtosis (leptokurtic shape) at small sample size. As sample size increases the skewness decreases trying to approach the zero level (skewness of standard normal) and kurtosis decreases trying to approach the kurtosis of standard normal. The empirical distribution fits standard normal starting at size 160 and larger than this at significance level 5% and 1% with associated P-value as shown in the table. $H_0=1$ means reject the null hypothesis that states the parameter distribution follows the standard normal distribution. While $H_0=0$ means fail to reject the null hypothesis. See following Figures (21-24)

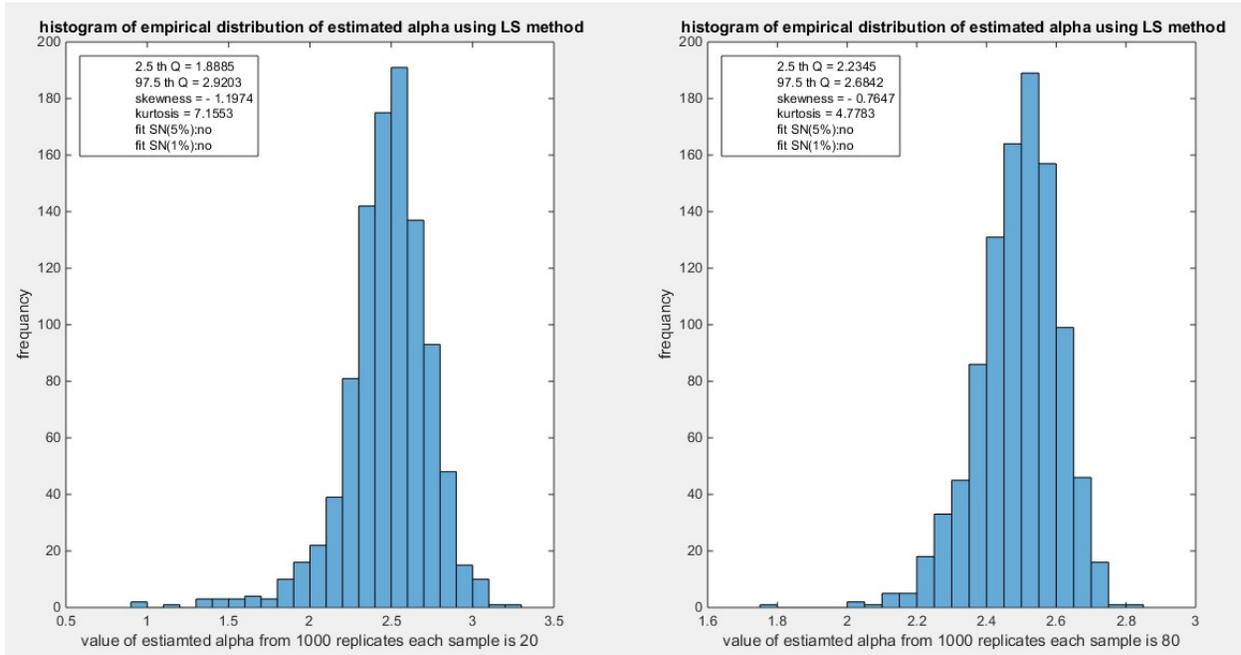

Fig. 21 shows the histogram of the empirical distribution of the estimated alpha from the 1000 replicates with sample size (n=20) on the left and (n=80) on the right using LS method

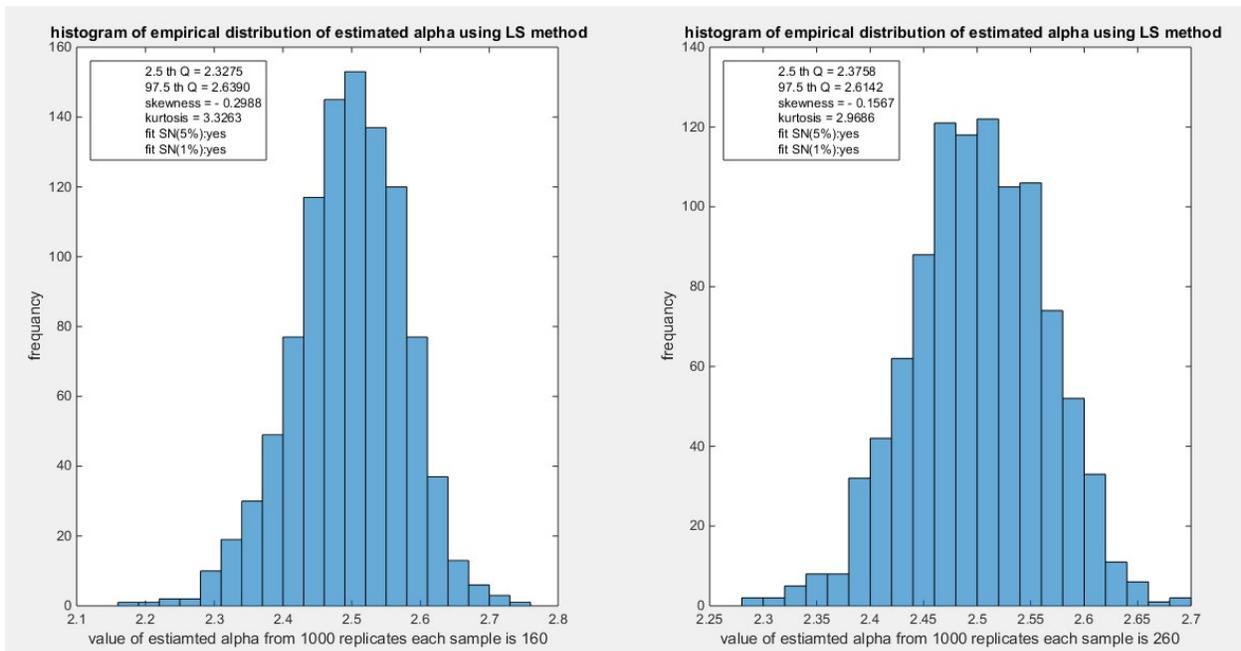

Fig.22 shows the histogram of the empirical distribution of the estimated alpha from the 1000 replicates with sample size (n=160) on the left and (n=260) on the right using LS method

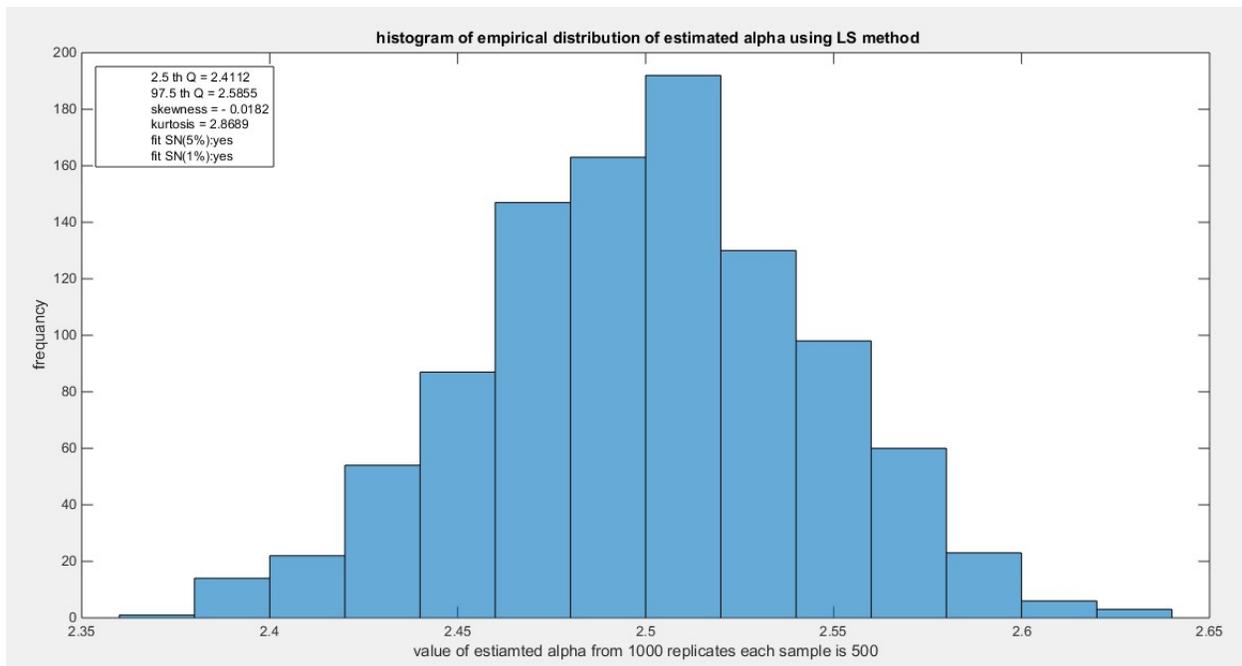

Fig.23 shows the histogram of the empirical distribution of the estimated alpha from the 1000 replicates with sample size (n =500) using LS method

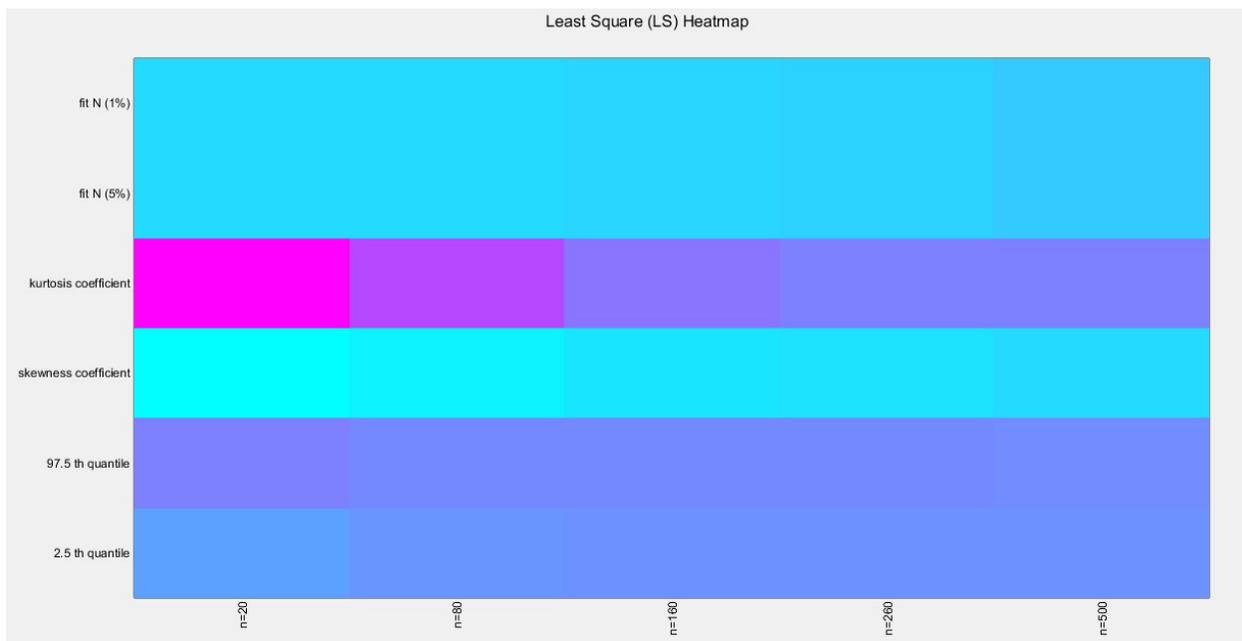

Fig.24 shows the heat map of the indices of the empirical distribution of the estimated alpha using (LS) method and how these indices change with changing the sample size from 20 to 500. (p values are shown)

Table (7): characteristics of empirical distribution of estimated alpha using percentile

| PERCENTILE | n=20 | n=80 | n=160 | n=260 | n=500 |
|---|---|---|---|---|---|
| 2.5 Q | 1.5756 | 2.0858 | 2.1745 | 2.2574 | 2.3239 |
| 97.5 Q | 3.0543 | 2.8417 | 2.7483 | 2.7020 | 2.6533 |
| Skewness | -0.2107 | 0.0299 | -0.1196 | 0.0412 | 0.0296 |
| Kurtosis | 2.6405 | 2.8171 | 2.8136 | 2.8258 | 3.0752 |
| Fit N (5%) | $H_0$=1 (0.0035) | $H_0$=0 (0.5) | $H_0$=0 (0.4261) | $H_0$=0 (0.5) | $H_0$=0 (0.3554) |
| Fit N (1%) | $H_0$=1 (0.0035) | $H_0$=0 (0.5) | $H_0$=0 (0.4261) | $H_0$=0 (0.5) | $H_0$=0 (0.3554) |

The empirical distribution of the estimated parameter alpha using percentile method is shown in Table 7. Each column represents a specific sample size with 1000 replicates in each size. Each column depicts the characteristics of the empirical distribution of the estimated alpha. The 2.5 th quantile and the 97.5 th quantile of the 1000 values of the estimated parameter in each sample shows that as the sample sizes increase the 2.5 th quantile rises while the 97.5 th quantile decreases. In other words, the distance between the two quantiles decreases as the sample size increases and this is reflected on the confidence interval (CI). As the sample size increases the CI becomes narrower. As sample size increases the distribution exhibits oscillation between a mild left skewness and a mild right skewness. In contrast, it exhibits gradual mild increase in negative excess kurtosis (platykurtic shape) as sample size rises and it approaches mesokurtic shape (kertosis coefficient=3) of the standard normal at sample size 500. The empirical distribution fits standard normal starting at size 80 and larger than this size at significance level 5% and 1% with associated P-value as shown in the table. $H_0$=1 means reject the null hypothesis that states the parameter distribution follows the standard normal distribution. While $H_0$=0 means fail to reject the null hypothesis. See following Figures (25-28)

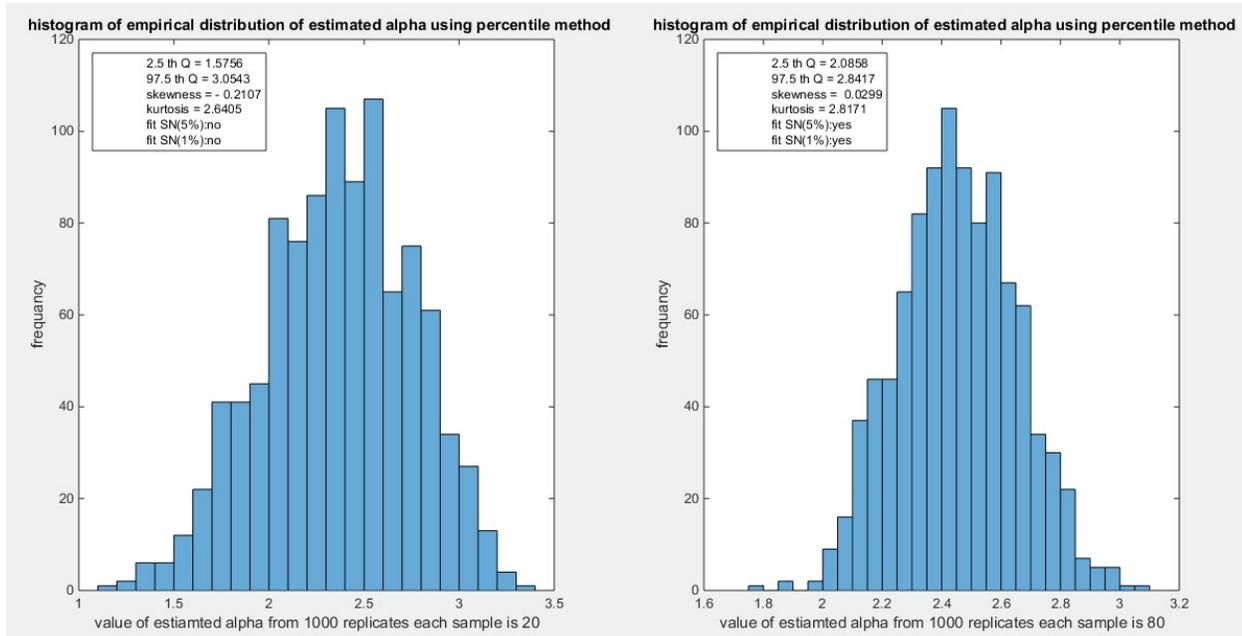

Fig. 25 shows the histogram of the empirical distribution of the estimated alpha from the 1000 replicates with sample size (n=20) on the left and (n=80) on the right using percentile method

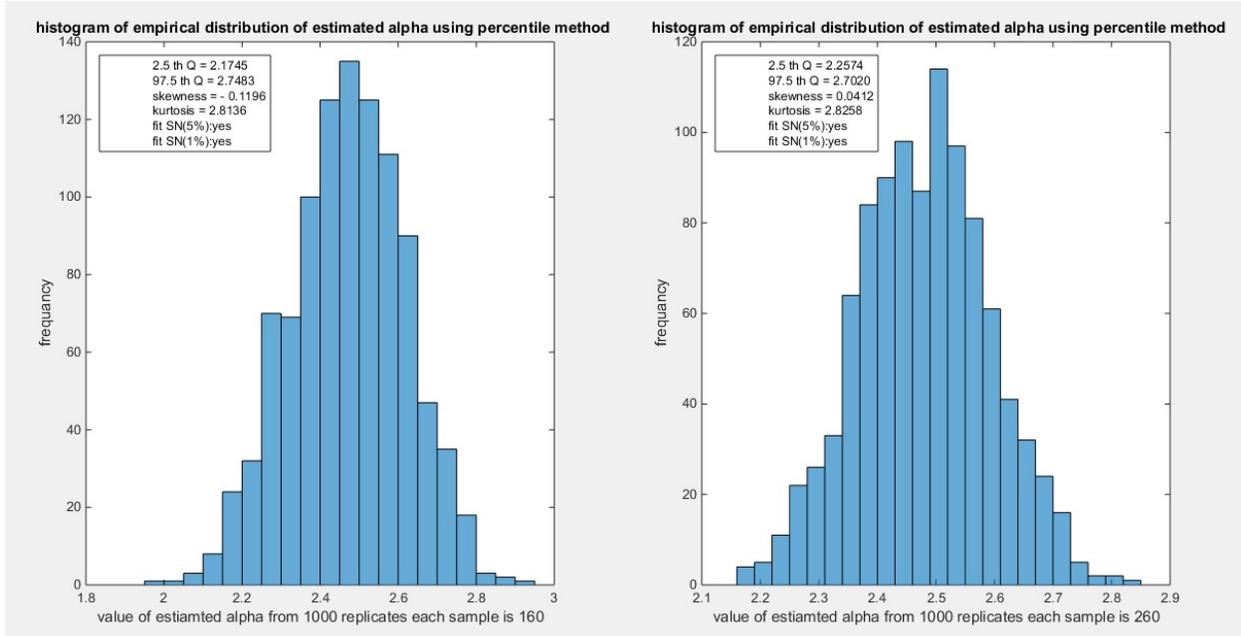

Fig. 26 shows the histogram of the empirical distribution of the estimated alpha from the 1000 replicates with sample size (n=160) on the left and (n=260) on the right using percentile method

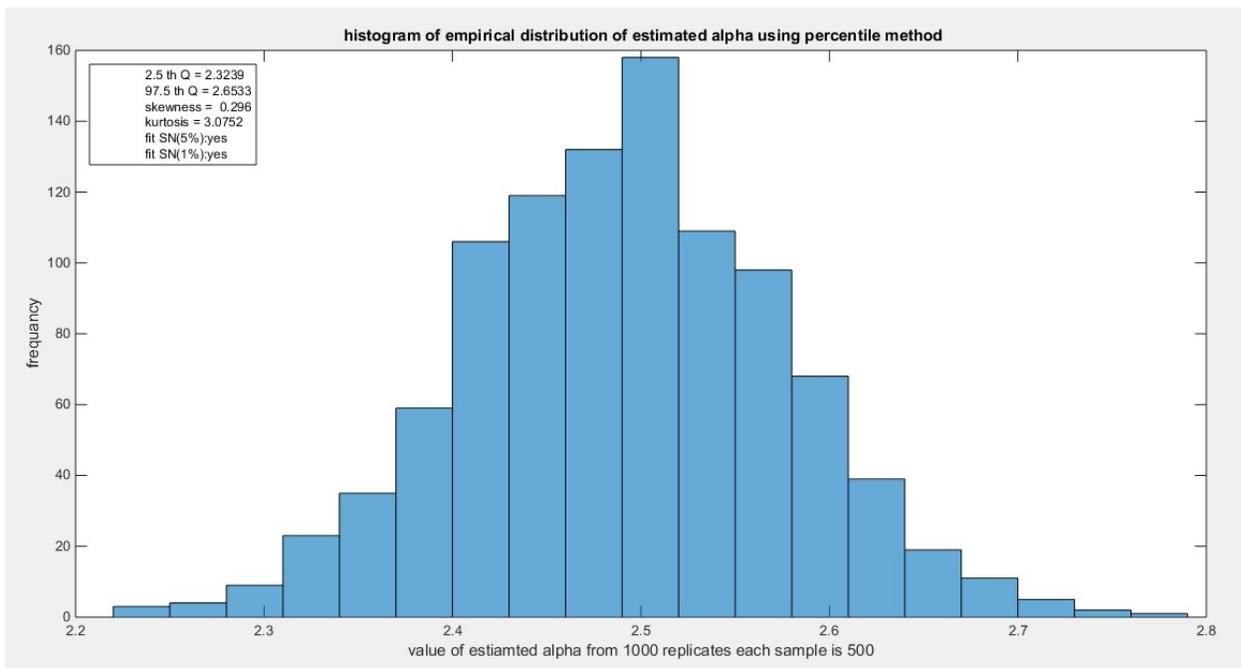

Fig. 27 shows the histogram of the empirical distribution of the estimated alpha from the 1000 replicates with sample size (n =500) using percentile method

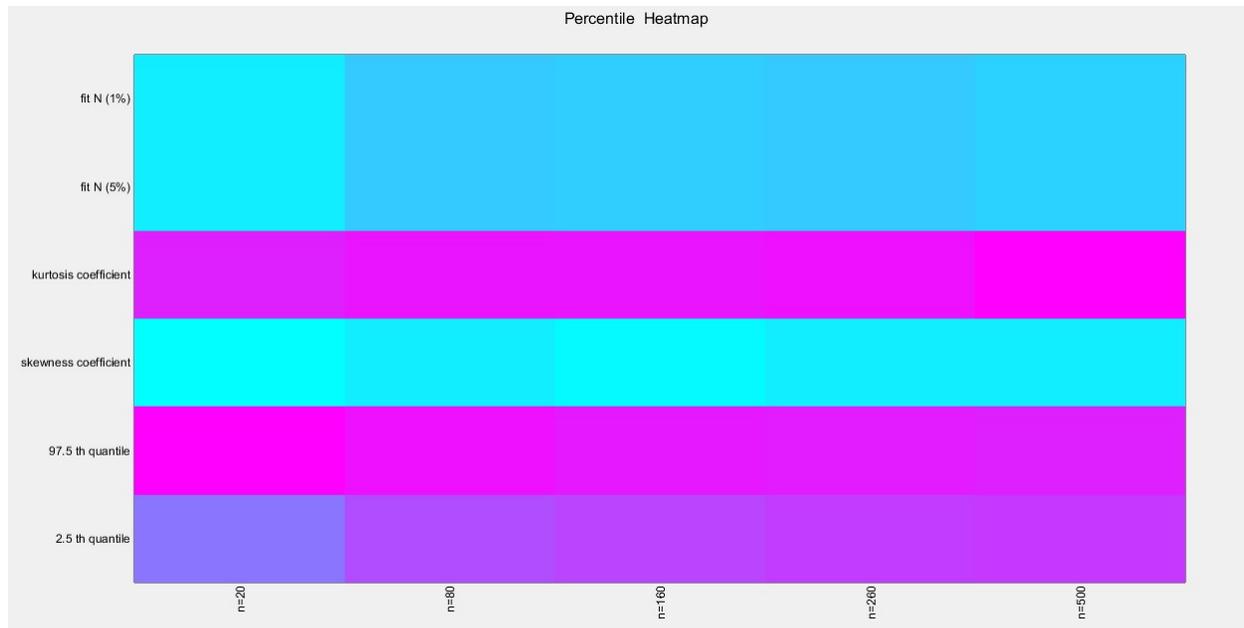

Fig. 28 shows the heat map of the indices of the empirical distribution of the estimated alpha using percentile method and how these indices change with changing the sample size from 20 to 500. (p values are shown)

Table (8): characteristics of empirical distribution of estimated alpha using WLS

| WLS | n=20 | n=80 | n=160 | n=260 | n=500 |
|---|---|---|---|---|---|
| 2.5 Q | 1.9476 | 2.2571 | 2.343 | 2.383 | 2.4152 |
| 97.5 Q | 2.8815 | 2.6832 | 2.6321 | 2.6121 | 2.5834 |
| Skewness | -0.8475 | -0.5139 | -0.1645 | -0.0862 | 0.0423 |
| Kurtosis | 5.3083 | 3.6886 | 3.1475 | 2.8469 | 2.9564 |
| Fit N (5%) | $H_0=1$ (0.001) | $H_0=1$ (0.001) | $H_0=0$ (0.5) | $H_0=0$ (0.5) | $H_0=0$ (0.5) |
| Fit N (1%) | $H_0=1$ (0.001) | $H_0=1$ (0.001) | $H_0=0$ (0.5) | $H_0=0$ (0.5) | $H_0=0$ (0.5) |

The empirical distribution of the estimated parameter alpha using WLS method is shown in Table 8. Each column represents a specific sample size with 1000 replicates in each size. Each column depicts the characteristics of the empirical distribution of the estimated alpha. The 2.5 th quantile and the 97.5 th quantile of the 1000 values of the estimated parameter in each sample shows that as the sample size increases the 2.5 th quantile rises while the 97.5 th quantile decreases. In other words, the distance between the two quantiles decreases as the sample size increases and this is reflected on the confidence interval (CI). As the sample size increases the CI becomes narrower. The distribution exhibits a mild left skewness and a high positive excess kurtosis (leptokurtic shape) at small sample size. As sample size increases the skewness decreases trying to approach the zero level (skewness of standard normal) and kurtosis decreases trying to approach the kurtosis of standard normal. The empirical distribution fits standard normal starting at size 160 and larger than this size at significance level 5% and 1% with associated P-value as shown in the table. $H_0=1$ means reject the null hypothesis that states the parameter distribution follows the standard normal distribution. While $H_0=0$ means fail to reject the null hypothesis. See the following Figures (29-32)

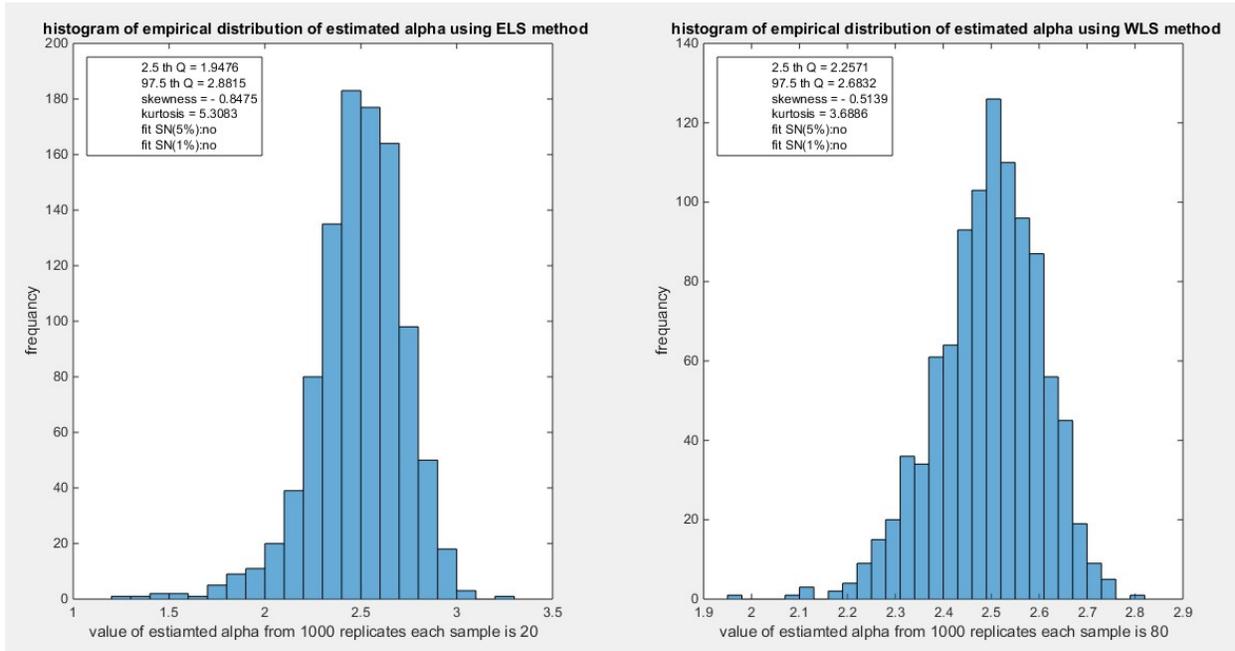

Fig. 29 shows the histogram of the empirical distribution of the estimated alpha from the 1000 replicates with sample size (n=20) on the left and (n=80) on the right using WLS method

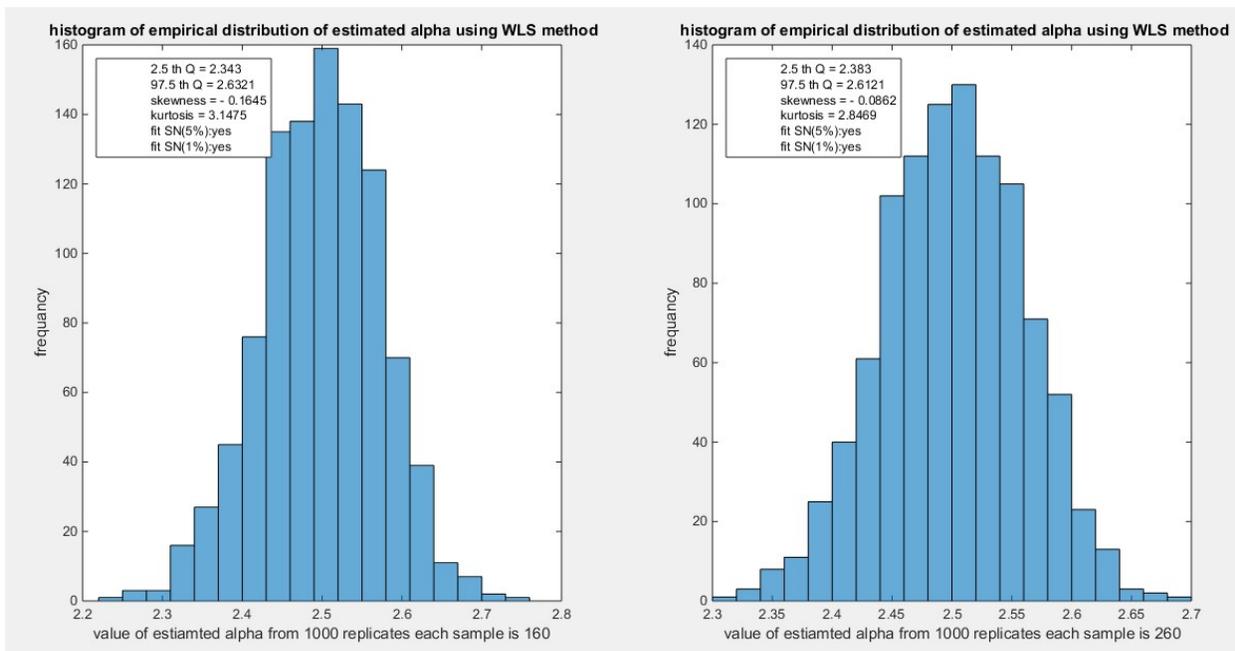

Fig. 30 shows the histogram of the empirical distribution of the estimated alpha from the 1000 replicates with sample size (n=160) on the left and (n=260) on the right using WLS method

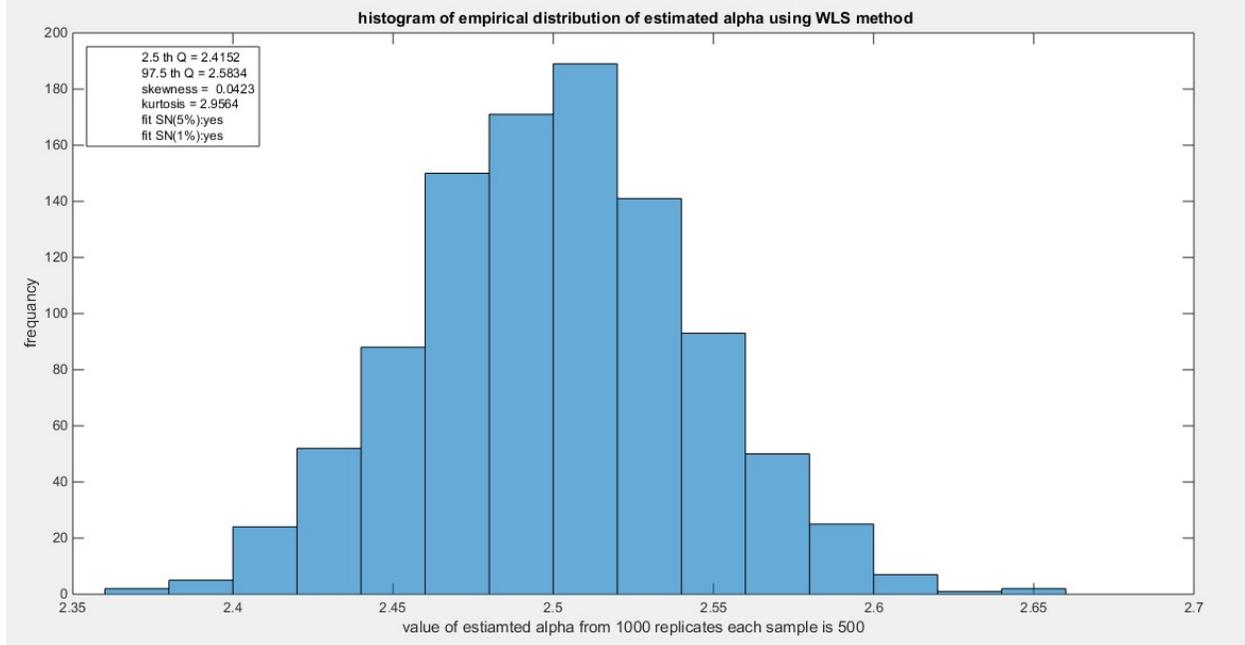

Fig. 31 shows the histogram of the empirical distribution of the estimated alpha from the 1000 replicates with sample size (n =500) using WLS method

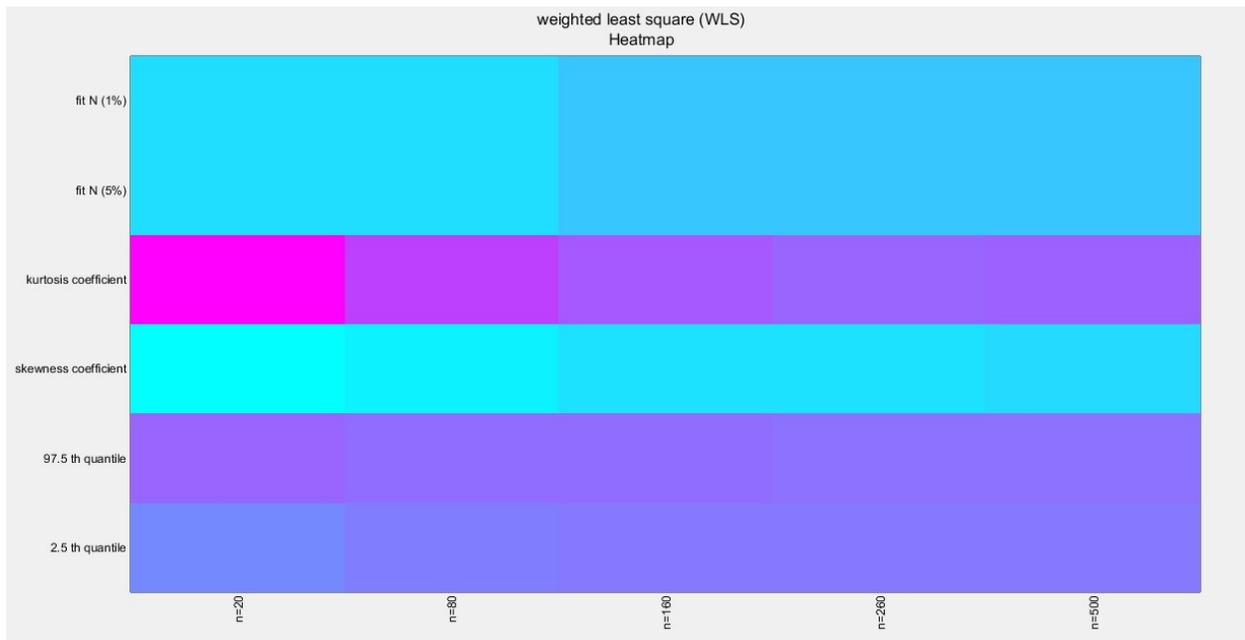

Fig.32 shows the heat map of the indices of the empirical distribution of the estimated alpha using (WLS) method and how these indices change with changing the sample size from 20 to 500. (p values are shown)

# ALPHA LEVEL =0.5

A simulation study is conducted utilizing the following sample sizes

$n = (20, 80, 160, 260, 500)$, and replicate N=1000 times. Various methods of estimation are utilized and compared with one another. The parameter alpha value chosen is $\alpha = (0.5)$.

Table (9): shows the SE from the 1000 replicates for each method.

| MEAN | MOM | MLE | MPS | AD | PERC | CVM | LS | WLS |
|---|---|---|---|---|---|---|---|---|
| n=20 | 0.4988 | 0.4888 | 0.4994 | 0.4925 | 0.5108 | 0.4921 | 0.4932 | 0.4961 |
| n=80 | 0.4989 | 0.4965 | 0.5002 | 0.4974 | 0.5033 | 0.4975 | 0.4978 | 0.4983 |
| n=160 | 0.4992 | 0.498 | 0.5001 | 0.4985 | 0.5016 | 0.4987 | 0.4989 | 0.4991 |
| n=260 | 0.4997 | 0.4991 | 0.5006 | 0.4995 | 0.5013 | 0.4996 | 0.4996 | 0.4997 |
| n=500 | 0.4999 | 0.4995 | 0.5004 | 0.4997 | 0.5008 | 0.4998 | 0.4999 | 0.4999 |

Table 9 shows that as sample size increases, the estimated parameter approaches the true value. The different methods are approximately equally efficient to approach the true value regardless the sample size. Figure 33 visually illustrates these results.

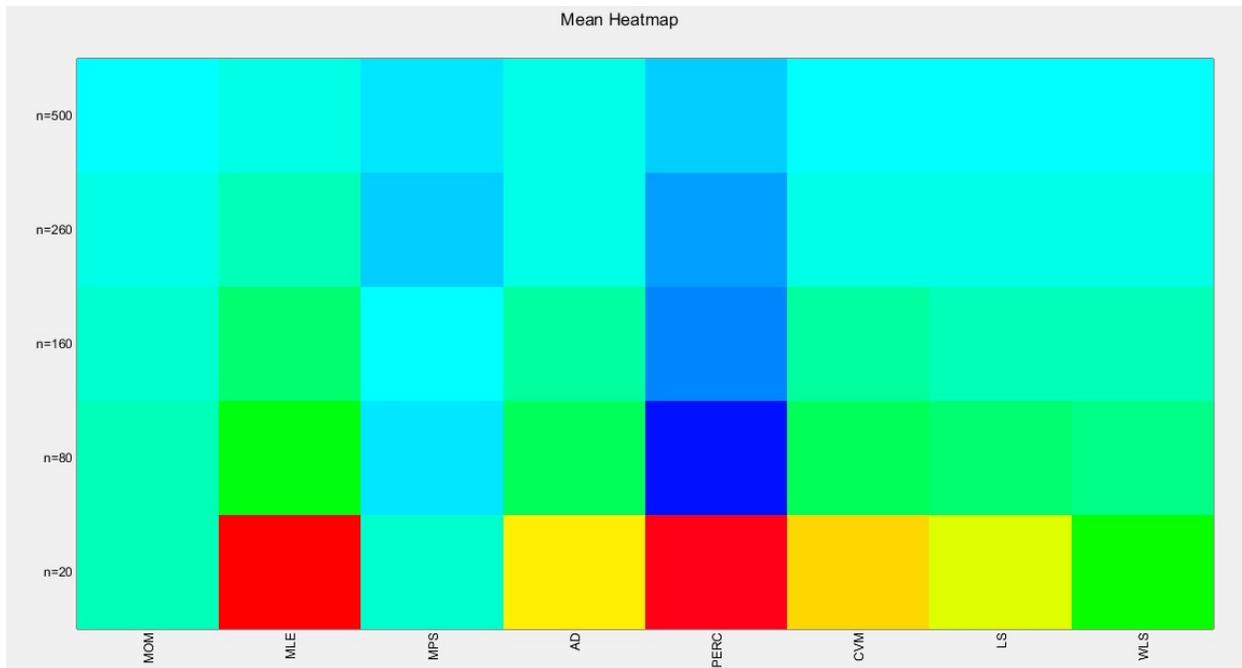

Fig. 33 shows the Heat-map for the mean of the estimated alpha parameter from running the simulation using different estimation methods with alpha value 0.5.

Table (10): shows the SE from the 1000 replicates for each method.

| SE | MOM | MLE | MPS | AD | PERC | CVM | LS | WLS |
|---|---|---|---|---|---|---|---|---|
| n=20 | 0.0013 | 0.0015 | 0.0014 | 0.0015 | 0.0013 | 0.0019 | 0.0019 | 0.0016 |
| n=80 | 0.00065 | 0.00068 | 0.00066 | 0.00069 | 0.00067 | 0.00074 | 0.00074 | 0.00069 |
| n=160 | 0.00046 | 0.00047 | 0.00046 | 0.00049 | 0.00046 | 0.00051 | 0.00051 | 0.00049 |
| n=260 | 0.00035 | 0.00035 | 0.00035 | 0.00037 | 0.00035 | 0.00038 | 0.00039 | 0.00037 |
| n=500 | 0.00026 | 0.00027 | 0.00026 | 0.00027 | 0.00027 | 0.00028 | 0.00028 | 0.00027 |

Table 10 shows that as sample size increases the standard error (SE) decreases. CVM and LS methods have the highest SE at all different sample sizes. The MOM & MPS have the lowest SE at n=500. MOM and MPS have nearly equal results at different sample sizes. This is also true as regards the pair of AD and WLS methods and the pair of CVM and LS methods. Figure 34 visually demonstrates these results.

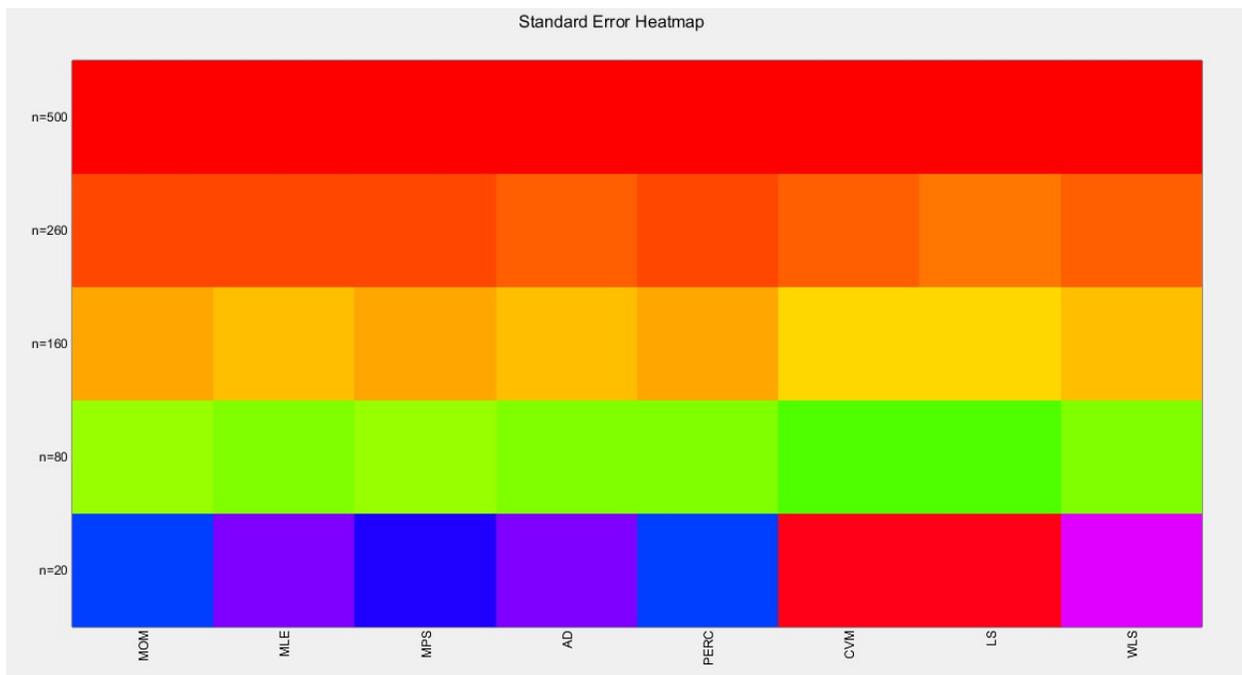

Fig. 34 shows the Heat-map for the SE of the estimated alpha parameter from running the simulation using different estimation methods with alpha value 0.5.

Table (11): shows the SE from the 1000 replicates for each method.

| AAB | MOM | MLE | MPS | AD | PERC | CVM | LS | WLS |
|---|---|---|---|---|---|---|---|---|
| n=20 | 0.0332 | 0.0361 | 0.0337 | 0.0367 | 0.0344 | 0.0418 | 0.0418 | 0.0378 |
| n=80 | 0.0166 | 0.0171 | 0.0167 | 0.0176 | 0.0172 | 0.0184 | 0.0184 | 0.0175 |
| n=160 | 0.0117 | 0.0119 | 0.0118 | 0.0122 | 0.0118 | 0.0127 | 0.0127 | 0.0122 |
| n=260 | 0.0085 | 0.0086 | 0.0085 | 0.0092 | 0.0088 | 0.0096 | 0.0097 | 0.0091 |
| n=500 | 0.0066 | 0.0066 | 0.0066 | 0.0068 | 0.0068 | 0.0070 | 0.0071 | 0.0068 |

Table 11 shows that as sample size increases the average absolute bias (AAB) decreases. CVM and LS methods have the highest AAB at all different sample sizes. The MOM, MLE, & MPS have the lowest AAB at n=500. The MOM, MLE, & MPS have nearly equal results at different sample sizes. This is also true as regards the pair of AD and WLS methods and the pair of CVM and LS methods. Figure 35 visually demonstrates these results.

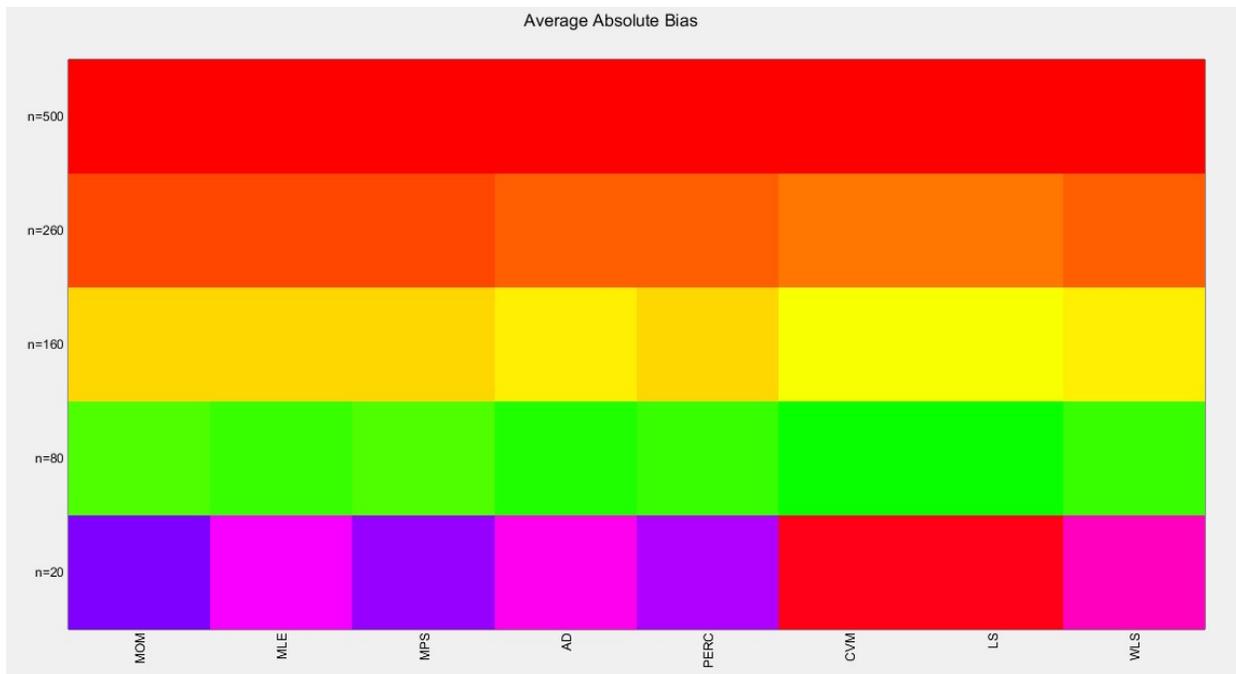

Fig. 35 shows the Heat-map for the AAB of the estimated alpha parameter from running the simulation using different estimation methods with alpha value 0.5.

Table (12): shows the MSE from the 1000 replicates for each method.

| MSE | MOM | MLE | MPS | AD | PERC | CVM | LS | WLS |
|---|---|---|---|---|---|---|---|---|
| n=20 | 0.0017 | 0.0025 | 0.002 | 0.0024 | 0.0019 | 0.0037 | 0.0037 | 0.0026 |
| n=80 | 0.00043 | 0.00047 | 0.00044 | 0.00049 | 0.000455 | 0.00055 | 0.00056 | 0.00049 |
| n=160 | 0.0002 | 0.00022 | 0.00022 | 0.00024 | 0.00021 | 0.00026 | 0.00026 | 0.00024 |
| n=260 | 0.00012 | 0.00012 | 0.00012 | 0.00014 | 0.00013 | 0.00015 | 0.00015 | 0.00013 |
| n=500 | 0.000068 | 0.000071 | 0.000070 | 0.000075 | 0.000071 | 0.00008 | 0.00008 | 0.000074 |

Table (13): shows the MRE from the 1000 replicates for each method.

| MRE | MOM | MLE | MPS | AD | PERC | CVM | LS | WLS |
|---|---|---|---|---|---|---|---|---|
| n=20 | 0.0664 | 0.0722 | 0.0674 | 0.0735 | 0.0689 | 0.0836 | 0.0836 | 0.0756 |
| n=80 | 0.0333 | 0.0342 | 0.0333 | 0.0351 | 0.0344 | 0.0368 | 0.0367 | 0.0351 |
| n=160 | 0.0233 | 0.0238 | 0.0235 | 0.0244 | 0.0237 | 0.0254 | 0.0254 | 0.0243 |
| n=260 | 0.017 | 0.0171 | 0.0170 | 0.0183 | 0.0176 | 0.0192 | 0.0194 | 0.0183 |
| n=500 | 0.0131 | 0.0133 | 0.0132 | 0.013 | 0.0135 | 0.014 | 0.0141 | 0.0136 |

The tables indicate that increasing the sample size leads to a decrease in the SE, AAB, MSE and MRE indices. For each method used, the indices decrease as the sample size increases. The values obtained from the MLE and MPS methods are nearly equal, especially with larger sample sizes (n=260 and n=500). Similarly, the AD and WLS methods yield comparable results. Additionally, the CVM and LS methods show approximately equal indices as the sample size increases. Overall, the methods demonstrate consistent results regarding the estimation values. Figure 36-37 display these findings

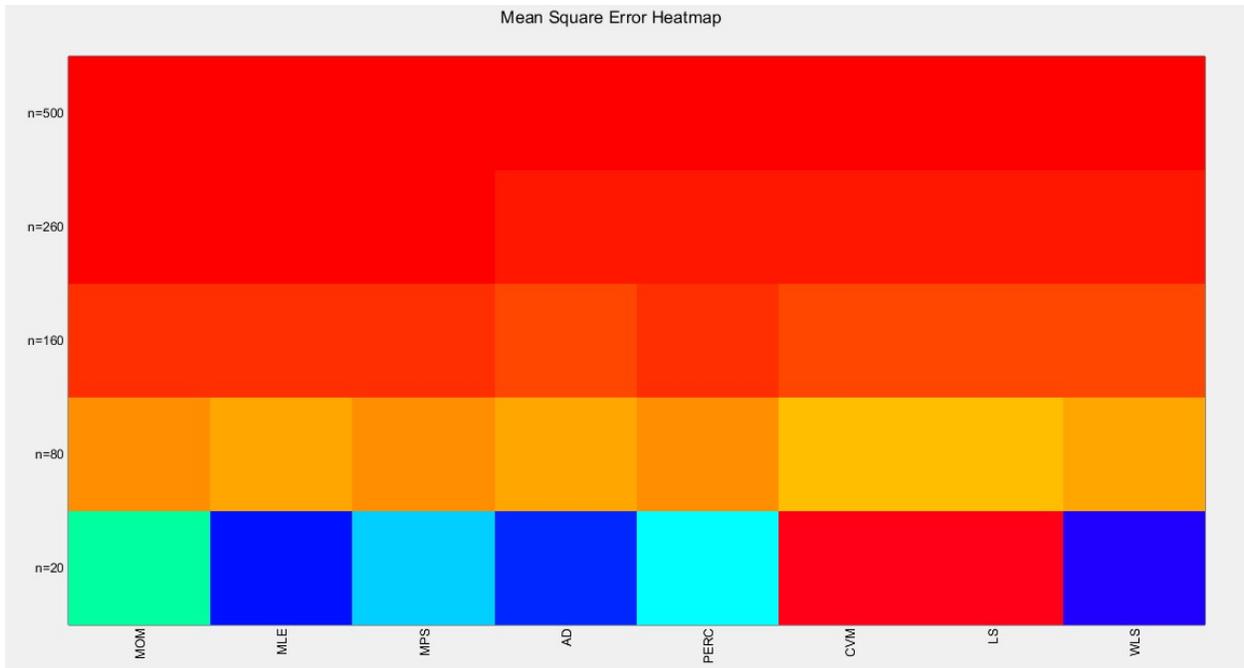

Fig. 36 shows the Heat-map for the MSE of the estimated alpha parameter from running the simulation using different estimation methods with alpha value 0.5.

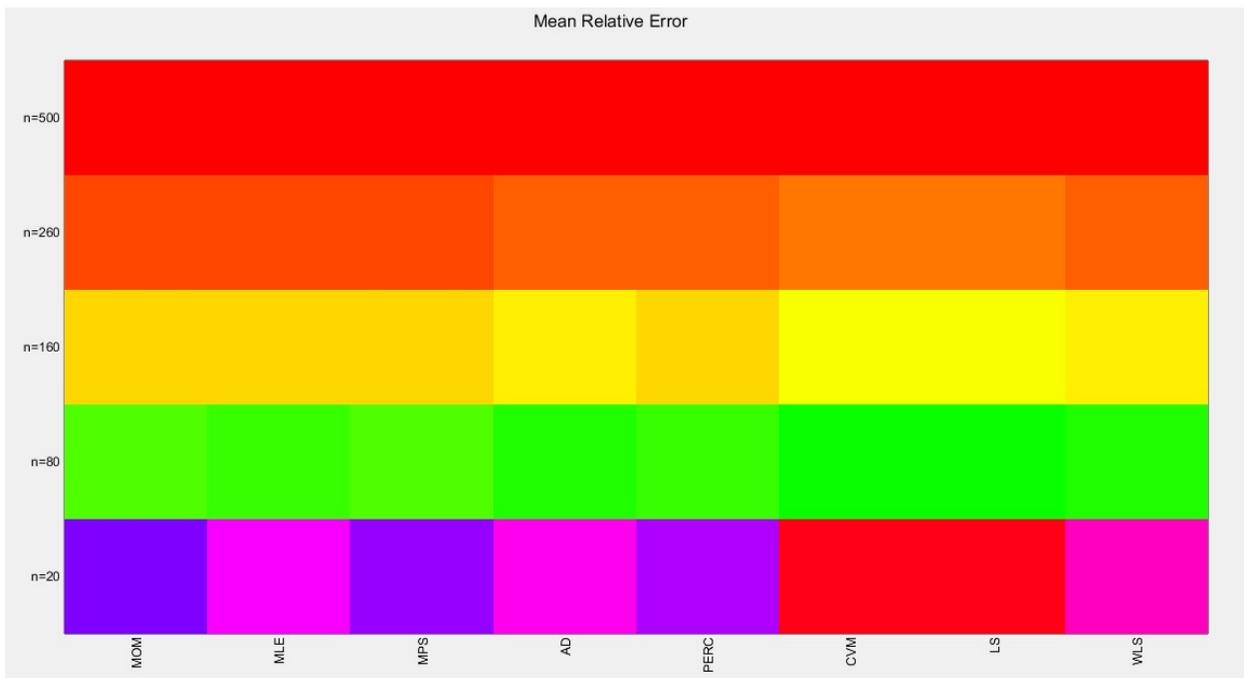

Fig. 37 shows the Heat-map for the MRE of the estimated alpha parameter from running the simulation using different estimation methods with alpha value 0.5.

Table (14): characteristics of empirical distribution of estimated alpha using MOM

| MOM | n=20 | n=80 | n=160 | n=260 | n=500 |
|---|---|---|---|---|---|
| 2.5 Q | 0.424 | 0.4607 | 0.4704 | 0.4776 | 0.4830 |
| 97.5 Q | 0.5801 | 0.5424 | 0.5277 | 0.5221 | 0.5156 |
| Skewness | 0.1049 | 0.0987 | 0.0338 | -0.0750 | -0.1145 |
| Kurtosis | 2.892 | 2.9217 | 2.9082 | 3.2618 | 3.0662 |
| Fit N (5%) | $H_0=0$ (0.1041) | $H_0=0$ (0.5) | $H_0=0$ (0.2235) | $H_0=1$ (0.0222) | $H_0=0$ (0.3337) |
| Fit N (1%) | $H_0=0$ (0.1041) | $H_0=0$ (0.5) | $H_0=0$ (0.2235) | $H_0=0$ (0.0222) | $H_0=0$ (0.3337) |

The empirical distribution of the estimated parameter alpha using MOM is shown in Table 14. Each column represents a specific sample size with 1000 replicates in each size. The 2.5 th quantile and the 97.5 th quantile of the 1000 values of the estimated parameter in each sample shows that as the sample size increases the 2.5 quantile rises while the 97.5 quantile decreases. In other words, the distance between the two quantiles decreases as the sample size increases and this is reflected on the confidence interval (CI). As the sample size increases the CI becomes narrower. The distribution exhibits a mild right skewness and a mild negative excess kurtosis (platykurtic shape) at small sample size. As sample size increases the skewness decreases trying to approach the zero level (skewness of standard normal) and kurtosis increases trying to approach the kurtosis of standard normal. The empirical distribution does not fit standard normal at size n=260 and at 5% significance level, otherwise, it fits standard normal distribution at different sample sizes and at significance level 5% and 1% with associated P-value as shown in the table. $H_0=1$ means reject the null hypothesis that states the parameter distribution follows the standard normal distribution. While $H_0=0$ means fail to reject the null hypothesis. See the following Figures (38-41)

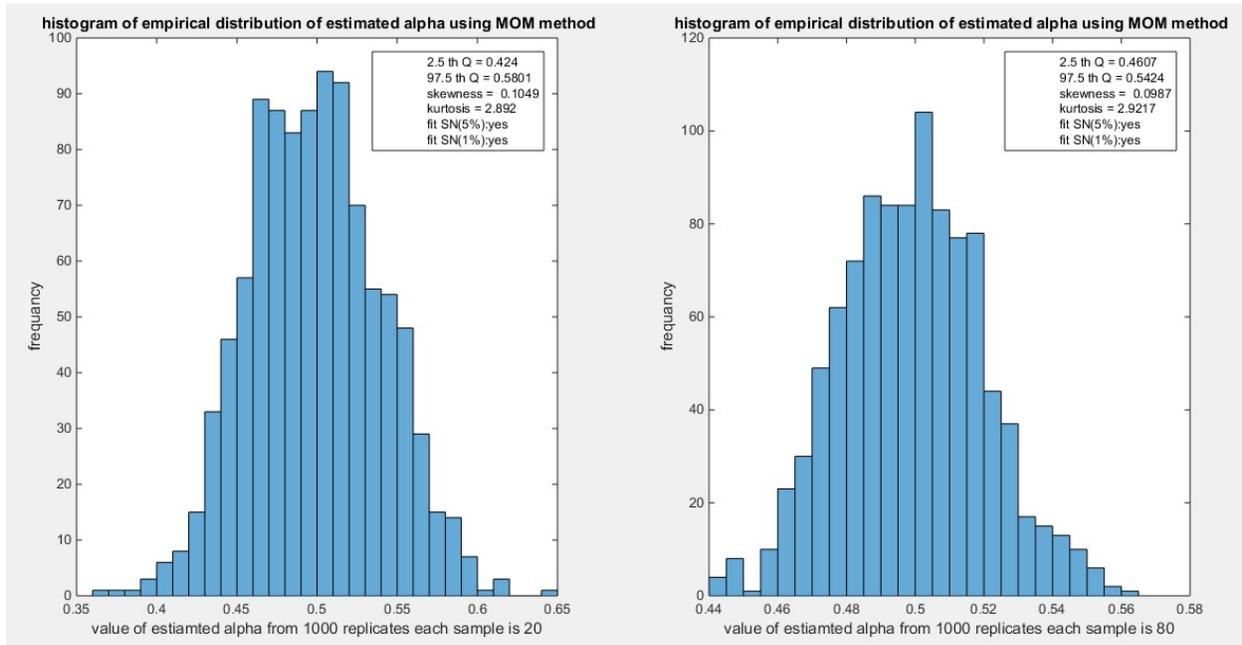

Fig. 38 shows the histogram of the empirical distribution of the estimated alpha from the 1000 replicates with sample size (n=20) on the left and (n=80) on the right using MOM method

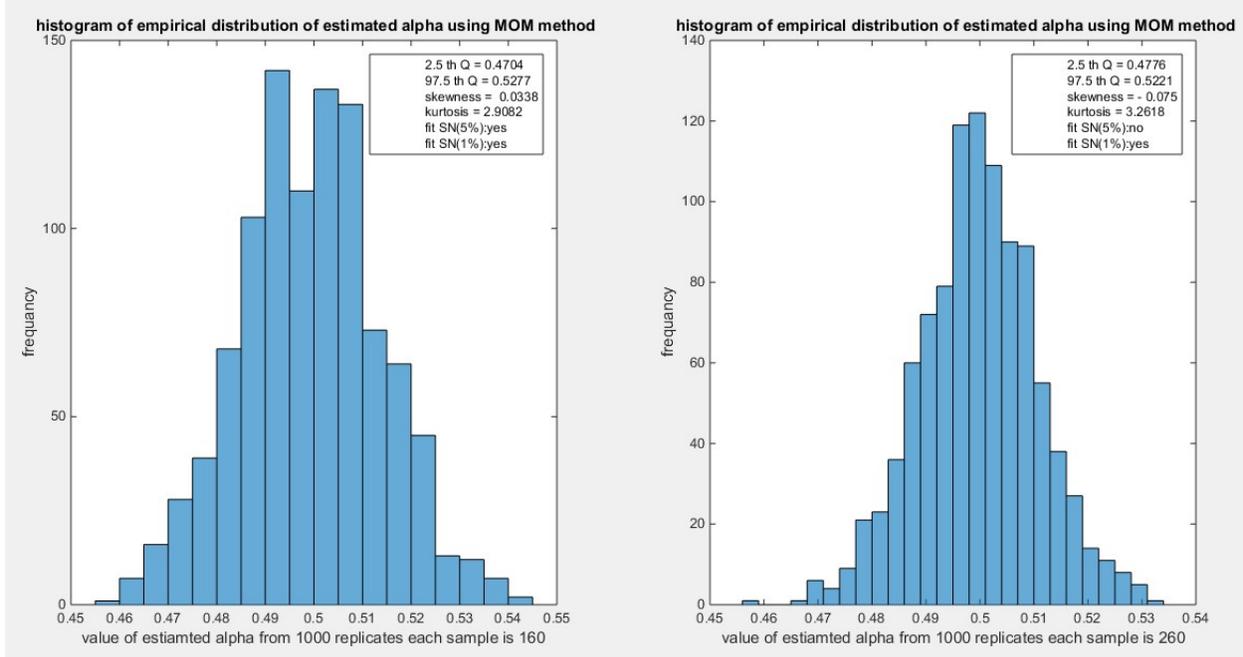

Fig. 39 shows the histogram of the empirical distribution of the estimated alpha from the 1000 replicates with sample size (n=160) on the left and (n=260) on the right using MOM method

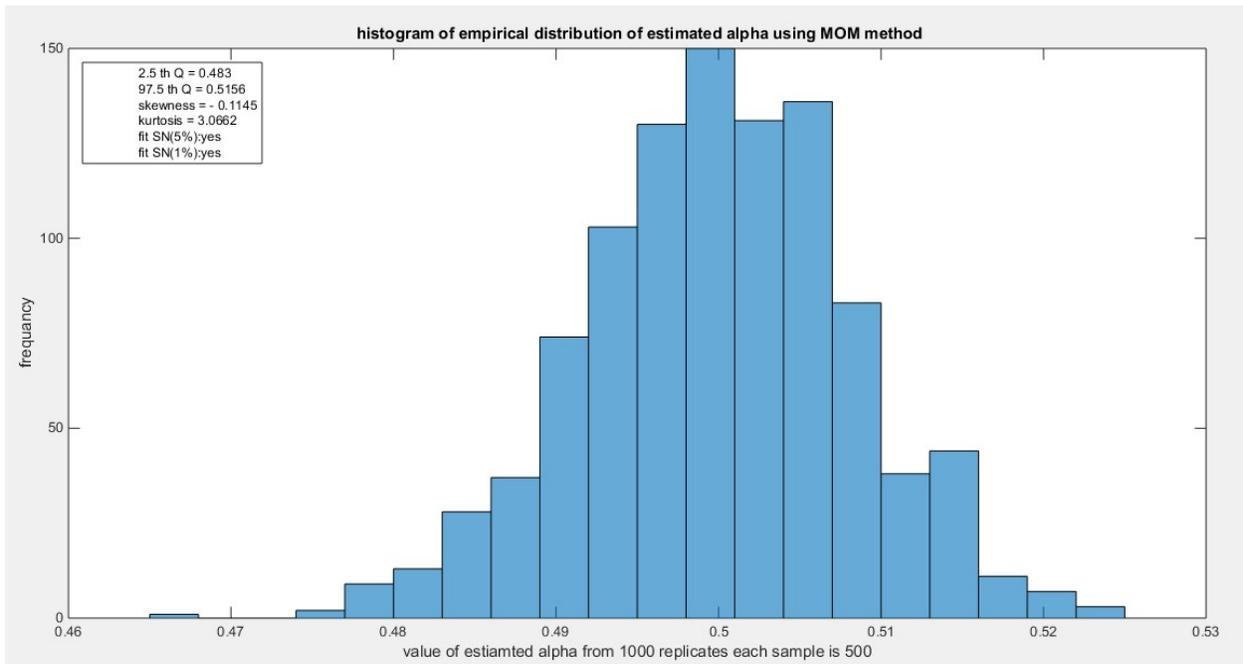

Fig. 40 shows the histogram of the empirical distribution of the estimated alpha from the 1000 replicates with sample size (n =500) using MOM method

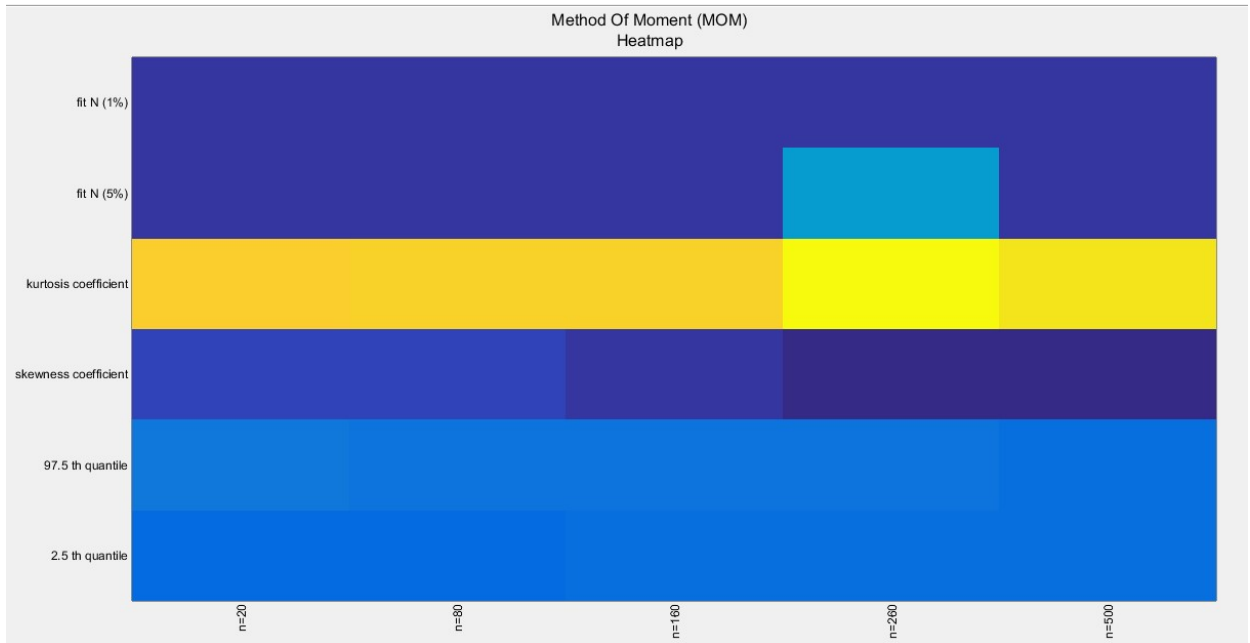

Fig. 41 shows the heat map of the indices of the empirical distribution of the estimated alpha using (MOM) method and how these indices change with changing the sample size from 20 to 500.( h0 values reported)

Table (15): characteristics of empirical distribution of estimated alpha using MLE

| MLE | n=20 | n=80 | n=160 | n=260 | n=500 |
|---|---|---|---|---|---|
| 2.5 Q | 0.381 | 0.4508 | 0.465 | 0.474 | 0.4814 |
| 97.5 Q | 0.5569 | 0.5344 | 0.524 | 0.5192 | 0.5144 |
| Skewness | -1.6493 | -0.5355 | -0.3951 | -0.4555 | -0.3762 |
| Kurtosis | 9.6769 | 3.4636 | 3.1770 | 3.6191 | 3.3215 |
| Fit N (5%) | $H_0=1$ (0.001) | $H_0=1$ (0.001) | $H_0=1$ (0.001) | $H_0=1$ (0.001) | $H_0=1$ (0.0098) |
| Fit N (1%) | $H_0=1$ (0.001) | $H_0=1$ (0.001) | $H_0=1$ (0.001) | $H_0=1$ (0.001) | $H_0=1$ (0.0098) |

The empirical distribution of the estimated parameter alpha using MLE is shown in Table 15. Each column represents a specific sample size with 1000 replicates in each size. The 2.5 th quantile and the 97.5 th quantile of the 1000 values of the estimated parameter in each sample shows that as the sample size increases the 2.5 quantile rises while the 97.5 quantile decreases. In other words, the distance between the two quantiles decreases as the sample size increases and this is reflected on the confidence interval (CI). As the sample size increases the CI becomes narrower. The distribution exhibits a moderate left skewness and a high positive excess kurtosis (leptokurtic shape) at small sample size. As sample size increases the skewness decreases trying to approach the zero level (skewness of standard normal) and kurtosis decreases trying to approach the kurtosis of standard normal. The empirical distribution does not fit standard normal at any size and at either significance level 5% or 1% with associated P-value as shown in the table. The distribution may fit the standard normal at sample sizes more than 500. $H_0=1$ means reject the null hypothesis that states the parameter distribution follows the standard normal distribution. While $H_0=0$ means fail to reject the null hypothesis. See the following Figures (42-45).

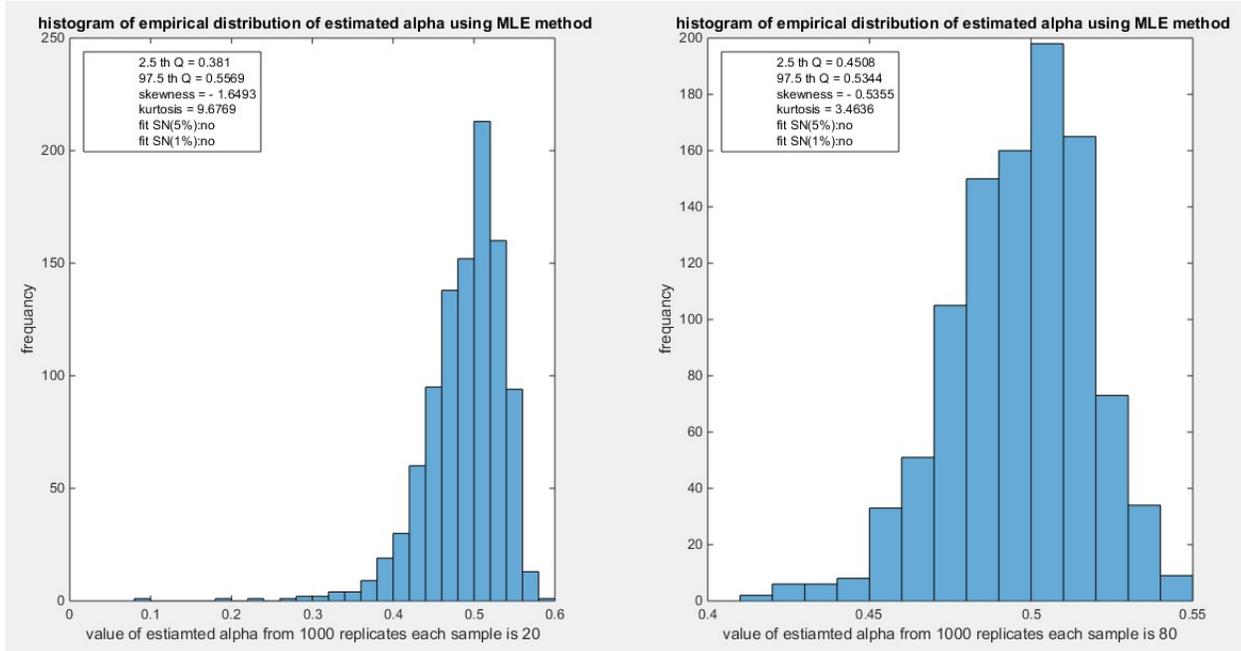

Fig. 42 shows the histogram of the empirical distribution of the estimated alpha from the 1000 replicates with sample size (n=20) on the left and (n=80) on the right using MLE method

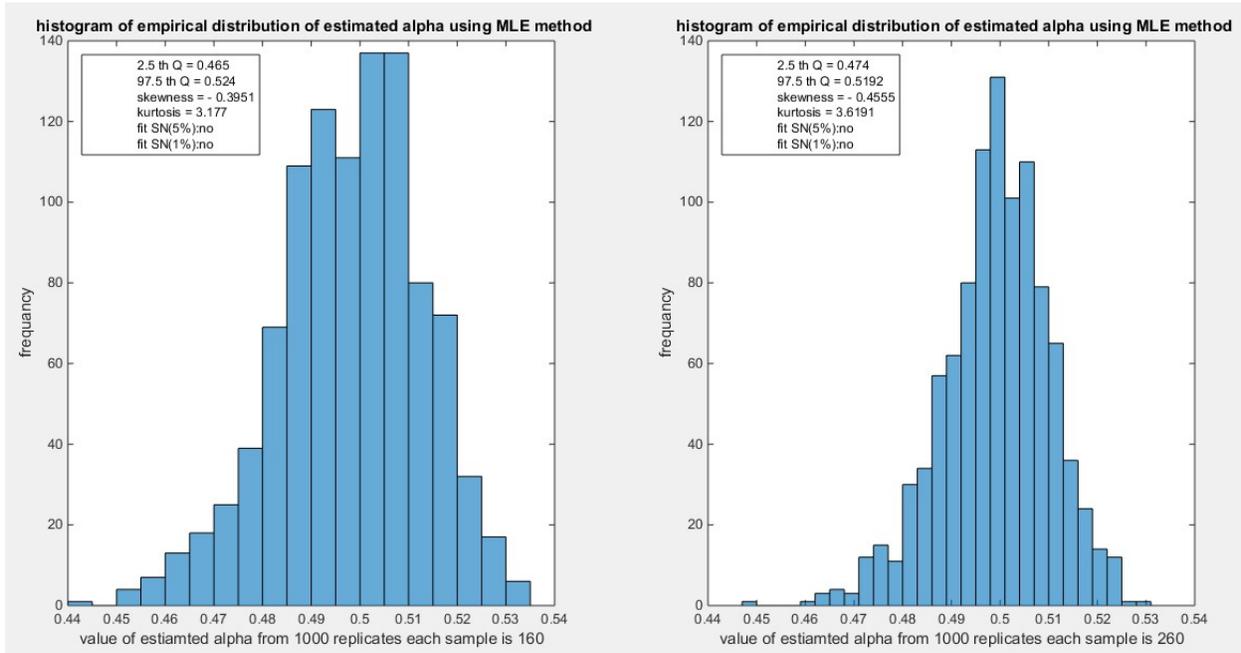

Fig. 43 shows the histogram of the empirical distribution of the estimated alpha from the 1000 replicates with sample size (n=160) on the left and (n=260) on the right using MLE method

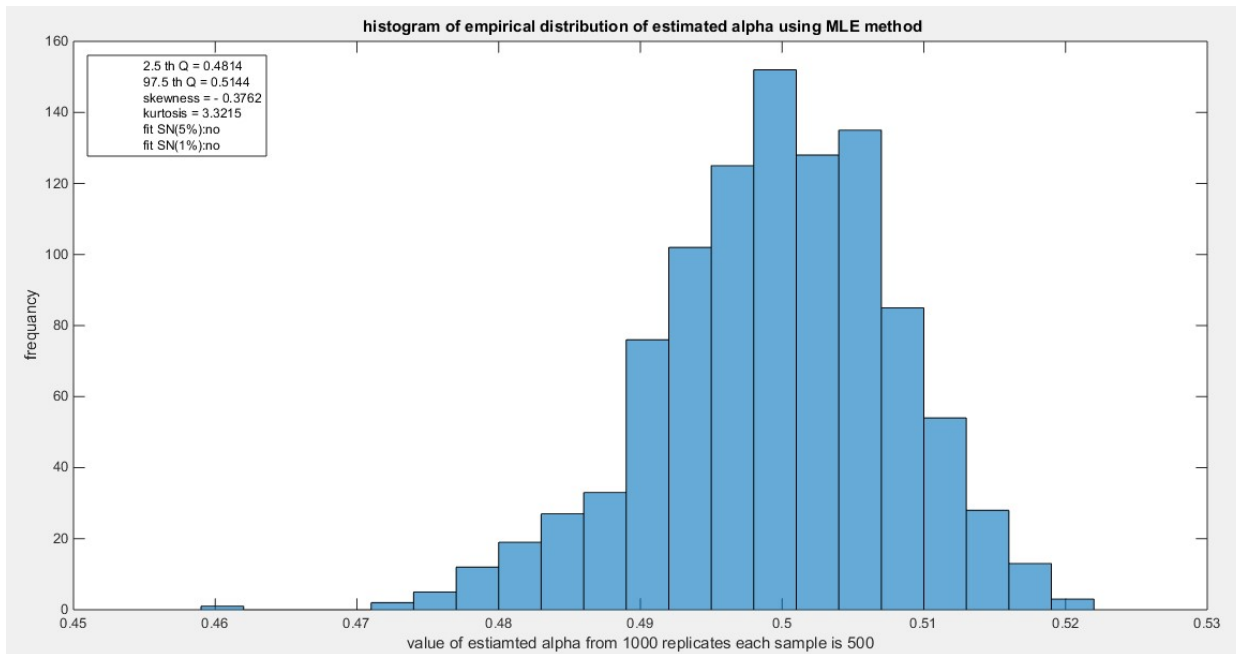

Fig. 44 shows the histogram of the empirical distribution of the estimated alpha from the 1000 replicates with sample size (n =500) using MLE method

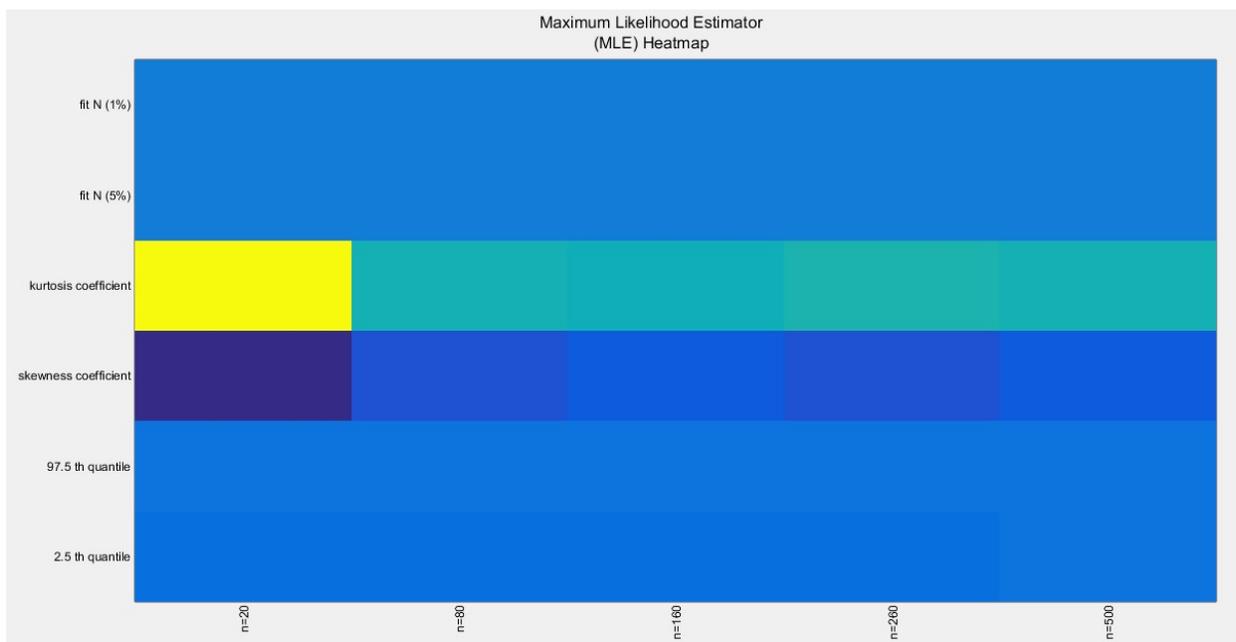

Fig. 45 shows the heat map of the indices of the empirical distribution of the estimated alpha using (MLE) method and how these indices change with changing the sample size from 20 to 500. (h value is shown)

Table (16): characteristics of empirical distribution of estimated alpha using MPS

| MPS | n=20 | n=80 | n=160 | n=260 | n=500 |
|---|---|---|---|---|---|
| 2.5 Q | 0.399 | 0.4557 | 0.4671 | 0.4757 | 0.4823 |
| 97.5 Q | 0.5631 | 0.537 | 0.5262 | 0.5203 | 0.5152 |
| Skewness | -1.5438 | -0.5296 | -0.3834 | -0.4521 | -0.3794 |
| Kurtosis | 8.894 | 3.4477 | 3.1806 | 3.6063 | 3.3356 |
| Fit N (5%) | $H_0=1$ (0.001) | $H_0=1$ (0.001) | $H_0=1$ (0.001) | $H_0=1$ (0.001) | $H_0=1$ (0.0147) |
| Fit N (1%) | $H_0=1$ (0.001) | $H_0=1$ (0.001) | $H_0=1$ (0.001) | $H_0=1$ (0.001) | $H_0=0$ (0.0147) |

The empirical distribution of the estimated parameter alpha using MPS is shown in Table 16. Each column represents a specific sample size with 1000 replicates in each size. The 2.5 th quantile and the 97.5 th quantile of the 1000 values of the estimated parameter in each sample shows that as the sample sizes increases the 2.5 quantile rises while the 97.5 quantile decreases. In other words, the distance between the quantiles decreases as the sample size increases and this is reflected on the confidence interval (CI). As the sample size increases the CI becomes narrower. The distribution exhibits a moderate left skewness and high positive excess kurtosis (leptokurtic shape) at small sample size. As sample size increases the skewness decreases trying to approach the zero level (skewness of standard normal) and kurtosis decreases trying to approach the kurtosis of standard normal. The empirical distribution does not fit the standard normal at different sample sizes except at sample size 500 and at significance level 1% with associated P-value as shown in the table. $H_0=1$ means reject the null hypothesis that states the parameter distribution follows the standard normal distribution. While $H_0=0$ means fail to reject the null hypothesis. See following Figures (46-49).

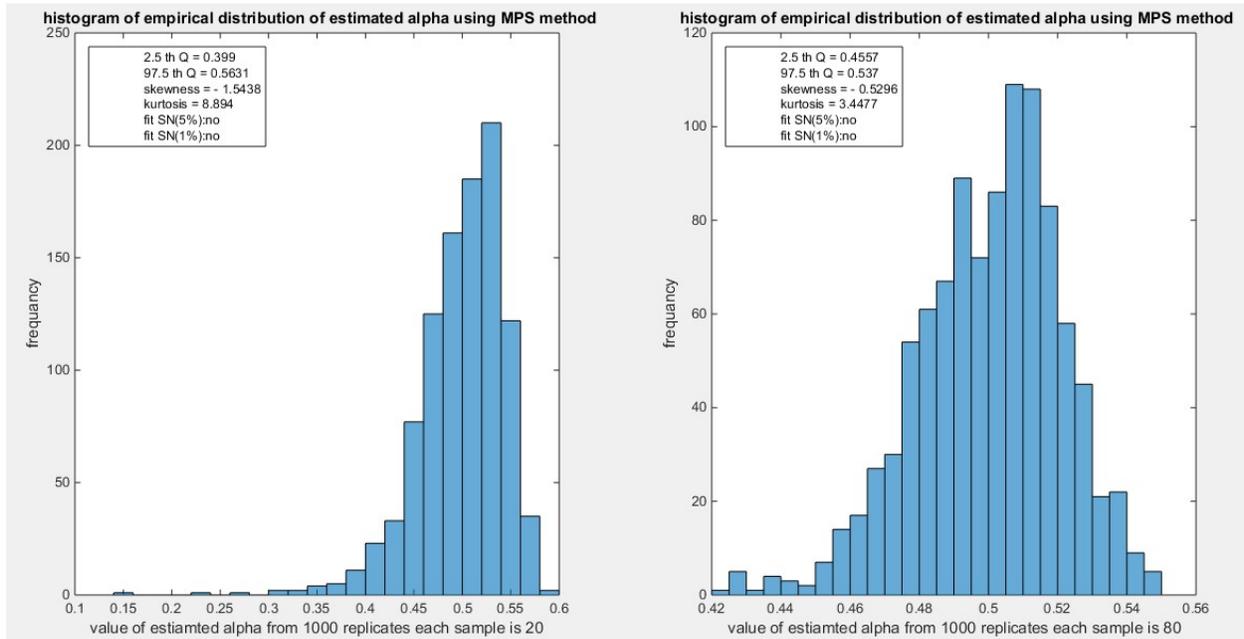

Fig. 46 shows the histogram of the empirical distribution of the estimated alpha from the 1000 replicates with sample size (n=20) on the left and (n=80) on the right using MPS method

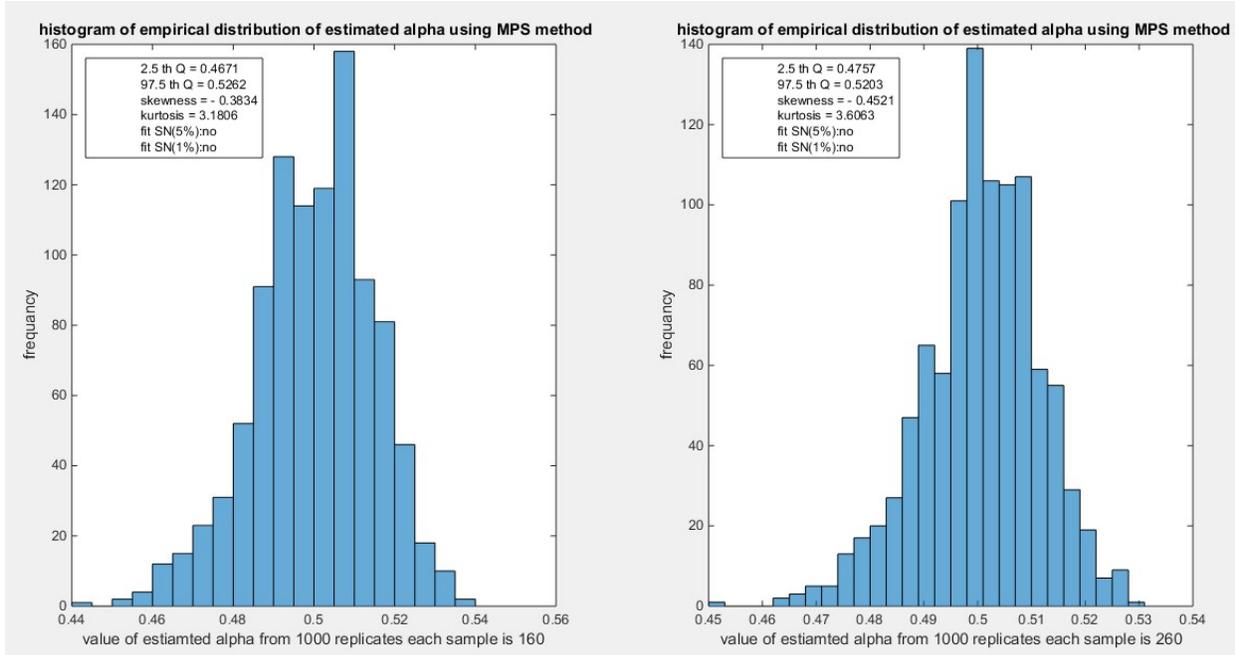

Fig. 47 shows the histogram of the empirical distribution of the estimated alpha from the 1000 replicates with sample size (n=160) on the left and (n=260) on the right using MPS method

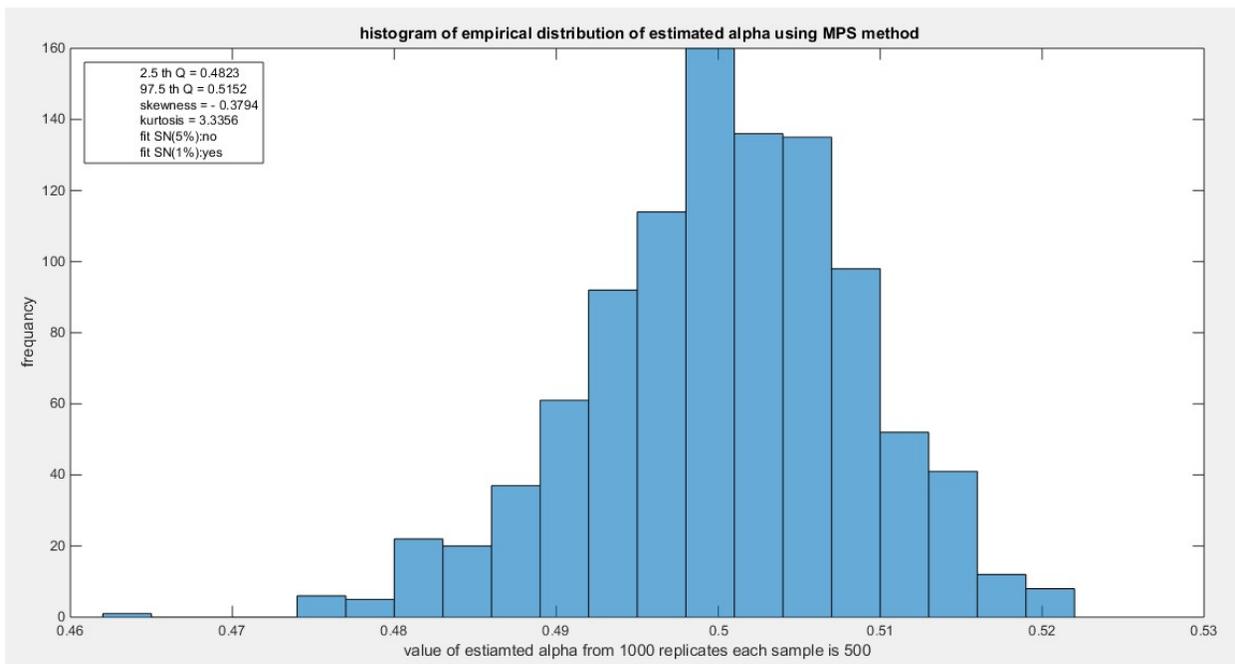

Fig. 48 shows the histogram of the empirical distribution of the estimated alpha from the 1000 replicates with sample size (n =500) using MPS method

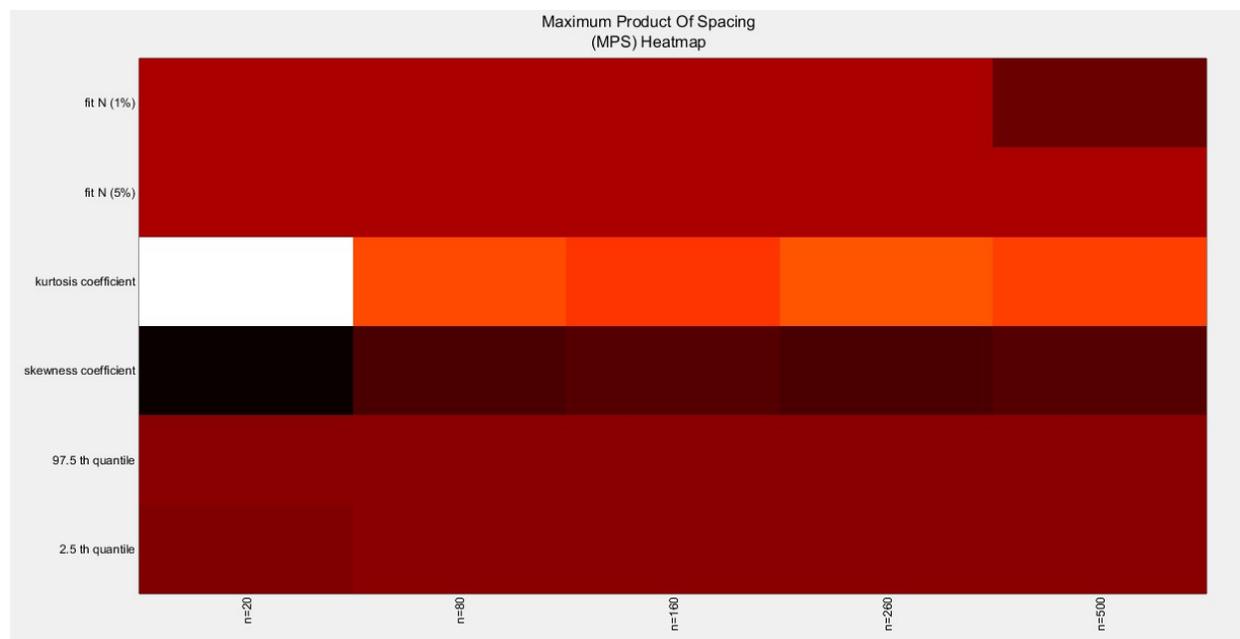

Fig. 49 shows the heat map of the indices of the empirical distribution of the estimated alpha using (MPS) method and how these indices change with changing the sample size from 20 to 500.(h value shown)

Table (17): characteristics of empirical distribution of estimated alpha using AD

| AD | n=20 | n=80 | n=160 | n=260 | n=500 |
|---|---|---|---|---|---|
| 2.5 Q | 0.3812 | 0.4523 | 0.4651 | 0.4748 | 0.4813 |
| 97.5 Q | 0.5622 | 0.537 | 0.5278 | 0.5220 | 0.5154 |
| Skewness | -1.3644 | -0.3869 | -0.312 | -0.4046 | -0.3145 |
| Kurtosis | 7.2624 | 3.0839 | 3.0796 | 3.5645 | 3.3159 |
| Fit N (5%) | $H_0=1$ (0.001) | $H_0=1$ (0.0011) | $H_0=1$ (0.0025) | $H_0=1$ (0.0115) | $H_0=1$ (0.0418) |
| Fit N (1%) | $H_0=1$ (0.001) | $H_0=1$ (0.0011) | $H_0=1$ (0.0025) | $H_0=0$ (0.0115) | $H_0=0$ (0.0418) |

The empirical distribution of the estimated parameter alpha using AD is shown in Table 17. Each column represents a specific sample size with 1000 replicates in each size. The 2.5 th quantile and the 97.5 th quantile of the 1000 values of the estimated parameter in each sample shows that as the sample sizes increases the 2.5 quantile rises while the 97.5 quantile decreases. In other words, the distance between the two quantiles decreases as the sample size increases and this is reflected on the confidence interval (CI). As the sample size increases the CI becomes narrower. The distribution exhibits a moderate left skewness and a moderate positive excess kurtosis (leptokurtic shape) at small sample size. As sample size increases the skewness decreases trying to approach the zero level (skewness of standard normal) and kurtosis decreases trying to approach the kurtosis of standard normal. The empirical distribution fits standard normal starting at size 260 and larger than this at significance level 1% with associated P-value as shown in the table. $H_0=1$ means reject the null hypothesis that states the parameter distribution follows the standard normal distribution. While $H_0=0$ means fail to reject the null hypothesis. See the following Figures (50-53)

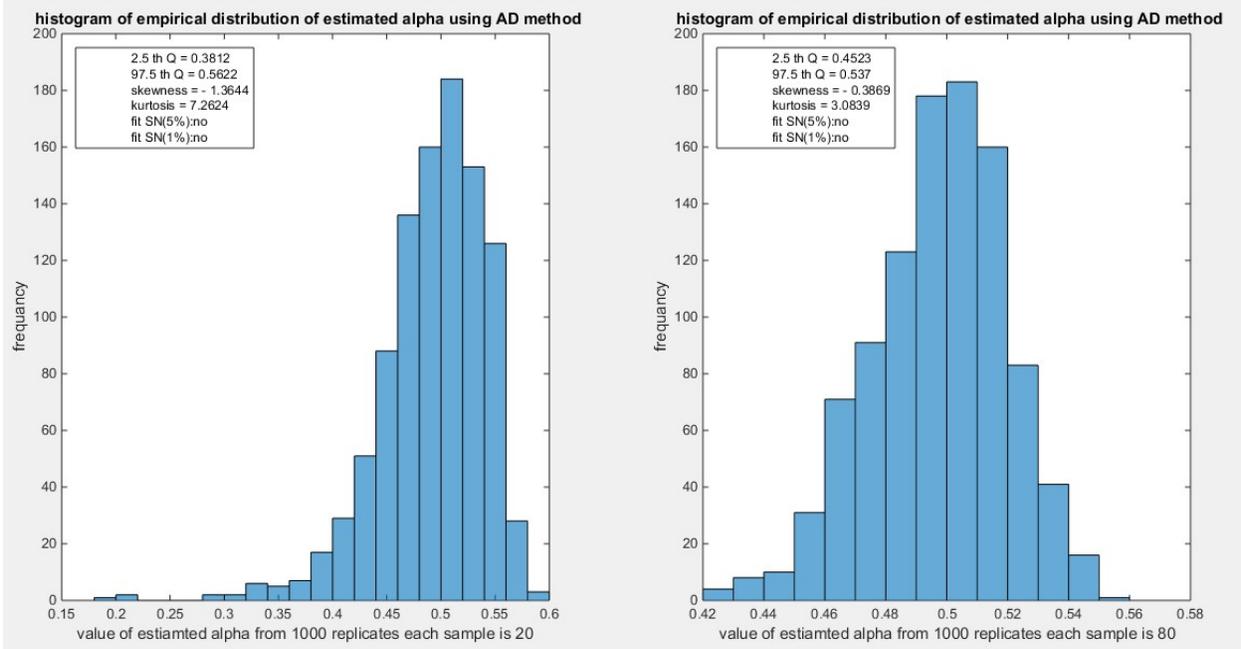

Fig. 50 shows the histogram of the empirical distribution of the estimated alpha from the 1000 replicates with sample size (n=20) on the left and (n=80) on the right using AD method

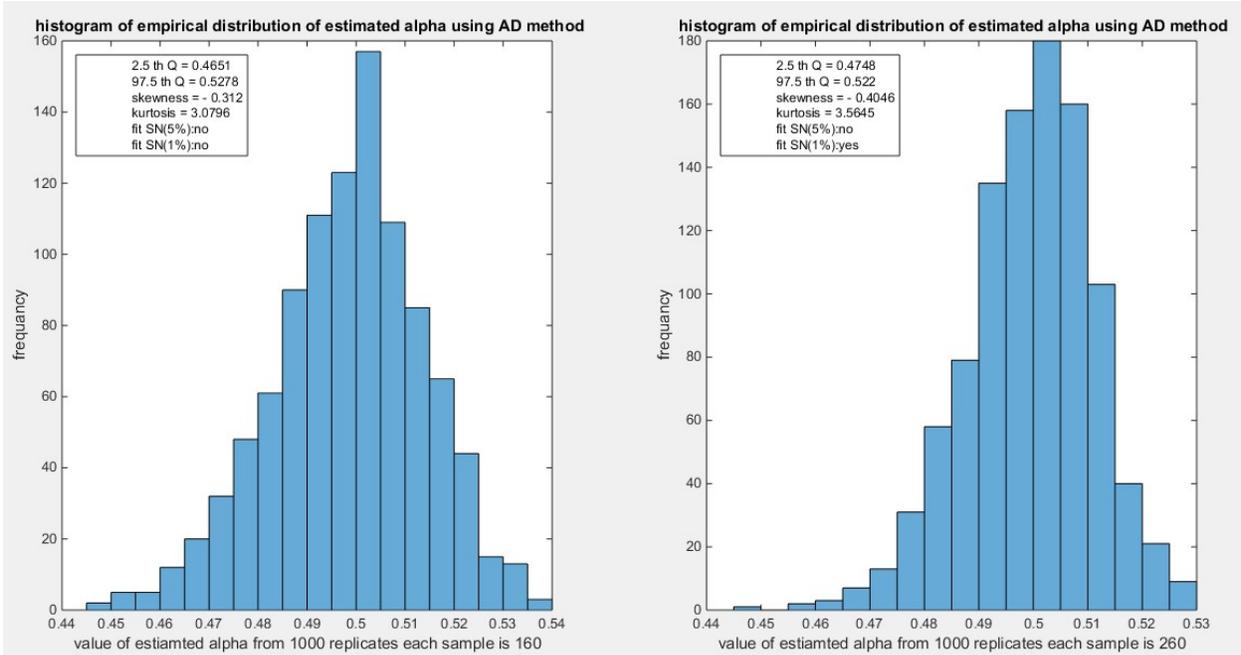

Fig. 51 shows the histogram of the empirical distribution of the estimated alpha from the 1000 replicates with sample size (n=20) on the left and (n=80) on the right using AD method

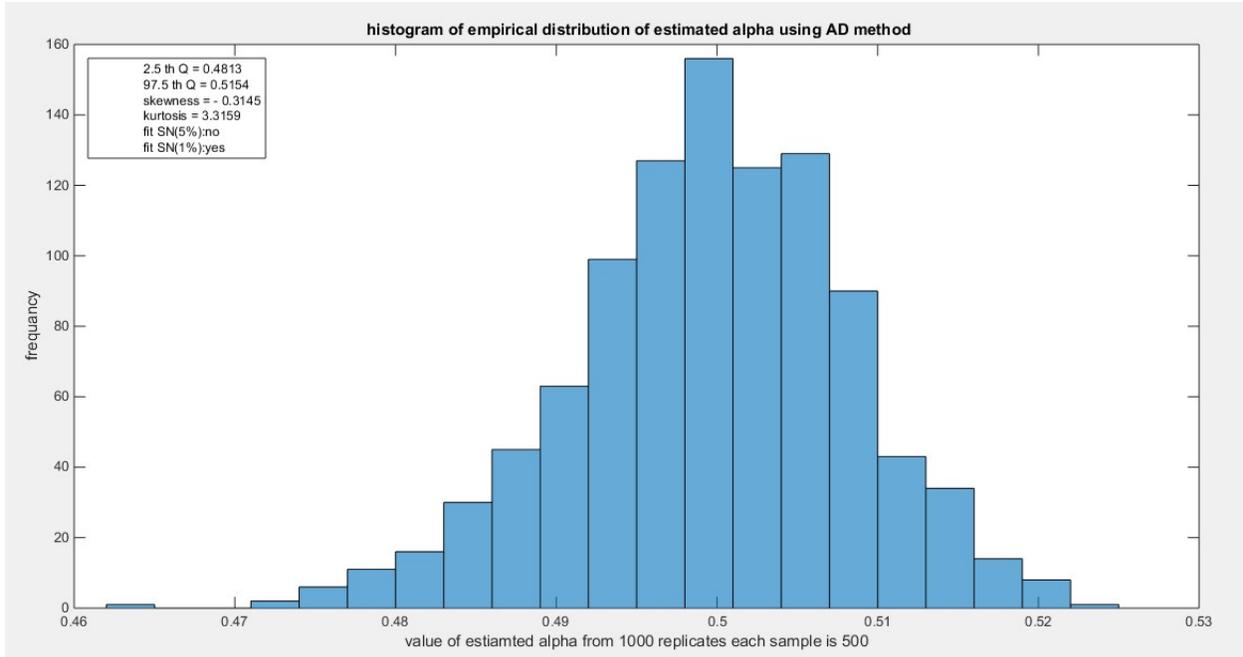

Fig. 52 shows the histogram of the empirical distribution of the estimated alpha from the 1000 replicates with sample size (n =500) using AD method

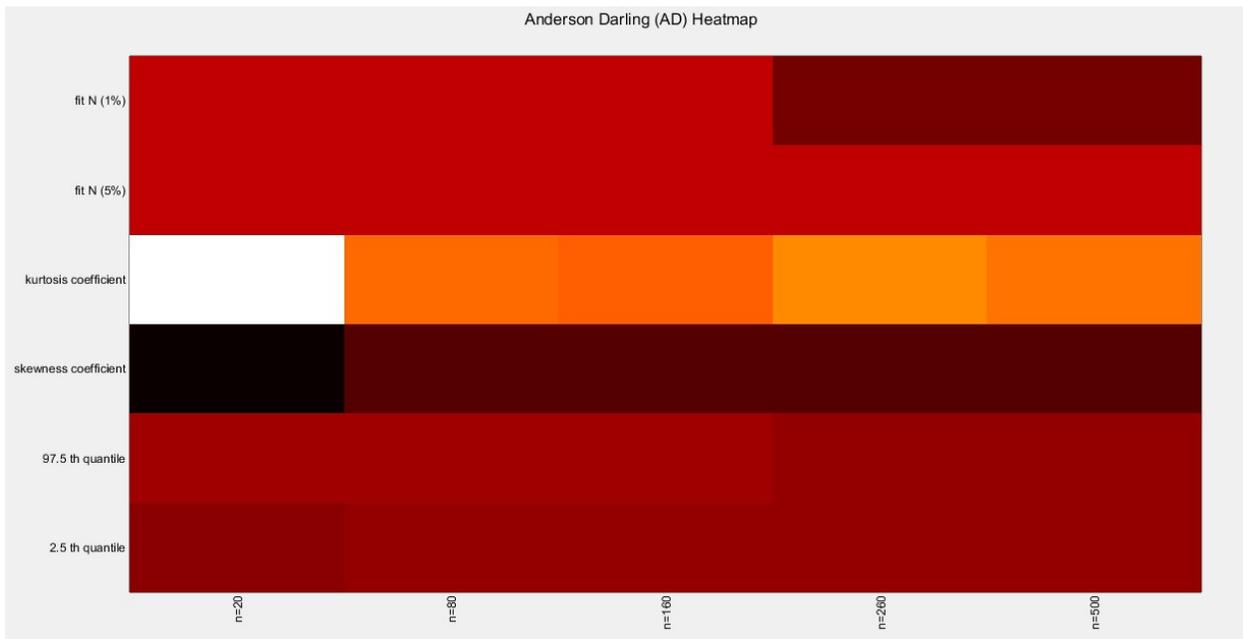

Fig. 53 shows the heat map of the indices of the empirical distribution of the estimated alpha using (AD) method and how these indices change with changing the sample size from 20 to 500. (h value shown)

Table (18): characteristics of empirical distribution of estimated alpha using CVM

| CVM | n=20 | n=80 | n=160 | n=260 | n=500 |
|---|---|---|---|---|---|
| 2.5 Q | 0.3675 | 0.45 | 0.465 | 0.4745 | 0.481 |
| 97.5 Q | 0.5804 | 0.5408 | 0.5292 | 0.5219 | 0.5166 |
| Skewness | -2.2301 | -0.349 | -0.2116 | -0.2663 | -0.2296 |
| Kurtosis | 15.7352 | 3.4125 | 3.1182 | 3.2822 | 3.3211 |
| Fit N (5%) | $H_0=1$ (0.001) | $H_0=1$ (0.001) | $H_0=1$ (0.0447) | $H_0=1$ (0.0369) | $H_0=0$ (0.1806) |
| Fit N (1%) | $H_0=1$ (0.001) | $H_0=1$ (0.001) | $H_0=0$ (0.0447) | $H_0=0$ (0.0369) | $H_0=0$ (0.1806) |

The empirical distribution of the estimated parameter alpha using CVM is shown in Table 18. Each column represents a specific sample size with 1000 replicates in each size. Each column depicts the characteristics of the empirical distribution of the estimated alpha. The 2.5 th quantile and the 97.5 th quantile of the 1000 values of the estimated parameter in each sample shows that as the sample size increases the 2.5 th quantile rises while the 97.5 th quantile decreases. In other words, the distance between the quantiles decreases as the sample size increases and this is reflected on the confidence interval (CI). As the sample size increases the CI becomes narrower. The distribution exhibits a moderate left skewness and a high positive excess kurtosis (leptokurtic shape) at small sample size. As sample size increases the skewness decreases trying to approach the zero level (skewness of standard normal) and kurtosis decreases trying to approach the kurtosis of standard normal. The empirical distribution fits standard normal starting at size 160 and larger than this at significance level 5% and 1% with associated P-value as shown in the table. $H_0=1$ means reject the null hypothesis that states the parameter distribution follows the standard normal distribution. While $H_0=0$ means fail to reject the null hypothesis. See the following Figures (54-57)

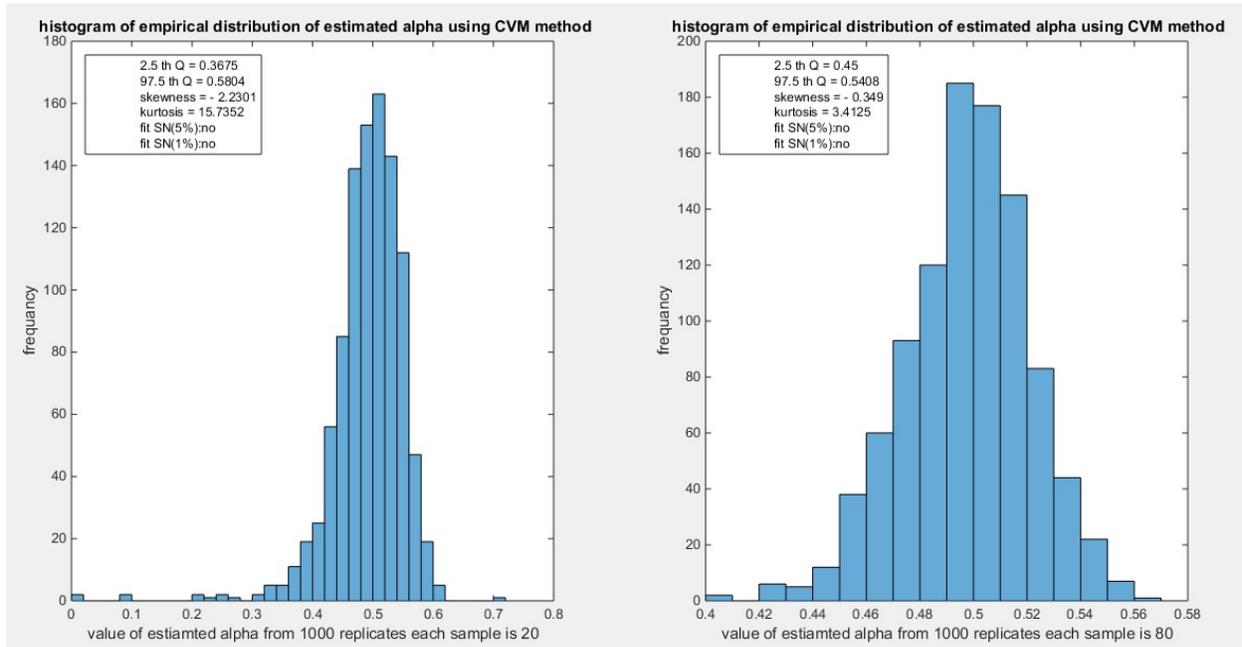

Fig. 54 shows the histogram of the empirical distribution of the estimated alpha from the 1000 replicates with sample size (n=20) on the left and (n=80) on the right using CVM method

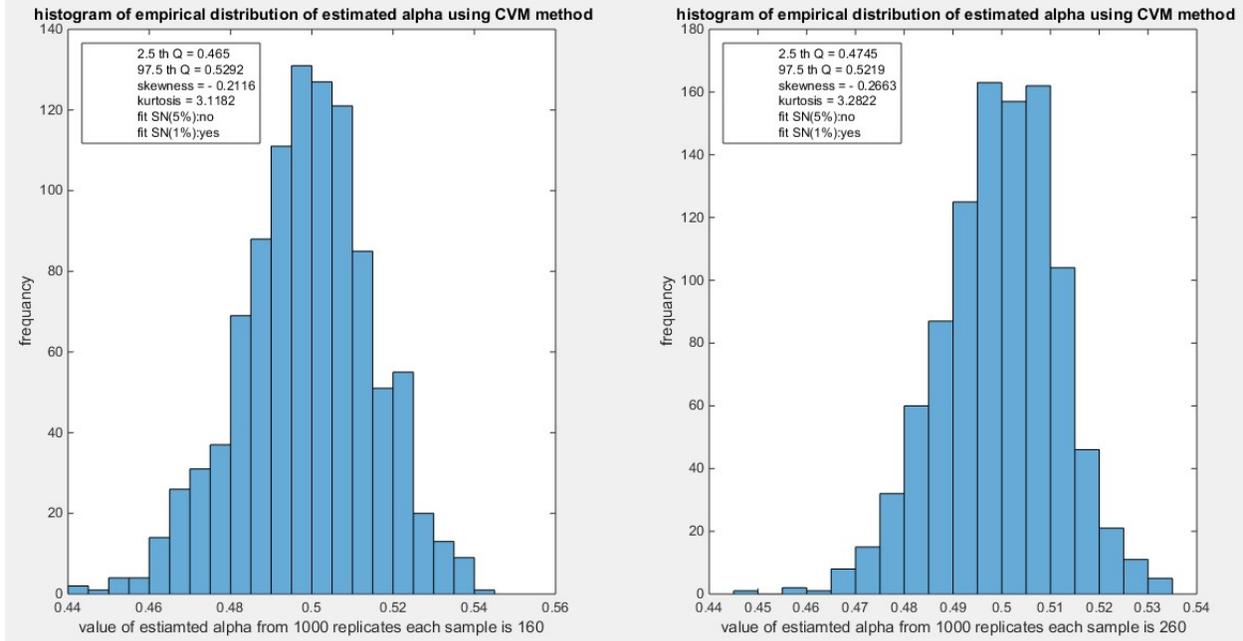

Fig. 55 shows the histogram of the empirical distribution of the estimated alpha from the 1000 replicates with sample size (n=160) on the left and (n=260) on the right using CVM method

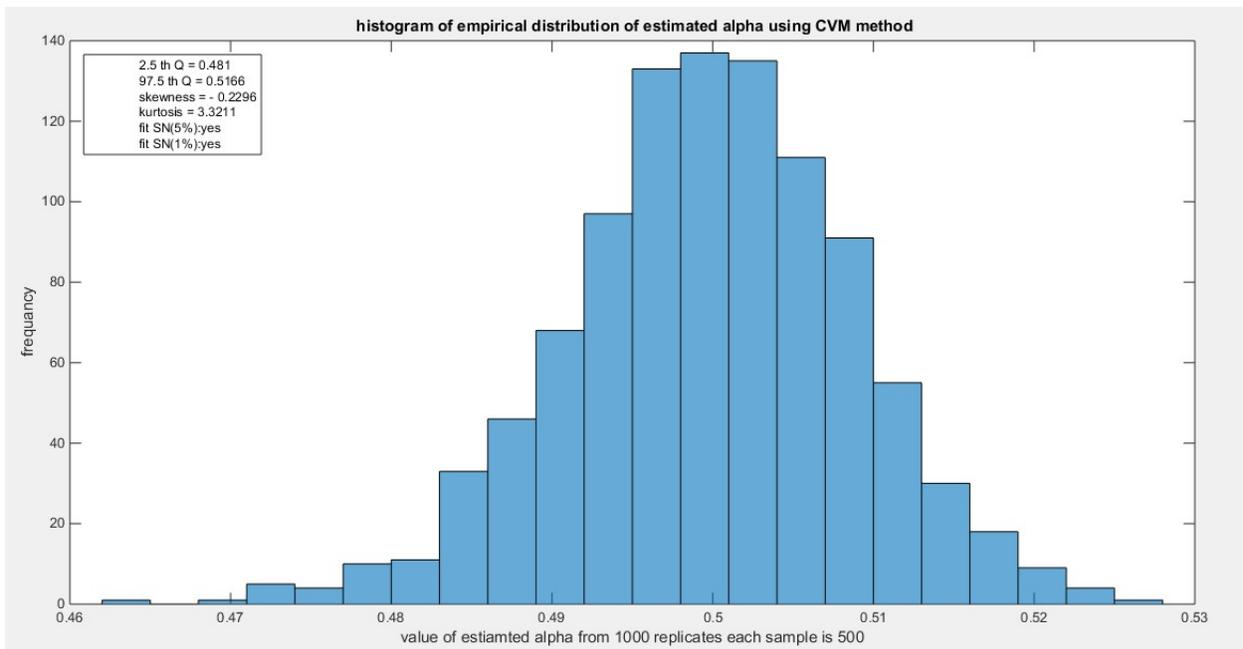

Fig.56 shows the histogram of the empirical distribution of the estimated alpha from the 1000 replicates with sample size (n =500) using CVM method

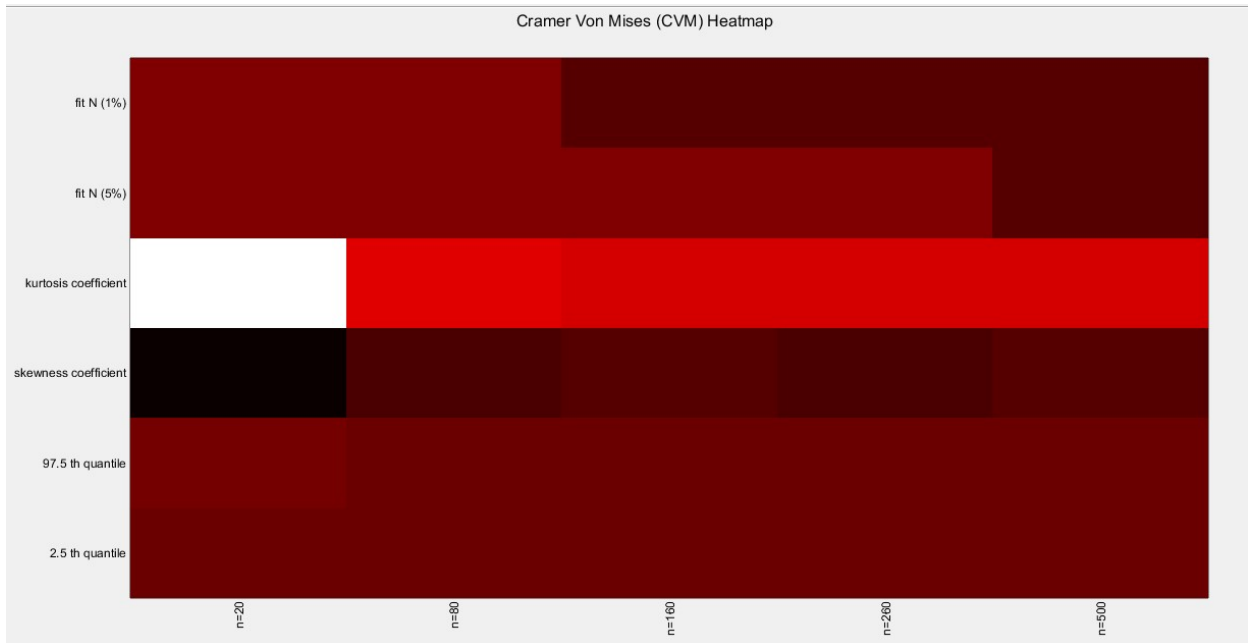

Fig. 57 shows the heat map of the indices of the empirical distribution of the estimated alpha using (cvm) method and how these indices change with changing the sample size from 20 to 500. (h value show)

Table (19): characteristics of empirical distribution of estimated alpha using LS

| LS | n=20 | n=80 | n=160 | n=260 | n=500 |
|---|---|---|---|---|---|
| 2.5 Q | 0.3691 | 0.4503 | 0.4651 | 0.4741 | 0.4809 |
| 97.5 Q | 0.5821 | 0.5411 | 0.5297 | 0.5218 | 0.5167 |
| Skewness | -2.1837 | -0.3467 | -0.2108 | -0.3274 | -0.2461 |
| Kurtosis | 15.4081 | 3.4104 | 3.1182 | 3.4525 | 3.3134 |
| Fit N (5%) | $H_0=1$ (0.001) | $H_0=1$ (0.001) | $H_0=1$ (0.0405) | $H_0=1$ (0.0213) | $H_0=0$ (0.1329) |
| Fit N (1%) | $H_0=1$ (0.001) | $H_0=1$ (0.001) | $H_0=0$ (0.0405) | $H_0=0$ (0.0213) | $H_0=0$ (0.1329) |

The empirical distribution of the estimated parameter alpha using LS is shown in Table 19. Each column represents a specific sample size with 1000 replicates in each size. Each column depicts the characteristics of the empirical distribution of the estimated alpha. The 2.5 th quantile and the 97.5 th quantile of the 1000 values of the estimated parameter in each sample shows that as the sample size increases the 2.5 th quantile rises while the 97.5 th quantile decreases. In other words, the distance between the quantiles decreases as the sample size increases and this is reflected on the confidence interval (CI). As the sample size increases the CI becomes narrower. The distribution exhibits a moderate left skewness and a high positive excess kurtosis (leptokurtic shape) at small sample size. As sample size increases the skewness decreases trying to approach the zero level (skewness of standard normal) and kurtosis decreases trying to approach the kurtosis of standard normal. The empirical distribution fits standard normal starting at size 160 and larger than this at significance level 5% and 1% with associated P-value as shown in the table. $H_0=1$ means reject the null hypothesis that states the parameter distribution follows the standard normal distribution. While $H_0=0$ means fail to reject the null hypothesis. See the following Figures (58-61)

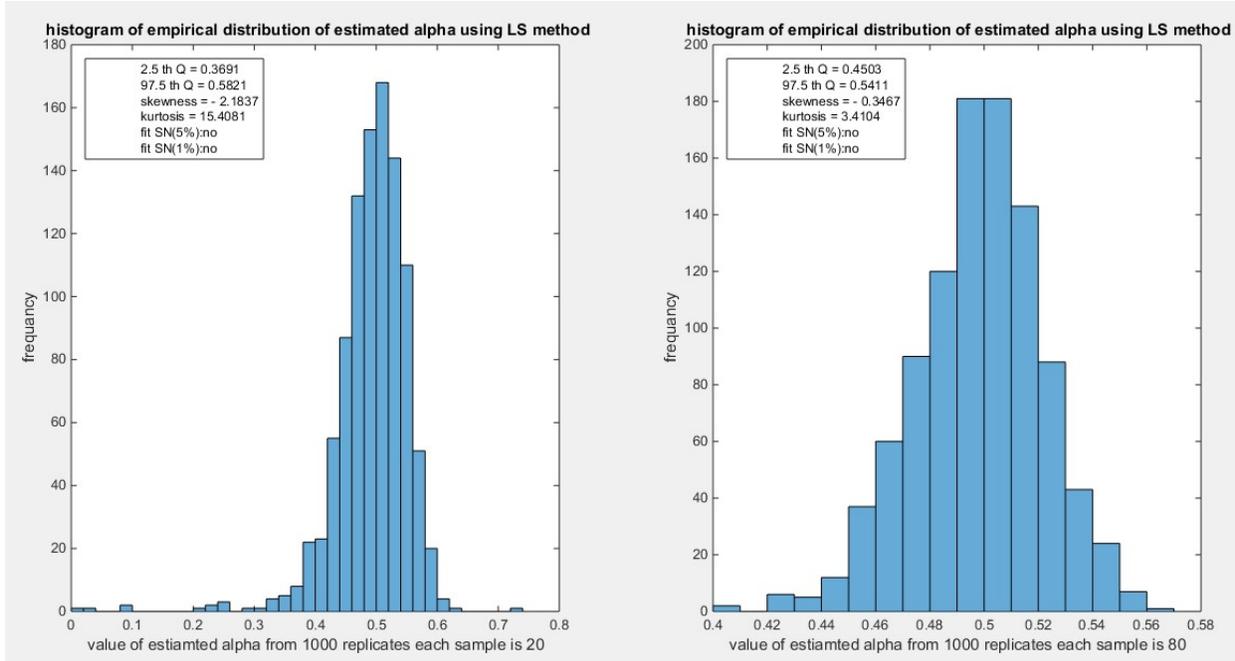

Fig. 58 shows the histogram of the empirical distribution of the estimated alpha from the 1000 replicates with sample size (n=20) on the left and (n=80) on the right using LS method

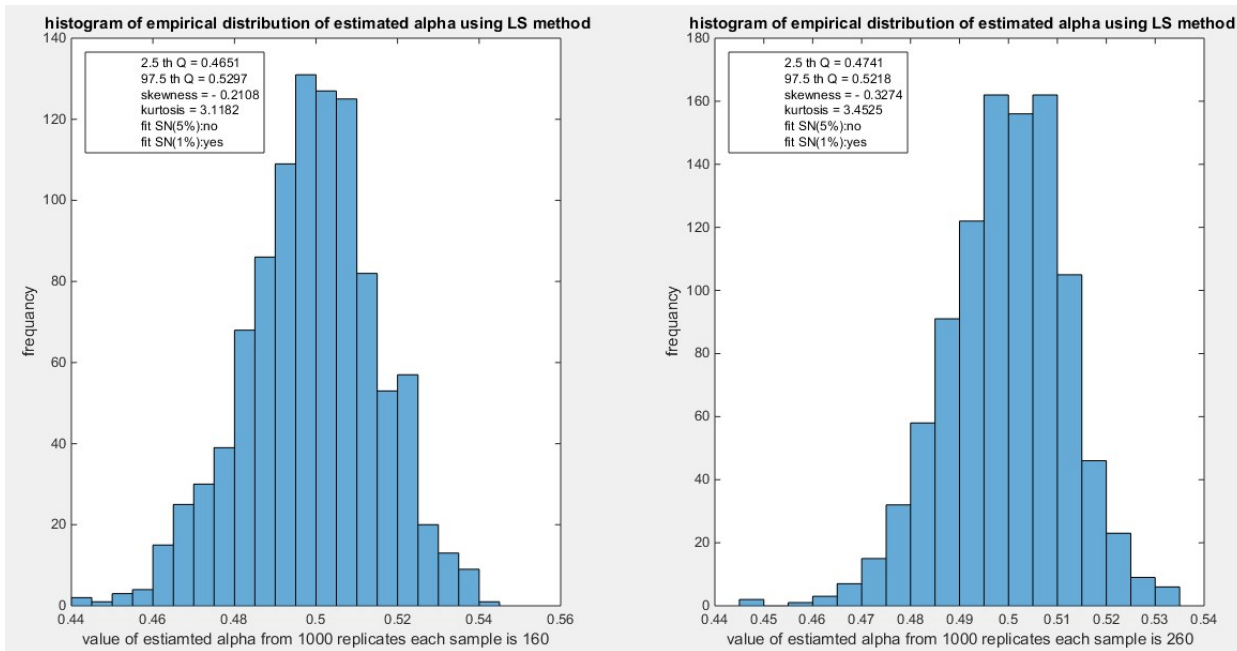

Fig. 59 shows the histogram of the empirical distribution of the estimated alpha from the 1000 replicates with sample size (n=160) on the left and (n=260) on the right using LS method

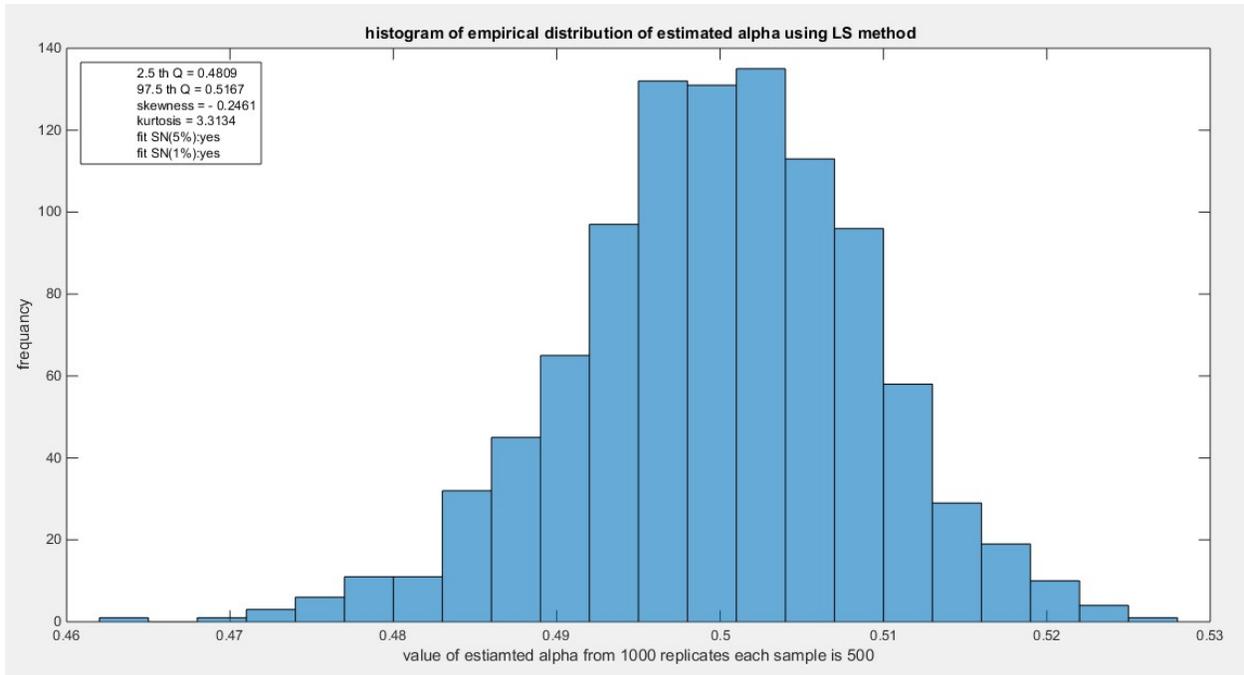

Fig. 60 shows the histogram of the empirical distribution of the estimated alpha from the 1000 replicates with sample size (n =500) using LS method

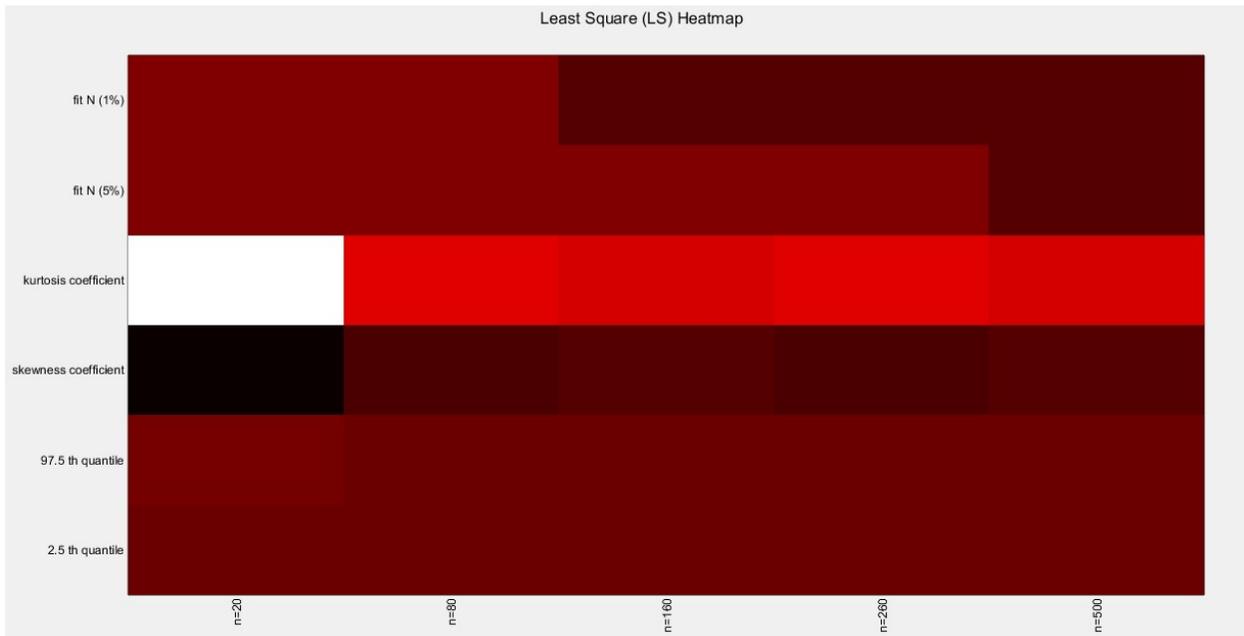

Fig. 61 shows the heat map of the indices of the empirical distribution of the estimated alpha using (LS) method and how these indices change with changing the sample size from 20 to 500. (h value is shown)

Table (20): characteristics of empirical distribution of estimated alpha using percentile

| PERCENTILE | n=20 | n=80 | n=160 | n=260 | n=500 |
|---|---|---|---|---|---|
| 2.5 Q | 0.4386 | 0.463 | 0.4736 | 0.4782 | 0.4836 |
| 97.5 Q | 0.6012 | 0.5473 | 0.5304 | 0.5246 | 0.5172 |
| Skewness | 0.3592 | 0.1697 | 0.0348 | -0.0114 | -0.1044 |
| Kurtosis | 3.1173 | 2.9065 | 2.8661 | 3.1032 | 2.9806 |
| Fit N (5%) | $H_0=1$ (0.0078) | $H_0=1$ (0.0299) | $H_0=1$ (0.046) | $H_0=0$ (0.1917) | $H_0=0$ (0.5) |
| Fit N (1%) | $H_0=1$ (0.0078) | $H_0=0$ (0.0299) | $H_0=0$ (0.046) | $H_0=0$ (0.1917) | $H_0=0$ (0.5) |

The empirical distribution of the estimated parameter alpha using percentile method is shown in Table 20. Each column represents a specific sample size with 1000 replicates in each size. Each column depicts the characteristics of the empirical distribution of the estimated alpha. The 2.5 th quantile and the 97.5 th quantile of the 1000 values of the estimated parameter in each sample shows that as the sample sizes increases the 2.5 th quantile rises while the 97.5 th quantile decreases. In other words, the distance between the quantiles decreases as the sample size increases and this is reflected on the confidence interval (CI). As the sample size increases the CI becomes narrower. As sample size increases the distribution exhibits oscillation between a mild left skewness and a mild right skewness. In contrast, it exhibits gradual mild decrease in kurtosis (platykurtic shape) as sample size rises and it approaches mesokurtic shape (kertosis coefficient=3) of the standard normal at sample size 500. The empirical distribution fits standard normal starting at size 80 and larger than this size at significance level 5% and 1% with associated P-value as shown in the table. $H_0=1$ means reject the null hypothesis that states the parameter distribution follows the standard normal distribution. While $H_0=0$ means fail to reject the null hypothesis. See figures (62-65)

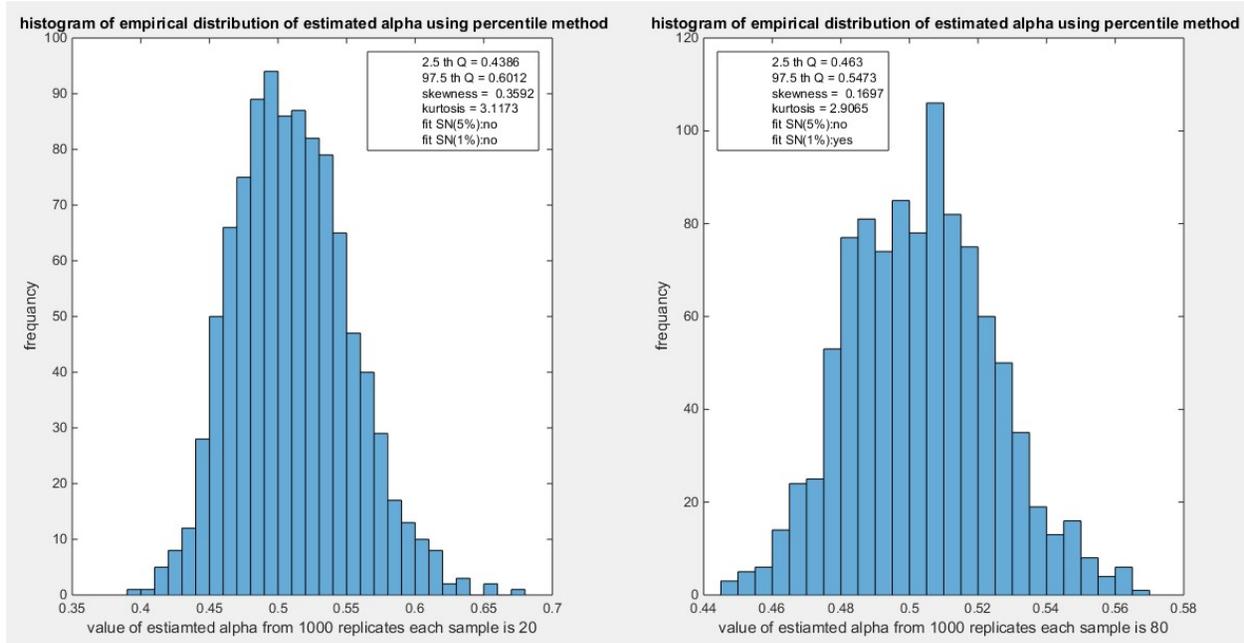

Fig. 62 shows the histogram of the empirical distribution of the estimated alpha from the 1000 replicates with sample size (n=20) on the left and (n=80) on the right using percentile method

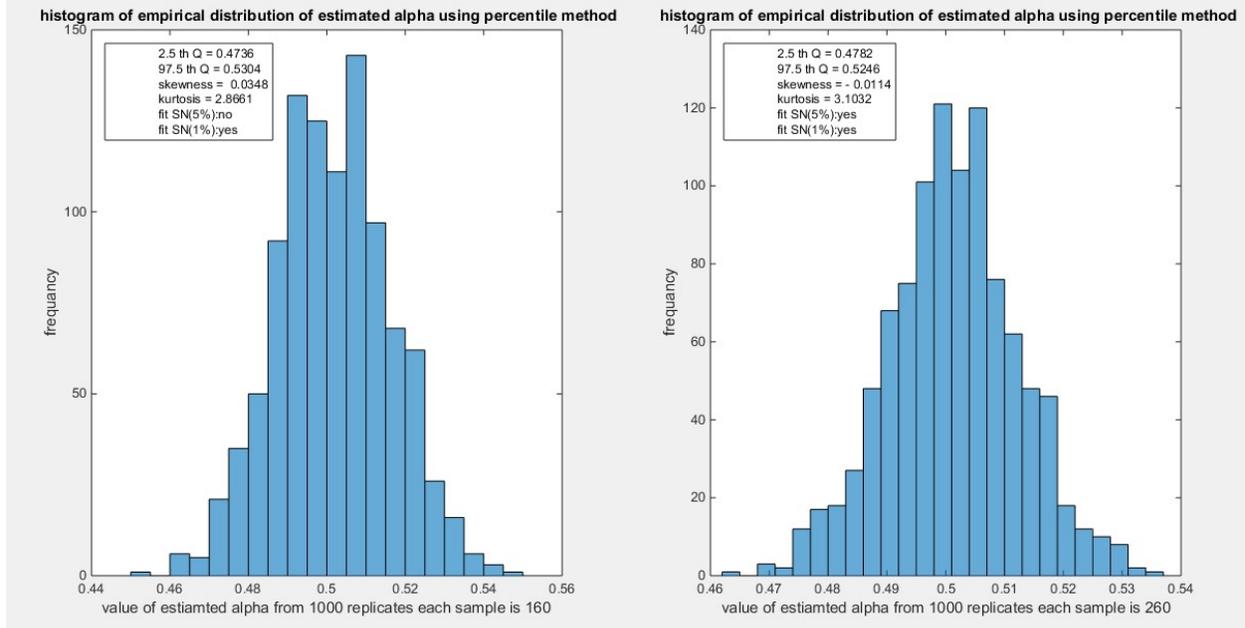

Fig. 63 shows the histogram of the empirical distribution of the estimated alpha from the 1000 replicates with sample size (n=160) on the left and (n=260) on the right using percentile method

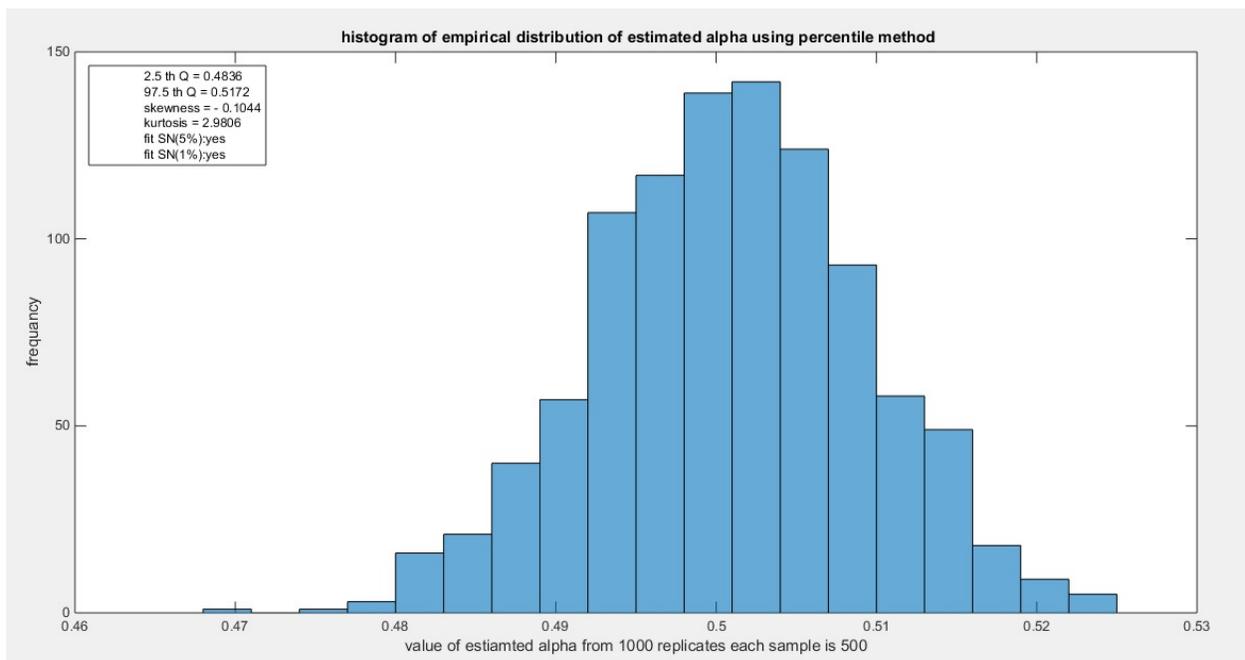

Fig. 64 shows the histogram of the empirical distribution of the estimated alpha from the 1000 replicates with sample size (n =500) using percentile method

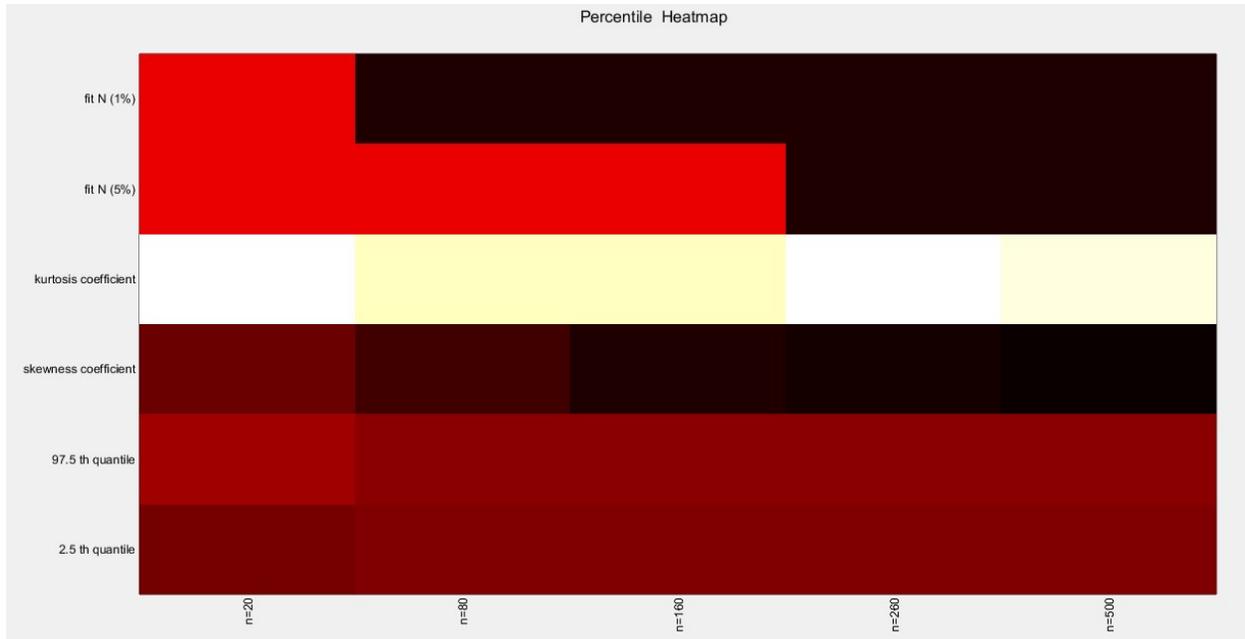

Fig. 65 shows the heat map of the indices of the empirical distribution of the estimated alpha using (percentile) method and how these indices change with changing the sample size from 20 to 500. (h value is shown)

Table (21): characteristics of empirical distribution of estimated alpha using WLS

| WLS | n=20 | n=80 | n=160 | n=260 | n=500 |
|---|---|---|---|---|---|
| 2.5 Q | 0.3885 | 0.4547 | 0.4676 | 0.4756 | 0.4816 |
| 97.5 Q | 0.5784 | 0.5406 | 0.5299 | 0.5222 | 0.5162 |
| Skewness | -1.2701 | -0.1687 | -0.113 | -0.2175 | -0.1815 |
| Kurtosis | 8.7862 | 3.1044 | 3.0195 | 3.338 | 3.1696 |
| Fit N (5%) | $H_0=1$ (0.001) | $H_0=0$ (0.2563) | $H_0=0$ (0.296) | $H_0=0$ (0.182) | $H_0=0$ (0.1715) |
| Fit N (1%) | $H_0=1$ (0.001) | $H_0=0$ (0.2563) | $H_0=0$ (0.296) | $H_0=0$ (0.182) | $H_0=0$ (0.1715) |

The empirical distribution of the estimated parameter alpha using WLS method is shown in Table 21. Each column represents a specific sample size with 1000 replicates in each size. Each column depicts the characteristics of the empirical distribution of the estimated alpha. The 2.5 th quantile and the 97.5 th quantile of the 1000 values of the estimated parameter in each sample shows that as the sample sizes increases the 2.5 th quantile rises while the 97.5 th quantile decreases. In other words, the distance between the quantiles decreases as the sample size increases and this is reflected on the confidence interval (CI). As the sample size increases the CI becomes narrower. The distribution exhibits mild left skewness and high positive excess kurtosis (leptokurtic shape) at small sample size. As sample size increases the skewness decreases trying to approach the zero level (skewness of standard normal) and kurtosis decreases trying to approach the kurtosis of standard normal. The empirical distribution fits standard normal starting at size 80 and larger than this size at significance level 5% and 1% with associated P-value as shown in the table. $H_0=1$

means reject the null hypothesis that states the parameter distribution follows the standard normal distribution. While H₀=0 means fail to reject the null hypothesis. See the following figures (66-69)

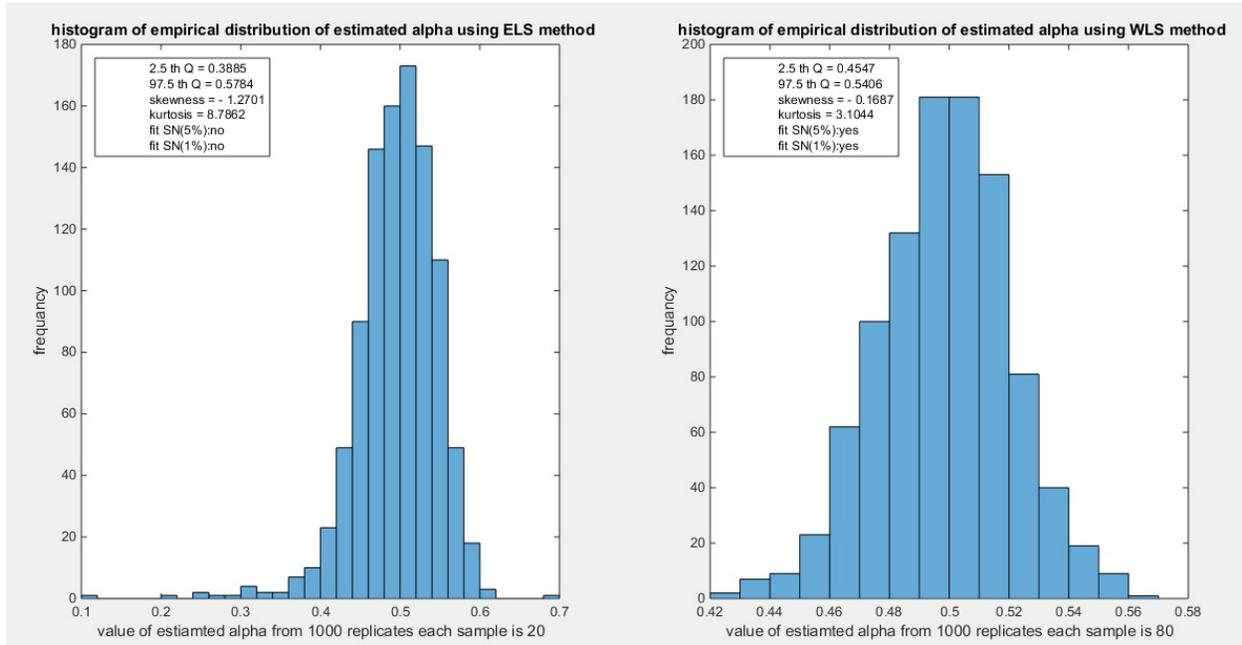

Fig. 66 shows the histogram of the empirical distribution of the estimated alpha from the 1000 replicates with sample size (n=20) on the left and (n=80) on the right using percentile method

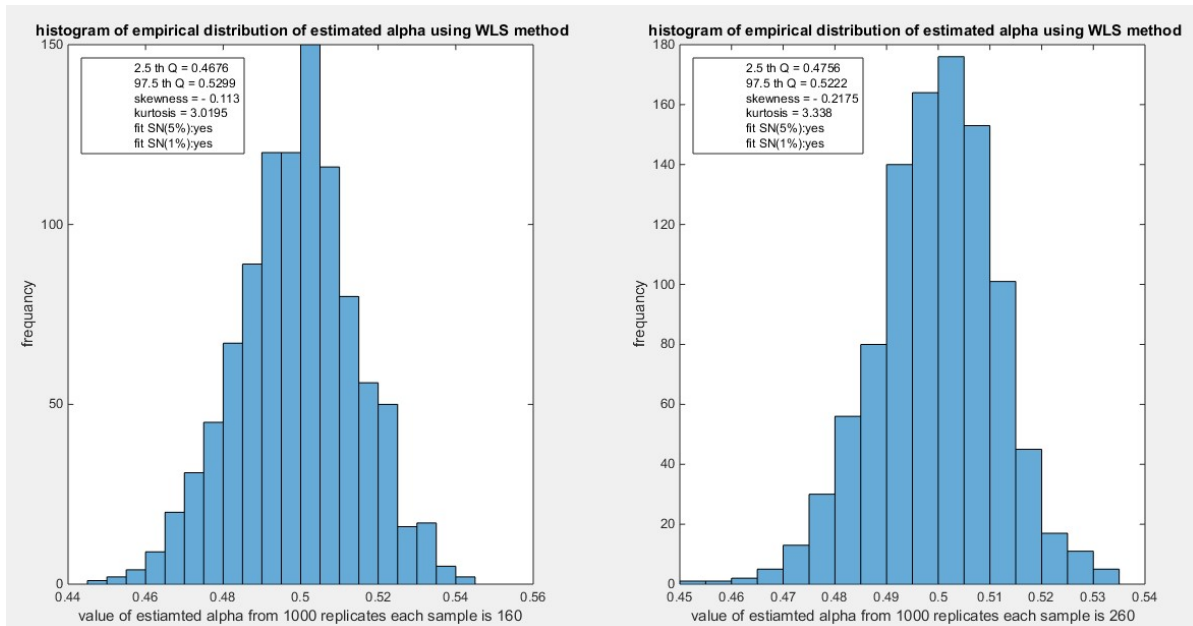

Fig.67 shows the histogram of the empirical distribution of the estimated alpha from the 1000 replicates with sample size (n=160) on the left and (n=260) on the right using percentile method

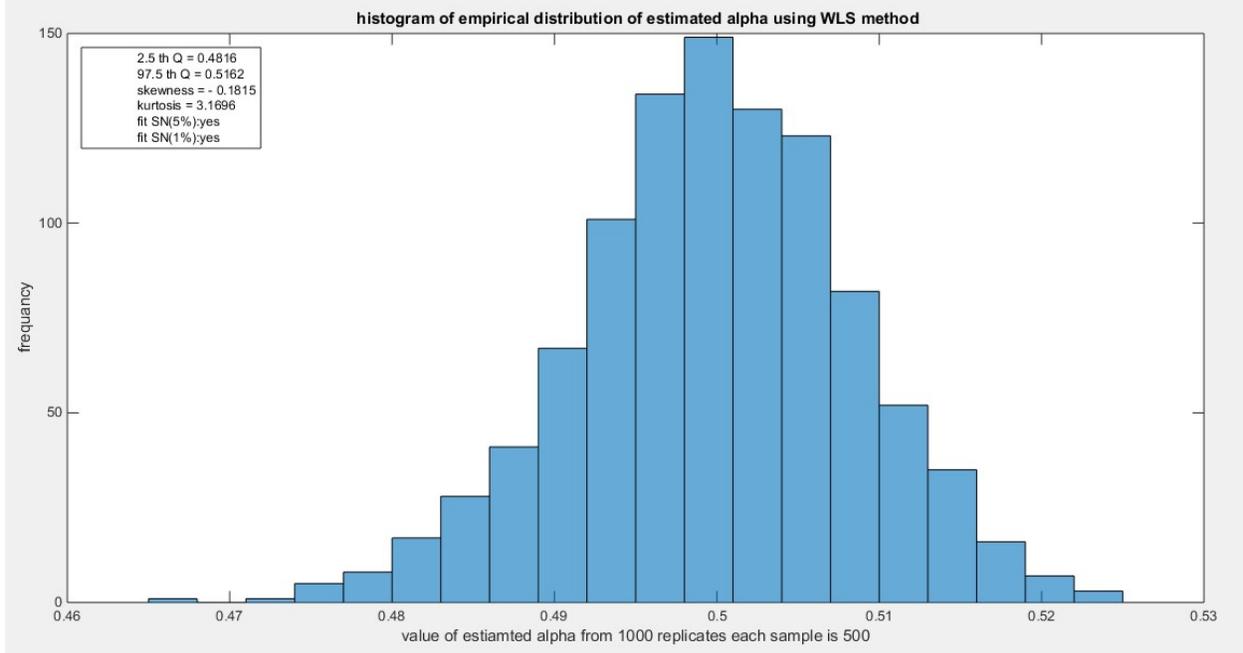

Fig.68 shows the histogram of the empirical distribution of the estimated alpha from the 1000 replicates with sample size (n =500) using percentile method

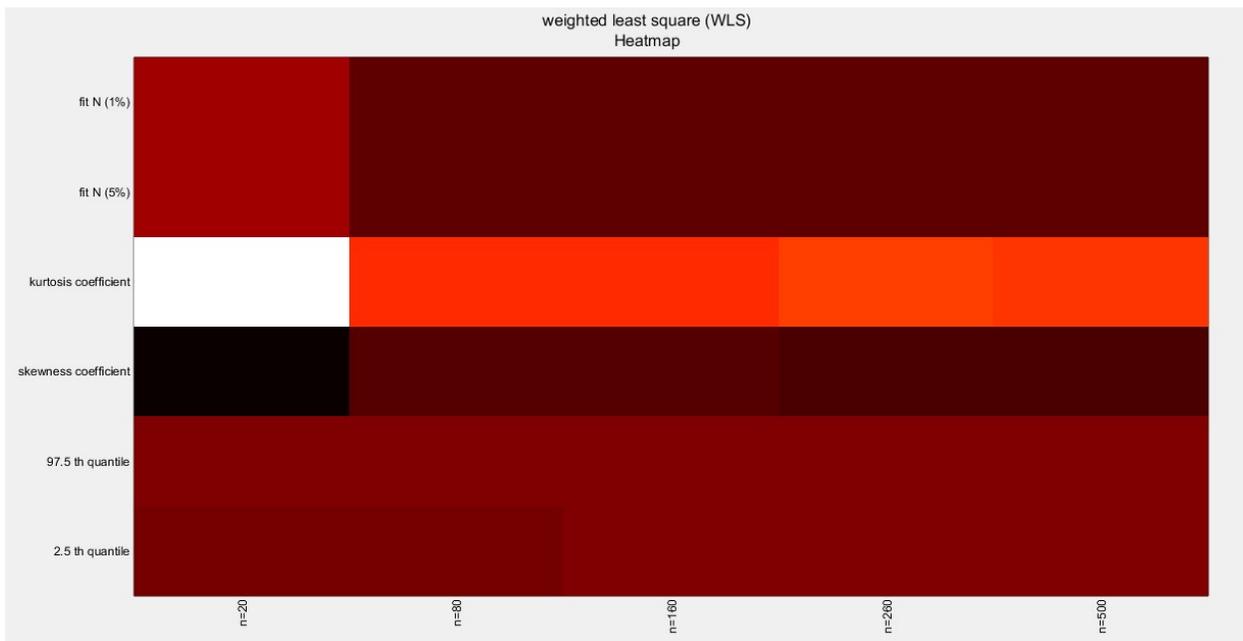

Fig.69 shows the heat map of the indices of the empirical distribution of the estimated alpha using (percentile) method and how these indices change with changing the sample size from 20 to 500.

# Supplementary materials (section 3)

**4.1. Analysis of the first data set** see table (1) & table (2). See figures (1-5).

Table (1): Descriptive statistics of the first data set:

| min | mean | std | skewness | kurtosis | $25_{perc}$ | $50_{perc}$ | $75_{perc}$ | max |
|---|---|---|---|---|---|---|---|---|
| 0.001 | 0.0345 | 0.056 | 2.5981 | 10.9552 | 0.0032 | 0.007 | 0.0455 | 0.259 |

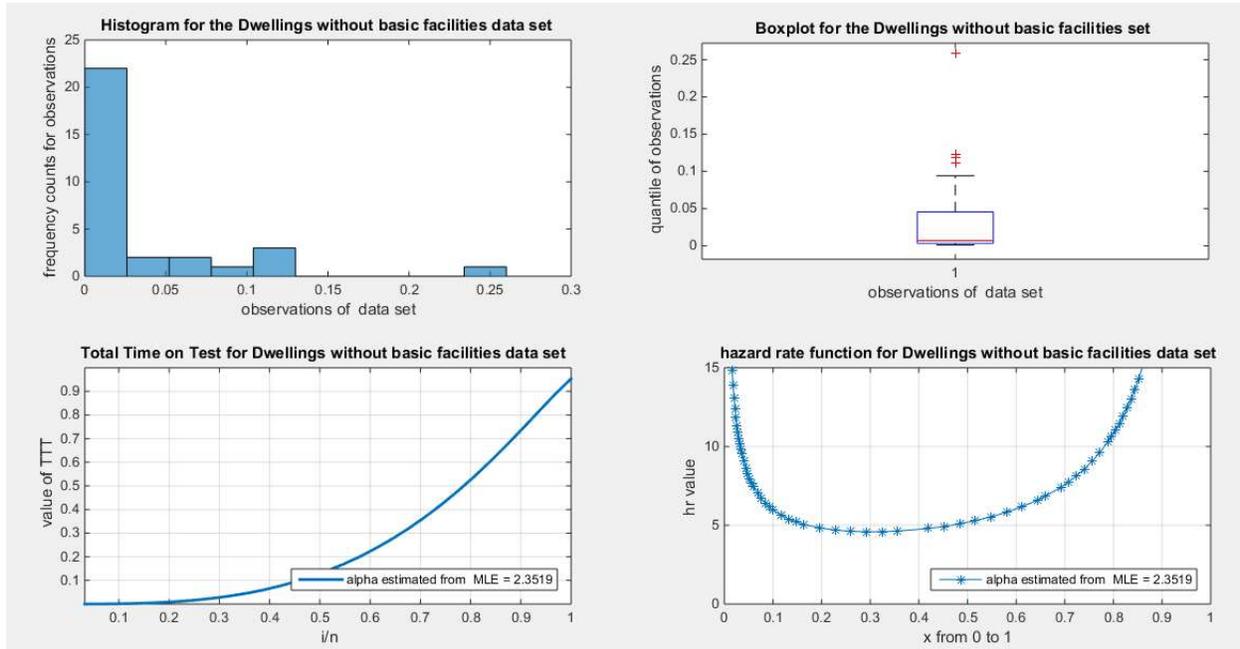

Fig.1 shows the histogram with right skewness and associated boxplot with extreme values located in the upper tail (around 4 values). The TTT plot shows convex shape which supports decreased failure rate that is obvious in the shape of the hazard function on the right lower graph. The bathtub shape clearly seen in the hazard shape function is not reflected on the TTT plot drawn used the first or the second approach.

This data set exhibits right skewness. After fitting the MBUR distribution to the data, the scaled TTT plot shows convex shape denoting decreased failure rate. The TTT plot used here is graphed by the second approach that depends on the theoretical quantile. The hazard rate function graph also depends on the estimated parameter after fitting the MBUR to the data. Although the hazard function has a bathtub shape, the scaled TTT plot does not show the well-known shape of convexity followed by concavity. The scaled TTT plot shows only the initial convexity. This is even true when the scaled TTT plot was graphed using the empirical or first approach as shown in figure (1). In figure (2), the scaled TTT plot was drawn using the first approach with convexity reflecting the decreased hazard rate function.

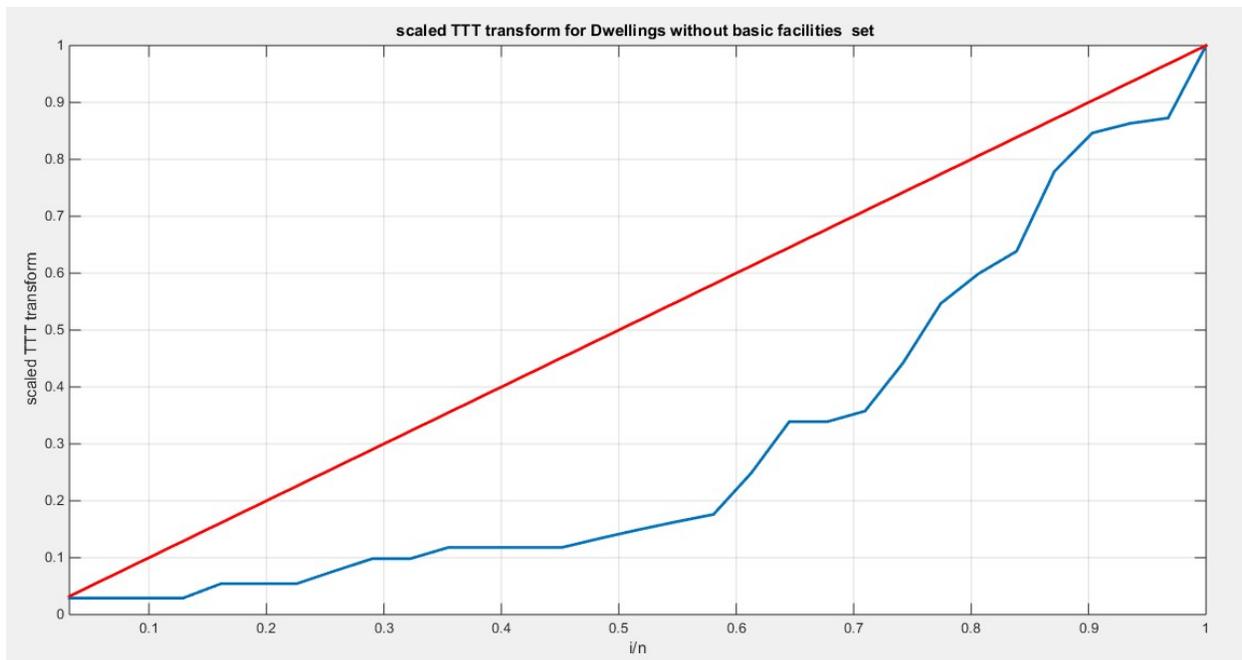

Fig. 2 shows the scaled TTT plot for the Dwellings without basic facilities using the first approach before fitting MBUR to the data.

Five distributions are evaluated to see whether they fit the data or not. The results of the fitting and validation indices are shown in table (2). The Kumaraswamy is the first to fit the data followed by Beta distribution and lastly MBUR. The Topp-Leone and Unit Lindley failed to fit the data. Figure (3) shows the empirical CDFs vs. the theoretical CDFs for the different competitors. Figure (4) shows the fitted PDFs for the different competitors. Figure (5) shows the QQ plot and the log-likelihood for the fitted MBUR. Although MBUR fitted the data, this is not obviously visually reflected in QQ plot especially in the upper tail. The Maximum log-likelihood value is attained at the estimated value of the parameter. The P-values for the estimators of alpha and beta parameters of the Beta distribution and Kumaraswamy distributions are significant $(p < 0.001)$.

P-values for the estimators of alpha of the MBUR distribution is significant $(p < 0.001)$. Figure (6) shows the QQ plot and PP plot for other distributions.

Table (2): Estimators and validation indices for the First data set

|  | Beta | | Kumaraswamy | | MBUR | Topp-Leone | Unit-Lindley |
|---|---|---|---|---|---|---|---|
| theta | $\alpha = 0.5086$ | | $\alpha = 0.6013$ | | 2.3519 | 0.2571 | 26.1445 |
|  | $\beta = 14.036$ | | $\beta = 8.5999$ | | | | |
| Var | .0323 | .6661 | .0086 | .2424 | 0.023 | 0.0021 | 20.5623 |
|  | .6661 | 22.1589 | .2424 | 9.228 | | | |
| SE | 0.03227 | | 0.01666 | | 0.0272 | 0.0082 | 0.8144 |
|  | 0.8455 | | 0.5456 | | | | |
| AIC | -153.5535 | | -155.8979 | | -143.3057 | -133.593 | -140.592 |
| CAIC | -153.1249 | | -155.4693 | | -143.1678 | -133.4551 | -140.454 |
| BIC | -150.6855 | | -153.0299 | | -141.8717 | -132.159 | -139.158 |
| HQIC | -152.6186 | | -154.963 | | -142.8382 | -133.1255 | -140.1243 |
| LL | 78.7767 | | 79.9489 | | 72.6528 | 67.7965 | 71.2959 |
| K-S Value | 0.2052 | | 0.1742 | | 0.2034 | 0.2818 | 0.3762 |
| H$_0$ | Fail to reject | | Fail to reject | | Fail to reject | reject | reject |
| P-value | 0.1271 | | 0.271 | | 0.1336 | 0.0114 | 0.000189 |
| AD | 1.288 | | 0.9566 | | 2.8789 | 4.6021 | 7.439 |
| CVM | 0.2419 | | 0.1659 | | 0.4976 | 0.8749 | 1.1683 |

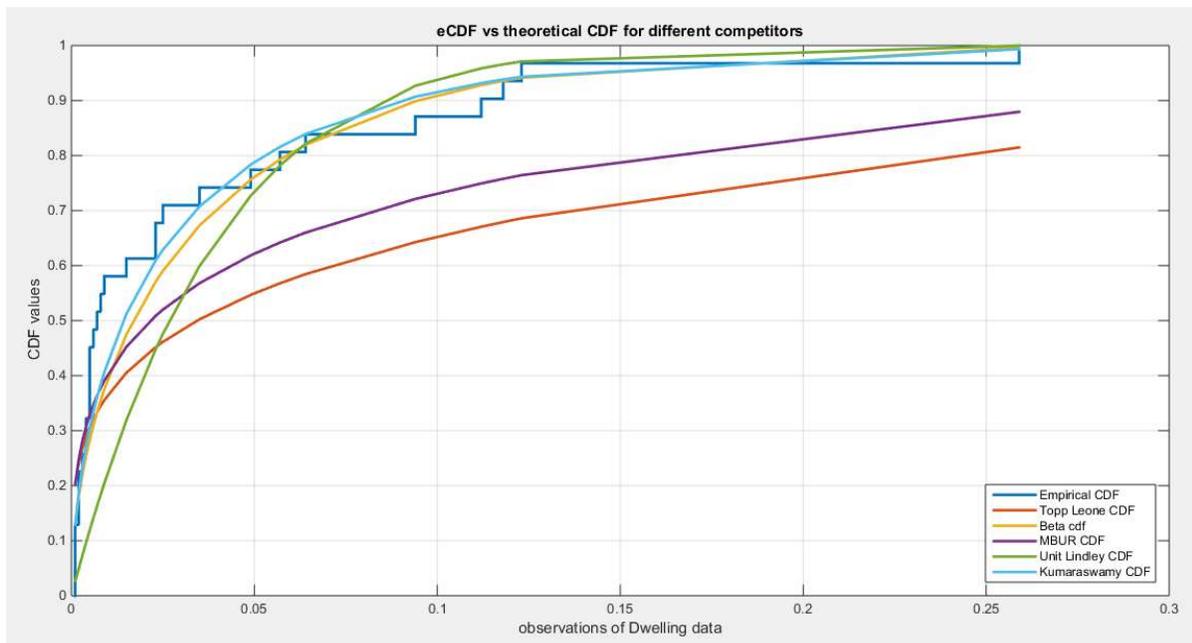

Fig. 3 shows the eCDF vs. theoretical CDF of the 5 distributions for the 1st data set (Dwellings without basic facilities).

As shown from the analysis, 3 distributions better fit the data than the others. These are Beta, Kumaraswamy, and MBUR distributions. This is because the K-S test failed to reject the null hypothesis, $H_0$, which hypothesized that the data being from the test distribution. Kumaraswamy has the highest negative values of NLL, AIC, corrected AIC, BIC, and HQIC values in comparison with the values obtained from the Beta and the MBUR distributions. MBUR distribution is the third distribution to fit the data after Kumaraswamy and Beta distribution in this order. Kumaraswamy is the distribution that has the smallest values of K-S test, AD test, and CVM test, followed by Beta then MBUR distribution. The one parameter distribution has done a good job to fit the data after the two parameters distributions; Kumaraswamy and Beta. And it fits the data better than the other one parameter distribution unit Lindely and Topp Leone which fail to fit the data.

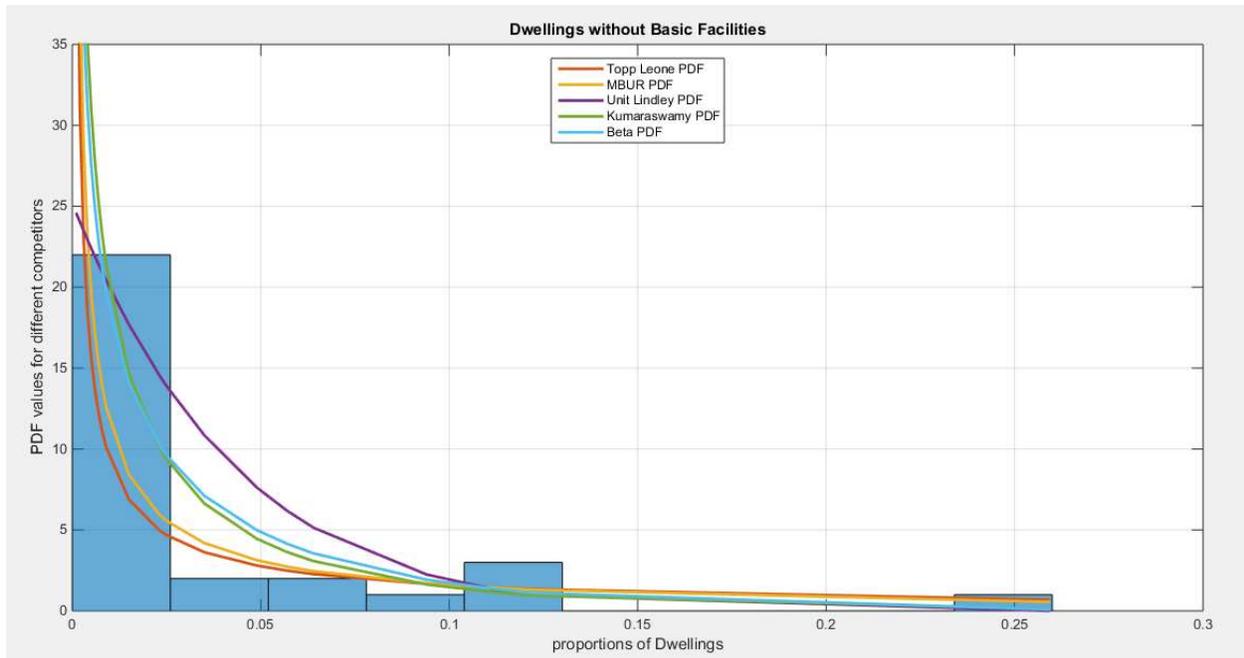

Fig. 4 shows the fitted PDFs for the different competitors.

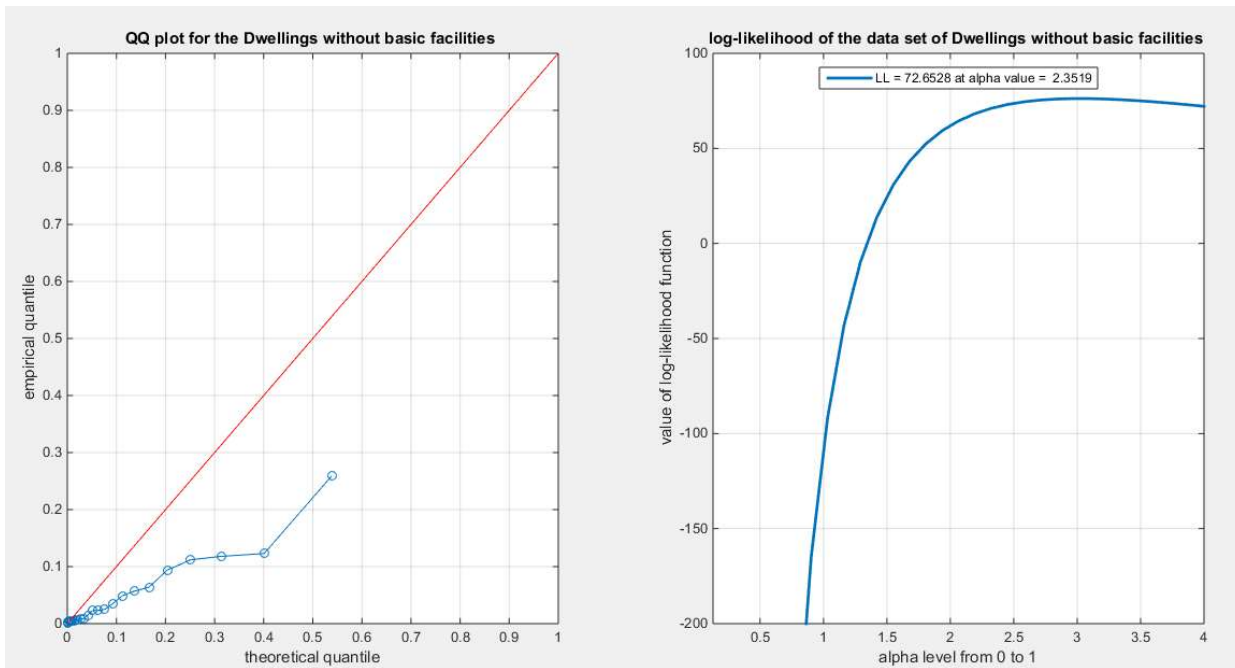

Fig.5 shows the QQ plot for Dwellings without basic facilities data set, on the left and the log-likelihood on the right, for the fitted MBUR distribution.

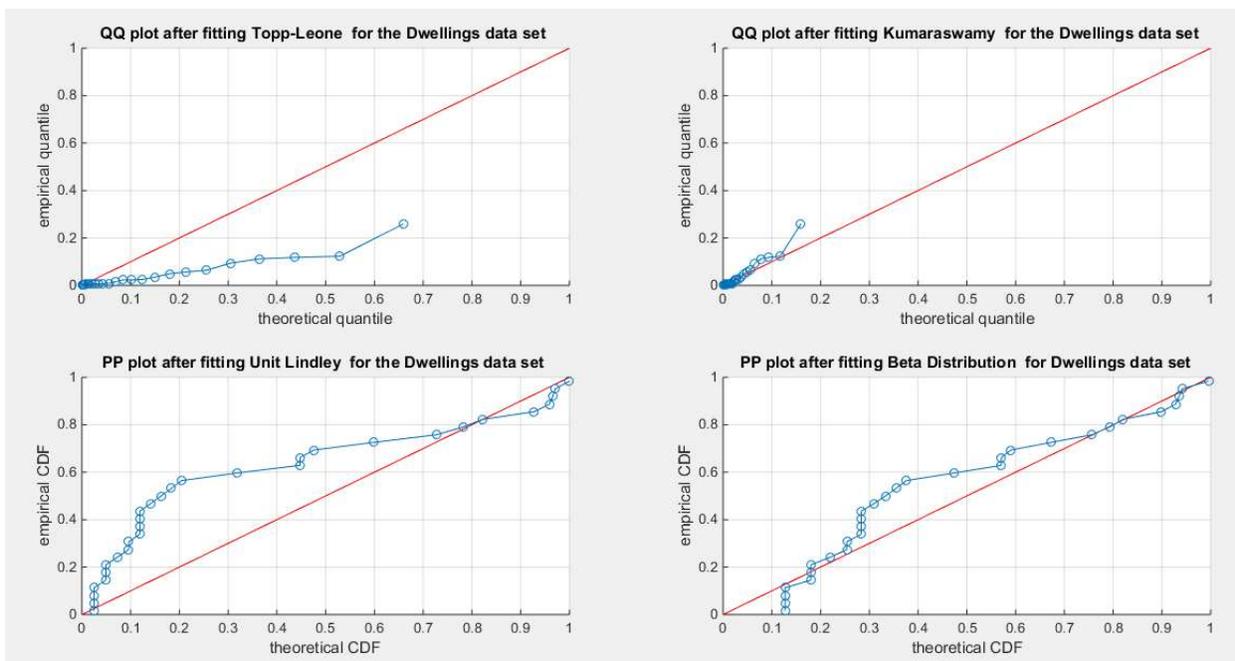

Fig.6 shows the QQ plot after fitting both Topp-Leone and Kumaraswamy distributions. The PP plot after fitting both Beta and Unit Lindley distributions are also seen.

### 4.3. Analysis of the third data set.  See Table (3) & Table (4). See figures (7-12)

Table (3): Descriptive statistics of the third data set

| min | mean | std | skewness | kurtosis | 25p | 50p | 75p | max |
|---|---|---|---|---|---|---|---|---|
| 0.42 | 0.7894 | 0.1504 | -1.3554 | 3.9461 | 0.75 | 0.84 | 0.895 | 0.94 |

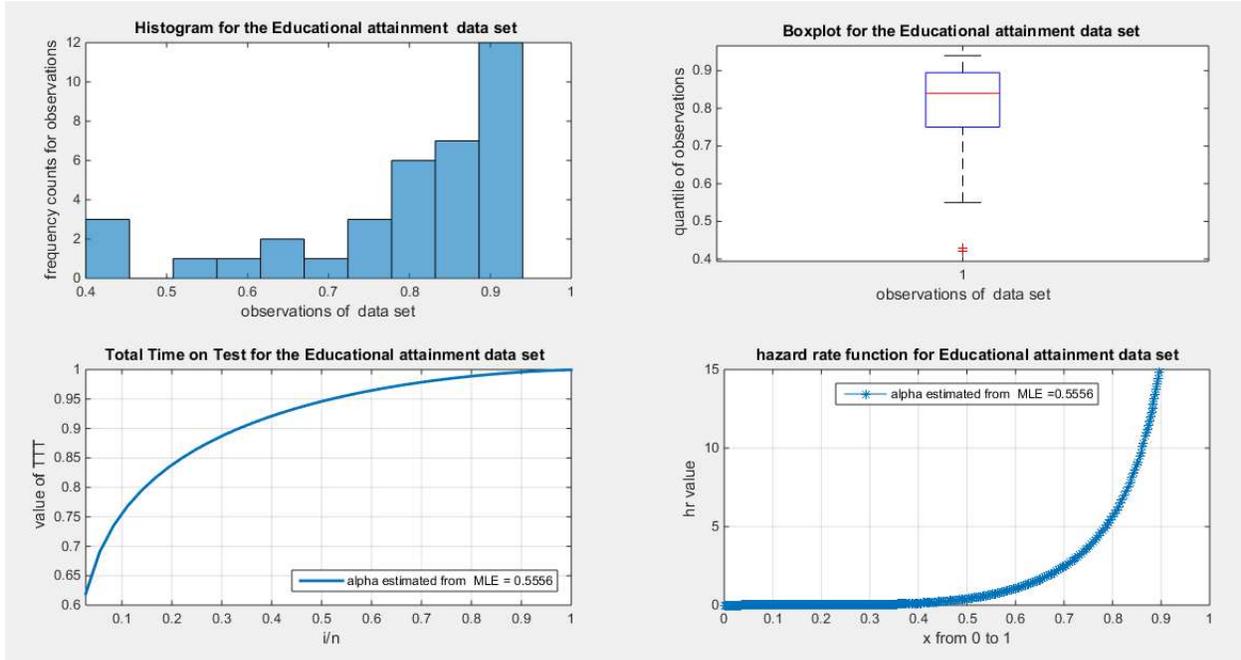

Fig.7 shows the histogram with left skewness and associated boxplot with two outliers or extreme values on the lower tail of the distribution. The TTT plot shows concave shape which supports increased failure rate that is obvious in the shape of the hazard function on the right lower graph.

The Unit Lindley distribution fitted the educational attainment data set followed by MBUR, Kumaraswamy and lastly Beta distribution. Topp-Leone did not fit the data. The MBUR is the second distribution to fit the data after UL because it has the second lowest indices (second largest negative values) as regards NLL, AIC, AIC corrected, BIC, HQIC. However, UL has the lowest values of KS test, AD test and CVM tests followed by Kumaraswamy, Beta and then MBUR.

The scaled TTT plot whether graphed by any approach, it is concave and so anticipating increased hazard rate function as shown in the lower 2 graphs of figure (7), and in figure (8). Table (4) shows the estimators for different distributions and the associated validity indices. Figure (9) illustrates the eCDFs and theoretical CDFs while Figure (10) shows the fitted PDFs. Figure (11) shows QQ plot after fitting BMUR distribution depicting an overall good alignment with some defective alignment at the lower end of the distribution. Figure (12) shows QQ plot and PP plot for other distributions.

P-values for the estimators of theta of the Unit Lindley distribution is significant ($p < 0.001$).P-values for the estimators of alpha of the MBUR distribution is significant ($p < 0.001$).

The P-values for the estimators of alpha and beta parameters of the Beta distribution and Kumaraswamy distributions are significant ($p < 0.001$).

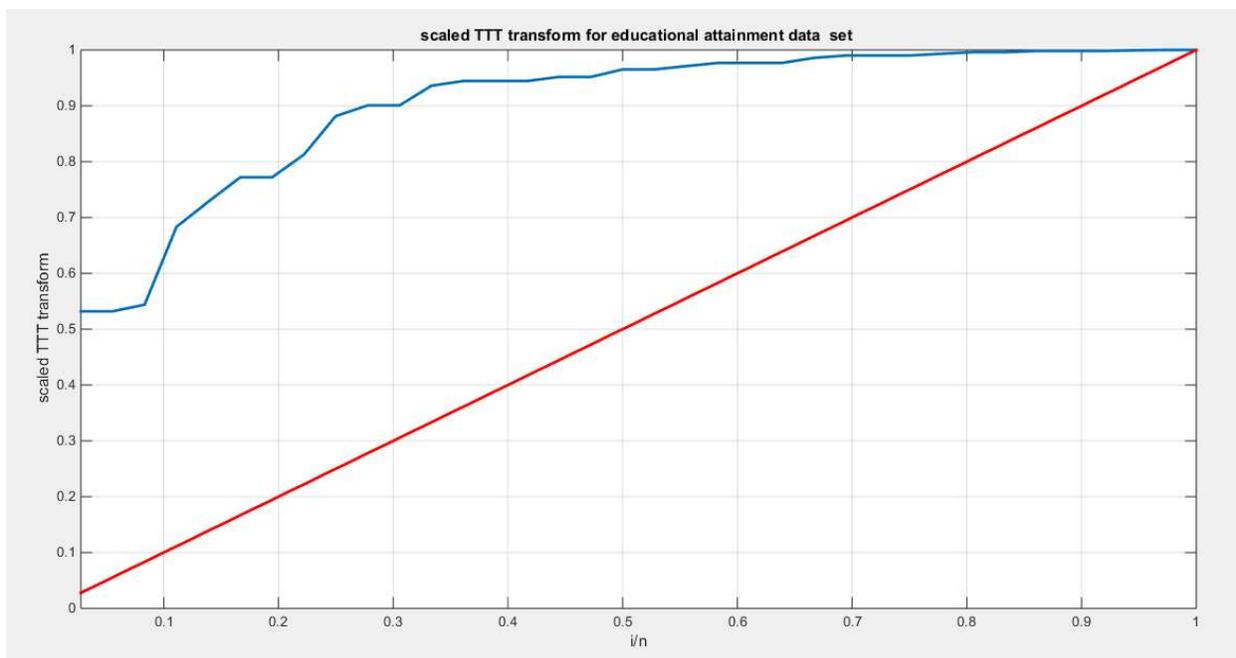

Fig. 8 shows the scaled TTT plot for the Educational attainment data set with a concave shape supporting the increased hazard rate as reflected in the shape of the hazard function.

Table (4): Estimators and validation indices for the Third data set

|  | Beta | | Kumaraswamy | | MBUR | Topp-Leone | Unit-Lindley |
|---|---|---|---|---|---|---|---|
| theta | $\alpha = 6.7222$ | | $\alpha = 6.0746$ | | 0.5556 | 13.4254 | 0.2905 |
|  | $\beta = 1.8405$ | | $\beta = 2.1284$ | | | | |
| Var | 3.4283 | 1.0938 | 1.3854 | 0.5232 | 0.0011 | 5.0067 | 0.0012 |
|  | 1.0938 | 0.416 | 0.5232 | 0.3234 | | | |
| SE | 0.3086 | | 0.1962 | | 0.0055 | 0.373 | 0.0058 |
|  | 0.1075 | | 0.0948 | | | | |
| AIC | -46.6152 | | -47.5937 | | -48.8713 | -40.5725 | -56.9322 |
| CAIC | -46.2516 | | -47.23 | | -48.7537 | -40.4548 | -56.8145 |
| BIC | -43.4482 | | -44.4266 | | -47.2878 | -38.9889 | -55.3487 |
| HQIC | -45.5098 | | -46.4883 | | -48.3186 | -40.0198 | -56.3795 |
| LL | 25.3076 | | 25.7968 | | 25.4357 | 21.2862 | 29.4661 |
| K-S | 0.1453 | | 0.1390 | | 0.1468 | 0.2493 | 0.0722 |
| H$_0$ | Fail to reject | | Fail to reject | | Fail to reject | Reject | Fail to reject |
| P-value | 0.2055 | | 0.2411 | | 0.1979 | 0.0062 | 0.8300 |
| AD | 1.2191 | | 1.1724 | | 1.2694 | 4.0059 | 0.2977 |
| CVM | 0.2041 | | 0.192 | | 0.2133 | 0.7501 | 0.0430 |

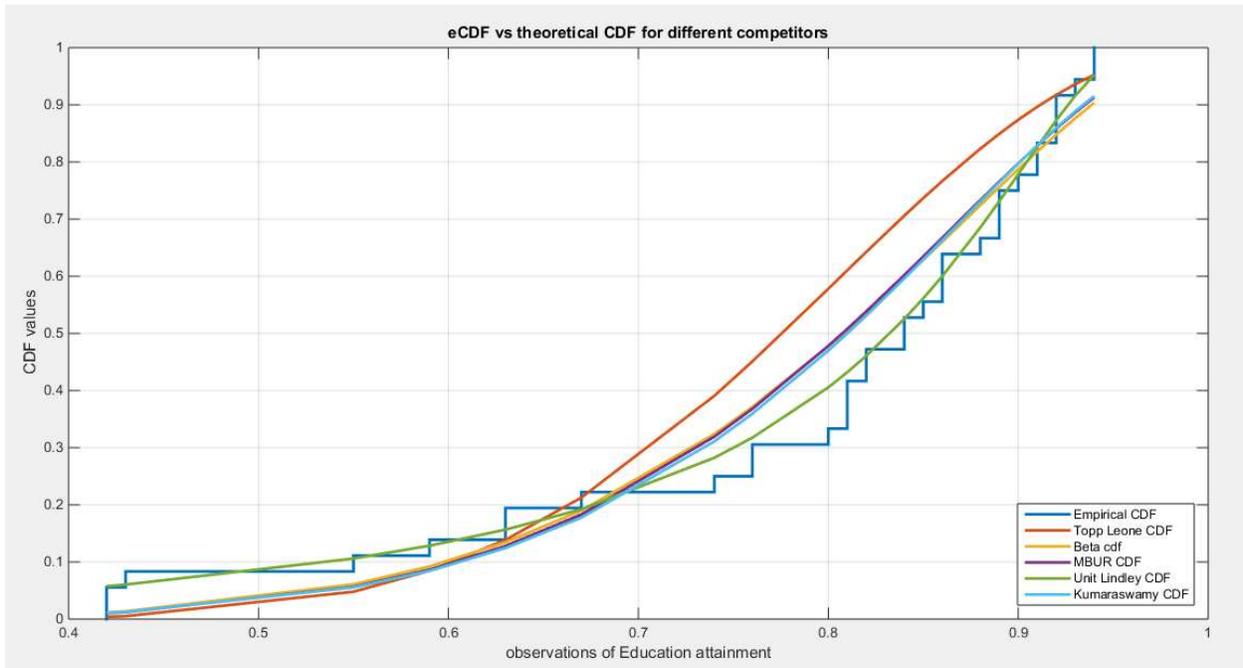

Fig. 9 shows the eCDFs vs. theoretical CDFs of the 5 distributions for the 3rd data set (Educational Attainment).

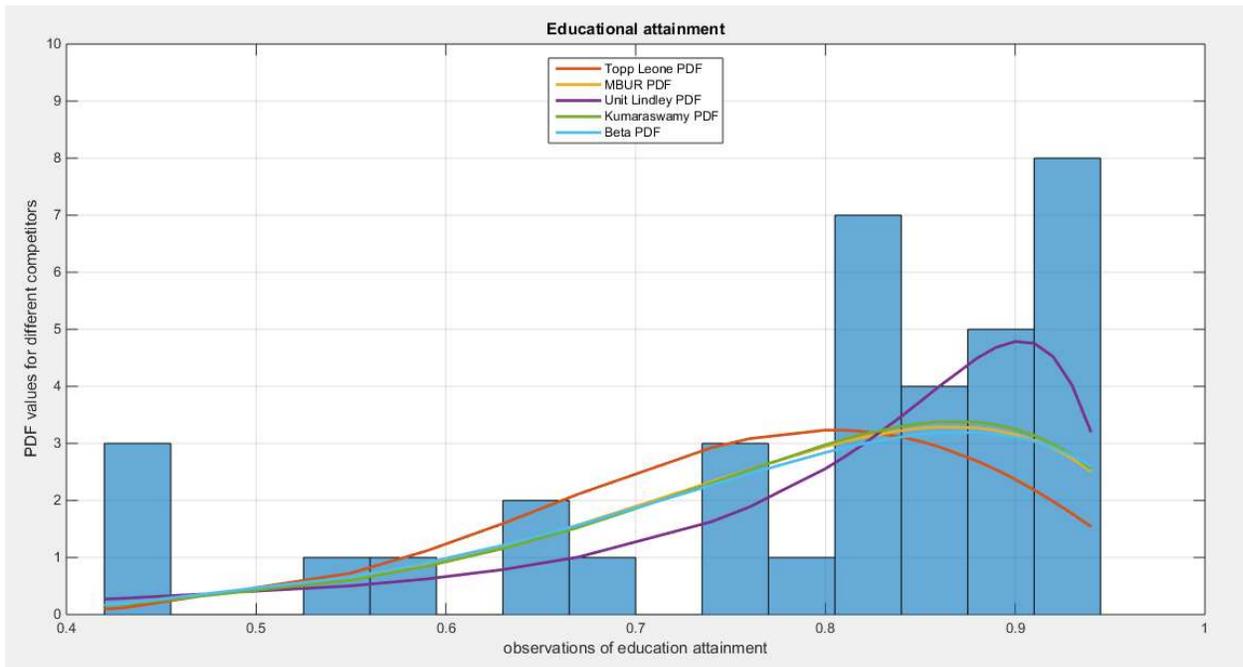

Fig.10 shows the PDFs of the fitted competitors

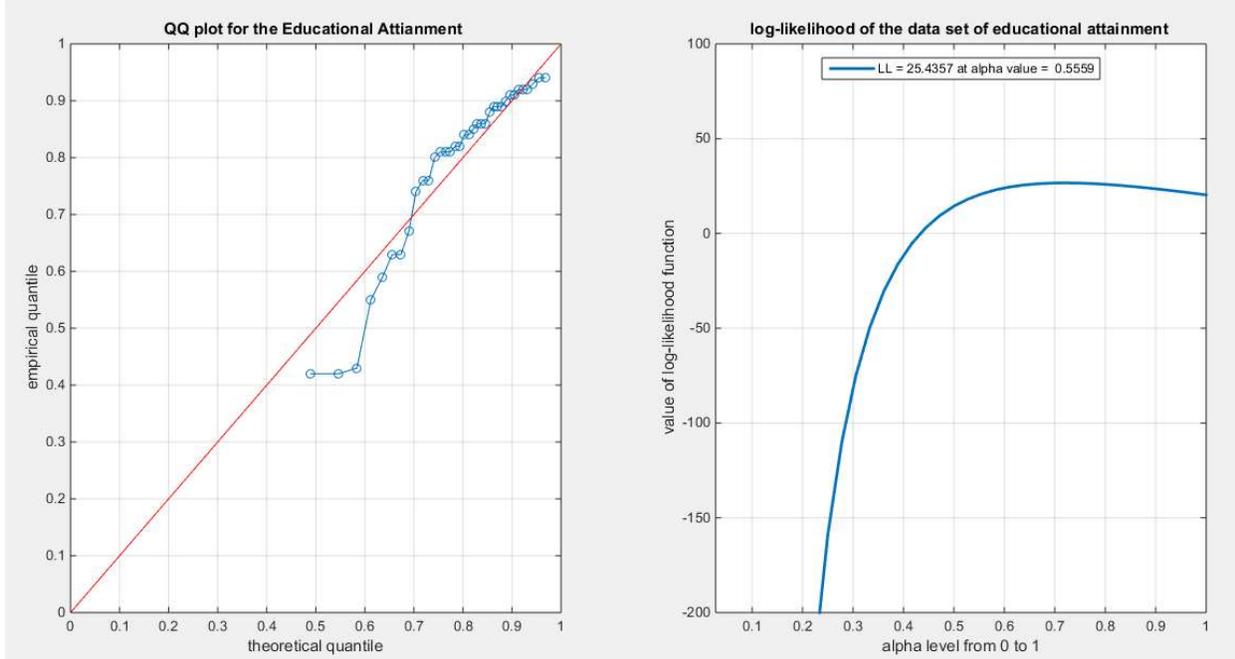

Fig. 11 shows the QQ plot for educational attainment data set, on the left hand side of the graph and the log-likelihood on the right after fitting MBUR distributions.

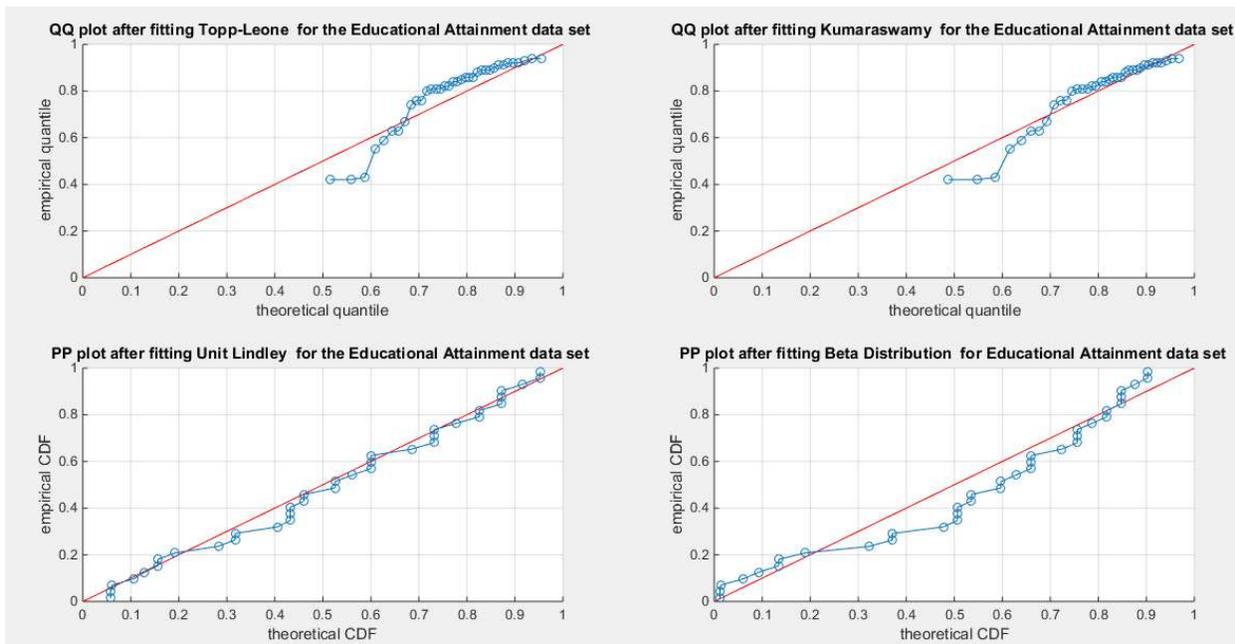

Fig. 12 shows the QQ plot after fitting both Topp-Leone and Kumaraswamy distributions. The PP plot after fitting both Beta and Unit Lindley distributions are also seen.

## 4.4. Analysis of the fourth data set. See Table (5) & Table (6). See figures (13-18).

Table (5): Descriptive statistics of the fourth data set

| min | mean | std | skewness | kurtosis | 25p | 50p | 75p | max |
|---|---|---|---|---|---|---|---|---|
| 0.26 | 0.4225 | 0.1244 | 1.1625 | 4.2363 | 0.33 | 0.405 | 0.465 | 0.74 |

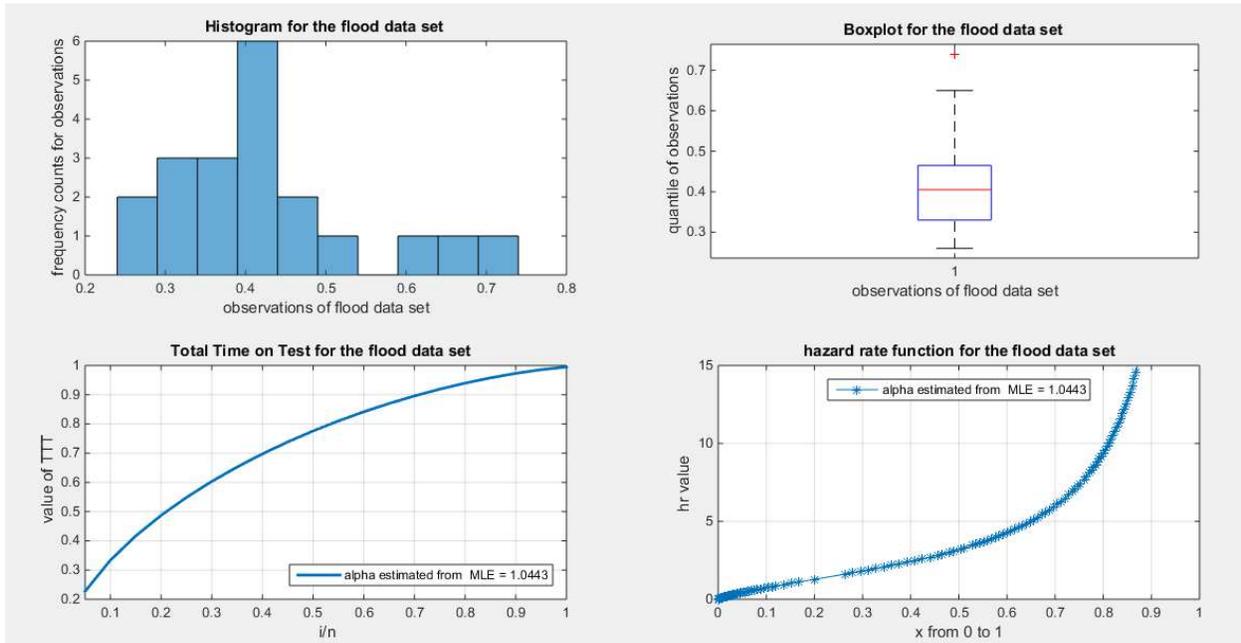

Fig.13 shows the histogram with right skewness and associated boxplot with one outlier or extreme value in the upper tail of the distribution. The TTT plot shows concave shape which supports increase failure rate that is obvious in the shape of the hazard function on the right lower graph

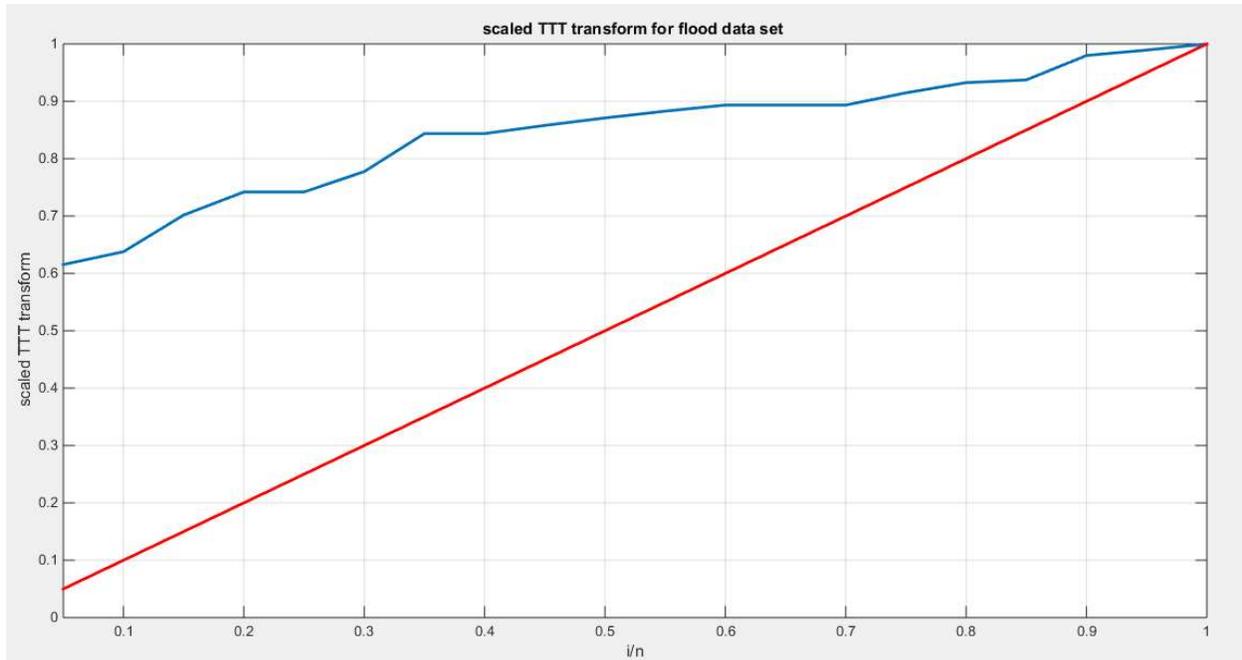

Fig. 14 shows the scaled TTT plot for the flood data set with a concave shape supporting the increased hazard rate as reflected in the shape of the hazard function.

The scaled TTT plot, in figure (13) using the theoretical approach and in figure (14) using the empirical approach, shows concave shape supporting increased hazard rate function.

The Beta distribution is the best among all distributions to fit the flood data set, this distribution is followed by Kumaraswamy, Unit Lindley and lastly by MBUR distribution. Topp-Leone did not fit the data as $H_0$ hypothesis is rejected. The MBUR has the least negative values of NLL, AIC, AIC corrected and BIC values than the other distributions. Beta distribution is the distribution that has the smallest KS-test, AD test and CVM test followed by the Kumaraswamy distribution, UL and then MBUR distribution. Although MBUR is the last distribution to fit the data, it can still fit the data while Topp-Leaone cannot. In other words, even it is not the best, but it is among the distributions that can fit the data. Table (6) shows these indices. Figure (15) shows eCDF and theoretical CDFs while figure (16) depicts the fitted PDFs for different competitors. The QQ plot do not visually show perfect alignment for MBUR in figure (17). In figure (18), the QQ plot and PP plot of other distributions are shown

The P-values for the estimators of alpha and beta parameters of the Beta distribution and Kumaraswamy distributions are significant $(p < 0.001)$.

P-values for the estimators of alpha of the MBUR distribution is significant $(p < 0.001)$.

P-values for the estimators of theta of the Unit Lindley distribution is significant $(p < 0.001)$.

Table (6): Estimators and validation indices for the Fourth data set

| | Beta | | Kumaraswamy | | MBUR | Topp-Leone | Unit-Lindley |
|---|---|---|---|---|---|---|---|
| theta | $\alpha = 6.8318$ | | $\alpha = 3.3777$ | | 1.0443 | 2.2413 | 1.6268 |
| | $\beta = 9.2376$ | | $\beta = 12.0057$ | | | | |
| Var | 7.22 | 7.2316 | 0.3651 | 2.8825 | 0.007 | 0.2512 | 0.0819 |
| | 7.2316 | 8.0159 | 2.8825 | 29.963 | | | |
| SE | 0.6008 | | 0.1351 | | 0.0187 | 0.1121 | 0.0639 |
| | 0.6331 | | 1.2239 | | | | |
| AIC | -24.3671 | | -21.9465 | | -10.9233 | -12.7627 | -12.3454 |
| CAIC | -23.6613 | | -21.2407 | | -10.7011 | -12.5405 | -12.1231 |
| BIC | -22.3757 | | -19.9551 | | -9.9276 | -11.767 | -11.3496 |
| HQIC | -23.9784 | | -21.5578 | | -10.7289 | -12.5684 | -12.151 |
| LL | 14.1836 | | 12.9733 | | 6.4617 | 7.3814 | 7.1727 |
| K-S | 0.2063 | | 0.2175 | | 0.3202 | 0.3409 | 0.2625 |
| H₀ | Fail to reject | | Fail to reject | | Fail to reject | Reject | Fail to reject |
| P-value | 0.3174 | | 0.2602 | | 0.0253 | 0.0141 | 0.0311 |
| AD | 0.7302 | | 0.9365 | | 2.7563 | 2.9131 | 2.3153 |
| CVM | 0.1242 | | 0.1653 | | 0.531 | 0.5857 | 0.4428 |

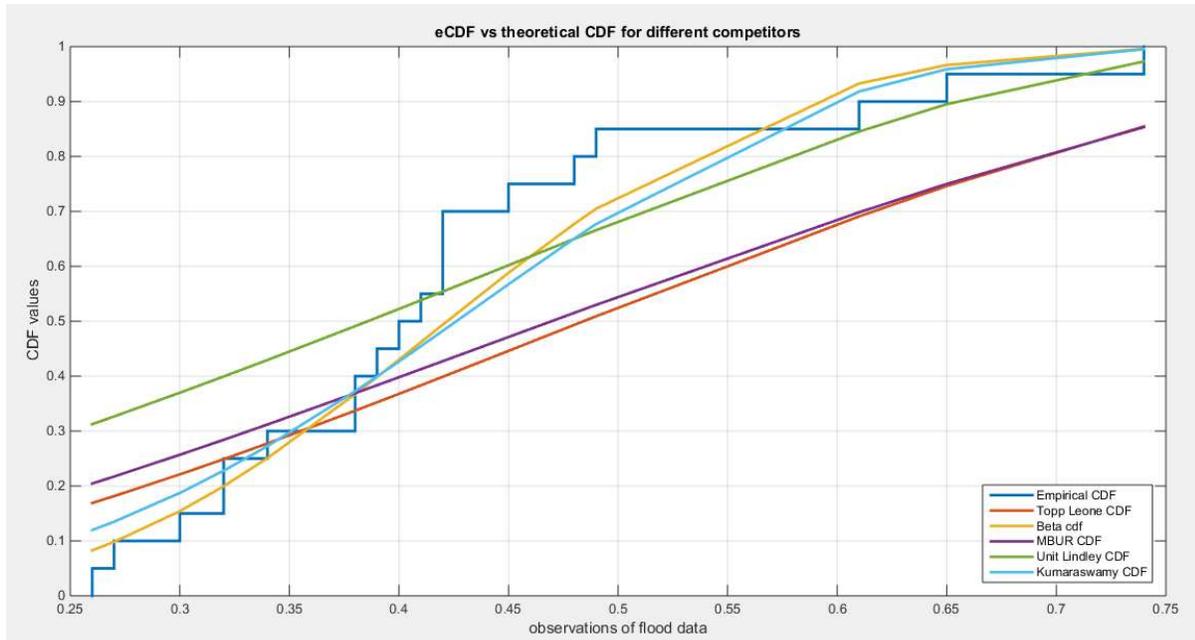

Fig. 15 shows the e CDF vs. theoretical CDF of the 5 distributions for the 4th data set (Flood Data).

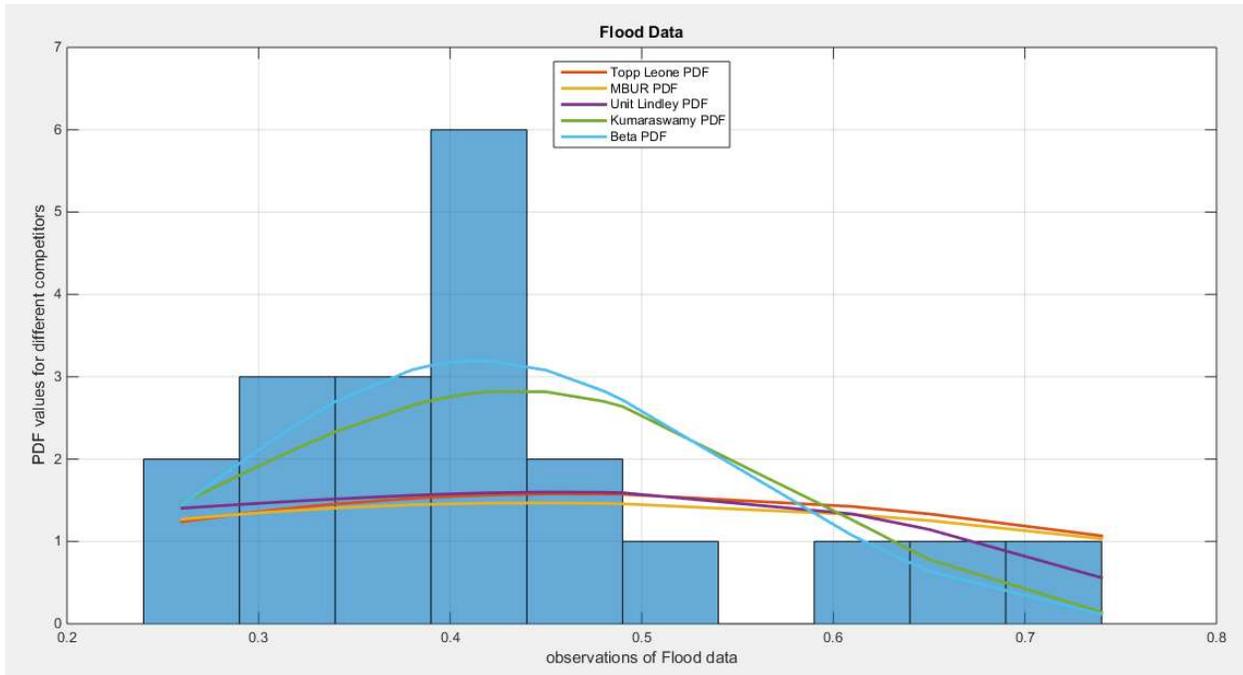

Fig 16 shows PDFs of fitted distributions

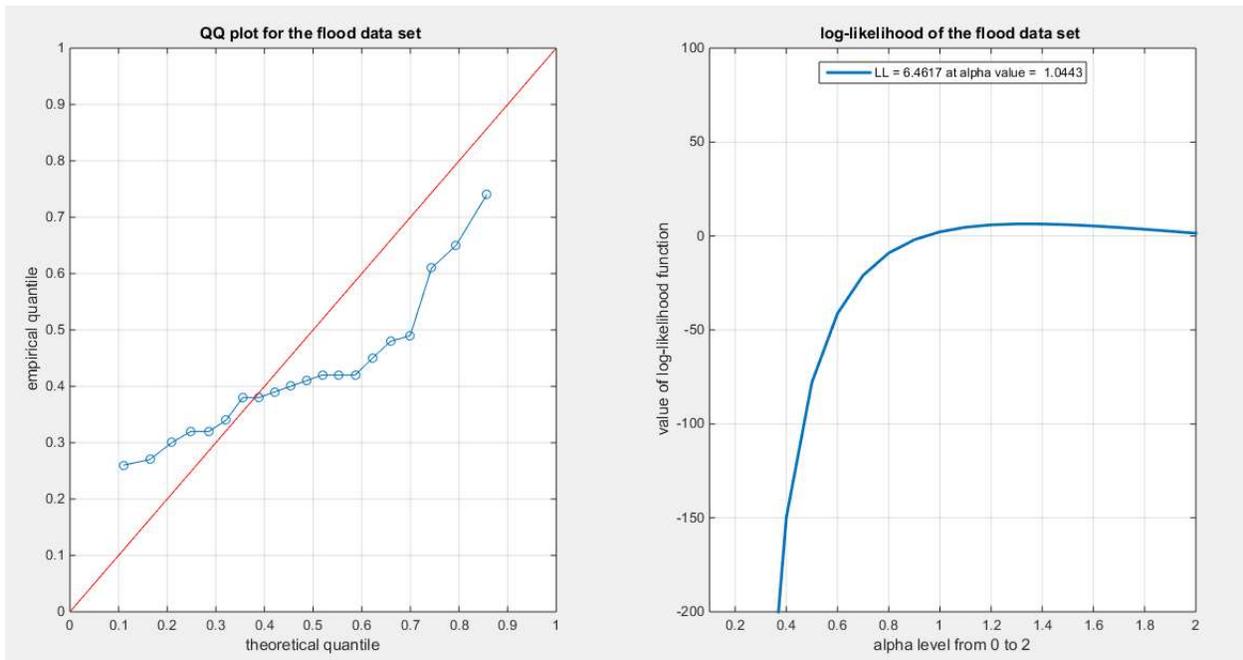

Fig. 17 shows the QQ plot for flood data set, on the left hand side of the graph and the log-likelihood on the right after fitting MBUR distribution.

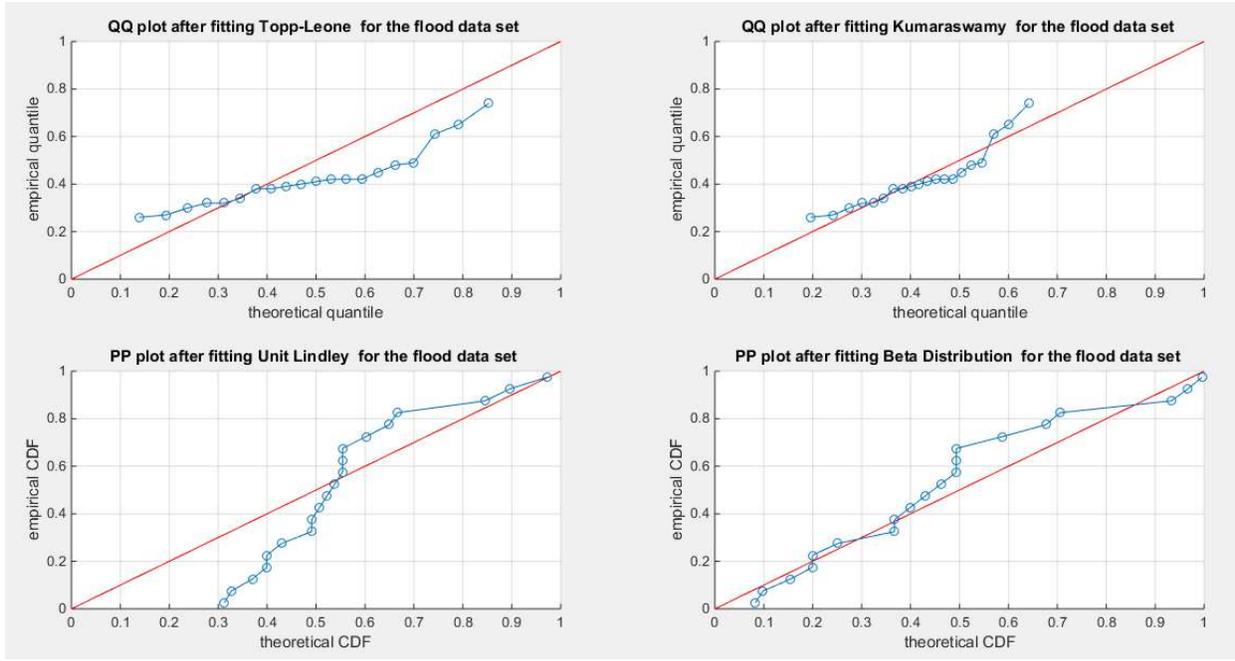

Fig.18 shows the QQ plot after fitting both Topp-Leone and Kumaraswamy distributions. The PP plot after fitting both Beta and Unit Lindley distributions are also seen.